\documentclass[12pt]{article}
\usepackage{epsf}
\usepackage{epsfig}
\usepackage{axodraw}
\usepackage{fancy}
\usepackage{graphics}
\advance \headheight by 3.0truept       

\begin{document}

\thispagestyle{empty}
\bigskip
{\Large\bf \centerline{\em Postdoctoral Working Report }
\centerline{Researches on Higgs and FCNC Physics}}
\bigskip
\normalsize

\vskip 0.5cm

\centerline{\bf Shou Hua Zhu}  \centerline{\sl  Institute of
Theoretical Physics, Academia Sinica,
      P.O.Box 2735,}
\centerline{\sl Beijing 100080, P.R.China}
\bigskip

\begin{abstract}

In this report, instead to give comprehensive review on two
important research fields during my first term postdoctoral
working period: Higgs and FCNC physics, I will collect part of my
recently works on it. Charged Higgs is the distinguished signature
of new physics, in this report, I review my two works on charged
Higgs associated production with top quark and $W$ boson at
hardron colliders. Our researches show that these two charged
Higgs production mechanisms are important channel not only in
finding charged Higgs, but also in studying the quantum structure
of new physics. Flavor Changing Neutral Current (FCNC) processes
are forbidden at tree level in the Standard Model (SM), so they
act as the ground to test quantum structure of the SM and also
very important channel in finding new physics beyond the SM. In
this report, I focus on the studies on FCNC processes on linear
colliders and in B-factories.

\vfill

\end{abstract}

\newpage
\tableofcontents
\newpage

\part{Higgs Physics}

\it
\begin{center}
\begin{minipage}{5in} At the moment, because of lack of
imagination, one cannot do much more than try to calculate effects
due to the Higgs system in order to make comparisons with
experiments results.

\begin{flushright}
--adapted from M. Veltman 1997 "Reflections on the Higgs system".
\end{flushright}
\end{minipage}
\end{center}

\rm \vskip 1cm
\section{Preface}
The standard model (SM) \cite{ld2} gives an excellent theoretical
description of the strong and electro-weak interaction. This
theory which is based on an $SU(3) \times SU(2) \times U(1)$ gauge
group, has been proven extraordinarily robust. Albeit its success,
the SM still has one part untested which is the mechanism of
electroweak symmetry spontaneous breaking (EWSB), through which
the gauge bosons and fermions gain their masses. In the SM, EWSB
is realized through one fundamental scalar field - Higgs field.
After EWSB, the physical world is left with one neutral Higgs
boson. The mass of Higgs boson is not predicted by the theory and
can only be determined by high energy experiments.

Although the SM is robust, there are theoretical aspects of the
SM, e.g. triviality \cite{ld3} and naturalness \cite{ld4} etc.,
which suggested the need for new physics. In addition, there are
certain open questions within the SM, such as too many free
parameters, origin of CP violation and flavor problem etc., whose
answer can only be found by invoking physics beyond the SM.

Among various new physics, supersymmetry (SUSY)\cite{ld5},
especially minimal supersymmetrical standard model (MSSM), is the
most elegant candidate. In order to preserve the SUSY and keep
theory anomaly free, in the MSSM, there should be introduced two
Higgs doublets to break the electroweak symmetry. After SUSY
breaking and EWSB, there are five physical Higgs bosons: three
neutral Higgs and two charged Higgs bosons. To find the Higgs
bosons predicted by the SM and the MSSM and study theirs
properties are the primary goal of present and next generation
colliders for both theoretical and high energy experimental
scientists.

The Higgs masses are not predicted by the SM and MSSM (in the
MSSM, there is a theoretical upper limit for lightest Higgs boson
$\leq 140$ GeV), which can only be determined by experiments.The
results coming from direct search for the Higgs in the process
$e^+e^- \rightarrow ZH$ at LEP 200 are, for the SM Higgs boson
\cite{ld6} $$ m_H > 107.7 GeV~~(95\% ~{\rm C.L.}) $$ which is
compatible with the result of the SM fit of all precision data
\cite{ld7} $$ M_H = \left(76^{\scriptstyle +85}_{\scriptstyle
-47}\right)~{\rm GeV} $$ or $$ M_H < 262~{\rm GeV}~~(95\% ~{\rm
C.L.})~; $$ for MSSM CP-even Higgs boson ($\tan\beta >1$) $$ m_H>
85.2 ~~(95\% ~{\rm C.L.}). $$

At upgraded Fermilab Tevatron (Run III), the mass of SM Higgs
boson can be pushed up to $\sim 180 GeV$ combined subprocesses
$qq^\prime \rightarrow WH$ and $gg \rightarrow H$ \cite{ld8}. And,
it is commonly thought that the combination of large hadron
collider (LHC) and next linear collider (NLC) will cover the mass
range of the Higgs boson up to 1 TeV or so.

         In the
above, I have given a brief general review of this field. In the
following, I shortly describe our works related to this topic:

\begin{itemize}

\item{\bf The Higgs boson production in $\gamma \gamma$ collisions at NLC
}\cite{ld9}

High energy $\gamma \gamma$ collision is the collision mode
realized at the NLC with almost the same center-of-mass energy and
luminosity, and provide more clean place to study the properties
of the Higgs  bosons. In the framework of the MSSM, we studied the
Higgs boson production in $\gamma \gamma$ collisions at NLC.
Especially, the light neutral Higgs boson pair production involves
many Feynman diagrams arising both from general two-Higgs-doublet
model particles and the supersymmetrical virtual particles. We
found the total cross section for the Higgs boson pair production
is sensitive to the model parameters, such as $\tan\beta$, triple
soft breaking terms $A_t, A_b$ and the Higgs boson masses etc.

\item{\bf Higgs boson associated production with $W$ at hadron colliders}
\cite{ld10}

Before the LHC comes to operate, to discuss the Higgs discovery
potential at present collider Fermilab Tevatron is an urgent task.
With the integrated luminosity $30 fb^{-1}$, through the process
$P\bar P \rightarrow q q^\prime \rightarrow WH$ followed by $H
\rightarrow b \bar b$ and $W \rightarrow \ell \bar \nu$, Tevatron
can find the mass of the Higgs boson up to $125 GeV$. In these
works, we have studied the Yukawa corrections arising from the top
loop as well as the leading electroweak corrections including the
Higgs contributions besides the top contributions. And we found
that in the SM, the corrections are small and at most few percent;
however, in the MSSM, the corrections could reach tens of percent
in the favorable parameters space.

\item{\bf
Higgs boson discovery potential through $bg$ channel} \cite{ld11}

For hadron colliders, especially LHC, the gluon distribution grows
rapidly, it may play an important role in producing Higgs boson in
particular for large $\tan\beta$ because the couplings of
down-type quarks with Higgs can be enhanced in this case. We study
the Higgs boson discovery potential through $bg$ channel for both
neutral and charged Higgs bosons. Indeed, we found that it is
possible to find the SUSY neutral Higgs boson at Tevatron if
$\tan\beta \geq 10$. For charged Higgs production, we also
calculated Yukawa correction, and found the magnitude of the
radiative corrections can exceed -20\% and not sensitive to the
mass of charged Higgs boson, these effects could be observable in
the experiments.

\item{\bf
Higgs boson production at NLC} \cite{ld12}

We have also studied the Higgs production at the NLC associated
with W boson or heavy quarks. The WH production is the
loop-induced process, and we found that the cross section can
reach 1 fb, but decease rapidly with the increment of the Higgs
mass. The $t\bar tH$ and $b \bar b H$ production have been
considered by many groups, we re-study this process under the
framework of M-theory. Our results show that the cross sections
are sensitive to the model parameters.

\item{\bf Radiative Higgs boson decay beyond the standard model}
\cite{ld13}

At LHC, the decay mode $H \rightarrow \gamma \gamma$ is used in
searching Higgs boson for the intermediate mass Higgs boson.
However, this searching strategy is suffered by the low decay rate
of this mode. In this work, we study the possibility of using $H
\rightarrow f\bar f \gamma$ in searching intermediate mass Higgs
boson where $f$ represent light fermions. Our study shows that, at
least, this channel can be used as discriminant between SM and
MSSM for a wide range of parameter space.

\end{itemize}

In the first section of this part, the supersymmetric electroweak
correction for $bg \rightarrow t H^-$ at hardron colliders will be
presented in details; in second section, we will study the process
of $b\bar b \rightarrow W^-H^+$ at CERN Large Hardron Collider in
supersymmetrical model.

\newpage

\section{Supersymmetric Electroweak Corrections to Charged Higgs
 Boson Production in Association with a Top Quark  at Hadron
Colliders }

\begin{footnotesize}
\begin{center}\begin{minipage}{5in}
\baselineskip=0.25in
\begin{center} ABSTRACT \end{center}

We calculate the $O(\alpha_{ew}m_{t(b)}^{2}/m_{W}^{2})$ and
$O(\alpha_{ew} m_{t(b)}^4/m_W^4)$ supersymmetric electroweak
corrections to the cross section for the charged Higgs boson
production in association with a top quark at the Tevatron and the
LHC. These corrections arise from the quantum effects which are
induced by potentially large Yukawa couplings from the Higgs
sector and the chargino-top(bottom)-sbottom(stop) couplings,
neutralino-top(bottom)-stop(sbottom) couplings and charged
Higgs-stop-sbottom couplings. They can decrease or increase the
cross section depending on $\tan\beta$ but are not very sensitive
to the mass of the charged Higgs boson for high $\tan\beta$. At
low $\tan\beta(=2)$ the corrections decrease the total cross
sections significantly, which exceed $-12\%$ for $m_{H^{\pm}}$
below $300GeV$ at both the Tevatron and the LHC, but for
$m_{H^{\pm}}>300GeV$ the corrections can become very small at the
LHC. For high $\tan\beta(=10,30)$ these corrections can decrease
or increase the total cross sections, and the magnitude of the
corrections are at most a few percent
at both the Tevatron and the LHC.

\end{minipage}\end{center}
\end{footnotesize}

\subsection{Introduction}

   There has been a great deal of interest in the charged Higgs bosons
appearing in the two-Higgs-doublet models(THDM)[1], particularly
the minimal supersymmetric standard model(MSSM)[2], which predicts
the existence of three neutral and two charged Higgs bosons $h, H,
A,$ and $H^{\pm}$. When the Higgs boson of the Standard Model(SM)
has a mass below 130-140 Gev and the h boson of the MSSM is in the
decoupling limit (which means that $H^{\pm}$ is too heavy anyway
to be possibly produced), the lightest neutral Higgs boson may be
difficult to distinguish from the neutral Higgs boson of the
standard model(SM). But charged Higgs bosons carry a distinctive
signature of the Higgs sector in the MSSM. Therefore, the search
for charged Higgs bosons is very important for probing the Higgs
sector of the MSSM and, therefore, will be one of the prime
objectives of the CERN Large Hadron Collider(LHC). At the LHC the
integrated luminosity is expected to reach $L=100 fb^{-1}$ per
year in the second phase. Recently, several studies of charged
Higgs boson production at hadron colliders have appeared in the
literature[3,4,5]. For a relatively light charged Higgs boson,
$m_{H^{\pm}}< m_t - m_b$, the dominate production processes at the
LHC are $gg\rightarrow t \bar t$ and $q\bar q\rightarrow t\bar t$
followed by the decay sequence $t\rightarrow bH^+\rightarrow b\tau
^+\nu_{\tau}$[6]. For a heavier charged Higgs boson the dominate
production process is $gb\rightarrow tH^-$[7,8,9]. Previous
studies showed that the search for heavy charged Higgs bosons with
$m_{H^{\pm}}>m_t + m_b$ at a hadron collider is seriously
complicated by QCD backgrounds due to processes such as
$gb\rightarrow t\bar tb, g\bar b\rightarrow t\bar t\bar b$, and
$gg\rightarrow t\bar tb\bar b$, as well as others process[8].
However, recent analyses[10,11] indicate that the decay mode
$H^+\rightarrow \tau ^+\nu$ provides an excellent signature for a
heavy charged Higgs boson in searches at the LHC. The discovery
region for $H^{\pm}$ is far greater than had been thought for a
large range of the $(m_{H^{\pm}}, \tan \beta)$ parameter space,
extending beyond $m_{H^{\pm}}\sim 1 TeV$ and down to at least
$\tan\beta \sim 3$, and potentially to $\tan\beta \sim 1.5$,
assuming the latest results for the SM parameters and parton
distribution functions as well as using kinematic selection
techniques and the tau polarization analysis[11]. Of course, it is
just a theoretical analysis and no experimental simulation has
been performed to make the statement very reliable so far.

   The one-loop radiative corrections to $H^-t$ associated
production have not been calculated, although this production
process has been studied extensively at tree-level[7,8,9]. In this
paper we present the calculations of the
$O(\alpha_{ew}m_{t(b)}^{2}/m_{W}^{2})$ supersymmetric(SUSY)
electroweak corrections to this associated $H^-t$ production
process at both the Fermilab Tevatron and the LHC in the MSSM.
These corrections arise from the quantum effects which are induced
by potentially large Yukawa couplings from the Higgs sector and
the chargino-top(bottom)-sbottom(stop) couplings, neutralino-
top(bottom)-stop(sbottom) couplings and charged Higgs-stop-sbottom
couplings which will contribute at the
$O(\alpha_{ew}m_{t(b)}^{4}/m_{W}^{4})$ to the self-energy of the
charged Higgs boson. In order to get a reliable estimate this
process has to be merged with the related gluon splitting
contribution $gg\rightarrow H^-t\bar{b}$. This leads to a
suppression by about $50\%$ at LO[12]. However, the complete
one-loop QCD corrections are probably more important, but not yet
available.

\subsection{Calculations and formulas}

 The tree-level
amplitude for $gb\rightarrow tH^-$ is
\begin{equation}
M_{0}=M_{0}^{(s)}+M_{0}^{(t)},
\end{equation}
where $M_{0}^{(s)}$ and $M_{0}^{(t)}$ represent the amplitudes
arising from diagrams in Fig.1$(a)$ and Fig.1$(b)$, respectively.
Explicitly,

\begin{eqnarray}
M_{0}^{(s)}&=&\frac{igg_{s}}{\sqrt{2}m_{W}(\hat{s}
-m_{b}^{2})}\overline{u}(p_{t})[2m_{t}\cot\beta p_{b}^{\mu}P_{L}
+2m_{b}\tan\beta p_{b}^{\mu}P_{R} -m_{t}\cot\beta
\gamma^{\mu}{\not{k}}P_{L} \nonumber \\ & &  -m_{b}\tan\beta
\gamma^{\mu}{\not{k}} P_{R}]u(p_{b})\varepsilon_{\mu}(k)
T_{ij}^{a},
\end{eqnarray}
and
\begin{eqnarray}
M_{0}^{(t)}&=& \frac{igg_{s}}{\sqrt{2}m_{W}(\hat{t}
-m_{t}^{2})}\overline{u}(p_{t})[2m_{t}\cot\beta p_{t}^{\mu}P_{L}
+2m_{b}\tan\beta p_{t}^{\mu}P_{R} -m_{t}\cot\beta
\gamma^{\mu}{\not{k}} P_{L} \nonumber \\ & &  -m_{b}\tan\beta
\gamma^{\mu}{\not{k}}P_{R}]u(p_{b})\varepsilon_{\mu}(k)T_{ij}^{a},
\end{eqnarray}
where $T^{a}$ are the $SU(3)$ color matrices and $\hat{s}$ and
$\hat{t}$ are the subprocess Mandelstam variables defined by $$
\hat{s}=(p_{b}+k)^2=(p_t+p_{H^-})^2,$$ and
$$\hat{t}=(p_t-k)^2=(p_{H^-}-p_b)^2. $$ Here the
Cabbibo-Kobayashi-Maskawa matrix element $V_{CKM}[bt]$ has been
taken to be unity.

The SUSY electroweak corrections of order
$O(\alpha_{ew}m_{t(b)}^{2}/m_{W}^{2})$
and $O(\alpha_{ew} m_{t(b)}^4/m_W^4)$ to the process
$gb\rightarrow H^-t$ arise from the Feynman diagrams shown in
Figs.1(c)-1(v) and Fig.2. We carried out the calculation in the
t'Hooft-Feynman gauge and used dimensional
reduction, which preserves supersymmetry, for
regularization of the ultraviolet divergences
in the virtual loop
corrections using the on-mass-shell renormalization scheme[13], in
which the fine-structure constant $\alpha_{ew}$ and physical
masses are chosen to be the renormalized parameters, and finite
parts of the counterterms are fixed by the renormalization
conditions. The coupling constant $g$ is related to the input
parameters $e, m_W,$ and $m_Z$ by $g^2= e^2/s_w^2$ and
$s_w^2=1-m_w^2/m_Z^2$. The parameter $\beta$ in the MSSM we are
considering must also be renormalized. Following the analysis of
ref.[14], this renormalization constant was fixed by the
requirement that the on-mass-shell $H^{+}\bar l\nu_l$ coupling
remain the same form as in Eq.(2) of ref.[14] to all orders of
perturbation theory. Taking into account the
$O(\alpha_{ew}m_{t(b)}^{2}/m_{W}^{2})$ Yukawa corrections, the
renormalized amplitude for the process $gb\rightarrow tH^{-}$ can
be written as
\begin{eqnarray}
M_{ren}&=& M_{0}^{(s)} +M_{0}^{(t)} +\delta M^{V_{1}(s)} +\delta
M^{V_{1}(t)} +\delta M^{s(s)} +\delta M^{s(t)} +\delta
M^{V_{2}(s)} \nonumber \\ & & +\delta M^{V_{2}(t)} +\delta
M^{b(s)} +\delta M^{b(t)} \equiv M_{0}^{(s)} +M_{0}^{(t)}
+\sum_{l} \delta M^{l},
\end{eqnarray}
where $\delta M^{V_1(s)},\delta M^{V_1(t)}, \delta M^{s(s)},
\delta M^{s(t)},\delta M^{V_2(s)}, \delta M^{V_2(t)},\delta
M^{b(s)}$, and $\delta M^{b(t)}$ represent the corrections to the
tree diagrams arising, respectively, from the $gbb$ vertex diagram
Fig.1(c)-1(d), the $gtt$ vertex diagram Fig.1(f)-1(g), the bottom
quark self-energy diagram Fig.1(i), the top quark self-energy
diagram Fig.1(k), the $btH^-$ vertex diagrams Figs.1(m)-1(n) and
Figs.1(p)-1(q), including their corresponding counterterms
Fig.1(e), Fig.1(h), Fig.1(j), Fig.1(l), Fig.1(o), and Fig.1(r),
and the box diagrams Figs.1$(s)-1(v)$. $\sum_{l} \delta M^{l}$
then represents the sum of the contributions to the Yukawa
corrections from all the diagrams in Figs.1(c)-1(v). The explicit
form of $\delta M^{l}$ can be expressed as
\begin{eqnarray}\label{}
\delta M^{l}&=& -\frac{ig^{3}g_{s} T_{ij}^{a}}
{4\sqrt{2}\times16\pi^{2}m_{W}} C^{l}\overline{u}
(p_{t})\{f_{1}^{l} \gamma^{\mu}P_{L} +f_{2}^{l} \gamma^{\mu}P_{R}
+f_{3}^{l}p_{b}^{\mu}P_{L} +f_{4}^{l}p_{b}^{\mu}P_{R}
+f_{5}^{l}p_{t}^{\mu}P_{L} \nonumber \\ & &
+f_{6}^{l}p_{t}^{\mu}P_{R} +f_{7}^{l}\gamma^{\mu}{\not{k}}P_{L}
+f_{8}^{l} \gamma^{\mu}{\not{k}}P_{R}
+f_{9}^{l}p_{b}^{\mu}{\not{k}}P_{L}
+f_{10}^{l}p_{b}^{\mu}{\not{k}}P_{R}
+f_{11}^{l}p_{t}^{\mu}{\not{k}}P_{L} \nonumber \\ & &
+f_{12}^{l}p_{t}^{\mu}{\not{k}}P_{R}\}u(p_{b})
\varepsilon_{\mu}(k),
\end{eqnarray}
where the $C^{l}$ are coefficients that depend on $\hat{s},
\hat{t}$, and the masses, and the $f_{i}^{l}$ are form factors;
both the coefficients $C^{l}$ and the form factors $f_{i}^{l}$ are
given explicitly in Appendix A. The corresponding amplitude
squared is
\begin{equation}
\overline{\sum}|M_{ren}|^{2}=\overline{\sum}|M_{0}^{(s)}
+M_{0}^{(t)}|^{2} +2Re\overline{\sum}[(\sum_{l}\delta M^{l})
(M_{0}^{(s)} +M_{0}^{(t)})^{\dag}],
\end{equation}
where
\begin{eqnarray}
\overline{\sum}|M_{0}^{(s)} +M_{0}^{(t)}|^{2}&=&
\frac{g^{2}g_{s}^{2}}{2N_{C}m_{W}^{2}}
\{\frac{1}{(\hat{s}-m_{b}^{2})^{2}}[(m_{t}^{2}\cot^{2}\beta
+m_{b}^{2}\tan^{2}\beta)(p_{b}\cdot kp_{t}\cdot k \nonumber
\\ &-& m_{b}^{2}p_{t}\cdot k +2p_{b}\cdot kp_{b}\cdot
p_{t}-m_{b}^{2}p_{b}\cdot p_{t}) +2m_{b}^{2}m_{t}^{2}(p_{b}\cdot k
- m_{b}^{2})] \nonumber \\ &+&
\frac{1}{(\hat{t}-m_{t}^{2})^{2}}[(m_{t}^{2}\cot^{2}\beta
+m_{b}^{2}\tan^{2}\beta)(p_{b}\cdot kp_{t}\cdot k
+m_{t}^{2}p_{b}\cdot k \nonumber \\ &-& m_{t}^{2}p_{b}\cdot p_{t})
+2m_{b}^{2}m_{t}^{2}(p_{t}\cdot k -m_{t}^{2})]
+\frac{1}{(\hat{s}-m_{b}^{2})(\hat{t}-m_{t}^{2})} \nonumber \\
&\times& [(m_{t}^{2}\cot^{2}\beta
+m_{b}^{2}\tan^{2}\beta)(2p_{b}\cdot kp_{t}\cdot k +2p_{b}\cdot
kp_{b}\cdot p_{t} -2(p_{b}\cdot p_{t})^{2} \nonumber \\ &-&
m_{b}^{2}p_{t}\cdot k +m_{t}^{2}p_{b}\cdot k)
+2m_{b}^{2}m_{t}^{2}(p_{t}\cdot k -p_{b}\cdot k -2p_{b}\cdot
p_{t})]\},
\end{eqnarray}
\begin{eqnarray}
\overline{\sum}\delta M^{l}(M_{0}^{(s)})^{\dag} &=&
-\frac{g^{4}g_{s}^2}{64N_{C}\times 16\pi^{2}
m_{W}^{2}(\hat{s}-m_{b}^{2})}
C^{l}\sum_{i=1}^{12}h_{i}^{(s)}f_{i}^{l},
\end{eqnarray}
and
\begin{eqnarray}
\overline{\sum}\delta M^{l}(M_{0}^{(t)})^{\dag} &=&
-\frac{g^{4}g_{s}^2} {64N_{C}\times
16\pi^{2}m_{W}^{2}(\hat{t}-m_{t}^{2})} C^{l}\sum_{i=1}^{12}
h_{i}^{(t)}f_{i}^{l}.
\end{eqnarray}
Here the color factor $N_{C}=3$ and $h_{i}^{(s)}$ and
$h_{i}^{(t)}$ are scalar functions whose explicit expressions are
given in Appendix B.

The cross section for the process $gb\rightarrow tH^{-}$ is
\begin{equation}
\hat{\sigma} =\int_{\hat{t}_{min}}^{\hat{t}_{max}}\frac{1}{16\pi
\hat{s}^2} \overline{\Sigma}|M_{ren}|^{2}d\hat{t}
\end{equation}
with
\begin{eqnarray*}
\hat{t}_{min} &=& \frac{m_{t}^{2} +m_{H^{-}}^{2} -\hat{s}}{2}
-\frac{1}{2}\sqrt{(\hat{s} -(m_{t} +m_{H^{-}})^{2})(\hat{s}
-(m_{t} -m_{H^{-}})^{2})},
\end{eqnarray*}
and
\begin{eqnarray*}
\hat{t}_{max} &=& \frac{m_{t}^{2} +m_{H^{-}}^{2} -\hat{s}}{2}
+\frac{1}{2}\sqrt{(\hat{s} -(m_{t} +m_{H^{-}})^{2})(\hat{s}
-(m_{t} -m_{H^{-}})^{2})}.
\end{eqnarray*}
The total hadronic cross section for $pp\rightarrow gb\rightarrow
tH^{-}$ can be obtained by folding the subprocess cross section
$\hat{\sigma}$ with the parton luminosity:
\begin{equation}
\sigma(s) =\int_{(m_{t} +m_{H^{-}})/\sqrt{s}}^{1}dz \frac{dL}{dz}
\hat{\sigma}(gb\rightarrow tH^{-} \ \ {\rm at} \ \ \hat{s}
=z^{2}s).
\end{equation}
Here $\sqrt{s}$ and $\sqrt{\hat{s}}$ are the CM energies of the
$pp$ and $gb$ states , respectively, and $dL/dz$ is the parton
luminosity, defined as
\begin{equation}
\frac{dL}{dz} =2z\int_{z^{2}}^{1}
\frac{dx}{x}f_{b/P}(x,\mu)f_{g/P} (z^{2}/x,\mu),
\end{equation}
where $f_{b/P}(x,\mu)$ and $f_{g/P}(z^{2}/x,\mu)$ are the bottom
quark and gluon parton distribution functions.

\subsection{Numerical results and conclusion}

In the following we present some numerical results for charged
Higgs boson production in association with a top quark at both the
Tevatron and the LHC. In our numerical calculations the SM
parameters were taken to be $m_W=80.41 GeV$, $m_Z=91.187 GeV$,
$m_t=176 GeV$, $\alpha_s(m_Z)=0.119$, and
$\alpha_{ew}(m_Z)={1\over 128.8}$[15]. And we used the running b
quark mass $\approx 3 GeV$ and the one-loop relations[16] from the
MSSM between the Higgs boson masses $m_{h,H,A,H^{\pm}}$ and the
parameters $\alpha$ and $ \beta$, and chose $m_{H^{\pm}}$ and
$\tan \beta$ as the two independent input parameters. And we used
the CTEQ5M[17] parton distributions throughout the calculations.
Other MSSM parameters were determined as follows:

(i) For the parameters $M_1,M_2$, and $\mu$ in the chargino and
neutralino matrix, we put $M_2=300GeV$ and then used the relation
$M_1=(5/3)(g'^2/g^2)M_2\simeq 0.5M_2$[2] to determine $M_1$. We
also put $\mu = -100 GeV$ except the numerical calculations as
shown in Fig.6(b), where $\mu$ is a variable.

(ii) For the parameters $m^2_{\tilde{Q},\tilde{U},\tilde{D}}$ and
$A_{t,b}$ in squark mass matrices

\begin{eqnarray}
M^2_{\tilde{q}} =\left(\begin{array}{cc} M_{LL}^2 & m_q M_{LR}\\
m_q M_{RL} & M_{RR}^2 \end{array} \right)
\end{eqnarray}
with
\begin{eqnarray}
M_{LL}^2 =m_{\tilde{Q}}^2 +m_q^2 +m_Z^2\cos 2\beta(I_q^{3L}
-e_q\sin^2\theta_W), \nonumber
\\ M_{RR}^2 =m_{\tilde{U},\tilde{D}}^2 +m_q^2 +m_Z^2
\cos 2\beta e_q\sin^2\theta_W, \nonumber
\\ M_{LR} =M_{RL} =\left(\begin{array}{ll} A_t -\mu\cot\beta &
(\tilde{q} =\tilde{t}) \\ A_b -\mu\tan\beta & (\tilde{q}
=\tilde{b}) \end{array} \right),
\end{eqnarray}
to simplify the calculation we assumed $m^2_{\tilde{Q}}
=m^2_{\tilde{U}} =m^2_{\tilde{D}}$ and $A_t=A_b$, and we put
$m_{\tilde{Q}}=500GeV$ and $A_t=200GeV$. But in the numerical
calculations of Fig.6(a) $A_t=A_b$ are the variables.

Some typical numerical calculations of the tree-level total cross
sections and the $O(\alpha_{ew}m^2_{t(b)}/m^2_W)$ SUSY electroweak
corrections as the functions of the charged Higgs boson mass,
$A_t=A_b$ and $\mu$, respectively, for three representative values
of $\tan\beta$ are given in Figs.3-6.

Figures 3(a) and 4(a) show that the tree-level total cross
sections as a function of the charged Higgs boson mass for three
representative values of $\tan\beta$. For $m_{H^{\pm}}=200GeV$ the
total cross sections at the Tevatron are at most only $0.7$ fb and
$0.1$ fb for $\tan\beta=2,30$ and $10$, respectively, and for
$m_{H^{\pm}}=300GeV$ the total cross sections are smaller than
$0.15$ fb for all three values of $\tan\beta$. However, at the LHC
the total cross sections are much larger: the order of thousands
of fb for $m_{H^{\pm}}$ in the range $100$ to $240 GeV$ and
$\tan\beta=2$ and $30$; and they are hundreds of fb for the
intermediate value $\tan\beta=10$. When the charged Higgs boson
mass becomes heavy($<500$ GeV), the total cross sections still are
larger than $100$ fb and $10$ fb for $\tan\beta=2,30$ and $10$,
respectively. For low $\tan\beta$ the top quark contribution is
enhanced while for high $\tan\beta$ the bottom quark contribution
becomes large. These results are smaller than ones given in
ref.[8,9] because we used the running b quark mass $\approx 3 GeV$
in the numerical calculations. We have confirmed that if we chose
$m_b=4.5 GeV$, our results will agree with ref.[8,9].

In Figs. 3(b) and 4(b) we show the corrections to the total cross
sections relative to the tree-level values as a function of
$m_{H^{\pm}}$ for $\tan\beta=2,10,$ and $30$. For $\tan\beta=2$
the corrections decrease the total cross sections significantly,
which exceed $-13\%$ for $m_{H^{\pm}}$ below $300GeV$ at the both
Tevatron and the LHC. But the corrections decrease as
$m_{H^{\pm}}$ increase. For example, as shown in Fig.4(b), the
corrections range between $-13\%\sim 0\%$ when $m_{H^{\pm}}$
increase from $300 GeV$ to $1 TeV$ at the LHC. For high
$\tan\beta(=10,30)$ these corrections become smaller, which can
decrease or increase the total cross sections depending on
$\tan\beta$, and the magnitude of the corrections are at most a
few percent for a wide range of the charged Higgs boson mass at
both the Tevatron and the LHC.

In Fig.5 we present the Yukawa correction from the Higgs sector
and the genuine SUSY electroweak correction from the couplings
involving the genuine SUSY particles(the chargino, neutralino and
squark) for $\tan\beta=30$ at the LHC, respectively. One can see
that the Yukawa correction and the genuine SUSY electroweak
correction have opposite signs, and thus cancel to some extent.
The former decrease the total cross sections, which can range
between $-8\%\sim -4\%$ for $m_{H^{\pm}}$ below $300GeV$, but the
latter increase the total cross sections, which range between
$10\%\sim 7\%$ for $m_{H^{\pm}}$ in the same range. In such a case
the combined effects just are about $2\%\sim 3\%$.

Figs.6(a) and 6(b) give the corrections as the functions of
$A_t=A_b$ and $\mu$ for $m_{H^{\pm}}=300 GeV$ at the LHC,
respectively, assuming $\tan\beta=2,10$ and $30$. From Figs.6(a)
and 6(b) one sees that the corrections increase or decrease slowly
with increasing $A_t=A_b$ and the magnitude of $\mu$,
respectively, for $\tan\beta=30,10$, and the corrections are not
very sensitive to both $A_t=A_b$ and $\mu$ for $\tan\beta=2$,
where the corrections are always about $-12\%$ and $-13\%$,
respectively. In general for large values of $A_t$ and small
values of $\tan\beta$ or large values of $\mu$ and $\tan\beta$,
one can get much larger corrections since the charged Higgs
boson-stop-sbottom couplings become stronger. For $\tan\beta=30$,
comparing Fig.4(b) with Fig.6(b), we can see that the corrections
indeed become larger as the values of $\mu$ increase. But for
$\tan\beta=2$ from Fig.4(a) and Fig.6(a) we found that the
corrections almost have no change when $A_t=A_b$ become larger.
Obviously the effects from the stronger couplings have been
suppressed by the decoupling effects because with an increase of
$A_t=A_b$ all the squark masses are still heavy, which almost is
same as discussed in Ref.[18].

In conclusion, we have calculated the
$O(\alpha_{ew}m_{t(b)}^{2}/m_{W}^{2})$ and $O(\alpha_{ew}
m_{t(b)}^4/m_W^4)$ SUSY electroweak corrections to the cross
section for the charged Higgs boson production in association with
a top quark at the Tevatron and the LHC. These corrections
decrease or increase the cross section depending on $\tan\beta$
but are not very sensitive to the mass of the charged Higgs boson
for high $\tan\beta$. At low $\tan\beta(=2)$ the corrections
decrease the total cross sections significantly, which exceed
$-12\%$ for $m_{H^{\pm}}$ below $300GeV$ at both the Tevatron and
the LHC, but for $m_{H^{\pm}}>300GeV$ the corrections can become
very small at the LHC. For high $\tan\beta(=10,30)$ these
corrections can decrease or increase the total cross sections, and
the magnitude of the corrections are at most a few percent
at both the Tevatron and the LHC.

{\small
\subsection{Appendix A}

The coefficients $C^l$ and form factors $f^l_i$ are the following:
\begin{eqnarray*}
C^{V_{1}(s)} &=& \frac{m_{b}^{2}} {m_{W}^{2}(\hat{s}-m_{b}^{2})},
\ \ \ \ C^{V_{1}(t)}= \frac{m_{t}^{2}} {m_{W}^{2}(\hat{t}
-m_{t}^{2})},\ \ \ \ C^{s(s)} =\frac{m_{b}^{2}}{m_{W}^{2}(\hat{s}
-m_{b}^{2})^{2}}, \\ C^{s(t)} &=& \frac{m_{t}^{2}}{m_{W}^{2}
(\hat{t}-m_{t}^{2})^{2}}, \ \ \ \ C^{V_{2}(s)} = \frac{m_bm_t}
{m_W^2(\hat{s}-m_{b}^{2})},\hspace{1.4cm} C^{V_{2}(t)}
=\frac{m_bm_t} {m_W^2(\hat{t}-m_{t}^{2})}, \\ C^{b(s)} &=&
C^{b(t)} = \frac{m_tm_b}{m_W^2},
\end{eqnarray*}
\begin{eqnarray*}
   f_{1}^{V_{1}(s)} &=& \eta^{(1)}[m_{b}(g_{2}^{V_{1}(s)}
-g_{3}^{V_{1}(s)}) -2p_{b}\cdot k g_{6}^{V_{1}(s)}],
\\ f_{2}^{V_{1}(s)} &=& \eta^{(2)}[m_{b}(g_{3}^{V_{1}(s)}
-g_{2}^{V_{1}(s)}) -2p_{b}\cdot k g_{7}^{V_{1}(s)}],
\\ f_{3}^{V_{1}(s)} &=& \eta^{(2)}[2(g_{1}^{V_{1}(s)}
+g_{2}^{V_{1}(s)}) +m_{b}(g_{4}^{V_{1}(s)} +g_{5}^{V_{1}(s)})
+2p_{b}\cdot k g_{8}^{V_{1}(s)}],
\\ f_{4}^{V_{1}(s)} &=& \eta^{(1)}[2(g_{1}^{V_{1}(s)}
+g_{3}^{V_{1}(s)}) +m_{b}(g_{4}^{V_{1}(s)} +g_{5}^{V_{1}(s)})
+2p_{b}\cdot k g_{9}^{V_{1}(s)}],
\\ f_{7}^{V_{1}(s)} &=& \eta^{(2)} [-(g_{1}^{V_{1}(s)}
+g_{2}^{V_{1}(s)}) +m_{b}(g_{6}^{V_{1}(s)} +g_{7}^{V_{1}(s)})],
\\ f_{8}^{V_{1}(s)} &=& \eta^{(1)} [-(g_{1}^{V_{1}(s)}
+g_{3}^{V_{1}(s)}) +m_{b}(g_{6}^{V_{1}(s)} +g_{7}^{V_{1}(s)})],
\\ f_{9}^{V_{1}(s)} &=& \eta^{(1)} [g_{4}^{V_{1}(s)}
+2g_{6}^{V_{1}(s)} +m_{b}(g_{8}^{V_{1}(s)} -g_{9}^{V_{1}(s)})],
\\ f_{10}^{V_{1}(s)} &=& \eta^{(2)} [g_{5}^{V_{1}(s)}
+2g_{7}^{V_{1}(s)} +m_{b}(g_{9}^{V_{1}(s)} -g_{8}^{V_{1}(s)})],
\\ f_{1}^{V_{2}(s)} &=& 2p_{b}\cdot k g_{3}^{V_{2}(s)},
\hspace{3.4cm} f_{2}^{V_{2}(s)} = 2p_{b}\cdot k g_{4}^{V_{2}(s)},
\\ f_{3}^{V_{2}(s)} &=& 2g_{1}^{V_{2}(s)}
+2m_{t}\cot\beta(\delta\Lambda_{L}^{(1)} +\delta\Lambda_{L}^{(2)}
+\delta\Lambda_{L}^{(3)}) -2m_{t}g_{3}^{V_{2}(s)}
+2m_{b}g_{4}^{V_{2}(s)},
\\ f_{4}^{V_{2}(s)} &=& 2g_{2}^{V_{2}(s)}
+2m_{b}\tan\beta(\delta\Lambda_{R}^{(1)} +\delta\Lambda_{R}^{(2)}
+\delta\Lambda_{R}^{(3)}) +2m_{b}g_{3}^{V_{2}(s)}
-2m_{t}g_{4}^{V_{2}(s)},
\\ f_{7}^{V_{2}(s)} &=& -\frac{1}{2}f_{3}^{V_{2}(s)},
\hspace{4.0cm} f_{8}^{V_{2}(s)} = -\frac{1}{2}f_{4}^{V_{2}(s)},
\\ f_{1}^{V_{2}(t)} &=& 2p_{t}\cdot k g_{3}^{V_{2}(t)},
\hspace{3.6cm} f_{2}^{V_{2}(t)} = 2p_{t}\cdot k g_{4}^{V_{2}(t)},
\\ f_{5}^{V_{2}(t)} &=& 2g_{1}^{V_{2}(t)}
+2m_{t}\cot\beta(\delta\Lambda_{L}^{(1)} +\delta\Lambda_{L}^{(2)}
+\delta\Lambda_{L}^{(3)}) -2m_{t}g_{3}^{V_{2}(t)}
+2m_{b}g_{4}^{V_{2}(t)},
\\ f_{6}^{V_{2}(t)} &=& 2g_{2}^{V_{2}(t)} +2m_{b}\tan\beta
(\delta\Lambda_{R}^{(1)} +\delta\Lambda_{R}^{(2)}
+\delta\Lambda_{R}^{(3)}) +2m_{b}g_{3}^{V_{2}(t)}
-2m_{t}g_{4}^{V_{2}(t)},
\\ f_{7}^{V_{2}(t)} &=& -\frac{1}{2}f_{5}^{V_{2}(t)},
\hspace{4.0cm} f_{8}^{V_{2}(t)} = -\frac{1}{2}f_{6}^{V_{2}(t)},
\\ f_{1}^{s(s)} &=& 2\eta^{(1)}p_{b}\cdot k[g_{1}^{s(s)}
+m_{b}(g_{2}^{s(s)} +g_{3}^{s(s)})],
\\ f_{2}^{s(s)} &=& 2\eta^{(2)}p_{b}\cdot k[g_{5}^{s(s)}
+m_{b}(g_{2}^{s(s)} +g_{4}^{s(s)})],
\\ f_{3}^{s(s)} &=& 2\eta^{(2)}[m_{b}(g_{1}^{s(s)} +g_5^{s(s)})
+2(m_{b}^{2} +p_{b}\cdot k) g_{2}^{s(s)} +(m_{b}^{2} +2p_{b}\cdot
k) g_{3}^{s(s)} +m_{b}^{2}g_{4}^{s(s)}],
\\ f_{4}^{s(s)} &=& 2\eta^{(1)}[m_{b}(g_{1}^{s(s)} +g_5^{s(s)})
+2(m_{b}^{2} +p_{b}\cdot k) g_{2}^{s(s)} +m_{b}^{2}g_{3}^{s(s)}
+(m_{b}^{2} +2p_{b}\cdot k) g_{4}^{s(s)}],
\\ f_{7}^{s(s)} &=& -\frac{1}{2}f_{3}^{s(s)}, \hspace{4.0cm}
f_{8}^{s(s)} = -\frac{1}{2}f_{4}^{s(s)},
\\ f_{1}^{b(s)} &=& \sum_{i=H^{0},h^{0},G^{0},A^{0}}
\eta_{i}^{(3)}\{\eta^{(2)}[2m_{b}(-3D_{312} +(1-\zeta_{i})D_{27})
+m_{b}^{3}(D_{0} +D_{12} -D_{22}
  \\ & & -D_{32}) -m_{t}^{2}m_{b}(D_{23} +2D_{39}) -2m_{b}p_{b}
\cdot k(2D_{36} +D_{24} +\zeta_{i}(D_{0} +D_{12}))
  \\ & & +2m_{b}p_{t}\cdot k(D_{25} +D_{310}) +2m_{b}p_{b}\cdot
p_{t}(D_{26} +2D_{38})] +\eta^{(1)}[2m_{t}(-3D_{313} +(1
  \\ & & +\zeta_{i})D_{27}) -m_{t}^{3}(D_{33} +(1+\zeta_{i})
D_{23}) +m_{b}^{2}m_{t}(D_{13} -2D_{38} +(1 +\zeta_{i})(D_{0}
  \\ & & -D_{22})) +2m_{t}p_{b}\cdot k(D_{13} -D_{310}
-(1 +\zeta_{i})(D_{12} +D_{24})) +2m_{t}p_{t}\cdot k(2D_{37}
  \\ & & +(1 +\zeta_{i})D_{25}) +2m_{t}p_{b}\cdot p_{t}(2D_{39}
+(1+\zeta_{i})D_{26})]\}
  \\ & &(-k,-p_{b},p_{t},m_{b},m_{b},m_{i},m_{t})
  \\ & & -\frac{8\sqrt{2}m_W}{\sin{2\beta}}\sum_{i,j,k}N_{k4}N_{k3}
^{\ast}R_i(b)R_j(t)\sigma_{ij}D_{27}(-k,-p_b,p_t, m_{\tilde{b}_i},
m_{\tilde{b}_i},m_{\tilde\chi_k^0},m_{\tilde{t}_j}),
\\ f_{2}^{b(s)} &=& f_{1}^{b(s)}(\eta^{(1)}\leftrightarrow
\eta^{(2)}, L_l \leftrightarrow R_l, N_{kl} \leftrightarrow
N_{kl}^\ast),
\\ f_{3}^{b(s)} &=& \sum_{i=H^{0},h^{0},G^{0},A^{0}}
\eta_{i}^{(3)}\{\eta^{(1)}[-4D_{27} +2m_{b}^{2}(D_{22} -D_{0}
-(1-\zeta_{i})(D_{12} +D_{22}))
  \\ & & +2m_{t}^{2}(D_{23}-(1+\zeta_{i})D_{26}) +4p_{t}\cdot
k(D_{26} -D_{25})] +\eta^{(2)}2m_{t}m_{b}(1 +\zeta_{i})(D_{22}
  \\ & & -D_{12} -D_{26})\}(-k,-p_{b},p_{t},m_{b},m_{b},m_{i},m_{t})
  \\ & & -\frac{8\sqrt{2}m_W}{\sin{2\beta}}\sum_{i,j,k}\sigma_{ij}
[-m_tN_{k4}N_{k3}^{\ast}R_i(b)R_j(t)D_{26}
+m_bN_{k4}^{\ast}N_{k3}L_i(b)L_j(t)(D_{12}
  \\ & & +D_{22}) +m_{\tilde\chi_k^0}N_{k4}^{\ast}N_{k3}^{\ast}
R_i(b)L_j(t)D_{12}] (-k,-p_b,p_t,m_{\tilde{b}_i},
m_{\tilde{b}_i},m_{\tilde\chi_k^0},m_{\tilde{t}_j}),
\\ f_{4}^{b(s)} &=& f_{3}^{b(s)}(\eta^{(1)}\leftrightarrow
\eta^{(2)}, L_l \leftrightarrow R_l, N_{kl} \leftrightarrow
N_{kl}^\ast),
\\ f_{5}^{b(s)} &=& \sum_{i=H^{0},h^{0},G^{0},A^{0}}
\eta_{i}^{(3)}\{\eta^{(1)}[12D_{313} +2m_{b}^{2}(2D_{38} -D_{13}
+(1-\zeta_{i})(D_{13}
  \\ & & +D_{26})) +2m_{t}^{2}(D_{33}+(1+\zeta_{i})D_{23})
+4p_{b}\cdot k(D_{25} +D_{310}) -4p_{t}\cdot k(D_{23}
  \\ & & +2D_{37})-4p_{t}\cdot p_{b}(D_{23} +2D_{39})]
+\eta^{(2)}2m_{t}m_{b}(1 +\zeta_{i})(D_{13} +D_{23}
  \\ & & -D_{26})\}(-k,-p_{b},p_{t},m_{b},m_{b},m_{i},m_{t})
  \\ & & +\frac{8\sqrt{2}m_W}{\sin{2\beta}}\sum_{i,j,k}\sigma_{ij}
[-m_tN_{k4}N_{k3}^{\ast}R_i(b)R_j(t)D_{23}
+m_bN_{k4}^{\ast}N_{k3}L_i(b)L_j(t)(D_{13}
  \\ & & +D_{26})
+m_{\tilde\chi_k^0}N_{k4}^{\ast}N_{k3}^{\ast}R_i(b)L_j(t)D_{13}]
(-k,-p_b,p_t,m_{\tilde{b}_i},m_{\tilde{b}_i},m_{\tilde\chi_k^0},
m_{\tilde{t}_j}) ,
\\ f_{6}^{b(s)} &=& f_{5}^{b(s)}(\eta^{(1)}\leftrightarrow
\eta^{(2)}, L_l \leftrightarrow R_l, N_{kl} \leftrightarrow
N_{kl}^\ast),
\\ f_{7}^{b(s)} &=& \sum_{i=H^{0},h^{0},G^{0},A^{0}}
\eta_{i}^{(3)}\{\eta^{(1)}[6(D_{27} -D_{311}) +m_{b}^{2}(D_{11}
-2D_{12} -2D_{22}
  \\ & & -2D_{36} +(1+\zeta_{i})(D_{0} +D_{12}))-m_{t}^{2}(2D_{23}
+2D_{37} +(1+\zeta_{i})D_{13}) -2p_{b}\cdot k(D_{12}
  \\ & & +2D_{24} +2D_{34}) +2p_{t}\cdot k(D_{13} +2D_{25}
+2D_{35}) +2p_{t}\cdot p_{b}(D_{13} +2D_{26}
  \\ & & +D_{310})] +\eta^{(2)}m_{t}m_{b}(1 +\zeta_{i})(D_{12}
-D_{13} -D_{0})\}(-k,-p_{b},p_{t},m_{b},m_{b},m_{i},m_{t}),
\\ f_{8}^{b(s)} &=& f_{7}^{b(s)}(\eta^{(1)}\leftrightarrow
\eta^{(2)}),
\\ f_{9}^{b(s)} &=& \sum_{i=H^{0},h^{0},G^{0},A^{0}}
\eta_{i}^{(3)}\{\eta^{(1)}2m_{t}[-D_{13} -D_{26}
+(1+\zeta_{i})(D_{12} +D_{24})]
  \\ & & +\eta^{(2)}2m_{b}[-D_{22} +D_{24} +\zeta_{i}(D_{0}
+2D_{12} +D_{24})]\} (-k,-p_{b},p_{t},m_{b},m_{b},m_{i},m_{t})
  \\ & & -\frac{8\sqrt{2}m_W}{\sin{2\beta}}\sum_{i,j,k}\sigma_{ij}
N_{k4}N_{k3}^{\ast}R_i(b)R_j(t)(D_{12}
  \\ & & +D_{24})(-k,-p_b,p_t,m_{\tilde{b}_i},m_{\tilde{b}_i},
m_{\tilde\chi_k^0}, m_{\tilde{t}_j}),
\\ f_{10}^{b(s)} &=& f_{9}^{b(s)}(\eta^{(1)}\leftrightarrow
\eta^{(2)}, L_l \leftrightarrow R_l, N_{kl} \leftrightarrow
N_{kl}^\ast),
\\ f_{11}^{b(s)} &=& \sum_{i=H^{0},h^{0},G^{0},A^{0}}
\eta_{i}^{(3)}\{\eta^{(1)}2m_{t}[D_{23} -(1+\zeta_{i})D_{25}]
-\eta^{(2)}2m_{b}[-D_{26} +D_{25}
  \\ & & +\zeta_{i}(D_{13}
+D_{25})]\}(-k,-p_{b},p_{t},m_{b},m_{b},m_{i},m_{t})
  \\ & & +\frac{8\sqrt{2}m_W}{\sin{2\beta}}\sum_{i,j,k}\sigma_{ij}
N_{k4}N_{k3}^{\ast}R_i(b)R_j(t)(D_{13}
  \\ & & +D_{25})(-k,-p_b,p_t,m_{\tilde{b}_i},m_{\tilde{b}_i},
m_{\tilde\chi_k^0}, m_{\tilde{t}_j}),
\\ f_{12}^{b(s)} &=& f_{11}^{b(s)}(\eta^{(1)}\leftrightarrow
\eta^{(2)}, L_l \leftrightarrow R_l, N_{kl} \leftrightarrow
N_{kl}^\ast),
\end{eqnarray*}
where $D_{0},D_{ij},D_{ijk}$ are the four-point Feynman integrals
[19]. The explicit forms of $\delta M^{V_1(t)}, \delta M^{s(t)},
\delta M^{b(t)}$ can be respectively obtained from $\delta
M^{V_1(s)}, \delta M^{s(s)}, \delta M^{b(s)}$ by the
transformation $U$ defined as $$p_{b}\rightarrow p_{t},\ \ \ \
\hat{s}\rightarrow \hat{t},\ \ \ \ k\rightarrow -k, \ \ \ \
\xi_{i}^{(1)}\rightarrow \xi_{i}^{(2)},\ \ \ \
\xi_{i}^{(3)}\rightarrow \xi_{i}^{(4)},\ \ \ \ \eta_i^{(1)}
\rightarrow \eta_i^{(2)},$$ $$m_{t}\leftrightarrow m_{b},\ \ \
\eta^{(1)}\leftrightarrow\eta^{(2)},\ \ \ \lambda_b
\leftrightarrow \lambda_t,\ \ \  m_{\tilde{t}_i} \leftrightarrow
m_{\tilde{b}_i},\ \ \  U_{i2} \leftrightarrow V_{i2}^\ast,\ \ \
N_{i3} \leftrightarrow N_{i4}^\ast,$$ $$L_i(b) \leftrightarrow
L_i(t), \ \ \ R_i(b) \leftrightarrow R_i(t), \ \ \ p_b^\mu
P_{L(R)} \leftrightarrow p_t^\mu P_{R(L)}, \ \ \ \gamma^\mu
\not{k}P_L \leftrightarrow \gamma^\mu \not{k}P_R.$$

All other form factors $f_i^l$ not listed above vanish. In the
above expressions we have used the following definitions:
\begin{eqnarray*}
\eta^{(1)} =m_{b}\tan\beta,\hspace{1.0cm} \eta^{(2)}
=m_{t}\cot\beta, & & \lambda_b =\frac{m_b}{\sqrt{2}m_W\cos\beta},
\hspace{1.0cm} \lambda_t =\frac{m_t}{\sqrt{2}m_W\sin\beta}
\\ L_1(q) =\cos\theta_q,\hspace{1.0cm} L_2(q) =-\sin\theta_q, & &
R_1(q) =\sin\theta_q,\hspace{1.0cm} R_2(q) =\cos\theta_q,
\\ \eta_{H^{0}}^{(1)} =\frac{\cos^{2}\alpha}{\cos^{2}\beta},
\hspace{1.5cm} \eta_{h^{0}}^{(1)} =\frac{\sin^{2}\alpha}
{\cos^{2}\beta}, & &\hspace{0.1cm} \eta_{A^{0}}^{(1)}
=\tan^{2}\beta,\hspace{1.5cm} \eta_{G^{0}}^{(1)} =1,
\\ \eta_{H^{0}}^{(2)} =\frac{\sin^{2}\alpha}{\sin^{2}\beta},
\hspace{1.5cm} \eta_{h^{0}}^{(2)} =\frac{\cos^{2}\alpha}
{\sin^{2}\beta},& & \hspace{0.1cm} \eta_{A^{0}}^{(2)}
=\cot^{2}\beta, \hspace{1.6cm} \eta_{G^{0}}^{(2)}=1,
\\ \eta_{H^{0}}^{(3)} =-\eta_{h^{0}}^{(3)} =\frac{\sin\alpha
\cos\alpha}{\sin\beta \cos\beta},\hspace{1.7cm} & &
\eta_{G^{0}}^{(3)} =-\eta_{A^{0}}^{(3)} =1,
\\ \xi_{H^{-}}^{(1)} =\frac{m_{t}^{2}}{m_{b}^{2}} \cot^{2}\beta,
\hspace{1.2cm}\xi_{G^{-}}^{(1)} =\frac{m_{t}^{2}}{m_{b}^{2}}, & &
\hspace{0.2cm}\xi_{H^{-}}^{(2)}
=\frac{m_{b}^{2}}{m_{t}^{2}}\tan^{2}\beta, \hspace{1.0cm}
\xi_{G^{-}}^{(2)} =\frac{m_{b}^{2}}{m_{t}^{2}},
\\ \xi_{H^{-}}^{(3)} =\tan^{2}\beta, \hspace{1.7cm}
\xi_{G^{-}}^{(3)} =1,\hspace{0.6cm} & &\hspace{0.2cm}
\xi_{H^{-}}^{(4)} =\cot^{2}\beta, \hspace{1.5cm} \xi_{G^{-}}^{(4)}
=1,
\end{eqnarray*}
$$\zeta_{H^{0}} = \zeta_{h^{0}} =\zeta_{H^{-}} =-\zeta_{A^{0}}
=-\zeta_{G^{0}} =-\zeta_{G^{-}} =1,$$
\begin{eqnarray*}
\sigma_{ij} &=& \frac{m_W}{\sqrt{2}}(\sin{2\beta}
-\frac{m_b^2\tan\beta +m_t^2\cot\beta}{m_W^2})L_i(b)L_j(t)
  \\ & & +\frac{m_tm_b}{\sqrt{2}m_W}(\tan\beta
+\cot\beta)R_i(b)R_j(t) -\frac{m_b}{\sqrt{2}m_W}(\mu
-A_b\tan\beta)R_i(b)L_j(t)
  \\ & & -\frac{m_t}{\sqrt{2}m_W}(\mu -A_t\cot\beta)L_i(b)R_j(t),
\\ g_{1}^{V_{1}(s)} &=& \sum_{i=H^{0},h^{0},G^{0},A^{0}}
\eta_{i}^{(1)}\{[\frac{1}{2} -2\overline{C}_{24} +m_{b}^{2}
(-2C_{11} +C_{12} -C_{21}+C_{23}) -\hat{s}(C_{12}
  \\ & & +C_{23})](-p_{b},-k,m_{i},m_{b},m_{b}) +[-F_{0} +F_{1}
+2m_{b}^{2}G_{1}
  \\ & & -(1+\zeta_{i})2m_{b}^{2}G_{0}](m_{b}^{2},m_{i},m_{b})\},
\\ g_{2}^{V_{1}(s)} &=& \sum_{i=H^{-},G^{-}} 2\{\xi_{i}^{(1)}
[\frac{1}{2} -2\overline{C}_{24} +m_{t}^{2}C_{0} +m_{b}^{2}(-C_{0}
-2C_{11} +C_{12} -C_{21} +C_{23})
  \\ & & -\hat{s}(C_{12} +C_{23})](-p_{b},-k,m_{i},m_{t},m_{t})
+[\xi_{i}^{(1)}(-F_{0} +F_{1}) -2m_{t}^{2}\zeta_{i}G_{0}
  \\ & & +m_{b}^{2}(\xi_{i}^{(1)}
+\xi_{i}^{(3)})(G_{1} -\zeta_{i}G_{0})](m_{b}^{2},m_{i},m_{t})\}
  \\ & & +\frac{4m_W^2}{m_b^2}\sum_{i,j}\{\lambda_b^2[R_j^2(b)
|N_{i3}|^2(-F_0 +F_1) +m_b^2|N_{i3}|^2(-G_0 +G_1)
-2m_bm_{\tilde\chi_i^0}
  \\ & & \times L_j(b)R_j(b)N_{i3}^{\ast2}G_0]
(m_b^2,m_{\tilde{b}_j},m_{\tilde\chi_i^0})
+[-2m_bm_{\tilde\chi_i^+}\lambda_b\lambda_t
L_j(t)R_j(t)V_{i2}^{\ast2}U_{i2}^{\ast2}G_0
  \\ & & +\lambda_t^2R_j^2(t)|V_{i2}|^2(-F_0 +F_1)
+m_b^2(\lambda_t^2R_j^2(t)|V_{i2}|^2
+\lambda_b^2L_j^2(t)|U_{i2}|^2)(-G_0
  \\ & & +G_1)](m_b^2,m_{\tilde{t}_j},m_{\tilde\chi_i^+})
-2\lambda_b^2R_j^2(b)|N_{i3}|^2\bar{C}_{24}
(-p_b,-k,m_{\tilde\chi_i^0},m_{\tilde{b}_j},m_{\tilde{b}_j})
  \\ & & -2\lambda_t^2R_j^2(t)|V_{i2}|^2\bar{C}_{24}
(-p_b,-k,m_{\tilde\chi_i^+},m_{\tilde{t}_j},m_{\tilde{t}_j})\} ,
\\ g_{3}^{V_{1}(s)} &=& g_{2}^{V_{1}(s)}(\xi_{i}^{(1)} \leftrightarrow
\xi_{i}^{(3)}, V_{i2} \leftrightarrow U_{i2}^\ast, N_{i3}
\leftrightarrow N_{i3}^\ast, L_j(b) \leftrightarrow R_j(b),
\lambda_bL_j(t) \leftrightarrow \lambda_tR_j(t)),
\\ g_{4}^{V_{1}(s)} &=& \sum_{i=H^{0},h^{0},G^{0},A^{0}}
\eta_{i}^{(1)}2m_{b} [C_{0} +2C_{11} +C_{21} +\zeta_{i}(C_{0}
+C_{11})] (-p_{b},-k,m_{i},m_{b},m_{b})
  \\ & & +\sum_{i=H^{-},G^{-}} 4m_{b}[\xi_{i}^{(3)}(C_0 +2C_{11}
+C_{21}) +\frac{m_{t}^{2}} {m_{b}^{2}}\zeta_{i}(C_{0} +C_{11})]
(-p_{b},-k,m_{i},m_{t},m_{t})
  \\ & & +\frac{8m_W^2}{m_b^2}\sum_{i,j}\{\lambda_b^2
[m_{\tilde\chi_i^0}L_j(b)R_j(b)N_{i3}^{\ast2} (C_0 +C_{11})
-m_bL_j^2(b)|N_{i3}|^2 (C_{11}
  \\ & & +C_{21})](-p_b,-k,m_{\tilde\chi_i^0},m_{\tilde{b}_j},
m_{\tilde{b}_j})
  \\ & & +[m_{\tilde\chi_i^+}\lambda_b \lambda_tL_j(t)R_j(t)
V_{i2}^{\ast}U_{i2}^{\ast}(C_0 +C_{11}) -m_b\lambda_b^2L_j^2(t)
|U_{i2}|^2 (C_{11}
  \\ & & +C_{21})](-p_b,-k,m_{\tilde\chi_i^+},m_{\tilde{t}_j},
m_{\tilde{t}_j})\} ,
\\ g_{5}^{V_{1}(s)} &=& g_{4}^{V_{1}(s)}(\xi_{i}^{(1)} \leftrightarrow
\xi_{i}^{(3)}, V_{i2} \leftrightarrow U_{i2}^\ast, N_{i3}
\leftrightarrow N_{i3}^\ast, L_j(b) \leftrightarrow R_j(b),
\lambda_bL_j(t) \leftrightarrow \lambda_tR_j(t)),
\\ g_{6}^{V_{1}(s)}
&=& -\sum_{i=H^{0},h^{0},G^{0},A^{0}} \eta_{i}^{(1)}m_{b} (C_{0}
+C_{11} +\zeta_{i}C_{0})(-p_{b},-k,m_{i},m_{b},m_{b})
  \\ & & -\sum_{i=H^{-},G^{-}} 2m_{b}[\xi_{i}^{(3)}(C_{0} +C_{11})
+\frac{m_{t}^{2}} {m_{b}^{2}}\zeta_{i}C_{0}]
(-p_{b},-k,m_{i},m_{t},m_{t}),
\\ g_{7}^{V_{1}(s)} &=& g_{6}^{V_{1}(s)}(\xi_{i}^{(1)} \leftrightarrow
\xi_{i}^{(3)}),
\\ g_{8}^{V_{1}(s)} &=& \sum_{i=H^{0},h^{0},G^{0},A^{0}}
2\eta_{i}^{(1)}(C_{12} +C_{23}) (-p_{b},-k,m_{i},m_{b},m_{b})
  \\ & & +\sum_{i=H^{-},G^{-}} 4\xi_{i}^{(1)}(C_{12} +C_{24})
(-p_{b},-k,m_{i},m_{t},m_{t})
  \\ & & -\frac{8m_W^2}{m_b^2}\sum_{i,j}\{\lambda_b^2R_j^2(b)
|N_{i3}|^2(C_{12} +C_{23})
(-p_b,-k,m_{\tilde\chi_i^0},m_{\tilde{b}_j},m_{\tilde{b}_j})
  \\ & & +\lambda_t^2R_j^2(t)|V_{i2}|^2(C_{12} +C_{23})
(-p_b,-k,m_{\tilde\chi_i^+},m_{\tilde{t}_j},m_{\tilde{t}_j})\},
\\ g_{9}^{V_{1}(s)} &=& g_{8}^{V_{1}(s)}(\xi_{i}^{(1)} \leftrightarrow
\xi_{i}^{(3)}, V_{i2} \leftrightarrow U_{i2}^\ast, N_{i3}
\leftrightarrow N_{i3}^\ast, L_j(b) \leftrightarrow R_j(b),
\lambda_bL_j(t) \leftrightarrow \lambda_tR_j(t)),
\\ g_{1}^{V_{2}(s)} &=& \sum_{i=H^{0},h^{0},G^{0},A^{0}}
\eta_{i}^{(3)}\{\eta^{(1)}[-\frac{1}{2} +4\overline{C}_{24}
+m_{t}^{2}(C_{0} +2C_{11} +\zeta_{i}(C_{0} +C_{11})
  \\ & & +C_{21} -C_{12} -C_{23}) +m_{H^{-}}^{2}(C_{22} -C_{23})
+\hat{s}(C_{12} +C_{23})] +\eta^{(2)}m_{b}m_{t}[\zeta_{i}C_{11}
  \\ & & +(1 +\zeta_i)C_{0}]\}(-p_{t},-p_{H^{-}},m_{i},
m_{t},m_{b}) +\frac{4\sqrt{2}m_W}{\sin{2\beta}}\sum_{i,j,k}
[m_tR_i(b)R_j(t) N_{k3}^{\ast}N_{k4}
  \\ & & \times (-C_{11} +C_{12})+m_{\tilde\chi_k^0}L_j(t)R_i(b)
N_{k3}^{\ast}N_{k4}^{\ast} C_0]\sigma_{ij}(-p_t,-p_{H^-},
m_{\tilde\chi_k^0},m_{\tilde{b}_i}, m_{\tilde{t}_j}),
\\ g_{2}^{V_{2}(s)} &=& g_{1}^{V_{2}(s)}(\eta^{(1)}\leftrightarrow
\eta^{(2)}, L_l \leftrightarrow R_l, N_{kl} \leftrightarrow
N_{kl}^\ast),
\\ g_{3}^{V_{2}(s)} &=& \sum_{i=H^{0},h^{0},G^{0},A^{0}}
\eta_{i}^{(3)}\{\eta^{(1)}m_{t}[C_{0} +C_{11} +\zeta_{i}(C_{0}
+C_{12})] +\eta^{(2)}\zeta_{i} m_{b}C_{12}\}
  \\ & & (-p_{t},-p_{H^{-}},m_{i},m_{t},m_{b})
  \\ & & -\frac{4\sqrt{2}m_W}{\sin{2\beta}}\sum_{i,j,k}R_i(b)R_j(t)
N_{k3}^{\ast}N_{k4}\sigma_{ij}C_{12} (-p_t,-p_{H^-},
m_{\tilde\chi_k^0}, m_{\tilde{b}_i},m_{\tilde{t}_j}),
\\ g_{4}^{V_{2}(s)} &=& g_{3}^{V_{2}(s)}(\eta^{(1)}\leftrightarrow
\eta^{(2)}, L_l \leftrightarrow R_l, N_{kl} \leftrightarrow
N_{kl}^\ast),
\\ g_{1}^{V_{2}(t)} &=& \sum_{i=H^{0},h^{0},G^{0},A^{0}}
\eta_{i}^{(3)}\{\eta^{(1)}[-\frac{1}{2} +4\overline{C}_{24}
+m_{b}^{2}(C_{0} +2C_{11} +\zeta_{i}(C_{0} +C_{11})
  \\ & & +C_{21} -C_{12} -C_{23}) +m_{H^{-}}^{2}(C_{22}
-C_{23}) +\hat{t}(C_{12} +C_{23})]
  \\ & & +\eta^{(2)}m_{b}m_{t}[C_{0} +\zeta_{i}(C_{0} +C_{11})]\}
(-p_{b},p_{H^{-}},m_{i},m_{b},m_{t})
  \\ & & +\frac{4\sqrt{2}m_W}{\sin{2\beta}}\sum_{i,j,k}
[m_bL_i(b)L_j(t)N_{k3}^{\ast}N_{k4}(-C_{11} +C_{12})
  \\ & & +m_{\tilde\chi_k^0}L_j(t)R_i(b)N_{k3}^{\ast}N_{k4}^{\ast}
C_0]\sigma_{ij}(-p_b,p_{H^-},m_{\tilde\chi_k^0},
m_{\tilde{b}_i},m_{\tilde{t}_j}),
\\ g_{2}^{V_{2}(t)} &=&
g_{1}^{V_{2}(t)}(\eta^{(1)}\leftrightarrow \eta^{(2)}, L_l
\leftrightarrow R_l, N_{kl} \leftrightarrow N_{kl}^\ast),
\\ g_{3}^{V_{2}(t)} &=& -\sum_{i=H^{0},h^{0},G^{0},A^{0}}
\eta_{i}^{(3)}\{\eta^{(1)}m_{b}[C_{0} +C_{11} +\zeta_{i}(C_{0}
+C_{12})] +\eta^{(2)}\zeta_{i} m_{t}C_{12}\}
  \\ & & (-p_{b},p_{H^{-}},m_{i},m_{b},m_{t})
  \\ & & +\frac{4\sqrt{2}m_W}{\sin{2\beta}}\sum_{i,j,k}R_i(b)R_j(t)
N_{k3}^{\ast}N_{k4}\sigma_{ij}C_{12} (-p_b,p_{H^-},
m_{\tilde\chi_k^0}, m_{\tilde{b}_i},m_{\tilde{t}_j}) ,
\\ g_{4}^{V_{2}(t)} &=& g_{3}^{V_{2}(t)}(\eta^{(1)}\leftrightarrow
\eta^{(2)}, L_l \leftrightarrow R_l, N_{kl} \leftrightarrow
N_{kl}^\ast),
\\ g_{1}^{s(s)} &=& \sum_{i=H^{0},h^{0},G^{0},A^{0}}m_{b}
\eta_{i}^{(1)}\{-\zeta_{i}F_{0}(p_{b}+k,m_{i},m_{b})
+[\zeta_{i}F_{0} -2m_{b}^{2}(1 +\zeta_{i})G_{0}
  \\ & & +2m_{b}^{2}G_{1}](m_{b}^{2},m_{i},m_{b})\}
+\sum_{i=H^{-},G^{-}}2m_{b}\{-\frac{m_{t}^{2}}{m_{b}^{2}}\zeta_{i}
F_{0}(p_{b}+k,m_{i},m_{t})
  \\ & & +[-2m_{t}^{2}\zeta_{i}G_{0} +m_{b}^{2} (\xi_{i}^{(1)}
+\xi_{i}^{(3)})(G_{1} -\zeta_{i}G_{0}) +\zeta_{i}
\frac{m_{t}^{2}}{m_{b}^{2}}F_{0}] (m_{b}^{2},m_{i},m_{t})\}
  \\ & & +\frac{4m_W^2}{m_b^2}\sum_{i,j}\{-m_{\tilde\chi_i^0}
\lambda_b^2 L_j(b)R_j(b)N_{i3}^{\ast2}F_0(p_b+k,m_{\tilde{b}_j},
m_{\tilde\chi_i^0})
  \\ & & -m_{\tilde\chi_i^+}\lambda_b\lambda_tL_j(b)R_j(b)
V_{i2}^{\ast} U_{i2}^{\ast}F_0(p_b+k,m_{\tilde{t}_j},
m_{\tilde\chi_i^+}) +[m_b^3\lambda_b^2|N_{i3}|^2(-G_0 +G_1)
  \\ & & -m_{\tilde\chi_i^0}\lambda_b^2
L_j(b)R_j(b)N_{i3}^{\ast2}(2m_b^2G_0 -F_0)](m_b^2,m_{\tilde{b}_j},
m_{\tilde\chi_i^0}) +[m_b^3(\lambda_b^2L_j^2(t)|U_{i2}|^2
  \\ & & +\lambda_t^2R_j^2(t)|V_{i2}|^2)(-G_0 +G_1)
-m_{\tilde\chi_i^+}\lambda_b\lambda_t
L_j(t)R_j(t)V_{i2}^{\ast}U_{i2}^{\ast}(2m_b^2G_0
  \\ & & -F_0)](m_b^2,m_{\tilde{t}_j}, m_{\tilde\chi_i^+})\},
\\ g_{2}^{s(s)} &=& \sum_{i=H^{0},h^{0},G^{0},A^{0}}
\eta_{i}^{(1)}(-F_{0} +F_{1})(p_{b}+k,m_{i},m_{b}),
\\ g_{3}^{s(s)} &=& \sum_{i=H^{0},h^{0},G^{0},A^{0}}\eta_{i}^{(1)}
[F_{0} -F_{1} -2m_{b}^{2}G_{1} +2(1 +\zeta_{i})m_{b}^{2}G_{0}]
(m_{b}^{2},m_{i},m_{b})
  \\ & & +\sum_{i=H^{-},G^{-}}2\{\xi_{i}^{(1)}(-F_{0} +F_{1})
(p_{b}+k,m_{i},m_{t}) -[\xi_{i}^{(1)}(-F_{0}+F_{1})
  \\ & &-2\zeta_{i}m_{t}^{2}G_{0} +m_{b}^{2}(\xi_{i}^{(1)}
+\xi_{i}^{(3)}) (G_{1} -\zeta_{i}G_{0})](m_{b}^{2},m_{i},m_{t})\}
  \\ & & -\frac{4m_W^2}{m_b^2}\sum_{i,j}\{\lambda_b^2[R_j^2(b)
|N_{i3}|^2(-F_0 +F_1) +|N_{i3}|^2m_b^2(-G_0 +G_1)
  \\ & & -2m_bm_{\tilde\chi_i^0}L_j(b)R_j(b)N_{i3}^{\ast2}G_0]
(m_b^2,m_{\tilde b_j},m_{\tilde\chi_i^0})
  \\ & & +[\lambda_t^2R_j^2(t)|V_{i2}|^2(-F_0 +F_1)
+m_b^2(\lambda_t^2R_j^2(t)|V_{i2}|^2
+\lambda_b^2L_j^2(t)|U_{i2}|^2)(G_1 -G_0)
  \\ & & -2m_bm_{\tilde\chi_i^+}L_j(t)R_j(t)\lambda_b
\lambda_tV_{i2}^{\ast} U_{i2}^{\ast} G_0] (m_b^2,m_{\tilde
t_j},m_{\tilde\chi_i^+})
  \\ & & -\lambda_b^2R_j^2(b)|N_{i3}|^2 (-F_0 +F_1)
(p_b+k,m_{\tilde{b}_j}, m_{\tilde\chi_i^0})
  \\ & & -\lambda_t^2R_j^2(t)|V_{i2}|^2 (-F_0
+F_1)(p_b+k,m_{\tilde{t}_j}, m_{\tilde\chi_i^+})\},
\\ g_{4}^{s(s)} &=& g_{3}^{s(s)}(\xi_{i}^{(1)} \leftrightarrow
\xi_{i}^{(3)}, V_{i2} \leftrightarrow U_{i2}^\ast, N_{i3}
\leftrightarrow N_{i3}^\ast, L_j(b) \leftrightarrow R_j(b),
\lambda_bL_j(t) \leftrightarrow \lambda_tR_j(t)),
\\ g_{5}^{s(s)} &=& g_{1}^{s(s)}(N_{i3}^{\ast} \rightarrow N_{i3},
V_{i2}^{\ast} \rightarrow V_{i2},U_{i2}^{\ast} \rightarrow
U_{i2}),
\\ \delta\Lambda_{L}^{(1)} &=& \frac{4N_{c}} {3m_{W}^{2}}
(1-\cot^{2}\theta_{W})[2m_{t}^{2}(\ln{\frac{m_{t}^{2}}{\mu^{2}}}
-1) +m_{b}^{2} +m_{t}^{2} -\frac{5}{6}m_{W}^{2} +m_{b}^{2}F_{0}
  \\ & & +(m_{b}^{2}-m_{t}^{2}-2m_{W}^{2})F_{1}]
(m_{W}^{2},m_{b},m_{t}) +\frac{4N_{c}}{3m_{W}^{2}} \cot^{2}
\theta_{W}\{-\frac{5}{6}[(g_{V}^{b})^{2} +(g_{A}^{b})^{2}
  \\ & & +(g_{V}^{t})^{2} +(g_{A}^{t})^{2}]m_{Z}^{2}
+[((g_{V}^{t})^{2} +(g_{A}^{t})^{2})(2m_{t}^{2}
\ln{\frac{m_{t}^{2}}{\mu^{2}}} +m_{t}^{2}F_{0} -2m_{Z}^{2}F_{1})
  \\ & & -((g_{V}^{t})^{2} -(g_{A}^{t})^{2})3m_{t}^{2}F_{0}]
(m_{Z}^{2},m_{t},m_{t}) +[((g_{V}^{b})^{2} +(g_{A}^{b})^{2})
(2m_{b}^{2}\ln{\frac{m_{b}^{2}} {\mu^{2}}}
  \\ & & +m_{b}^{2}F_{0} -2m_{Z}^{2}F_{1})-((g_{V}^{b})^{2}
-(g_{A}^{b})^{2}) 3m_{b}^{2}F_{0}] (m_{Z}^{2},m_{b},m_{b})\}
+\frac{4N_{c}}{m_{W}^{2}} [(\cot^{2}\beta
  \\ & & -1)m_{t}^{2}F_{0} +(m_{t}^{2}-m_{b}^{2}
-2m_{t}^{2}\cot^{2}\beta)F_{1} +(m_{t}^{2}\cot^{2}\beta +m_{b}^{2}
\tan^{2}\beta
  \\ & & +2m_{b}^{2})m_{t}^{2}G_{0} -(m_{t}^{2}\cot^{2}\beta
+m_{b}^{2} \tan^{2}\beta)m_{H^{-}}^{2}G_{1}] (m_{H^{-}}^{2},m_{t},
m_{b})
  \\ & & +\sum_{i=H^{0},h^{0},G^{0},A^{0}} \frac{1}{2m_{W}^{2}}
\{m_{b}^{2}\eta_{i}^{(1)}[F_{1}-F_{0} -2m_{b}^{2}(G_0
+\zeta_{i}G_{0} -G_{1})] (m_{b}^{2},m_{i},m_{b})
  \\ & & -m_{t}^{2} \eta_{i}^{(2)}[-(1+2\zeta_{i})F_{0} +F_{1}
+2m_{t}^{2} (1+\zeta_{i})G_{0} -2m_{t}^{2}G_{1}]
(m_{t}^{2},m_{i},m_{t})\}
  \\ & & +\sum_{i=H^{-},G^{-}}\frac{1}{m_{W}^{2}}
\{m_{b}^{2}[\xi_{i}^{(1)}(-F_{0} +F_{1}) -2m_{t}^{2}\zeta_{i}G_{0}
+m_{b}^{2}(\xi_{i}^{(1)} +\xi_{i}^{(3)})
  \\ & & \times(G_{1} -\zeta_{i}G_{0})](m_{b}^{2},m_{i},m_{t})
-m_{t}^{2}[-\frac{2m_{b}^{2}}{m_{t}^{2}}\zeta_{i}F_{0}
+\xi_{i}^{(2)}(-F_{0} +F_{1}) +2m_{b}^{2}\zeta_{i}G_{0}
  \\ & & -m_{t}^{2}(\xi_{i}^{(2)} +\xi_{i}^{(4)})(G_{1}
-\zeta_{i}G_{0})](m_{t}^{2},m_{i},m_{b})\}
-2N_C\sum_{i,j}\{2\sigma_{ij}\sigma_{ij}G_0
  \\ & & +\frac{1}{m_W^2}L_i(b)L_j(t)[L_i(b)L_j(t) (\frac{m_b^2}
{\cos^2\beta} +\frac{m_t^2}{\sin^2\beta})\cos{2\beta}
  \\ & & +R_i(b)R_j(t)m_tm_b(\tan^2\beta
-\cot^2\beta)]\}(m_{H^-}^2,m_{\tilde{t}_j}, m_{\tilde{b}_i}) ,
\\ \delta\Lambda_{L}^{(2)} &=& -2\sum_{i,j}\{\lambda^2_t[-\frac
{2m_{\tilde{\chi}^0_i}}{m_t} L_j(t)R_j(t)N_{i4}^{\ast2}(F_0
-m_t^2G_0) +|N_{i4}|^2(R_j^2(t)(-F_0 +F_1)
  \\ & & -m_t^2(-G_0 +G_1))] (m_t^2,m_{\tilde{t}_j},
m_{\tilde{\chi}_i^0}) +[-\frac {2m_{\tilde{\chi}^+_i}}{m_t}
\lambda_b\lambda_t L_j(b)R_j(b)U_{i2}^\ast V_{i2}^\ast(F_0
  \\ & & -m_t^2G_0) +\lambda_b^2R_j^2(b)|U_{i2}|^2(-F_0 +F_1)
-m_t^2(\lambda_t^2L_j^2(b)|V_{i2}|^2
+\lambda_b^2R_j^2(b)|U_{i2}|^2)
  \\ & & \times(-G_0 +G_1)](m_t^2,m_{\tilde{b}_j},
m_{\tilde{\chi}_i^+})\},
\\ \delta\Lambda_{L}^{(3)} &=& 2\sum_{i,j}\{\lambda^2_b
[|N{i3}|^2(R_j^2(b)(-F_0 +F_1) +m_b^2(-G_0 +G_1))
-2m_bm_{\tilde{\chi}_i^0}L_j(b)R_j(b)
  \\ & & \times N_{i3}^{\ast2}G_0] (m_b^2,m_{\tilde{b}_j},
m_{\tilde{\chi}_i^0}) +[-2m_{\tilde{\chi}^+_i} m_b
\lambda_b\lambda_t L_j(t)R_j(t)U_{i2}^\ast V_{i2}^\ast G_0
  \\ & & +\lambda_b^2L_j^2(b)|U_{i2}|^2(-F_0 +F_1) +m_b^2
(\lambda_t^2R_j^2(t)|V_{i2}|^2 +\lambda_b^2L_j^2(t)|U_{i2}|^2)
  \\ & & \times(-G_0 +G_1)]
(m_b^2,m_{\tilde{t}_j},m_{\tilde{\chi}_i^+})\},
\\ \delta\Lambda_{R}^{(1)} &=& \sum_{i=H^{0},h^{0},G^{0},A^{0}}
\frac{1}{2m_{W}^{2}} \{m_{t}^{2}\eta_{i}^{(2)}[-F_{0} +F_{1}
-2m_{t}^{2}(G_0+\zeta_{i}G_{0} -G_{1})] (m_{t}^{2},m_{i},m_{t})
  \\ & & -m_{b}^{2} \eta_{i}^{(1)}[-F_{0} +F_{1} -2\zeta_{i}F_{0}
+2m_{b}^{2}(1+\zeta_{i})G_{0} -2m_{b}^{2}G_{1}]
(m_{b}^{2},m_{i},m_{b})\}
  \\ & & +\sum_{i=H^{-},G^{-}}\frac{1}{m_{W}^{2}}\{m_{t}^{2}
[\xi_{i}^{(2)} (-F_{0} +F_{1}) -2m_{b}^{2}\zeta_{i}G_{0}
+m_{t}^{2}(\xi_{i}^{(2)} +\xi_{i}^{(4)})(G_{1}
  \\ & & -\zeta_{i}G_{0})](m_{t}^{2},m_{i},m_{b})
-m_{b}^{2}[-\frac{2m_{t}^{2}}{m_{b}^{2}}\zeta_{i}F_{0}
+\xi_{i}^{(1)}(-F_{0} +F_{1}) +2m_{t}^{2}\zeta_{i}G_{0}
  \\ & & -m_{b}^{2}(\xi_{i}^{(1)} +\xi_{i}^{(3)})(G_{1}
-\zeta_{i}G_{0})](m_{b}^{2},m_{i},m_{t})\},
\\ \delta\Lambda_{R}^{(2)} &=& \delta\Lambda_{L}^{(2)}(U),
\hspace{4.0cm} \delta\Lambda_{R}^{(3)} =
\delta\Lambda_{L}^{(3)}(U).
\end{eqnarray*}
Here $C_{0},C_{ij}$ are the three-point Feynman integrals[19] and
$\overline{C}_{24}\equiv-\frac{1}{4}\Delta+C_{24}$, while
\begin{eqnarray*}
F_{n}(q,m_{1},m_{2})&=&\int_{0}^{1}dyy^{n}\ln{[\frac{-q^{2}y(1-y)
+m_{1}^{2}(1-y)+m_{2}^{2}y}{\mu^{2}}]},
\\ G_{n}(q,m_{1},m_{2})
&=&-\int_{0}^{1}dy\frac{y^{n+1}(1-y)} {-q^{2}y(1-y)
+m_{1}^{2}(1-y) +m_{2}^{2}y},
\end{eqnarray*}
and
\begin{eqnarray*}
g_{V}^{t} = \frac{1}{2}-\frac{4}{3}\sin^{2}\theta_{W}, \ \ \ \
g_{A}^{t}=\frac{1}{2}, \ \ \ \ \ \ \  g_{V}^{b} = -\frac{1}{2}
+\frac{2}{3}\sin^{2}\theta_{W}, \ \ \ \ g_{A}^{b} =-\frac{1}{2},
\end{eqnarray*}
which are the SM couplings of the top and bottom quarks to the Z
boson. The definitions of $\theta_q, U_{ij}, V_{ij}, N_{ij}, \mu,
A_{q}$ can be found in ref.[2].

\subsection{Appendix B}
\begin{eqnarray*}
h_{1}^{(i)} &=& 4m_{t}\eta^{(2)}(2p_{b}\cdot k -p^{(i)}\cdot p_b)
-4m_{b}\eta^{(1)}(p^{(i)}\cdot p_{t} +p_{t}\cdot k),
\\ h_{2}^{(i)} &=& h_{1}^{(i)}(\eta^{(1)}
\leftrightarrow \eta^{(2)}),
\\ h_{3}^{(i)} &=& 2\eta^{(2)}(2p_{b}\cdot kp_{b}\cdot p_{t}
-m_{b}^{2}p_{t}\cdot k -2p^{(i)}\cdot p_bp_{b}\cdot p_{t})
+2m_{b}m_{t}\eta^{(1)}(p_{b}\cdot k
  \\ & & -2p^{(i)}\cdot p_b),
\\ h_{4}^{(i)} &=& h_{3}^{(i)}(\eta^{(1)}
\leftrightarrow \eta^{(2)}),
\\ h_{5}^{(i)} &=& 2\eta^{(2)}(m_{t}^{2}p_{b}\cdot k
-2p^{(i)}\cdot p_tp_{b}\cdot p_{t})
+2m_{b}m_{t}m_{t}\eta^{(1)}(p_{t}\cdot k -2p^{(i)}\cdot p_t),
\\ h_{6}^{(i)} &=& h_{5}^{(i)}(\eta^{(1)}
\leftrightarrow \eta^{(2)}),
\\ h_{7}^{(i)} &=& 4\eta^{(2)}(p^{(i)}\cdot
p_bp_{t}\cdot k -p^{(i)}\cdot kp_{b}\cdot p_{t} -p_b\cdot
kp^{(i)}\cdot p_t -2p_{b}\cdot kp_{t}\cdot k)
  \\ & & -4m_{b}m_{t}\eta^{(1)} p^{(i)}\cdot k,
\\ h_{8}^{(i)} &=& h_{7}^{(i)}(\eta^{(1)}
\leftrightarrow \eta^{(2)}),
\\ h_{9}^{(i)} &=& 4m_{t}\eta^{(2)} p_{b}\cdot k(p_{b}\cdot k
-p^{(i)}\cdot p_b) -4m_{b}\eta^{(1)}p^{(i)}\cdot p_bp_{t}\cdot k,
\\ h_{10}^{(i)} &=& h_{9}^{(i)}(\eta^{(1)}
\leftrightarrow \eta^{(2)}),
\\ h_{11}^{(i)} &=& 4m_{t}\eta^{(2)} p_{b}\cdot k(p_{t}\cdot k
-p^{(i)}\cdot p_t) -4m_{b}\eta^{(1)} p_{t}\cdot kp^{(i)}\cdot p_t,
\\ h_{12}^{(i)} &=&h_{11}^{(i)}(\eta^{(1)}
\leftrightarrow\eta^{(2)}),
\end{eqnarray*}
where the index $i$ represents the two channels $s$ and $t$, and
$p^{(s)}=p_b$, $p^{(t)}=p_t$. \eject \baselineskip=0.25in {\LARGE
References} \vspace{0.2cm} }

\begin{itemize}
\item[{\rm[1]}] For a review, see J.Gunion, H. Haber, G. Kane, and
            S.Dawson, The Higgs Hunter's Guide(Addison-Wesley,
            New York,1990).
\item[{\rm[2]}] H.E. Haber and G.L. Kane, Phys. Rep. 117, 75(1985);
            J.F. Gunion and H.E. Haber, Nucl. Phys. {\bf B272}, 1(1986).
\item[{\rm[3]}] E.Eichten, I.Hinchliffe, K. Lane, and C. Quigg, Rev.
            Mod. Phys. 56, 579(1984); 1065(E)(1986); N.G. Deshpande, X.
Tata,
            and D. A. Dicus, Phys. Rev. {\bf D29}, 1527(1984); S.
Willenbrock,
            Phys. Rev. {\bf D35}, 173(1987); A. Krause, T.Plehn, M. Spria,
            and P. M. Zerwas, Nucl. Phys. {\bf B519}, 85(1998); J.Yi, M.
Wen-Gan,
            H.Liang, H. Meng, and Y. Zeng-Hui, J. Phys. G23, 385(1997);
            Erratum-ibid. G23, 1151(1997).
\item[{\rm[4]}] D.A.Dicus, J.L.Hewett, C.Kao and T.G.Rizzo, Phys. Rev. {\bf
D40}, 787(1989);
            A.A. Barrientos Bendez$\acute{u}$ and B.A. Kniehl, Phys. Rev.
            {\bf D59}, 015009(1999).
\item[{\rm[5]}] S. Moretti and K. Odagiri, Phys. Rev. {\bf
D59},055008(1999).
\item[{\rm[6]}] Z.Kunszt and F. Zwirner, Nucl. Phys. {\bf B385}, 3(1992),
            and references cited therein.
\item[{\rm[7]}] J.F. Gunion, H.E. Haber, F.E. Paige, W.-K. Tung, and
            S. Willenbrock, Nucl. Phys. {\bf B294},621(1987); R.M.
            Barnett, H.E. Haber, and D.E. Soper, ibid. B306,
            697(1988); F.I. Olness and W.-K. Tung, ibid. {\bf B308},
            813(1988).
\item[{\rm[8]}] V. Barger, R.J.N. Phillips, and D.P. Roy, Phys. Lett.
            {\bf B324}, 236(1994).
\item[{\rm[9]}] C.S. Huang and S.H. Zhu, Phys. Rev. {\bf D60},
            075012(1999).
\item[{\rm[10]}] K. Odagiri, hep-ph/9901432; Phys. Lett. {\bf B452},
327(1999).
\item[{\rm[11]}] D.P. Roy, Phys. Lett. {\bf B459}, 607(1999).
\item[{\rm[12]}] Francesca Borzumati, Jean-Loic Kneur, and Nir
            Polonsky, Phys. Rev. {\bf D60}, 115011(1999).
\item[{\rm[13]}] S. Sirlin, Phys. Rev. {\bf D22}, 971 (1980);
            W. J. Marciano and A. Sirlin,{\sl ibid.} {\bf 22}, 2695(1980);
            {\bf 31}, 213(E) (1985);
            A. Sirlin and W.J. Marciano, Nucl. Phys. {\bf B189}, 442(1981);
            K.I. Aoki et.al., Prog. Theor. Phys. Suppl. {\bf 73}, 1(1982).
\item[{\rm[14]}] A. Mendez and A. Pomarol, Phys.Lett.{\bf B279}, 98(1992).
\item[{\rm[15]}] Particle Data Group, C.Caso {\it et al}, Eur.Phys.J.C 3,
1(1998).
\item[{\rm[16]}] J.Gunion, A.Turski, Phys. Rev. {\bf D39}, 2701(1989);
            {\bf D40}, 2333(1990); J.R.Espinosa, M.Quiros, Phys. Lett. {\bf
            B266}, 389(1991); M.Carena, M.Quiros, C.E.M.Wagner, Nucl. Phys.
            {\bf B461}, 407(1996).
\item[{\rm[17]}] H.L. Lai, et al.(CTEQ collaboration), hep-ph/9903282.

\item[{\rm[18}] C.S.Li, R.J.Oakes, and J.M. Yang, Phys. Rev. {\bf D55},
5780(1997).

\item[{\rm[19]}] G.Passarino and M.Veltman, Nucl. Phys. {\bf B160},
151(1979);
            A.Axelrod, {\sl ibid.} {\bf B209}, 349 (1982); M.Clements {\sl
et al.}, Phys. Rev. {\bf D27}, 570 (1983).

\end{itemize}

\newpage

\vspace{-0.2cm}
\begin{picture}(120,120)(0,0)
\Gluon(5,100)(25,78){-2.5}{3} \ArrowLine(5,56)(25,78)
\ArrowLine(25,78)(75,78) \Vertex(25,78){1} \Vertex(75,78){1}
\ArrowLine(75,78)(95,100) \DashLine(75,78)(95,56){3}
\Text(95,105)[] {$t$} \Text(98,51)[] {$H^-$} \Text(5,105)[]{$g$}
\Text(5,51)[] {$b$} \Text(48,25)[]{$(a)$}
\end{picture}
\vspace{-0.2cm} \hspace{1.0cm}
\begin{picture}(120,120)(0,0)
\Gluon(5,98)(40,98){-2.5}{4} \ArrowLine(5,58)(40,58)
\ArrowLine(40,58)(40,98) \Vertex(40,98){1} \Vertex(40,58){1}
\ArrowLine(40,98)(75,98) \DashLine(40,58)(75,58){3}
\Text(80,101)[] {$t$} \Text(85,55)[] {$H^-$} \Text(0,101)[]{$g$}
\Text(0,55)[] {$b$} \Text(43,35)[]{$(b)$}
\end{picture}
\vspace{-0.2cm}  \hspace{0.9cm}
\begin{picture}(120,120)(0,0)
\Gluon(5,100)(25,78){-2.5}{3} \ArrowLine(5,56)(13.6,65.4)
\Line(13.6,65.4)(25,78) \ArrowLine(25,78)(75,78) \Vertex(25,78){1}
\Vertex(75,78){1} \ArrowLine(75,78)(95,100)
\DashLine(75,78)(95,56){3} \DashCArc(25,78)(17,-132,0){3}
\Text(95,105)[] {$t$} \Vertex(13.6,65.4){1} \Vertex(42,78){1}
\Text(98,51)[] {$H^-$} \Text(5,105)[]{$g$} \Text(5,51)[] {$b$}
\Text(48,35)[]{$(c)$}
\end{picture}\\
\vspace{-0.2cm}  \hspace{0.4cm}
\begin{picture}(120,120)(0,0)
\Gluon(5,100)(25,78){-2.5}{3} \ArrowLine(5,56)(13.6,65.4)
\DashLine(13.6,65.4)(25,78){3} \DashLine(25,78)(42,78){3}
\ArrowLine(42,78)(75,78) \Vertex(25,78){1} \Vertex(75,78){1}
\ArrowLine(75,78)(95,100) \DashLine(75,78)(95,56){3}
\CArc(25,78)(17,-132,0) \Text(95,105)[] {$t$}
\Vertex(13.6,65.4){1} \Vertex(42,78){1} \Text(98,51)[] {$H^-$}
\Text(5,105)[]{$g$} \Text(5,51)[] {$b$} \Text(48,35)[]{$(d)$}
\end{picture}
\vspace{-0.2cm} \hspace{0.9cm}
\begin{picture}(120,120)(0,0)
\Gluon(5,100)(20,83){-2.5}{3} \ArrowLine(5,56)(20,73)
\ArrowLine(30,78)(75,78) \Line(22,82)(28,74) \Line(22,74)(28,82)
\Vertex(75,78){1} \ArrowLine(75,78)(95,100)
\DashLine(75,78)(95,56){3} \Text(95,105)[] {$t$} \Text(98,51)[]
{$H^-$} \Text(5,105)[]{$g$} \Text(5,51)[] {$b$}
\Text(48,25)[]{$(e)$}
\end{picture}
\vspace{-0.2cm} \hspace{1.1cm}
\begin{picture}(120,120)(0,0)
\Gluon(5,98)(40,98){-2.5}{4} \ArrowLine(5,58)(40,58)
\Line(40,98)(57.5,98) \DashLine(57.5,98)(40,78){3}
\Line(40,98)(40,88) \ArrowLine(40,58)(40,88) \Vertex(40,98){1}
\Vertex(40,58){1} \ArrowLine(57.5,98)(75,98)
\DashLine(40,58)(75,58){3} \Vertex(57.5,98){1} \Vertex(40,78){1}
\Text(80,101)[] {$t$} \Text(85,55)[] {$H^-$} \Text(0,101)[]{$g$}
\Text(0,55)[] {$b$} \Text(43,35)[]{$(f)$}
\end{picture}\\
\vspace{-0.2cm} \hspace{0.4cm}
\begin{picture}(120,120)(0,0)
\Gluon(5,98)(40,98){-2.5}{4} \ArrowLine(5,58)(40,58)
\DashLine(40,98)(57.5,98){3} \Line(57.5,98)(40,78)
\DashLine(40,78)(40,98){3} \ArrowLine(40,58)(40,78)
\Vertex(40,98){1} \Vertex(40,58){1} \ArrowLine(57.5,98)(75,98)
\DashLine(40,58)(75,58){3} \Vertex(57.5,98){1} \Vertex(40,78){1}
\Text(80,101)[] {$t$} \Text(85,55)[] {$H^-$} \Text(0,101)[]{$g$}
\Text(0,55)[] {$b$} \Text(43,35)[]{$(g)$}
\end{picture}
\vspace{-0.2cm} \hspace{1.1cm}
\begin{picture}(120,120)(0,0)
\Gluon(5,98)(35,98){-2.5}{4} \ArrowLine(5,58)(40,58)
\ArrowLine(40,58)(40,93) \Vertex(40,58){1}
\ArrowLine(45,98)(75,98) \DashLine(40,58)(75,58){3}
\Line(37,102)(43,94) \Line(37,94)(43,102) \Text(80,101)[] {$t$}
\Text(85,55)[] {$H^-$} \Text(0,101)[]{$g$} \Text(0,55)[] {$b$}
\Text(43,35)[]{$(h)$}
\end{picture}
\vspace{-0.2cm} \hspace{0.8cm}
\begin{picture}(120,120)(0,0)
\Gluon(5,100)(25,78){-2.5}{3} \ArrowLine(5,56)(25,78)
\DashCArc(50,78)(17,0,180){3} \Line(25,78)(75,78)
\Vertex(25,78){1} \Vertex(75,78){1} \ArrowLine(75,78)(95,100)
\Vertex(33,78){1} \Vertex(67,78){1} \DashLine(75,78)(95,56){3}
\Text(95,105)[] {$t$} \Text(98,51)[] {$H^-$} \Text(5,105)[]{$g$}
\Text(5,51)[] {$b$} \Text(48,35)[]{$(i)$}
\end{picture}\\
\vspace{-0.2cm} \hspace{0.4cm}
\begin{picture}(120,120)(0,0)
\Gluon(5,100)(25,78){-2.5}{3} \ArrowLine(5,56)(25,78)
\Line(25,78)(45,78) \Line(55,78)(75,78) \Line(47,82)(53,74)
\Line(53,82)(47,74) \Vertex(25,78){1} \Vertex(75,78){1}
\ArrowLine(75,78)(95,100) \DashLine(75,78)(95,56){3}
\Text(95,105)[] {$t$} \Text(98,51)[] {$H^-$} \Text(5,105)[]{$g$}
\Text(5,51)[] {$b$} \Text(48,25)[]{$(j)$}
\end{picture}
\vspace{-0.3cm} \hspace{1.1cm}
\begin{picture}(120,120)(0,0)
\Gluon(5,98)(40,98){-2.5}{4} \ArrowLine(5,58)(40,58)
\Line(40,58)(40,98) \Vertex(40,98){1} \Vertex(40,58){1}
\DashCArc(40,78)(13,-90,90){3} \ArrowLine(40,98)(75,98)
\Vertex(40,91){1} \Vertex(40,65){1} \DashLine(40,58)(75,58){3}
\Text(80,101)[] {$t$} \Text(85,55)[] {$H^-$} \Text(0,101)[]{$g$}
\Text(0,55)[] {$b$} \Text(43,35)[]{$(k)$}
\end{picture}
\vspace{-0.3cm} \hspace{1.0cm}
\begin{picture}(120,120)(0,0)
\Gluon(5,98)(40,98){-2.5}{4} \ArrowLine(5,58)(40,58)
\Line(40,58)(40,73) \Line(40,83)(40,98) \Vertex(40,98){1}
\Vertex(40,58){1} \ArrowLine(40,98)(75,98) \Line(37,82)(43,74)
\Line(37,74)(43,82) \DashLine(40,58)(75,58){3} \Text(80,101)[]
{$t$} \Text(85,55)[] {$H^-$} \Text(0,101)[]{$g$} \Text(0,55)[]
{$b$} \Text(43,35)[]{$(l)$}
\end{picture}\\
\vspace{-0.3cm} \hspace{0.4cm}
\begin{picture}(120,120)(0,0)
\Gluon(5,100)(25,78){-2.5}{3} \ArrowLine(5,56)(25,78)
\ArrowLine(25,78)(75,78) \Vertex(25,78){1} \Vertex(75,78){1}
\DashCArc(75,78)(17,48,180){3} \ArrowLine(86.4,90.6)(95,100)
\Line(75,78)(86.4,90.6) \Vertex(58,78){1} \Vertex(86.4,90.6){1}
\DashLine(75,78)(95,56){3} \Text(95,105)[] {$t$} \Text(98,51)[]
{$H^-$} \Text(5,105)[]{$g$} \Text(5,51)[] {$b$}
\Text(48,35)[]{$(m)$}
\end{picture}
\vspace{-0.3cm} \hspace{1.0cm}
\begin{picture}(120,120)(0,0)
\Gluon(5,100)(25,78){-2.5}{3} \ArrowLine(5,56)(25,78)
\ArrowLine(25,78)(75,78) \Vertex(25,78){1} \Vertex(75,78){1}
\ArrowLine(75,78)(95,100) \DashLine(75,78)(95,56){3}
\DashCArc(75,78)(17,-180,-48){3} \Vertex(86.4,65.4){1}
\Vertex(58,78){1} \Text(95,105)[] {$t$} \Text(98,51)[] {$H^-$}
\Text(5,105)[]{$g$} \Text(5,51)[] {$b$} \Text(92,74)[]{\small
$\tilde{t}_j$} \Text(72,55)[]{\small $\tilde{b}_i$}
\Text(65,87)[]{\small $\tilde\chi_k^0$} \Text(48,35)[]{$(n)$}
\end{picture}
\vspace{-0.3cm} \hspace{1.0cm}
\begin{picture}(120,120)(0,0)
\Gluon(5,100)(25,78){-2.5}{3} \ArrowLine(5,56)(25,78)
\ArrowLine(25,78)(70,78) \Vertex(25,78){1}
\ArrowLine(80,82)(95,100) \DashLine(80,74)(95,56){3}
\Line(72,82)(78,74) \Line(72,74)(78,82) \Text(95,105)[] {$t$}
\Text(98,51)[] {$H^-$} \Text(5,105)[]{$g$} \Text(5,51)[] {$b$}
\Text(48,25)[]{$(o)$}
\end{picture}\\
\vspace{-0.3cm} \hspace{0.4cm}
\begin{picture}(120,120)(0,0)
\Gluon(5,98)(40,98){-2.5}{4} \ArrowLine(5,58)(22.5,58)
\Line(22.5,58)(40,58) \Line(40,58)(40,68) \ArrowLine(40,68)(40,98)
\Vertex(40,98){1} \DashLine(22.5,58)(40,78){3} \Vertex(40,58){1}
\Vertex(40,78){1} \Vertex(22.5,58){1} \ArrowLine(40,98)(75,98)
\DashLine(40,58)(75,58){3} \Text(80,101)[] {$t$} \Text(85,55)[]
{$H^-$} \Text(0,101)[]{$g$} \Text(0,55)[] {$b$}
\Text(43,35)[]{$(p)$}
\end{picture}
\vspace{-0.3cm}\hspace{1.1cm}
\begin{picture}(120,120)(0,0)
\Gluon(5,98)(40,98){-2.5}{4} \ArrowLine(5,58)(22.5,58)
\ArrowLine(40,78)(40,98) \Vertex(40,98){1} \Vertex(40,58){1}
\DashLine(40,58)(40,78){3} \Line(22.5,58)(40,78)
\ArrowLine(40,98)(75,98) \DashLine(22.5,58)(75,58){3}
\Vertex(40,78){1} \Vertex(22.5,58){1} \Text(80,101)[] {$t$}
\Text(85,55)[] {$H^-$} \Text(0,101)[]{$g$} \Text(0,55)[] {$b$}
\Text(34,50)[]{\small $\tilde{b}_i$} \Text(48,68)[]{\small
$\tilde{t}_j$} \Text(23,72)[]{\small $\tilde\chi_k^0$}
\Text(43,35)[]{$(q)$}
\end{picture}
\vspace{-0.3cm}\hspace{1.1cm}
\begin{picture}(120,120)(0,0)
\Gluon(5,98)(40,98){-2.5}{4} \ArrowLine(5,58)(35,58)
\ArrowLine(40,63)(40,98) \Vertex(40,98){1} \Line(37,62)(43,54)
\Line(37,54)(43,62) \ArrowLine(40,98)(75,98)
\DashLine(45,58)(75,58){3} \Text(80,101)[] {$t$} \Text(85,55)[]
{$H^-$} \Text(0,101)[]{$g$} \Text(0,55)[] {$b$}
\Text(43,35)[]{$(r)$}
\end{picture}\\
\vspace{-0.3cm} \hspace{0.4cm}
\begin{picture}(120,120)(0,0)
\Gluon(5,98)(32,98){-2.5}{3} \ArrowLine(5,58)(32,58)
\DashLine(32,58)(32,98){3} \Vertex(32,98){1} \Vertex(32,58){1}
\DashLine(32,98)(58,98){3} \DashLine(58,98)(85,98){3}
\Line(32,58)(58,58) \Line(58,98)(58,58) \DashLine(58,58)(85,58){3}
\Vertex(58,98){1} \Vertex(58,58){1} \Text(90,55)[] {$t$}
\Text(95,101)[] {$H^-$} \Text(0,101)[]{$g$} \Text(0,55)[] {$b$}
\Text(25,78)[]{\small $\tilde{b}_l$} \Text(65,78)[]{\small
$\tilde{t}_j$} \Text(45,110)[]{\small $\tilde{b}_i$}
\Text(45,50)[]{\small $\tilde\chi_k^0$} \Text(48,35)[]{$(s)$}
\end{picture}
\vspace{-0.3cm} \hspace{1.1cm}
\begin{picture}(120,120)(0,0)
\Gluon(5,98)(32,98){-2.5}{3} \ArrowLine(5,58)(32,58)
\ArrowLine(32,58)(32,98) \Vertex(32,98){1} \Vertex(32,58){1}
\ArrowLine(32,98)(58,98) \DashLine(58,98)(85,98){3}
\DashLine(32,58)(58,58){3} \ArrowLine(58,98)(58,58)
\ArrowLine(58,58)(85,58) \Vertex(58,98){1} \Vertex(58,58){1}
\Text(90,55)[] {$t$} \Text(95,101)[] {$H^-$} \Text(0,101)[]{$g$}
\Text(0,55)[] {$b$} \Text(25,78)[]{$b$} \Text(65,78)[]{$t$}
\Text(48,35)[]{$(t)$}
\end{picture}
\vspace{-0.3cm}\hspace{0.8cm}
\begin{picture}(120,120)(0,0)
\Gluon(5,98)(32,98){-2.5}{3} \DashLine(5,58)(32,58){3}
\ArrowLine(32,58)(32,98) \Vertex(32,98){1} \Vertex(32,58){1}
\ArrowLine(32,98)(58,98) \ArrowLine(58,98)(85,98)
\ArrowLine(58,58)(32,58) \ArrowLine(85,58)(58,58)
\DashLine(58,58)(58,98){3} \Vertex(58,98){1} \Vertex(58,58){1}
\Text(90,101)[] {$t$} \Text(90,55)[] {$b$} \Text(0,101)[]{$g$}
\Text(25,78)[]{$t$} \Text(45,50)[]{$b$} \Text(0,55)[] {$H^-$}
\Text(48,35)[]{$(u)$}
\end{picture}\\
\vspace{-0.3cm}\hspace{0.6cm}
\begin{picture}(120,120)(0,0)
\Gluon(5,98)(32,98){-2.5}{3} \DashLine(5,58)(32,58){3}
\DashLine(32,58)(32,98){3} \Vertex(32,98){1} \Vertex(32,58){1}
\DashLine(32,98)(58,98){3} \ArrowLine(58,98)(85,98)
\DashLine(58,58)(32,58){3} \ArrowLine(85,58)(58,58)
\Line(58,58)(58,98) \Vertex(58,98){1} \Vertex(58,58){1}
\Text(90,101)[] {$t$} \Text(90,55)[] {$b$} \Text(0,101)[]{$g$}
\Text(25,78)[]{\small $\tilde{t}_j$} \Text(45,50)[]{\small
$\tilde{b}_i$} \Text(0,55)[] {$H^-$} \Text(68,78)[]{\small
$\tilde{t}_l$} \Text(45,107)[]{\small
$\tilde\chi_k^0$}\Text(48,35)[]{$(v)$}
\end{picture}

{\small \begin{figure}[ht]
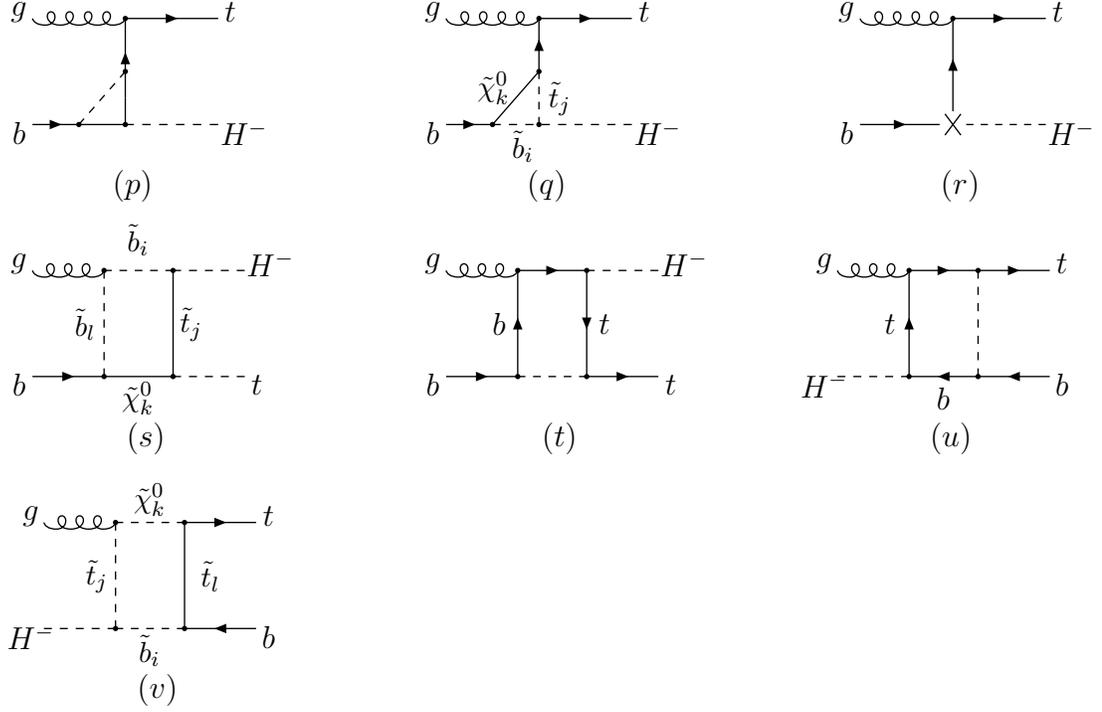
 \caption[]{ \small Feynman diagrams
contributing to $O(\alpha_{ew} m_{t(b)}^{2}/m_{W}^{2})$ Yukawa
corrections to $gb\rightarrow tH^{-}$: $(a)$ and $(b)$ are tree level
diagrams; $(c)-(v)$ are one-loop diagrams. The dashed lines
represent $H,h,A,H^{\pm},G^{0}$ and $G^{\pm}$ for diagrams $(c)$ and $(f)$;
$H,h,A$ and $G^{0}$ for diagrams $(m),(p),(t)$ and $(u)$;
$\tilde{t},\tilde{b},H,h,A,H^{\pm},G^{0}$ and $G^{\pm}$ for
$(i)$ and $(k)$,
where the solid lines represent charginos and
neutralinos if the dashed lines represent squarks.
For diagrams $(d)$ and $(g)$, the solid lines in the loop represent
$\tilde\chi^0$ and $\tilde\chi^+$
 and the dashed lines represent squarks.}
\end{figure}}

\newpage
\vspace{0.1cm} \hspace{0.1cm}
\begin{picture}(120,120)(0,0)
\ArrowLine(0,62)(25,62) \Line(25,62)(75,62)
\ArrowLine(75,62)(100,62) \DashCArc(50,62)(25,0,180){3}
\Vertex(25,62){1} \Vertex(75,62){1} \Text(5,53)[] {$t(b)$}
\Text(103,53)[] {$t(b)$} \Text(53,5)[]{$(a)$}
\end{picture}
\vspace{0.1cm}  \hspace{0.8cm}
\begin{picture}(120,120)(0,0)
\Photon(5,60)(33,60){2.5}{4} \Photon(77,60)(105,60){2.5}{4}
\ArrowArc(55,60)(22,0,180) \ArrowArc(55,60)(22,180,360)
\Vertex(33,60){1} \Vertex(77,60){1} \Text(5,50)[]{$W^{-}$}
\Text(108,50)[]{$W^{-}$} \Text(55,93)[]{$t$} \Text(55,28)[]{$b$}
\Text(55,5)[]{$(b)$}
\end{picture}
\vspace{0.1cm}  \hspace{0.8cm}
\begin{picture}(120,120)(0,0)
\Photon(5,60)(33,60){2.5}{4} \Photon(77,60)(105,60){2.5}{4}
\ArrowArc(55,60)(22,0,180) \ArrowArc(55,60)(22,180,360)
\Vertex(33,60){1} \Vertex(77,60){1} \Text(5,50)[]{$Z^{0}$}
\Text(108,50)[]{$Z^{0}$} \Text(55,93)[]{$t(b)$}
\Text(55,28)[]{$t(b)$} \Text(55,5)[]{$(c)$}
\end{picture}\\
\vspace{-0.5cm} \hspace{0.5cm}
\begin{picture}(120,120)(0,0)
\DashLine(5,80)(33,80){3} \Photon(77,80)(105,80){2.5}{4}
\ArrowArc(55,80)(22,0,180) \ArrowArc(55,80)(22,180,360)
\Vertex(33,80){1} \Vertex(77,80){1} \Text(5,70)[]{$H^{-}$}
\Text(108,70)[]{$W^{-}$} \Text(55,113)[]{$t$} \Text(55,48)[]{$b$}
\Text(55,25)[]{$(d)$}
\end{picture}
\vspace{-0.5cm} \hspace{1.0cm}
\begin{picture}(120,120)(0,0)
\DashLine(5,80)(33,80){3} \Photon(77,80)(105,80){2.5}{4}
\DashCArc(55,80)(22,0,360){3} \Vertex(33,80){1} \Vertex(77,80){1}
\Text(5,70)[]{$H^{-}$} \Text(108,70)[]{$W^{-}$}
\Text(55,113)[]{\small $\tilde{t}_j$} \Text(55,48)[]{\small
$\tilde{b}_i$} \Text(55,25)[]{$(e)$}
\end{picture}
\vspace{-0.5cm}  \hspace{1.0cm}
\begin{picture}(120,120)(0,0)
\DashLine(5,80)(33,80){3} \DashLine(77,80)(105,80){3}
\ArrowArc(55,80)(22,0,180) \ArrowArc(55,80)(22,180,360)
\Vertex(33,80){1} \Vertex(77,80){1} \Text(5,70)[]{$H^{-}$}
\Text(108,70)[]{$H^{-}$} \Text(55,113)[]{$t$} \Text(55,48)[]{$b$}
\Text(55,25)[]{$(f)$}
\end{picture}\\

\vskip 1cm
\vspace{-0.5cm}\hspace{0.1cm}
\begin{picture}(120,120)(0,0)
\DashLine(5,80)(33,80){3} \DashLine(77,80)(105,80){3}
\DashCArc(55,80)(22,0,360){3} \Vertex(33,80){1} \Vertex(77,80){1}
\Text(5,70)[]{$H^{-}$} \Text(108,70)[]{$H^{-}$}
\Text(55,113)[]{\small $\tilde{t}_j$} \Text(55,48)[]{\small
$\tilde{b}_i$} \Text(55,25)[]{$(g)$}
\end{picture}

{\small \begin{figure}[ht] \caption[]{ \small Self-energy
Feynman diagrams
contributing to renormalization constants: The dashed lines
represent $\tilde{t},\tilde{b},H,h,A,H^{\pm},G^{0}$ and $G^{\pm}$ for
diagram $(a)$,
where the solid lines represent charginos and
neutralinos if the dashed lines represent squarks. }
\end{figure}}

\begin{figure}
\epsfxsize=15 cm \centerline{ \epsffile{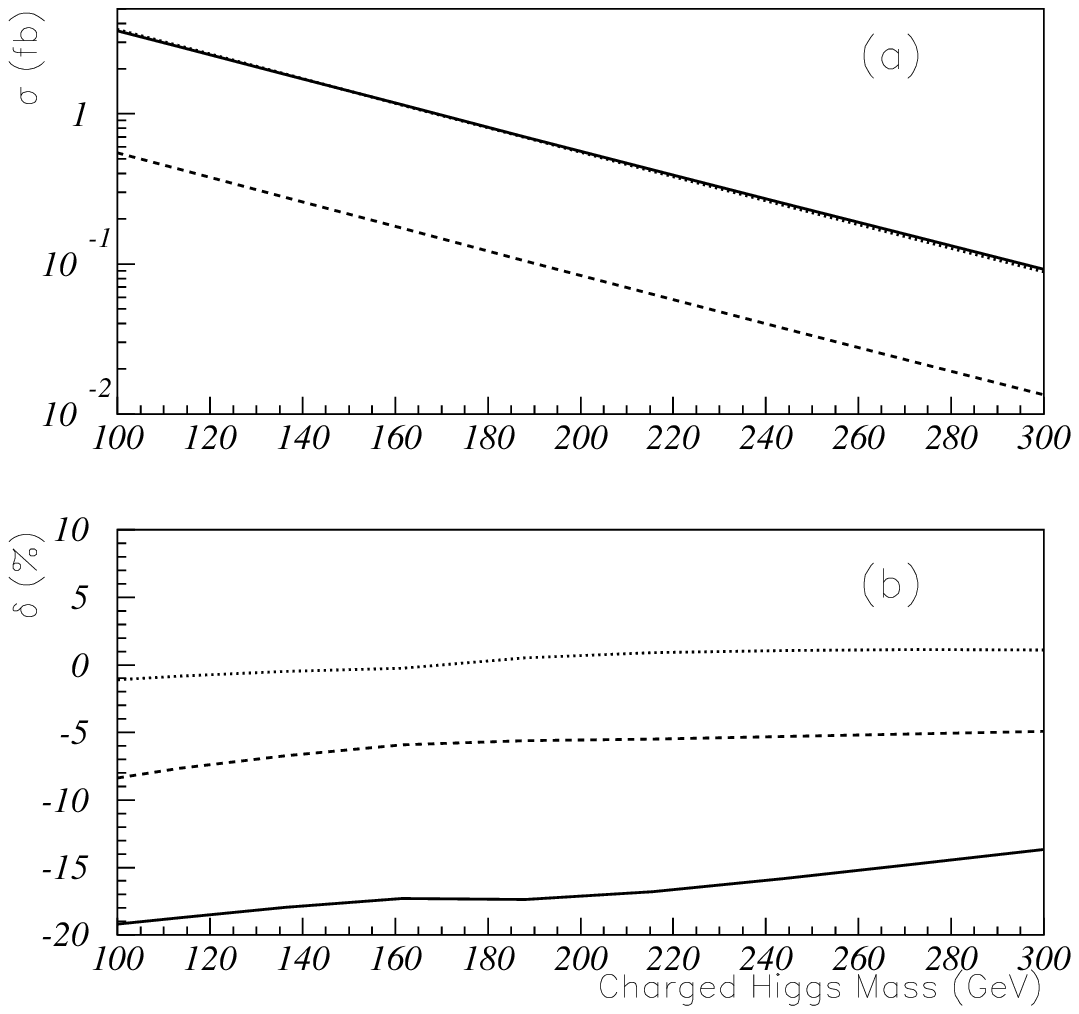 }}
\caption[]{
    The tree-level total cross sections (a) and relative one-loop
    corrections (b) versus $m_{H^{\pm}}$ at the Tevatron with
    $\sqrt{s}= 2$  TeV. The solid, dashed and dotted lines correspond to
    $\tan\beta=2,10$ and $30$, respectively.}
\end{figure}

\begin{figure}
\epsfxsize=15 cm \centerline{ \epsffile{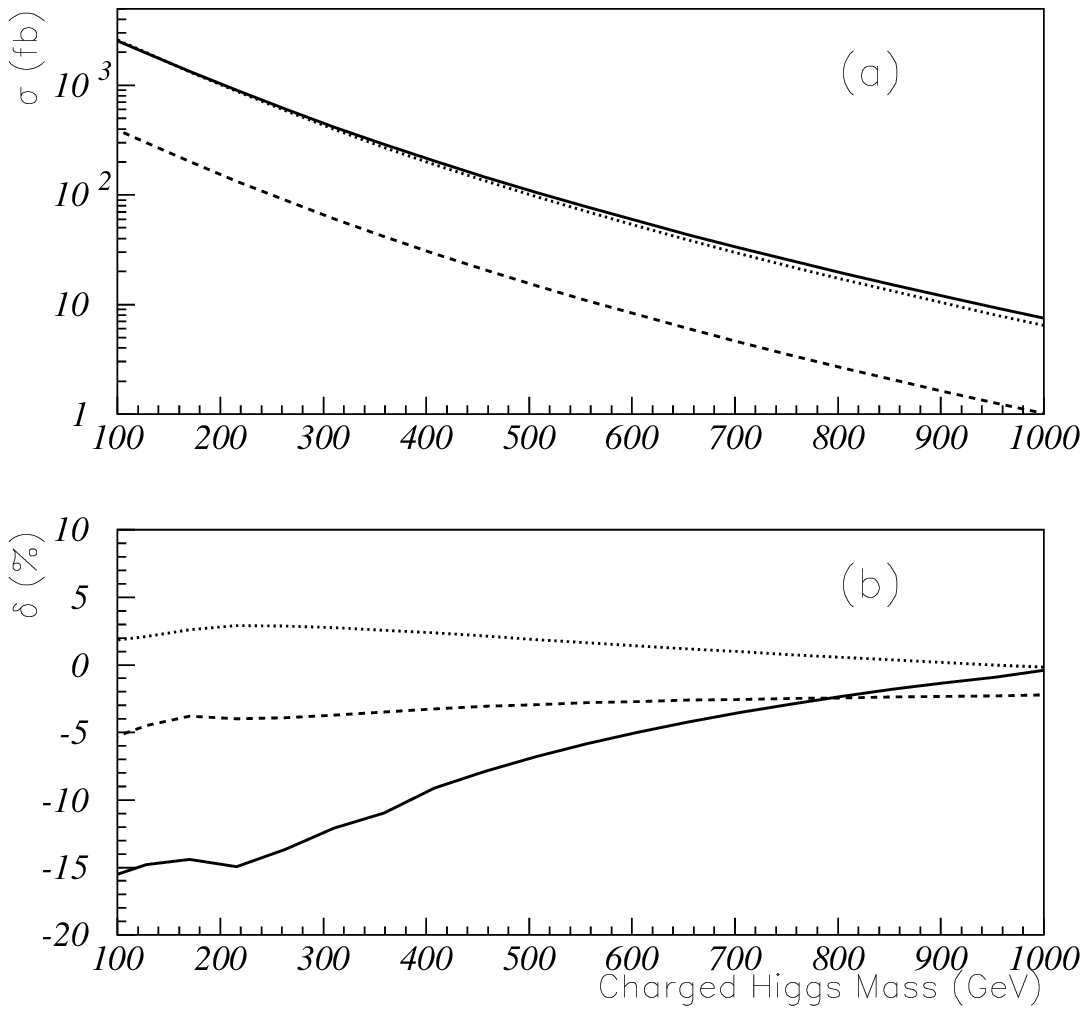 } }
\caption[]{
    The tree-level total cross sections (a) and relative one-loop
    corrections (b) versus $m_{H^{\pm}}$ at the LHC with
    $\sqrt{s}= 14$  TeV. The solid, dashed and dotted lines correspond to
    $\tan\beta=2,10$ and $30$, respectively.}
\end{figure}

\begin{figure}
\epsfxsize=15 cm \centerline{ \epsffile{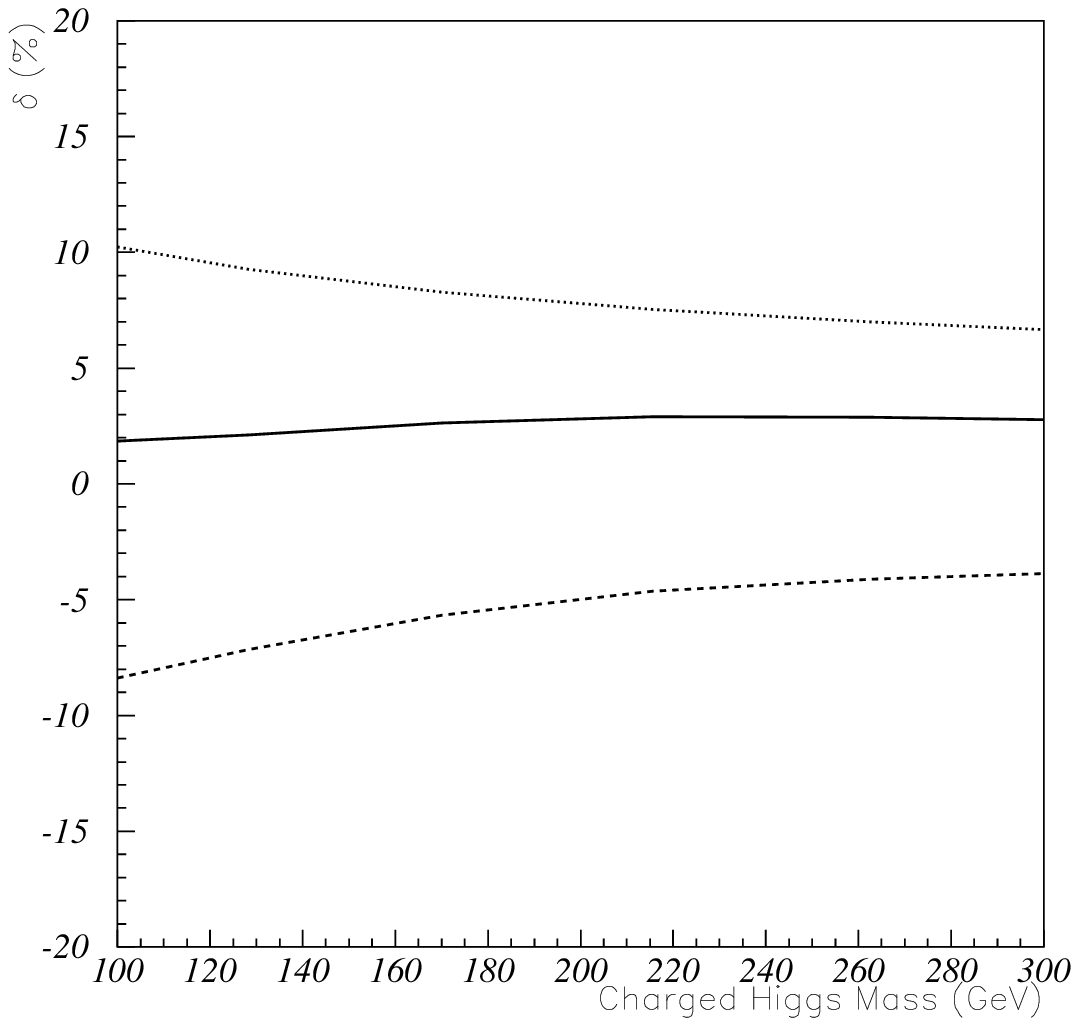 }}
\caption[]{ The radiative correction from top, bottom quarks
(dashed line)
 and genuine SUSY
particles (dotted line), as well as total contributions (solid line)
 when  $\tan\beta=30$
 at the LHC with
    $\sqrt{s}= 14$  TeV.}
\end{figure}

\begin{figure}
\epsfxsize=15 cm \centerline{ \epsffile{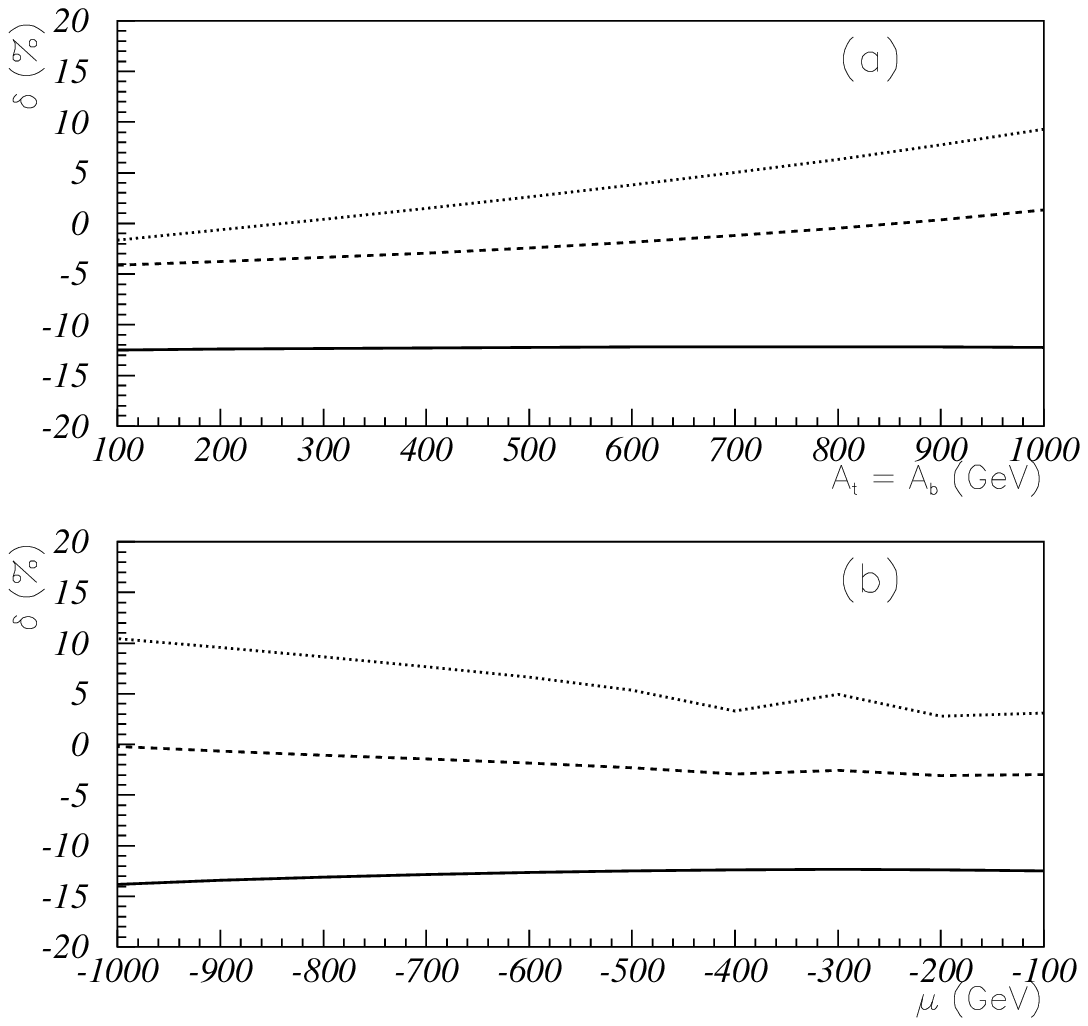 }}
\caption[]{
    Relative one-loop
    corrections versus $A_t$, $A_b$ (a) as well as
$\mu$ (b) at the LHC with
    $\sqrt{s}= 14$  TeV,
where $m_{H^{\pm}}=300 GeV$ and the solid,
dashed and dotted lines correspond to
    $\tan\beta=2,10$ and $30$, respectively.
For (a), $\mu=-100 GeV$, and for (b), $A_t=A_b=200 GeV$.}
\end{figure}



\setcounter{figure}{0} \setcounter{table}{0}
\setcounter{equation}{0}

\newpage
\section{ Supersymmetric Electroweak
Corrections to $W^{\pm}H^{\mp}$ Associated Production at the CERN
Large Hadron Collider}

\begin{footnotesize}
\begin{center}\begin{minipage}{5in}
\baselineskip=0.25in
\begin{center} ABSTRACT \end{center}

The $O(\alpha_{ew}m_{t(b)}^{2}/m_{W}^{2})$ and $O(\alpha_{ew}
m_{t(b)}^4/m_W^4)$ supersymmetric electroweak corrections to the
cross section for $W^{\pm}H^{\mp}$ associated production at the
LHC are calculated in the minimal supersymmetric standard model.
Those corrections arise from the quantum effects which are induced
by the Yukawa couplings from the Higgs sector and the
chargino-top(bottom)-sbottom(stop) couplings,
neutralino-top(bottom)-stop(sbottom) couplings and charged
Higgs-stop-sbottom couplings. The numerical results show that the
Yukawa corrections arising from the Higgs sector can decrease the
total cross sections significantly for low $\tan\beta(=1.5$ and
$2)$ when $m_{H^+}(<300)$GeV, which exceed $-12\%$. For high
$\tan\beta$ the Yukawa corrections become negligibly small. The
genuine supersymmetric electroweak corrections can increase or
decrease the total cross sections depending on the supersymmetric
parameters, which can exceed $-25\%$ for the favorable
supersymmetric parameter values. We also show that the genuine
supersymmetric electroweak corrections depend strongly on the
choice of $\tan\beta$, $A_t$, $M_{\tilde Q}$ and $\mu$. For large
values of $A_t$, or large values of $\mu$ and $\tan\beta$, one can
get much larger corrections. The corrections can become very
small, in contrast, for larger values of $M_{\tilde Q}$.

\end{minipage}\end{center}
\end{footnotesize}

\subsection{Introduction}

One of the most important objectives of the CERN Large Hadron
Collider (LHC) is the search for Higgs boson. In various
extensions of the Higgs sector of the standard model(SM), for
example, in the two-Higgs-doublet models(THDM)[1], particularly
the minimal supersymmetric standard model(MSSM)[2], there are
physical charged Higgs bosons, which do not belong to the spectrum
of the SM and therefore their discovery would be instant evidence
of new physics. In much of the parameter space preferred by the
MSSM, namely $m_{H^{\pm}}> m_W$ and $1< \tan\beta < m_t/m_b$[3,4],
the LHC will provide the greatest opportunity for the discovery of
charged Higgs boson. Previous studies have shown that for a
relatively light charged Higgs boson, $m_{H^{\pm}}< m_t - m_b$,
the dominate production processes at the LHC are $gg\rightarrow t
\bar t$ and $q\bar q\rightarrow t\bar t$ followed by the decay
sequence $t\rightarrow bH^+\rightarrow b\tau ^+\nu_{\tau}$[5], and
for a heavier charged Higgs boson the dominate production process
is $gb\rightarrow tH^-$[6,7,8]. Besides the processes mentioned
above, in Ref.[9] Dicus et al. also studied the production of a
charged Higgs boson in association with a $W$ boson via $b\bar b$
annihilation at the tree level and $gg$ fusion at one loop at
hadron colliders. Since the leptonic decays of $W$ boson would
serve as a spectacular trigger for the charged Higgs boson search,
these processes seem attractive. But the authors of Ref.[9] only
considered the case where the value of $\tan\beta$ to be in the
range $0.3-2.3$. Recently Barrientos Bendezu and Kniehl[10]
further studied these processes and presented theoretical
predictions for the $W^{\pm}H^{\mp}$ production cross section at
the LHC and Tevatron's Run II, where they generalize the analysis
of Ref.[9] for arbitrary values of $\tan\beta$ and to update it.
They found that the $W^{\pm}H^{\mp}$ production would have a
sizeable cross section and its signal should have a significant
rate at the LHC unless $m_{H^{\mp}}$ is very large.

As analyzed in Ref.[7,11], the search for heavy charged Higgs
bosons with $m_{H^+}>m_t + m_b$ at a hadron collider is seriously
complicated by QCD backgrounds. For example, the processes
suggested in Ref.[10] suffer from the irreducible background due
to top quark pair production, $q\bar q\rightarrow t\bar t$ and
$gg\rightarrow t\bar t$ with subsequent decay through the
intermediate state $b\bar bW^+W^-$, and heavy charged Higgs boson
produced in association with $W^{\pm}$ gauge bosons cannot be
resolved at the LHC, via semileptonic $W^+W^-$ decays, for charged
Higgs boson masses in the range between $2m_t$ and $600$GeV at
neither low nor high $\tan\beta$[11]. However, recent
analyses[12,13] have shown that the decay mode $H^+\rightarrow
\tau ^+\nu$, indeed dominant for light charged Higgs bosons below
the top threshold for any accessible $\tan\beta$[14], provides an
excellent signature for a heavy charged Higgs boson in searches at
the LHC. The discover region for $H^{\pm}$ is far greater than had
been thought for a large range of the $(m_{H^{\pm}}, \tan \beta)$
parameter space, extending beyond $m_{H^{\pm}}\sim 1$TeV and down
to at least $\tan\beta \sim 3$, and potentially to $\tan\beta \sim
1.5$, assuming the latest results for the SM parameters and parton
distribution functions as well as using kinematic selection
techniques and the tau polarization analysis[13]. Of course, it is
just a theoretical analysis and no experimental simulation has
been performed to make the statement very reliable so far.

Since the contributions to the $W^{\pm}H^{\mp}$ production cross
section due to $b\bar b$ annihilation at the tree level are
greater than ones due to $gg$ fusion which proceeds at one-loop,
it is important to calculate the one-loop radiative corrections to
the $W^{\pm}H^{\mp}$ production via $b\bar b$ annihilation for
more accurate theoretical predictions for the cross sections. In
this paper we present the calculations of the
$O(\alpha_{ew}m_{t(b)}^{2}/m_{W}^{2})$ and
$O(\alpha_{ew}m_{t(b)}^{4}/m_{W}^{4})$ supersymmetric(SUSY)
electroweak(EW) corrections to this $W^{\pm}H^{\mp}$ associated
production process at the LHC in the MSSM. These corrections arise
from the quantum effects which are induced by potentially large
Yukawa couplings from the Higgs sector and the
chargino-top(bottom)-sbottom(stop) couplings, neutralino-
top(bottom)-stop(sbottom) couplings and charged Higgs-stop-sbottom
couplings which will contribute at the
$O(\alpha_{ew}m_{t(b)}^{4}/m_{W}^{4})$ to the self-energy of the
charged Higgs boson. The relevant QCD corrections are expected to
be larger, but not yet available.

The arrangement of this paper is as follows. In Sec.II we give the
analytic results. In Sec.III we present some numerical examples
and discuss the implications of our results. Some notations used
in this paper and the lengthy expressions of the form factors are
summarized in Appendix A, B.

\subsection{Calculations and formulas}

The Feynman diagrams for the charged Higgs boson production via
$b(p_1)\bar b(p_2)\rightarrow W^{\pm}(k)H^{\mp}(p_3)$, which
include the SUSY EW corrections to the process, are shown in Fig.1
and Fig.2. We carried out the calculation in the t'Hooft-Feynman
gauge and used dimensional reduction, which preserves
supersymmetry, for regularization of the ultraviolet divergences
in the virtual loop corrections using the on-mass-shell
renormalization scheme[15], in which the fine-structure constant
$\alpha_{ew}$ and physical masses are chosen to be the
renormalized parameters, and finite parts of the counterterms are
fixed by the renormalization conditions. The coupling constant $g$
is related to the input parameters $e$, $m_W,$ and $m_Z$ via $g^2=
e^2/s_w^2$ and $s_w^2=1-m_W^2/m_Z^2$. As far as the parameters
$\beta$ and $\alpha$, for the MSSM we are considering, they have
to be renormalized, too. In the MSSM they are not independent.
Nevertheless, we follow the approach of Mendez and Pomarol[16] in
which they consider them as independent renormalized parameters
and fixed the corresponding renormalization constants by a
renormalization condition that the on-mass-shell $H^+\bar l \nu_l$
and $h \bar l l$ couplings keep the forms of Eq.(3) of Ref.[16] to
all order of perturbation theory.

We define the Mandelstam variables as
\begin{eqnarray}
\hat s =(p_1 +p_2)^2 =(k +p_3)^2, \nonumber \\ \hat t =(p_1 -k)^2
=(p_2 -p_3)^2, \nonumber \\ \hat u =(p_1 -p_3)^2 =(p_2 -k)^2.
\end{eqnarray}

The relevant renormalization constants are defined as
\begin{eqnarray}
& & m_{W0}^2 =m_W^2 +\delta m_W^2,\ \ \ m_{Z0}^2 =m_Z^2 +\delta
m_Z^2, \nonumber
\\ & & \tan\beta_0 =(1+\delta Z_\beta)\tan\beta, \nonumber
\\ & & \sin\alpha_0 =(1+\delta Z_\alpha)\sin\alpha, \nonumber
\\ & & W_0^{\pm \mu} =(1+\delta Z_W)^{1/2}W^{\pm\mu}
+iZ_{H^{\pm}W^{\pm}}^{1/2}\partial^\mu H^\mp, \nonumber
\\ & & H_0^\pm =(1+\delta Z_{H^\pm})^{1/2}H^\pm, \nonumber
\\ & & Z_0^\mu =(1+\delta Z_Z)^{1/2}Z^\mu +iZ_{ZA}^{1/2}\partial^\mu A,
\nonumber
\\ & & A_0 =(1+\delta Z_A)^{1/2}A, \nonumber
\\ & & H_0 =(1+\delta Z_H)^{1/2}H +Z_{Hh}^{1/2}h, \nonumber
\\ & & h_0 =(1+\delta Z_h)^{1/2}h +Z_{hH}^{1/2}H.
\end{eqnarray}

Taking into account the $O(\alpha_{ew}m_{t(b)}^{2}/m_{W}^{2})$ and
$O(\alpha_{ew}m_{t(b)}^{4}/m_{W}^{4})$ SUSY EW corrections, the
renormalized amplitude for $b\bar b\rightarrow W^{-}H^{+}$ can be
written as
\begin{eqnarray}
&& M_{ren} = M_0^{(s)} +M_0^{(t)} +[\delta \hat M^{V_1(s)} +\delta
\hat M^{S(s)} +\delta \hat M^{V_2(s)}](H_i) +[\delta \hat
M^{V_1(s)} \nonumber
\\ && \hspace{1.4cm} +\delta \hat M^{S(s)} +\delta \hat M^{V_2(s)}](A)
+\delta \hat M^{V_1(t)} +\delta \hat M^{S(t)} +\delta \hat
M^{V_2(t)} +\delta M^{box},
\end{eqnarray}
where $M_0^{(s)}$ and $M_0^{(t)}$ are the tree-level amplitudes
arising from Fig.1$(a)$ and Fig.1$(b)$, respectively, which are
given by
\begin{eqnarray}
&& M_0^{(s)} =
  -i\sum_{i}\frac{gh_b\alpha_{2i}\varphi_{11}}{\sqrt{2}(\hat s-m_{H_i}^2)}
  \sum_{j=1}^{4}M_j+\frac{igh_b\beta_{12}}{\sqrt{2}(\hat
  s-m_{A}^2)}(M_1-M_2+M_3-M_4)
\end{eqnarray}
and
\begin{eqnarray}
&& M_0^{(t)} =\frac{i g}{\sqrt{2}(\hat t-m_t^2)}
  (2h_b\beta_{12}M_2-h_bm_b\beta_{12}M_5+h_tm_t\beta_{11}M_6-h_b\beta_{12}M_{12}).
\end{eqnarray}
Here $h_b\equiv gm_b/\sqrt{2}m_W\cos\beta$ and $h_t\equiv
gm_t/\sqrt{2}m_W\sin\beta$ are the Yukawa couplings from the
bottom and top quarks, $p_1$ and $p_2$ denote the momentum of
incoming quarks $b$ and $\bar b$, respectively, while $k$ and
$p_3$ are used for the outgoing $W^-$ Boson and $H^+$ Boson,
respectively. The notations $\alpha_{ij}$, $\beta_{ij}$ and
$\varphi_{ij}$ used in the above expressions are defined in
Appendix A, and $H_i$ stands for Higgs Bosons $h$ with $i=1$ and
$H$ with $i=2$. $M_i$ are the standard matrix elements, which are
defined by
\begin{eqnarray}
&&M_1= \bar v(p_2) P_R u(p_1)p_1\cdot\varepsilon(k),\nonumber
\\&&M_2= \bar v(p_2) P_L u(p_1)p_1\cdot\varepsilon(k),\nonumber
\\&&M_3= \bar v(p_2) P_R u(p_1)p_2\cdot\varepsilon(k),\nonumber
\\&&M_4= \bar v(p_2) P_L u(p_1)p_2\cdot\varepsilon(k),\nonumber
\\&&M_5= \bar v(p_2) \not\varepsilon(k) P_R u(p_1),\nonumber
\\&&M_6= \bar v(p_2) \not\varepsilon(k) P_L u(p_1),\nonumber
\\&&M_7= \bar v(p_2) \not\varepsilon(k) P_R u(p_1)p_1\cdot\varepsilon(k),\nonumber
\\&&M_8= \bar v(p_2) \not\varepsilon(k) P_L u(p_1)p_1\cdot\varepsilon(k),\nonumber
\\&&M_9= \bar v(p_2) \not\varepsilon(k) P_R u(p_1)p_2\cdot\varepsilon(k),\nonumber
\\&&M_{10}= \bar v(p_2) \not\varepsilon(k) P_L u(p_1)p_2\cdot\varepsilon(k),\nonumber
\\&&M_{11}= \bar v(p_2) \not k\not\varepsilon(k) P_R u(p_1),\nonumber
\\&&M_{12}= \bar v(p_2) \not k\not\varepsilon(k) P_L u(p_1),
\end{eqnarray}
where $P_{L,R} \equiv (1\mp \gamma_5)/2$. The vertex and
self-energy corrections to the tree-level process are included in
$\delta \hat M^{V,S}$, which are given by
\begin{eqnarray} && \delta \hat M^{V_1(s)}(H_i)
=-\frac{igh_b}{\sqrt{2}} \{ \sum_{i=1,2}
\frac{\alpha_{2i}\varphi_{i1}}{\hat s -m_{H_i}^2}
  [\frac{\delta h_b}{h_b}+\frac{1}{2}\delta Z^b_L +\frac{1}{2}\delta Z_R^b
  +\frac{1}{2}\delta Z_{H_i}]
  \nonumber \\ && \hspace{2.1cm}
  +\frac{\sin(\beta -\alpha)\sin\alpha} {\hat s -m_H^2}(\tan\alpha
  \delta Z_\alpha +Z_{hH}^{1/2}) -\frac{\cos(\beta -\alpha)}
  {\hat s -m_h^2}(\sin\alpha\delta Z_\alpha
  \nonumber \\ && \hspace{2.1cm}
  -\cos\alpha Z_{Hh}^{1/2})\}\sum_{j=1}^{4}M_j +\delta M^{V_1(s)}(H),\nonumber
\\ && \delta \hat M^{V_1(s)}(A) =-\frac{igh_b\sin\beta}{\sqrt{2}(\hat
  s -m_A^2)}[\frac{\delta h_b}{h_b}+\cos^2\beta\delta Z_\beta
   +\frac{1}{2}\delta Z^b_L +\frac{1}{2}\delta Z_R^b +\frac{1}{2}
  \delta Z_{A}
    \nonumber \\ && \hspace{2.1cm}
  +\frac{im_W}{\tan\beta\cos\theta_W}Z^{1/2}_{hH}]
  (M_1-M_2+M_3-M_4) +\delta M^{V_1(s)}(A),\nonumber
\\ && \delta \hat M^{S(s)}(H_i) =\frac{igh_b}{\sqrt{2}}
  \sum_{i=1,2}\frac{\alpha_{2i}\varphi_{i1}}{(\hat s
  -m_{H_i}^2)^2}[\delta m_{H_i}^2 -(\hat s -m_{H_i}^2)\delta
  Z_{H_i} -(\hat s
  \nonumber \\ && \hspace{2.0cm}
  -m_H^2)Z_{Hh}^{1/2} -(\hat s -m_h^2)Z_{hH}^{1/2}]
  \sum_{j=1}^{4}M_j +\delta M^{S(s)}(H),\nonumber
\\ && \delta \hat M^{S(s)}(A) =\frac{igh_b\sin\beta}{\sqrt{2}(\hat s
  -m_A^2)}[\delta m_A^2 -(\hat s -m_A^2)\delta
  Z_A](M_1-M_2+M_3-M_4)
  \nonumber \\ && \hspace{2.0cm}
  +\delta M^{S(s)}(A),\nonumber
\\ && \delta \hat M^{V_2(s)}(H_i) =-\frac{igh_b}{\sqrt{2}}
  \{\sum_{i=1,2}\frac{\alpha_{2i}\varphi_{i1}}{\hat s
  -m_{H_i}^2}(\frac{\delta g}{g} +\frac{1}{2}\delta Z_{W^-} +\frac{1}{2}\delta Z_{H^+}
  +\frac{1}{2} Z_{H_i})
  \nonumber \\ && \hspace{2.1cm} -\frac{\cos\alpha\cos(\beta -\alpha)}
  {\hat s -m_H^2}(\sin\beta\cos\beta\delta Z_\beta
  -\tan\alpha\delta Z_\alpha -Z_{hH}^{1/2}
  \nonumber \\ && \hspace{2.1cm}
  +m_WZ_{HW}^{1/2})
  +\frac{\sin\alpha\sin(\beta -\alpha)}{\hat s -m_h^2}(\sin\beta
  \cos\beta\delta Z_\beta
  \nonumber \\ && \hspace{2.1cm}
  -\tan\alpha\delta Z_\alpha +Z_{Hh}^{1/2} +m_WZ_{HW}^{1/2})\}
  \sum_{j=1}^{4}M_j   +\delta M^{V_2(s)}(H),\nonumber
\\ && \delta \hat M^{V_2(s)}(A) =-\frac{igh_b\sin\beta}{\sqrt{2}(\hat
  s -m_A^2)}[\frac{\delta g}{g} +\frac{1}{2}\delta Z^A
  +\frac{1}{2}\delta Z_{H^+}
  \nonumber \\ && \hspace{2.0cm}
   +\frac{1}{2}\delta Z_{W^-}]  (M_1-M_2+M_3-M_4) +\delta M^{V_2(s)}(A),\nonumber
\\ && \delta \hat M^{V_1(t)} =\frac{i g}{\sqrt{2}(\hat t-m_t^2)}
  (2h_b\beta_{12}M_2-h_bm_b\beta_{12}M_5+h_tm_t\beta_{11}M_6
  \nonumber \\ && \hspace{1.5cm} -h_b\beta_{12}M_{12})
   (\frac{\delta g}{g} +\frac{1}{2}\delta Z^t_L +\frac{1}{2}\delta Z_L^b
   +\frac{1}{2}\delta Z_{W^-})
  +\delta M^{V_1(t)},\nonumber
\\ && \delta \hat M^{S(t)} =\frac{i g}{\sqrt{2}(\hat t -m_t^2)^2}[
  (2m_t^2 \frac{\delta m_t}{m_t} + m_t^2 \delta Z_L^t - \hat t
  \delta Z_L^t) (2h_b\beta_{12} M_2
  \nonumber \\ && \hspace{1.5cm} - h_bm_b\beta_{12}M_5 - h_t\beta_{12}M_{12}
  +\frac{1}{2}h_tm_t\beta_{11}M_6) +\frac{1}{2}(2\hat t\frac{\delta m_t}{m_t}
  \nonumber \\ && \hspace{1.5cm} + m_t^2\delta Z_R^t -\hat t \delta Z_R^t)
  h_tm_t\beta_{11}M_6]+ \delta M^{S(t)},\nonumber
\\ && \delta \hat M^{V_2(t)} =\frac{ig^2}{2m_W(\hat t -m_t^2)}
  [m_t^2\cot\beta(\frac{\delta h_t}{h_t} -\cos^2\beta\delta Z_\beta
  +\frac{1}{2}\delta Z^b_L +\frac{1}{2}\delta Z_R^t
    \nonumber \\ && \hspace{1.5cm}
  +\frac{1}{2}\delta Z_{H^+} +\frac{m_W}{\cot\beta}Z_{HW}^{1/2})
  M_6 +m_b\tan\beta(\frac{\delta h_b}{h_b}
  +\sin^2\beta\delta Z_\beta +\frac{1}{2}\delta Z^t_L
 \nonumber \\ && \hspace{1.5cm}
  +\frac{1}{2}\delta Z_R^b +\frac{1}{2}\delta Z_{H^+}
   -\frac{m_W}{\tan\beta}Z_{HW}^{1/2})
  (2M_2-M_{12} -m_bM_5)] +\delta M^{V_2(t)},
\end{eqnarray}
  with
 \begin{eqnarray}
 && \frac{\delta g}{g}= \frac{\delta e}{e}+\frac{1}{2}\frac{\delta
 m_Z^2}{m_Z^2}-\frac{1}{2}\frac{\delta m_Z^2-\delta
 m_W^2}{m_Z^2-m_W^2}\nonumber,
 \\&& \frac{\delta h_b}{h_b}=\frac{\delta g}{g}+\frac{\delta
 m_b}{m_b}-\frac{1}{2}\frac{\delta m_W^2}{m_W^2}+\cos^2\beta\delta
 Z_\beta\nonumber,
 \\&& \frac{\delta h_t}{h_t}=\frac{\delta g}{g}+\frac{\delta
 m_t}{m_t}-\frac{1}{2}\frac{\delta m_W^2}{m_W^2}-\sin^2\beta\delta
 Z_\beta\nonumber,
 \\&& \delta Z_\beta =-\frac{\delta g}{g}+ \frac{1}{2}\frac{\delta m_W^2}{m_W^2}
 -\frac{1}{2}\delta Z_{H^+} - \frac{m_W}{\tan \beta} Z_{HW}^{1/2}\nonumber,
 \\&& \delta Z_\alpha = -\frac{\delta g}{g}+ \frac{1}{2}\frac{\delta m_W^2}{m_W^2}
 -\frac{1}{2}\delta Z_h - \cot \alpha Z_{Hh}^{1/2} - \sin^2 \beta \delta
 Z_\beta.
 \end{eqnarray}
 The $\delta e/e$ appearing in Eq.(8) does not contain the
 $O(\alpha_{ew} m_{t(b)}^2/m_W^2)$ corrections and needs not be
 considered in our calculations. And $\delta M^{V_1(s)}(H_i)$, $\delta M^{V_1(s)}(A)$, $\delta
 M^{S(s)}(H_i)$, $\delta M^{S(s)}(A)$, $\delta M^{V_2(s)}(H_i)$,
$\delta M^{V_2(s)}(A)$, $\delta M^{V_1(t)}$, $\delta M^{S(t)}$,
$\delta M^{V_2(t)}$ and $\delta M^{box}$ represent the irreducible
corrections arising, respectively, from the $b\bar bH(h)$ vertex
diagrams shown in Fig.$1(c)-1(d)$, the $b\bar bA$ vertex diagrams
shown in Fig.$1(c)-1(d)$, the $H$ and $h$ boson self-energy
diagrams in Fig.$1(i)-1(k)$, the $A$ boson self-energy diagrams
shown in Fig.$1(i)-1(k)$, the $H(h)W^-H^+$ vertex diagrams shown
in Fig.$1(f)-1(h)$, the $AW^-H^+$ vertex diagrams shown in
Fig.$1(f)-1(h)$, the $btW^-$ vertex diagrams Fig.$1(l)-1(o)$, the
top quark self-energy diagrams Fig.$1(r)$, the $t\bar bH^+$ vertex
diagrams Fig.$1(p)-1(q)$, and the box diagrams Fig.$1(s)-1(x)$.
All above $\delta M^{V,S}$ and $\delta M^{box}$ can be written in
the form
\begin{eqnarray}
&& \delta M^{V,S,box} =i\sum_{i=1}^{12}f_i^{V,S,box}M_i,
\end{eqnarray}
where the $f_i^{V,S,box}$ are form factors, which are given
explicitly in Appendix B.

Calculating the self-energy diagrams in Fig.2, we can get the
explicit expressions of all the renormalization constants as
following:
\begin{eqnarray}
&& \frac{\delta m_t}{m_t} =\sum_i\frac{-h_t^2}{32\pi^2}
  [\alpha_{1i}^2(-B_0^{ttH_i} +B_1^{ttH_i})
  +\beta_{1i}^2(B_0^{ttA_i} +B_1^{ttA_i})]
 \nonumber \\ && \hspace{0.9cm}
 -\sum_i\frac{1}{32\pi^2m_t}[(h_t^2m_t
  \beta_{1i}^2 +h_b^2m_b\beta_{2i}^2)B_1^{tbH_i^+} +2h_bh_t
  \beta_{1i}\beta_{2i}B_0^{tbH_i^+}]
  \nonumber\\ && \hspace{0.9cm} +\sum_{i,j}\frac{h_t^2}{32\pi^2m_t}
  [m_t|N_{j4}|^2(B_0^{t\tilde t_i\tilde \chi_j^0} +B_1^{t\tilde
  t_i\tilde\chi_j^0}) +m_{\tilde\chi_j^0}\theta_{i1}^t
  \theta_{i2}^t(N_{j4}^2 +N_{j4}^{\ast 2})B_0^{t\tilde t_i\tilde
  \chi_j^0}]
  \nonumber\\ && \hspace{0.9cm} +\sum_{i,j}\frac{1}{32\pi^2m_t}\{m_t[h_t^2
  (\theta_{i1}^b)^2|V_{j1}|^2 +h_b^2(\theta_{i2}^b)^2|U_{j2}|^2]
  (B_0^{t\tilde b_i\tilde \chi_j^+} +B_1^{t\tilde b_i\tilde
  \chi_j^+})
  \nonumber\\ && \hspace{0.9cm} +h_bh_tm_{\tilde \chi_j^+} \theta_{i1}^b
  \theta_{i2}^b(U_{j2}V_{j2} +U_{j2}^\ast V_{j2}^\ast)B_0^{t
  \tilde b_i\tilde \chi_j^+}\}\nonumber,
\\ && \delta Z_L^t =\sum_i\frac{h_b^2\beta_{2i}^2}{16\pi^2}
  B_1^{tbH_i^+} -\sum_{i,j}\frac{h_t^2(\theta_{i2}^t)^2}{16\pi^2}
  |N_{j4}|^2(B_0^{t\tilde t_i \tilde\chi_j^0} +B_1^{t\tilde t_i
  \tilde\chi_j^0})
  \nonumber\\ && \hspace{0.9cm} -\sum_{i,j}\frac{h_b^2(\theta_{i2}^b)^2}
  {16\pi^2}|U_{j2}|^2(B_0^{t\tilde b_i \tilde\chi_j^+}
  +B_1^{t\tilde b_i \tilde\chi_j^+}) +\delta^t\nonumber,
\\ && \delta Z_R^t =\sum_i\frac{h_t^2\beta_{1i}^2}{16\pi^2}
  B_1^{tbH_i^+} -\sum_{i,j}\frac{h_t^2(\theta_{i1}^t)^2}{16\pi^2}
  |N_{j4}|^2(B_0^{t\tilde t_i \tilde\chi_j^0} +B_1^{t\tilde t_i
  \tilde\chi_j^0})
  \nonumber\\ && \hspace{0.9cm} -\sum_{i,j}\frac{h_t^2(\theta_{i1}^b)^2}
  {16\pi^2}|V_{j2}|^2(B_0^{t\tilde b_i \tilde\chi_j^+}
  +B_1^{t\tilde b_i \tilde\chi_j^+}) +\delta^t\nonumber,
\\ && \delta^t =\sum_i\frac{h_t^2}{32\pi^2}\{\alpha_{1i}^2
  [B_1^{ttH_i} -2m_t^2(B_0^{ttH_i} -B_1^{ttH_i})] +\beta_{1i}^2
  [B_1^{ttA_i} +2m_t^2(B_0^{ttA_i} +B_1^{ttA_i})]\}
  \nonumber\\ && \hspace{0.6cm} +\sum_i\frac{m_t}{16\pi^2}[m_t(h_t^2
  \beta_{1i}^2 +h_b^2 \beta_{2i}^2)B_0^{'tbH_i^+} +2h_bh_tm_b
  \beta_{1i}\beta_{2i} B_0^{'tbH_i^+}]
  \nonumber\\ && \hspace{0.6cm} -\sum_{i,j}\frac{h_t^2m_t}{16\pi^2}[m_t
  |N_{j4}|^2(B_0^{'t\tilde t_i\tilde \chi_j^0} +B_1^{'t\tilde t_i
  \tilde \chi_j^0}) +m_{\tilde\chi_j^0}\theta_{i1}^t\theta_{i2}^t
  (N_{j4}^2 +N_{j4}^{\ast 2})B_0^{'t\tilde t_i\tilde \chi_j^0}]
  \nonumber\\ && \hspace{0.6cm} -\sum_{i,j}\frac{m_t}{16\pi^2}\{m_t[h_t^2
  (\theta_{i1}^b)^2|V_{j1}|^2 +h_b^2(\theta_{i2}^b)^2|U_{j2}|^2]
  (B_0^{'t \tilde b_i\tilde \chi_j^+} +B_1^{'t \tilde b_i\tilde
  \chi_j^+})
  \nonumber\\ && \hspace{0.6cm} +h_bh_t m_{\tilde\chi_j^+}\theta_{i1}^b
  \theta_{i2}^b(U_{j2}V_{j2} +U_{j2}^\ast V_{j2}^\ast)B_0^{'t
  \tilde b_i\tilde \chi_j^+}\}\nonumber,
\\ && \delta m_W^2 =\frac{g^2}{16\pi^2}\{(m_b^2 -m_t^2)(1
  +\frac{m_b^2 -m_t^2 -2m_W^2}{2m_W^2}B_0^{0bt})-2m_t^2B_0^{0tt}
  \nonumber\\ && \hspace{1.0cm} -\frac{1}{2m_W^2}[(m_b^2 -m_t^2)^2 +(m_b^2
  +m_t^2)m_W^2]B_0^{Wbt}\}\nonumber,
\\&& \delta Z_W =\frac{g^2}{32\pi^2m_W^2}\{\frac{(m_b^2 -m_t^2)^2}
  {m_W^2}(B_0^{0bt} -B_0^{Wbt}) +[(m_b^2 -m_t^2)^2
  \nonumber\\ && \hspace{0.9cm} +(m_b^2 +m_t^2)m_W^2]B_0^{'Wbt}\}\nonumber,
\\&& \delta m_Z^2 =\frac{g^2s_W^2}{18c_W^2\pi^2}[\frac{m_b^2}{2}
  (3 -2s_W^2)(B_0^{Zbb} +B_0^{0bb}) -m_t^2(3 -4s_W^2)(B_0^{Ztt}
  -B_0^{0tt})]
  \nonumber\\ && \hspace{0.9cm} +\frac{g^2}{32c_W^2\pi^2}[m_b^2(B_0^{Zbb}
  -2B_0^{0bb}) -m_t^2(B_0^{Ztt} + 2 B_0^{0tt})]\nonumber,
\\&& \delta Z_{H^+} =\frac{3}{16\pi^2}[2(h_t^2\beta_{11}^2
  +h_b^2\beta_{21}^2)(B_1^{H^+bt} +m_b^2B_0^{'H^+bt}
  +m_{H^+}^2B_1^{'H^+bt})
  \nonumber\\ && \hspace{1.1cm} -4h_bh_tm_bm_t\beta_{11}\beta_{21}
  B_0^{'H^+bt} +\sum_{i,j,i',j'} (\theta_{ii'}^b)^2(\theta_{jj'}^t)^2(h_b
  \Theta_{i'j'1}^5 +h_t\Theta_{i'j'1}^6)^2B_0^{'H^+\tilde b_i
  \tilde t_j}]\nonumber,
\\&& \delta m_{H_k}^2 =\frac{3}{16\pi^2}\{-2h_t^2\alpha_{1k}^2
  [m_t^2(1 +B_0^{0tt} +2B_0^{H_ktt}) +m_{H_k}^2B_1^{H_ktt}]
  -2h_b^2\alpha_{2k}^2[m_b^2(1
  \nonumber\\ && \hspace{1.1cm} +B_0^{0bb} +2B_0^{H_kbb})
  +m_{H_k}^2B_1^{H_kbb}] +\sum_{i,j,i',j'} [(h_t\theta_{ii'}^t\theta_{jj'}^t
  \Theta_{i'j'k}^1)^2B_0^{H_k\tilde t_i\tilde t_j}
  \nonumber\\ && \hspace{1.1cm} +(h_b\theta_{ii'}^b \theta_{jj'}^b
  \Theta_{i'j'k}^2)^2 B_0^{H_k\tilde b_i\tilde b_j}]
  +\sum_i h_b^2 m_{\tilde b_i}^2 \alpha_{2k}^2(1+B_0^{0\tilde b_i\tilde b_i})
  \nonumber\\ && \hspace{1.1cm}+\sum_i h_t^2 m_{\tilde t_i}^2 \alpha_{1k}^2(1+B_0^{0\tilde t_i\tilde t_i})
  \}\nonumber,
\\&& \delta Z_{H_k} =\frac{3}{16\pi^2}\{2h_t^2\alpha_{1k}^2
  (B_1^{H_ktt} +2m_t^2B_0^{'H_ktt} +m_{H_k}^2B_1^{'H_ktt})
  +2h_b^2\alpha_{2k}^2(B_1^{H_kbb}
  \nonumber\\ && \hspace{1.1cm} +2m_b^2B_0^{'H_kbb}
  +m_{H_k}^2B_1^{'H_kbb}) +\sum_{i,j,i',j'} [(h_t\theta_{ii'}^t\theta_{jj'}^t
  \Theta_{i'j'k}^1)^2B_0^{'H_k\tilde t_i\tilde t_j}
  \nonumber\\ && \hspace{1.1cm} +(h_b\theta_{ii'}^b \theta_{jj'}^b
  \Theta_{i'j'k}^2)^2 B_0^{'H_k\tilde b_i\tilde b_j}]\}\nonumber,
\\&& \delta m_{A_k}^2 =\frac{3}{16\pi^2}\{2h_t^2\beta_{1k}^2
  [m_t^2(1 +B_0^{0tt}) +m_{A_k}^2B_1^{A_ktt}]
  +2h_b^2\beta_{2k}^2[m_b^2(1 +B_0^{0bb})
  \nonumber\\ && \hspace{1.1cm} +m_{A_k}^2B_1^{A_kbb}] -\sum_{i,j,i',j'} [(h_t
  \theta_{ii'}^t\theta_{jj'}^t\Theta_{i'j'k}^3)^2B_0^{A_k
  \tilde t_i\tilde t_j} +(h_b\theta_{ii'}^b \theta_{jj'}^b
  \Theta_{i'j'k}^4)^2 B_0^{A_k\tilde b_i\tilde b_j}]
  \nonumber\\ && \hspace{1.1cm} +\sum_i h_b^2 m_{\tilde b_i}^2 \beta_{2k}^2(1+B_0^{0\tilde b_i\tilde
  b_i})
  +\sum_i h_t^2 m_{\tilde t_i}^2 \beta_{1k}^2(1+B_0^{0\tilde t_i\tilde t_i})
  \}\nonumber,
\\&& \delta Z_{A_k} =\frac{3}{16\pi^2}\{2h_t^2\beta_{1k}^2
  (B_1^{A_ktt} +m_{A_k}^2B_1^{'A_ktt})
  +2h_b^2\beta_{2k}^2(B_1^{A_kbb}
  +m_{A_k}^2B_1^{'A_kbb})
  \nonumber\\ && \hspace{1.1cm} -\sum_{i,j,i',j'} [(h_t\theta_{ii'}^t\theta_{jj'}^t
  \Theta_{i'j'k}^3)^2B_0^{'A_k\tilde t_i\tilde t_j}
  +(h_b\theta_{ii'}^b \theta_{jj'}^b
  \Theta_{i'j'k}^4)^2 B_0^{'A_k\tilde b_i\tilde b_j}]\},\nonumber\\
&&Z_{H^+W}=\frac{-3g}{16\sqrt{2}\pi^2m_{H^+}^2m_W^2} [
(h_tm_t\beta_{11}+h_bm_b\beta_{12})
((m_b^2-m_t^2)(B_0^{0bt}-B_0^{H^+bt})-m_{H
^+}^2B_0^{H^+bt})\nonumber\\
&&\hspace{1.1cm}+\sum_{i,j,i',j'}\theta^b_{i1}\theta^b_{ii'}\theta^t_{j1}\theta^t_{jj'}
(h_b\Theta^5_{i'j'1}+h_t\Theta^6_{i'j'1}) (m_{\tilde
t_j}^2-m_{\tilde b_i}^2) (B_0^{0\tilde b_i \tilde t_j}
-B_0^{H^+\tilde b_i\tilde t_j})],\nonumber\\
&&Z_{AZ}=\frac{-i3gc_W}{16\sqrt{2}\pi^2m_W^2}
(h_tm_t\beta_{11}B_0^{Att}-h_bm_b\beta_{12}B_0^{Abb})\nonumber\\
&&\hspace{1.1cm}+\frac{igc_W}{32\pi^2m_{A}^2m_W^2}\sum_{i,j,i',j'}\{
h_b\theta^b_{ii'}\theta^b_{jj'} \Theta^4_{j'i'1}
[(3-2s_W^2)\theta^b_{i1}\theta^b_{j1}-2s_W^2\theta^b_{i2}\theta^b_{j2}]
(m_{\tilde b_i}^2-m_{\tilde b_j}^2) (B_0^{0\tilde b_i\tilde
b_j}\nonumber\\ &&\hspace{1.1cm}-B_0^{A\tilde b_i\tilde
b_j})-h_t\theta^t_{ii'}\theta^t_{jj'} \Theta^3_{j'i'1}
[(3-4s_W^2)\theta^t_{i1}\theta^t_{j1}-4s_W^2\theta^t_{i2}\theta^t_{j2}
](m_{\tilde t_i}^2-m_{\tilde t_j}^2) (B_0^{0\tilde t_i\tilde
t_j}-B_0^{A\tilde t_i\tilde t_j})\},\nonumber\\
&&Z_{hH}^{1/2}=\frac{3\alpha_{11}\alpha_{12}}{16\pi^2(m_{h}^2-m_{H}^2)}
[2m_b^2(1+B_0^{0bb}+2B_0^{Hbb})-2m_t^2(1+B_0^{0tt}+2B_0^{Htt})\nonumber\\
&&\hspace{1.1cm} -m_{H}^2(B_0^{Hbb}-B_0^{Htt})]\nonumber\\
&&\hspace{1.1cm}+\frac{3}{16\pi^2(m_{h}^2-m_{H}^2)}\sum_{i,j,i',j'}[
(h_b\theta^b_{ii'}\theta^b_{jj'})^2\Theta^2_{i'j'1}\Theta^2_{i'j'2}
B_0^{H\tilde b_i \tilde b_j}
+(h_t\theta^t_{ii'}\theta^t_{jj'})^2\Theta^1_{i'j'1}\Theta^1_{i'j'2}
B_0^{H\tilde t_i \tilde t_j} ]\nonumber\\
&&\hspace{1.1cm}-\frac{3\alpha_{11}\alpha_{12}}{16\pi^2(m_{h}^2-m_{H}^2)}\sum_i[
h_b^2m_{\tilde b_i}^2(1+B_0^{0\tilde b_i\tilde b_i})
+h_t^2m_{\tilde t_i}^2(1+B_0^{0\tilde t_i\tilde t_i})],\nonumber\\
&&Z_{Hh}^{1/2}=Z_{hH}^{1/2}|_{h\leftrightarrow H},
\end{eqnarray}
  with
\begin{eqnarray}
 && B_{n}^{ijk}=(-1)^n\{\frac{\Delta}{n+1}-\int_{0}^{1}dyy^{n}\ln{[\frac{m_i^{2}y(y-1)
+m_{j}^{2}(1-y)+m_{k}^{2}y}{\mu^{2}}]}\},
\\ && B_{n}^{'ijk}
=(-1)^n\int_{0}^{1}dy\frac{y^{n+1}(1-y)} {m_i^2 y(y-1)
+m_{j}^{2}(1-y) +m_{k}^{2}y}.
\end{eqnarray}
The notations $\theta^t_{ij}$ and $\theta^b_{ij}$ used in above
expressions are defined in Appendix A. $A_i$ stands for $A$ with
$i=1$ and $G^0$ with $i=2$. $H^+_i$ stands for $H^+$ with $i=1$
and $G^+$ with $i=2$. $\frac{\delta m_b}{m_b}$, $\delta Z_L^b$,
$\delta Z_R^b$ can be obtained, respectively, from $\frac{\delta
m_t}{m_t}$, $\delta Z_L^t$, $\delta Z_R^t$ by the transformation:
 $$h_b\leftrightarrow h_t,
m_b\leftrightarrow m_t, m_{\tilde b_i}\leftrightarrow m_{\tilde
t_i}, \alpha_{1i}\leftrightarrow \alpha_{2i},
\beta_{1i}\leftrightarrow \beta_{2i}, \theta_{ij}^b\leftrightarrow
\theta_{ij}^t, N_{i4}\rightarrow N_{i3}, U_{i2}\rightarrow
V_{i2}.$$

The corresponding amplitude squared is
\begin{eqnarray}
\overline{\sum}|M_{ren}|^{2} =\overline{\sum}|M_0^{(s)}
+M_0^{(t)}|^{2} +2Re\overline{\sum}[(\sum\delta M) (M_0^{(s)}
+M_0^{(t)})^{\dag}].
\end{eqnarray}

The cross section for the process $b\bar b\rightarrow
W^{\pm}H^{\mp}$ is
\begin{equation}
\hat{\sigma} =\int_{\hat{t}_{-}}^{\hat{t}_{+}}\frac{1}{16\pi
\hat{s}^2} \overline{\Sigma}|M_{ren}|^{2}d\hat{t}
\end{equation}
with
\begin{eqnarray}
\hat{t}_{\pm} &=& \frac{m_{W}^{2} +m_{H^{-}}^{2} -\hat{s}}{2} \pm
\frac{1}{2}\sqrt{(\hat{s} -(m_{W} +m_{H^{-}})^{2})(\hat{s} -(m_{W}
-m_{H^{-}})^{2})}.
\end{eqnarray}
The total hadronic cross section for $pp\rightarrow b\bar b
\rightarrow W^{\pm}H^{\mp}$ can be obtained by folding the
subprocess cross section $\hat{\sigma}$ with the parton
luminosity:
\begin{equation}
\sigma(s) =\int_{(m_{W} +m_{H^{-}})/\sqrt{s}}^{1}dz \frac{dL}{dz}
\hat{\sigma}(b\bar b\rightarrow W^{\pm}H^{\mp} \ \ {\rm at} \ \
\hat{s} =z^{2}s).
\end{equation}
Here $\sqrt{s}$ and $\sqrt{\hat{s}}$ are the CM energies of the
$pp$ and $b\bar b$ states , respectively, and $dL/dz$ is the
parton luminosity, defined as
\begin{equation}
\frac{dL}{dz} =2z\int_{z^{2}}^{1}
\frac{dx}{x}f_{b/P}(x,\mu)f_{\bar b/P} (z^{2}/x,\mu),
\end{equation}
where $f_{b/P}(x,\mu)$ and $f_{\bar b/P}(z^{2}/x,\mu)$ are the
bottom and anti-bottom quark parton distribution functions,
respectively.

\subsection{Numerical results and conclusion}

We now present some numerical results for the SUSY EW corrections
to $W^{\pm}H^{\mp}$ associated production at the LHC. The SM input
parameters in our calculations were taken to be
$\alpha_{ew}(m_Z)=1/128.8$, $m_W=80.375$GeV and
$m_Z=91.1867$GeV[17], and $m_t=175.6$GeV and $m_b=4.7$GeV, which
were taken according to Ref.[10] for comparison. We used the
CTEQ5M parton distributions throughout the calculations[18]. The
one-loop relations[19] between the Higgs boson masses
$M_{h,H,A,H^\mp}$ and the parameters $\alpha$ and $\beta$ in the
MSSM were used, and $m_{H^+}$ and $\beta$ were chosen as the two
independent input parameters. Other MSSM parameters were
determined as follows:

(i) For the parameters $M_1$, $M_2$ and $\mu$ in the chargino and
neutralino matrix, we take $M_2$ and $\mu$ as the input
parameters, and then used the relation
$M_1=(5/3)(g'^2/g^2)M_2\simeq 0.5M_2$[2] to determine $M_1$.

(ii) For the parameters $m^2_{\tilde{Q},\tilde{U},\tilde{D}}$ and
$A_{t,b}$ in squark mass matrices
\begin{eqnarray}
M^2_{\tilde{q}} =\left(\begin{array}{cc} M_{LL}^2 & m_q M_{LR}\\
m_q M_{RL} & M_{RR}^2 \end{array} \right)
\end{eqnarray}
with
\begin{eqnarray}
&&M_{LL}^2 =m_{\tilde{Q}}^2 +m_q^2 +m_Z^2\cos 2\beta(I_q^{3L}
-e_q\sin^2\theta_W), \nonumber
\\&& M_{RR}^2 =m_{\tilde{U},\tilde{D}}^2 +m_q^2 +m_Z^2
\cos 2\beta e_q\sin^2\theta_W, \nonumber
\\&& M_{LR} =M_{RL} =\left(\begin{array}{ll} A_t -\mu\cot\beta &
(\tilde{q} =\tilde{t}) \\ A_b -\mu\tan\beta & (\tilde{q}
=\tilde{b}) \end{array} \right),
\end{eqnarray}
to simplify the calculation we assumed $M_{\tilde Q}=M_{\tilde
U}=M_{\tilde D}$ and $A_t=A_b$, and we used $M_{\tilde Q}$ and
$A_t$ as the input parameters except the numerical calculations as
shown in Fig.6, where we took $m_{\tilde t_1}$, $m_{\tilde b_1}$
and $A_t=A_b$ as the input parameters.

Some typical numerical calculations of the Yukawa corrections and
the genuine SUSY EW corrections are given in Fig.3-4 and Fig.5-9,
respectively.

In Fig.3 we present the Yukawa corrections to the total cross
sections relative to the tree-level values as a function of
$m_{H^+}$ for $\tan\beta = 1.5, 2, 6$ and $30$. For $\tan\beta =
1.5$ and $2$ the corrections decrease the total cross sections
significantly, which exceed $-6\%$ for $m_{H^+}<500$GeV and
$-12\%$ for $m_{H^+}<300$GeV. For $\tan\beta(= 6)$ these
corrections also decrease the total cross sections, although
relatively smaller, which exceed $-2.5\%$ for $m_{H^+}<500$GeV and
exceed $-5\%$ for $m_{H^+}<250$GeV. But for high $\tan\beta(= 30)$
these corrections become positive, which increase the total cross
sections slightly. Note that there are the peaks at $m_{H^+} =
180.3$GeV, which arise from the singularity of the charged Higgs
boson wavefunction renormalization constant at the threshold point
$m_{H^+} = m_t+m_b$.

In Fig.4 we show the Yukawa corrections as a function of
$\tan\beta$ for $m_{H^+} = 100, 150, 200$ and $300$GeV. For
$\tan\beta<4$ the corrections reduce the total cross sections by
more than $10\%$ with $m_{H^+} = 100, 150$ and $200$GeV. With
$m_{H^+}=300$GeV the corrections are only significant for
$1<\tan\beta<5$. For high $\tan\beta (>10)$ the corrections become
negligibly small for all above $m_{H^+}$ values.

Fig.5 gives the genuine SUSY EW corrections as a function of
$m_{H^+}$ for $\tan\beta = 1.5, 2, 6$ and $30$, respectively,
assuming $M_2=300$GeV, $\mu=-100$GeV, $A_t=A_b=200$GeV, and
$M_{\tilde Q}=M_{\tilde U}=M_{\tilde D}=500$GeV. From this figure
one sees that the corrections are very small and negligible, which
is reasonable because the squark masses are now very large and
also the couplings of the charged Higgs boson-squarks are small
for the values of $A_{t,b}$, $M_{\tilde Q, \tilde U, \tilde D}$
and $\mu$ used in those numerical calculations. In contrast, in
Fig.6 when we take the lighter sqarks masses: $m_{\tilde
t_1}=100$GeV and $m_{\tilde b_1}=150$GeV, and put $A_t=A_b=1$TeV,
which are relatively larger, assuming $M_2=200$GeV, $\mu=100$GeV
and $M_{\tilde Q}=M_{\tilde U}$, the genuine SUSY EW corrections
are enhanced significantly, especially for low $\tan\beta(=1.5)$
and $m_{H^+}$ below $250$GeV, which can exceed $-30\%$. But when
$m_{H^+}>250$GeV the corrections increase the cross sections,
which can exceed $10\%$. For $\tan\beta=6$ and $30$ the
corrections are at most $10\%$ and become small with an increase
of $m_{H^+}$. The sharp dips at $m_{H^+}=250$GeV are again due to
the singularity of the charged Higgs boson wavefunction
renormalization constant at the threshold point $m_{H^+}
=m_{\tilde t_1}+m_{\tilde b_1}=250$GeV.

Fig.7, Fig.8 and Fig.9 give the genuine SUSY EW corrections versus
$A_t=A_b$, $M_{\tilde Q}=M_{\tilde U}=M_{\tilde D}$ and $\mu$,
respectively, for $\tan\beta= 1.5$ and $30$. In each figure we
fixed $m_{H^+} = 200$GeV and $M_2=300$GeV.

Fig.7 shows that the corrections are negative for $\tan\beta=1.5$
and positive for $\tan\beta=30$, assuming $M_{\tilde Q}=M_{\tilde
U}=M_{\tilde D}=400$GeV and $\mu=100$GeV. For both $\tan\beta=1.5$
and $30$ the magnitude of the corrections increases with
increasing $A_t=A_b$. When $A_t=A_b=1$TeV the corrections can
reach $-6\%$ and $7.5\%$ for $\tan\beta = 1.5$ and $30$,
respectively. Otherwise, when $A_t=A_b$ decrease to $100$GeV, the
corrections become negligibly small. This result is due to the
fact that large values of $A_t=A_b$ not only enhance the
couplings, but also give a large splitting between the masses of
$\tilde t_1 (\tilde b_1)$ and $\tilde t_2 (\tilde b_2)$, and in
consequence lighter $\tilde t_1$ and $\tilde b_1$.

Fig.8 also show that the corrections are negative for
$\tan\beta=1.5$ and positive for $\tan\beta=30$, assuming
$A_t=A_b=500$GeV and $\mu=100$GeV. When $M_{\tilde Q, \tilde U,
\tilde D} =250$GeV the corrections can reach $-3.6\%$ for
$\tan\beta = 1.5$ and $7.3\%$ for $\tan\beta = 30$. But the
magnitude of the corrections drops below one percent when
$M_{\tilde Q, \tilde U, \tilde D}$ increase to $750$GeV. This is
because for larger values of $M_{\tilde Q, \tilde U, \tilde D}$
the squarks have larger masses and their virtual effects decrease
due to the decoupling effects.

In Fig.9 we present the genuine SUSY EW corrections as a function
of $\mu$, assuming $A_t=A_b=500$GeV and $M_{\tilde Q}=M_{\tilde
U}=M_{\tilde D}=400$GeV. For $\tan\beta= 30$ the magnitude of the
corrections increase with an increase of $|\mu|$, which varies
from $0\%$ to $5\%$ when $|\mu|$ ranges between $0 \sim 500$GeV.
For $\tan\beta = 1.5$ the corrections are relatively small and
increase slowly from about $0\%$ to $3.5\%$ when $\mu$ ranges
between $-500$GeV$\sim 500$GeV. This result indicates that large
values of $\mu$ and $\tan\beta$ can enhance the corrections
significantly since the couplings become stronger.

In conclusion, we have calculated the $O(\alpha_{ew} m_{t(b)}^{2}
/ m_{W}^{2})$ and $O(\alpha_{ew} m_{t(b)}^4 / m_W^4)$ SUSY EW
corrections to the cross sections for $W^{\pm}H^{\mp}$ associated
production at the LHC in the MSSM. The numerical results show that
the Yukawa corrections arising from the Higgs sector can decrease
the total cross sections significantly for low $\tan\beta(=1.5$
and $2)$ when $m_{H^+}(<300)$GeV, which exceed $-12\%$. For high
$\tan\beta$ the Yukawa corrections become negligibly small. The
genuine SUSY EW corrections can increase or decrease the total
cross sections depending on the SUSY parameters, which can exceed
$-25\%$ for the favorable SUSY parameter values. We also show that
the genuine SUSY EW corrections depend strongly on the choice of
$\tan\beta$, $A_t$, $M_{\tilde Q}$ and $\mu$. For large values of
$A_t$, or large values of $\mu$ and $\tan\beta$, one can get much
larger corrections. The correcan become very small, in contrast,
for larger values of $M_{\tilde Q}$.

{\small
\subsection{Appendix A}

We present some notations used in this paper here. We introduce an
angle $\varphi=\beta-\alpha$, and for each angle $\alpha$,
$\beta$, $\varphi$, $\theta^t$ or $\theta^b$, we define
\begin{eqnarray*}
&&\alpha_{ij}=\left(\begin{array}{cc} \cos\alpha & \sin\alpha\\
-\sin\alpha & \cos\alpha\end{array} \right),
\beta_{ij}=\left(\begin{array}{cc} \cos\beta & \sin\beta\\
-\sin\beta & \cos\beta\end{array} \right),
\varphi_{ij}=\left(\begin{array}{cc} \cos\varphi & \sin\varphi\\
-\sin\varphi & \cos\varphi\end{array} \right),\nonumber\\
&&\theta_{ij}^t=\left(\begin{array}{cc} \cos\theta^t &
\sin\theta^t\\ -\sin\theta^t & \cos\theta^t\end{array} \right),
\theta_{ij}^b=\left(\begin{array}{cc} \cos\theta^b &
\sin\theta^b\\ -\sin\theta^b & \cos\theta^b\end{array} \right)
\end{eqnarray*}
We define six matrix $\Theta^i_{jkl}, i=1-6$ for the couplings
between squarks and Higgses:
\begin{eqnarray*}
&\Theta^1_{ij1}=&\frac{1}{\sqrt{2}} \left(\begin{array}{cc}
2m_t\cos\alpha &A_t\cos\alpha+\mu\sin\alpha\\
A_t\cos\alpha+\mu\sin\alpha& 2m_t\cos\alpha\end{array}
\right)\nonumber\\ &\Theta^1_{ij2}=&\frac{1}{\sqrt{2}}
\left(\begin{array}{cc} 2m_t\sin\alpha
&A_t\sin\alpha-\mu\cos\alpha\\ A_t\sin\alpha-\mu\cos\alpha&
2m_t\sin\alpha\end{array} \right)\\
&\Theta^2_{ij1}=&\frac{-1}{\sqrt{2}} \left(\begin{array}{cc}
2m_b\sin\alpha &A_b\sin\alpha+\mu\cos\alpha\\
A_b\sin\alpha+\mu\cos\alpha&
2m_b\sin\alpha\end{array}\right)\nonumber\\
&\Theta^2_{ij2}=&\frac{1}{\sqrt{2}} \left(\begin{array}{cc}
2m_b\cos\alpha &A_b\cos\alpha-\mu\sin\alpha\\
A_b\cos\alpha-\mu\sin\alpha& 2m_b\cos\alpha\end{array} \right)\\
&\Theta^3_{ij1}=&\frac{1}{\sqrt{2}} \left(\begin{array}{cc} 0
&A_t\cos\beta+\mu\sin\beta\\ -A_t\cos\beta-\mu\sin\beta&
0\end{array} \right)\nonumber\\
&\Theta^3_{ij2}=&\frac{1}{\sqrt{2}} \left(\begin{array}{cc} 0
&A_t\sin\beta-\mu\cos\beta\\ -A_t\sin\beta+\mu\cos\beta&
0\end{array} \right)\\ &\Theta^4_{ij1}=&\frac{1}{\sqrt{2}}
\left(\begin{array}{cc} 0 &A_b\sin\beta+\mu\cos\beta\\
-A_b\sin\beta-\mu\cos\beta& 0\end{array} \right)\nonumber\\
&\Theta^4_{ij2}=&\frac{1}{\sqrt{2}} \left(\begin{array}{cc} 0
&-A_b\cos\beta+\mu\sin\beta\\ A_b\cos\beta-\mu\sin\beta&
0\end{array} \right)\\ &\Theta^5_{ij1}=&\left(\begin{array}{cc}
m_b\sin\beta &0\\ A_b\sin\beta+\mu\cos\beta&m_t\sin\beta
\end{array} \right)\nonumber\\
&\Theta^5_{ij2}=&\left(\begin{array}{cc} -m_b\cos\beta &0\\
-A_b\cos\beta+\mu\sin\beta&0
\end{array} \right)\\
&\Theta^6_{ij1}=&\left(\begin{array}{cc} m_t\cos\beta
&A_t\cos\beta+\mu\sin\beta\\ 0&m_b\cos\beta
\end{array} \right)\nonumber\\
&\Theta^6_{ij2}=&\left(\begin{array}{cc} m_t\sin\beta
&A_t\sin\beta-\mu\cos\beta\\ 0&0
\end{array} \right)
\end{eqnarray*}

\subsection{Appendix B}

 The form factors defined in
Eq.(9) are the following:
\begin{eqnarray*}
&& f_1^{V_1(s)}(H) =
  \sum_{i,j}\frac{gh_b^3\alpha_{2i}^2\alpha_{2j}\varphi_{j1}}
  {32\sqrt{2}\pi^2(\hat s -m_{H_j}^2)}\{B_0^{bbH_i}
  +[4m_b^2C_0
  \\ && \hspace{1.8cm} +(4m_b^2 +\hat s)C_1](\hat
  s,m_b^2,m_b^2,m_b^2,m_b^2,m_{H_i}^2)\}
  \\ && \hspace{1.8cm} +\sum_{i,j}\frac{-gh_b^3\beta_{2i}^2\alpha_{2j}
  \varphi_{j1}} {32\sqrt{2}\pi^2(\hat s -m_{H_j}^2)}
  [B_0^{bbA_i} -(4m_b^2 -\hat s)C_1(\hat
  s,m_b^2,m_b^2,m_b^2,m_b^2,m_{A_i}^2)]
  \\ && \hspace{1.8cm} +\sum_{i,j}\frac{gh_t\alpha_{1j}\varphi_{j1}}
  {16\sqrt{2}\pi^2(\hat s -m_{H_j}^2)}
  \{-h_bh_t\beta_{1i}\beta_{2i}B_0^{btH_i^+}
  +[(h_t^2m_bm_t\beta_{1i}^2
   \\ && \hspace{1.8cm} +2h_bh_tm_t^2
  \beta_{1i}\beta_{2i} +h_b^2m_bm_t\beta_{2i}^2)C_0
  +(2h_t^2m_bm_t\beta_{1i}^2 +h_bh_t\hat s\beta_{1i}\beta_{2i}
   \\ && \hspace{1.8cm}  +2h_b^2m_bm_t\beta_{2i}^2)C_1] (\hat
  s,m_b^2,m_b^2,m_t^2,m_t^2,m_{H_i^+}^2)\}
  \\ && \hspace{1.8cm} +\sum_{i,j,k,l}\sum_{i',j'}\frac{gh_t\varphi_{l1}\theta_{ii'}^t
  \theta_{jj'}^t\Theta_{j'i'l}^1}{16\pi^2(\hat s -m_{H_l}^2)}
  [h_b^2m_b\theta_{i1}^t\theta_{j1}^t|U_{k2}|^2(C_0 +C_1 +C_2)
  \\ && \hspace{1.8cm} +m_{\tilde\chi_k^+}h_bh_t\theta_{i2}^t
  \theta_{j1}^tU_{k2}V_{k2} C_0 -h_t^2m_b\theta_{i2}^t
  \theta_{j2}^t|V_{j2}|^2C_1]
  (\hat s,m_b^2,m_b^2,m_{\hat t_i}^2,m_{\hat t_j}^2,m_{\tilde\chi_k^+})
  \\ && \hspace{1.8cm} +\sum_{i,j,k,l}\sum_{j',k'}\frac{gh_b^3\varphi_{i1}
  \theta_{jj'}^b\theta_{kk'}^bN_{l3}\Theta_{j'k'i}^2}{16\pi^2
  (\hat s -m_{H_i}^2)}[m_b\theta_{j1}^b\theta_{k1}^bN_{l3}^\ast
  (C_0 +C_1 +C_2)
  \\ && \hspace{1.8cm} -m_b\theta_{j2}^b\theta_{k2}^bN_{l3}^\ast
  C_1 +m_{\tilde\chi_l^0}\theta_{j1}^b \theta_{k2}^bN_{l3}C_0]
  (\hat s,m_b^2,m_b^2,m_{\hat b_j}^2,m_{\hat b_k}^2,m_{\tilde\chi_l^0}),
\\ && f_2^{V_1(s)}(H) =f_1^{V_1(s)}(H)(h_b\theta_{n1}^t
  \leftrightarrow h_t\theta_{n2}^t,\theta_{n1}^b \leftrightarrow
  \theta_{n2}^b,U_{n2} \leftrightarrow V_{n2}^\ast,N_{n3}
  \leftrightarrow N_{n3}^\ast), \\
&& f_3^{V_1(s)}(H) =f_1^{V_1(s)}(H),
\\ && f_4^{V_1(s)}(H) =f_2^{V_1(s)}(H);
\\&& f_i^{V_1(s)}(A)=f_i^{V_1(s)}(A)_a+f_i^{V_1(s)}(A)_b,
\end{eqnarray*}
where


All other form factors $f_i$ not listed above vanish.

Here $A_0$, $C_i$, $D_i$ and $D_{ij}$ are the one-, three- and
four-point Feynman integrals[20]. The definitions of $U_{ij}$,
$V_{ij}$, $N_{ij}$, $O^L_{ij}$ and $O^R_{ij}$ can be found in
Ref.[2]. }

{\LARGE References} \vspace{0.2cm}
\begin{itemize}
\item[{\rm[1]}] For a review, see J.Gunion, H. Haber, G. Kane, and
            S.Dawson, The Higgs Hunter's Guide(Addison-Wesley,
            New York,1990).
\item[{\rm[2]}] H.E. Haber and G.L. Kane, Phys. Rep. 117, 75(1985);
            J.F. Gunion and H.E. Haber, Nucl. Phys. {\bf B272}, 1(1986).
\item[{\rm[3]}] CMS Technical Proposal. CERN/LHC94-43 LHCC/P1, December 1994.
\item[{\rm[4]}] CDF Collaboration, Phys. Rev. Lett. 79, 35(1997);
                D0 Collaboration, Phys. Rev.Lett. 82, 4975(1999).
\item[{\rm[5]}] Z.Kunszt and F. Zwirner, Nucl. Phys. {\bf B385}, 3(1992),
            and references cited therein.
\item[{\rm[6]}] J.F. Gunion, H.E. Haber, F.E. Paige, W.-K. Tung, and
             S. Willenbrock, Nucl. Phys. {\bf B294},621(1987); R.M.
             Barnett, H.E. Haber, and D.E. Soper, ibid. B306,
             697(1988); F.I. Olness and W.-K. Tung, ibid. {\bf B308},
             813(1988).
\item[{\rm[7]}] V. Barger, R.J.N. Phillips, and D.P. Roy, Phys. Lett.
            {\bf B324}, 236(1994).
\item[{\rm[8]}] C.S. Huang and S.H. Zhu, Phys. Rev. {\bf D60},
                075012(1999).
                L.G. Jin, C.S.Li, R.J. Oakes, and S.H. Zhu, to appear in
                Eur.Phys.J.C.
                L.G. Jin, C.S.Li, R.J. Oakes, and S.H. Zhu, hep-ph/0003159.
\item[{\rm[9]}] D.A. Dicus, J.L.Hewett, C. Kao, and T.G. Rizzo, Phys. Rev.
                {\bf D40},789(1989).
\item[{\rm[10]}] A.A.Barrientos Bendezu and B.A. Kniehl, Phys. Rev.
                 {\bf D59}, 015009(1998).
\item[{\rm[11]}] S. Moretti and K. Odagiri, Phys. Rev. {\bf D59}, 055008(1999).
\item[{\rm[12]}] K. Odagiri, hep-ph/9901432; Phys. Lett. {\bf B452}, 327(1999).
\item[{\rm[13]}] D.P. Roy, Phys. Lett. {\bf B459}, 607(1999).
\item[{\rm[14]}] S. Raychaudhuri and D.P.Roy, Phys. Rev. {\bf D53},
                 4902(1996).
\item[{\rm[15]}] S. Sirlin, Phys. Rev. {\bf D22}, 971 (1980);
            W. J. Marciano and A. Sirlin,{\sl ibid.} {\bf 22}, 2695(1980);
            {\bf 31}, 213(E) (1985);
            A. Sirlin and W.J. Marciano, Nucl. Phys. {\bf B189}, 442(1981);
            K.I. Aoki et.al., Prog. Theor. Phys. Suppl. {\bf 73}, 1(1982).
\item[{\rm[16]}] A. Mendez and A. Pomarol, Phys.Lett.{\bf B279}, 98(1992).
\item[{\rm[17]}] Particle Data Group, C.Caso {\it et al}, Eur.Phys.J.C 3,
1(1998).
\item[{\rm[18]}] H.L. Lai, et al.(CTEQ collaboration), hep-ph/9903282.
\item[{\rm[19]}] J.Gunion, A.Turski, Phys. Rev. {\bf D39}, 2701(1989);
            {\bf D40}, 2333(1990); J.R.Espinosa, M.Quiros, Phys. Lett. {\bf
            B266}, 389(1991); M.Carena, M.Quiros, C.E.M.Wagner, Nucl. Phys.
            {\bf B461}, 407(1996).

\item[{\rm[20]}] G.Passarino and M.Veltman, Nucl. Phys. {\bf B160},
                 151(1979); A.Axelrod, {\sl ibid.} {\bf B209}, 349 (1982);
                 M.Clements {\sl et al.}, Phys. Rev. {\bf D27}, 570
                 (1983); A.Denner, Fortschr. Phys. {\bf 41}, 4 (1993); R.
                 Mertig {\sl et al.}, Comput. Phys. Commun. {\bf 64}, 345
                 (1991).
\end{itemize}

\eject

\begin{picture}(120,120)(0,0)
\DashLine(35,60)(85,60){3} \ArrowLine(15,80)(35,60)
\ArrowLine(35,60)(15,40) \Photon(85,60)(105,80){1.5}{5}
\DashLine(85,60)(105,40){3} \Vertex(35,60){1} \Vertex(85,60){1}
\Text(10,85)[]{$b$} \Text(10,35)[]{$\bar b$} \Text(60,70)[]{\small
$H,h,A$} \Text(115,85)[]{\small $W^-$} \Text(115,35)[]{\small
$H^+$} \Text(60,15)[]{$(a)$}
\end{picture}
\hspace{1.0cm}
\begin{picture}(120,120)(0,0)
\ArrowLine(60,80)(60,40) \ArrowLine(20,80)(60,80)
\ArrowLine(60,40)(20,40) \Photon(60,80)(100,80){1.5}{5}
\DashLine(60,40)(100,40){3} \Vertex(60,80){1} \Vertex(60,40){1}
\Text(15,85)[]{$b$} \Text(15,35)[]{$\bar b$} \Text(68,60)[]{$t$}
\Text(110,85)[]{\small $W^-$} \Text(110,35)[]{\small $H^+$}
\Text(60,15)[]{$(b)$}
\end{picture}
\hspace{1.0cm}
\begin{picture}(120,120)(0,0)
\DashLine(50,60)(85,60){3} \ArrowLine(15,80)(50,60)
\ArrowLine(50,60)(15,40) \DashLine(28,73)(28,47){3}
\Photon(85,60)(105,80){1.5}{5} \DashLine(85,60)(105,40){3}
\Vertex(50,60){1} \Vertex(85,60){1} \Vertex(28,73){1}
\Vertex(28,47){1} \Text(10,85)[]{$b$} \Text(10,35)[]{$\bar b$}
\Text(67.5,70)[]{\small $1$} \Text(115,85)[]{\small $W^-$}
\Text(115,35)[]{\small $H^+$} \Text(15,60)[]{\small $2;3$}
\Text(45,75)[]{$b;t$} \Text(45,45)[]{$b;t$} \Text(60,15)[]{$(c)$}
\end{picture}

\begin{picture}(120,120)(0,0)
\DashLine(50,60)(85,60){3} \ArrowLine(15,80)(28,73)
\DashLine(28,73)(50,60){3} \ArrowLine(28,47)(15,40)
\DashLine(50,60)(28,47){3} \Line(28,73)(28,47)
\Photon(85,60)(105,80){1.5}{5} \DashLine(85,60)(105,40){3}
\Vertex(50,60){1} \Vertex(85,60){1} \Vertex(28,73){1}
\Vertex(28,47){1} \Text(10,85)[]{$b$} \Text(10,35)[]{$\bar b$}
\Text(115,85)[]{\small $W^-$} \Text(115,35)[]{\small $H^+$}
\Text(67.5,70)[]{\small $1$} \Text(10,60)[]{\small $\tilde
\chi^0;\tilde \chi^+$} \Text(39,78)[]{$b;t$} \Text(39,42)[]{$b;t$}
\Text(60,15)[]{$(d)$}
\end{picture}
\hspace{1.0cm}
\begin{picture}(120,120)(0,0)
\DashLine(35,60)(65,60){3} \ArrowLine(15,80)(35,60)
\ArrowLine(35,60)(15,40) \ArrowLine(65,60)(85,80)
\ArrowLine(85,40)(65,60) \ArrowLine(85,80)(85,40)
\Photon(85,80)(105,80){1.5}{4} \DashLine(85,40)(105,40){3}
\Vertex(65,60){1} \Vertex(35,60){1} \Vertex(85,80){1}
\Vertex(85,40){1} \Text(10,85)[]{$b$} \Text(10,35)[]{$\bar b$}
\Text(115,85)[]{\small $W^-$} \Text(115,35)[]{\small $H^+$}
\Text(50,70)[]{\small $1$} \Text(75,80)[]{$b$} \Text(75,40)[]{$b$}
\Text(95,60)[]{$t$} \Text(60,15)[]{$(e)$}
\end{picture}
\hspace{1.0cm}
\begin{picture}(120,120)(0,0)
\DashLine(35,60)(65,60){3} \ArrowLine(15,80)(35,60)
\ArrowLine(35,60)(15,40) \ArrowLine(85,80)(65,60)
\ArrowLine(65,60)(85,40) \ArrowLine(85,40)(85,80)
\Photon(85,80)(105,80){1.5}{4} \DashLine(85,40)(105,40){3}
\Vertex(65,60){1} \Vertex(35,60){1} \Vertex(85,80){1}
\Vertex(85,40){1} \Text(10,85)[]{$b$} \Text(10,35)[]{$\bar b$}
\Text(115,85)[]{\small $W^-$} \Text(115,35)[]{\small $H^+$}
\Text(50,70)[]{\small $1$} \Text(75,80)[]{$t$} \Text(75,40)[]{$t$}
\Text(95,60)[]{$b$} \Text(60,15)[]{$(f)$}
\end{picture}

\begin{picture}(120,120)(0,0)
\DashLine(35,60)(65,60){3} \ArrowLine(15,80)(35,60)
\ArrowLine(35,60)(15,40) \DashLine(65,60)(85,80){3}
\DashLine(85,40)(65,60){3} \DashLine(85,80)(85,40){3}
\Photon(85,80)(105,80){1.5}{4} \DashLine(85,40)(105,40){3}
\Vertex(65,60){1} \Vertex(35,60){1} \Vertex(85,80){1}
\Vertex(85,40){1} \Text(10,85)[]{$b$} \Text(10,35)[]{$\bar b$}
\Text(115,85)[]{\small $W^-$} \Text(115,35)[]{\small $H^+$}
\Text(50,70)[]{\small $2$} \Text(75,80)[]{$\tilde b$}
\Text(75,40)[]{$\tilde b$} \Text(95,60)[]{$\tilde t$}
\Text(60,15)[]{$(g)$}
\end{picture}
\hspace{1.0cm}
\begin{picture}(120,120)(0,0)
\DashLine(35,60)(65,60){3} \ArrowLine(15,80)(35,60)
\ArrowLine(35,60)(15,40) \DashLine(85,80)(65,60){3}
\DashLine(65,60)(85,40){3} \DashLine(85,40)(85,80){3}
\Photon(85,80)(105,80){1.5}{4} \DashLine(85,40)(105,40){3}
\Vertex(65,60){1} \Vertex(35,60){1} \Vertex(85,80){1}
\Vertex(85,40){1} \Text(10,85)[]{$b$} \Text(10,35)[]{$\bar b$}
\Text(115,85)[]{\small $W^-$} \Text(115,35)[]{\small $H^+$}
\Text(50,70)[]{\small $2$} \Text(75,80)[]{$\tilde t$}
\Text(75,40)[]{$\tilde t$} \Text(95,60)[]{$\tilde b$}
\Text(60,15)[]{$(h)$}
\end{picture}
\hspace{1.0cm}
\begin{picture}(120,120)(0,0)
\DashLine(35,60)(47,60){3} \DashLine(73,60)(85,60){3}
\ArrowArc(60,60)(13,180,0) \ArrowArc(60,60)(13,360,180)
\ArrowLine(15,80)(35,60) \ArrowLine(35,60)(15,40)
\Photon(85,60)(105,80){1.5}{5} \DashLine(85,60)(105,40){3}
\Vertex(35,60){1} \Vertex(85,60){1} \Vertex(47,60){1}
\Vertex(73,60){1} \Text(10,85)[]{$b$} \Text(10,35)[]{$\bar b$}
\Text(115,85)[]{\small $W^-$} \Text(115,35)[]{\small $H^+$}
\Text(41,70)[]{\small $1$} \Text(79,70)[]{\small $1$}
\Text(60,85)[]{$t;b$} \Text(60,35)[]{$t;b$} \Text(60,15)[]{$(i)$}
\end{picture}

\begin{picture}(120,120)(0,0)
\DashLine(35,60)(47,60){3} \DashLine(73,60)(85,60){3}
\DashCArc(60,60)(13,0,360){3} \ArrowLine(15,80)(35,60)
\ArrowLine(35,60)(15,40) \Photon(85,60)(105,80){1.5}{5}
\DashLine(85,60)(105,40){3} \Vertex(35,60){1} \Vertex(85,60){1}
\Vertex(47,60){1} \Vertex(73,60){1} \Text(10,85)[]{$b$}
\Text(10,35)[]{$\bar b$} \Text(115,85)[]{\small $W^-$}
\Text(115,35)[]{\small $H^+$} \Text(60,85)[]{$\tilde t;\tilde b$}
\Text(60,35)[]{$\tilde t;\tilde b$} \Text(41,70)[]{\small $2$}
\Text(79,70)[]{\small $1$} \Text(60,15)[]{$(j)$}
\end{picture}
\hspace{1.0cm}
\begin{picture}(120,120)(0,0)
\DashLine(35,60)(85,60){3} \DashCArc(60,70)(10,0,360){3}
\ArrowLine(15,80)(35,60) \ArrowLine(35,60)(15,40)
\Photon(85,60)(105,80){1.5}{5} \DashLine(85,60)(105,40){3}
\Vertex(35,60){1} \Vertex(85,60){1} \Vertex(60,60){1}
\Text(10,85)[]{$b$} \Text(10,35)[]{$\bar b$}
\Text(115,85)[]{\small $W^-$} \Text(115,35)[]{\small $H^+$}
\Text(60,90)[]{$\tilde t;\tilde b$} \Text(41,70)[]{\small $2$}
\Text(79,70)[]{\small $1$} \Text(60,15)[]{$(k)$}
\end{picture}
\hspace{1.0cm}
\begin{picture}(120,120)(0,0)
\Line(60,80)(60,66) \ArrowLine(60,66)(60,40)
\ArrowLine(20,80)(40,80) \Line(40,80)(60,80)
\ArrowLine(60,40)(20,40) \Photon(60,80)(100,80){1.5}{5}
\DashLine(60,40)(100,40){3} \DashLine(40,80)(60,60){3}
\Vertex(60,80){1} \Vertex(60,40){1} \Vertex(60,60){1}
\Vertex(40,80){1} \Text(15,85)[]{$b$} \Text(15,35)[]{$\bar b$}
\Text(68,60)[]{$t$} \Text(110,85)[]{\small $W^-$}
\Text(110,35)[]{\small $H^+$} \Text(45,65)[]{\small $2$}
\Text(50,90)[]{$b$} \Text(60,15)[]{$(l)$}
\end{picture}

\begin{picture}(120,120)(0,0)
\DashLine(60,80)(60,60){3} \ArrowLine(60,60)(60,40)
\ArrowLine(20,80)(40,80) \DashLine(40,80)(60,80){3}
\ArrowLine(60,40)(20,40) \Photon(60,80)(100,80){1.5}{5}
\DashLine(60,40)(100,40){3} \ArrowLine(40,80)(60,60)
\Vertex(60,80){1} \Vertex(60,40){1} \Vertex(60,60){1}
\Vertex(40,80){1} \Text(15,85)[]{$b$} \Text(15,35)[]{$\bar b$}
\Text(110,85)[]{\small $W^-$} \Text(110,35)[]{\small $H^+$}
\Text(40,65)[]{$\tilde \chi^0$} \Text(50,90)[]{$\tilde b$}
\Text(67,70)[]{$\tilde t$} \Text(67,50)[]{$t$}
\Text(60,15)[]{$(m)$}
\end{picture}
\hspace{1.0cm}
\begin{picture}(120,120)(0,0)
\Line(60,80)(60,66) \ArrowLine(60,66)(60,40)
\ArrowLine(20,80)(60,80) \ArrowLine(60,40)(20,40)
\DashLine(60,80)(80,80){3} \DashLine(80,80)(60,60){3}
\Photon(80,80)(100,80){1.5}{3} \DashLine(60,40)(100,40){3}
\Vertex(60,80){1} \Vertex(60,40){1} \Vertex(80,80){1}
\Vertex(60,60){1} \Text(15,85)[]{$b$} \Text(15,35)[]{$\bar b$}
\Text(110,85)[]{\small $W^-$} \Text(110,35)[]{\small $H^+$}
\Text(53,50)[]{$t$} \Text(70,90)[]{\small $1;3$}
\Text(75,62)[]{\small $3;1$} \Text(50,65)[]{$b;t$}
\Text(60,15)[]{$(n)$}
\end{picture}
\hspace{1.0cm}
\begin{picture}(120,120)(0,0)
\DashLine(60,80)(60,60){3} \ArrowLine(60,60)(60,40)
\ArrowLine(20,80)(60,80) \ArrowLine(60,40)(20,40)
\Line(60,80)(80,80) \Line(80,80)(60,60)
\Photon(80,80)(100,80){1.5}{3} \DashLine(60,40)(100,40){3}
\Vertex(60,80){1} \Vertex(60,40){1} \Vertex(80,80){1}
\Vertex(60,60){1} \Text(15,85)[]{$b$} \Text(15,35)[]{$\bar b$}
\Text(110,85)[]{\small $W^-$} \Text(110,35)[]{\small $H^+$}
\Text(53,50)[]{$t$} \Text(70,88)[]{\small
$\tilde\chi^0;\tilde\chi^+$} \Text(48,70)[]{\small $\tilde
b;\tilde t$} \Text(85,67)[]{\small $\tilde\chi^+;\tilde\chi^0$}
\Text(60,15)[]{$(o)$}
\end{picture}

\begin{picture}(120,120)(0,0)
\ArrowLine(60,80)(60,53) \Line(60,53)(60,40)
\DashLine(60,60)(40,40){3} \ArrowLine(20,80)(60,80)
\Line(60,40)(40,40) \ArrowLine(40,40)(20,40)
\Photon(60,80)(100,80){1.5}{5} \DashLine(60,40)(100,40){3}
\Vertex(60,80){1} \Vertex(60,40){1} \Vertex(60,60){1}
\Vertex(40,40){1} \Text(15,85)[]{$b$} \Text(15,35)[]{$\bar b$}
\Text(68,60)[]{$t$} \Text(110,85)[]{\small $W^-$}
\Text(110,35)[]{\small $H^+$} \Text(45,53)[]{\small $2$}
\Text(60,15)[]{$(p)$}
\end{picture}
\hspace{1.0cm}
\begin{picture}(120,120)(0,0)
\ArrowLine(60,80)(60,60) \DashLine(60,60)(60,40){3}
\Line(60,60)(40,40) \ArrowLine(20,80)(60,80)
\DashLine(60,40)(40,40){3} \ArrowLine(40,40)(20,40)
\Photon(60,80)(100,80){1.5}{5} \DashLine(60,40)(100,40){3}
\Vertex(60,80){1} \Vertex(60,40){1} \Vertex(60,60){1}
\Vertex(40,40){1} \Text(15,85)[]{$b$} \Text(15,35)[]{$\bar b$}
\Text(68,70)[]{$t$} \Text(110,85)[]{\small $W^-$}
\Text(110,35)[]{\small $H^+$} \Text(68,50)[]{$\tilde t$}
\Text(50,32)[]{$\tilde b$} \Text(47,56)[]{$\tilde\chi^0$}
\Text(60,15)[]{$(q)$}
\end{picture}
\hspace{1.0cm}
\begin{picture}(120,120)(0,0)
\ArrowLine(60,80)(60,40) \ArrowLine(20,80)(60,80)
\DashCArc(60,60)(13,-90,90){3} \ArrowLine(60,40)(20,40)
\Photon(60,80)(100,80){1.5}{5} \DashLine(60,40)(100,40){3}
\Vertex(60,80){1} \Vertex(60,40){1} \Vertex(60,73){1}
\Vertex(60,47){1} \Text(15,85)[]{$b$} \Text(15,35)[]{$\bar b$}
\Text(110,85)[]{\small $W^-$} \Text(110,35)[]{\small $H^+$}
\Text(60,15)[]{$(r)$}
\end{picture}

\begin{picture}(120,120)(0,0)
\ArrowLine(15,80)(50,80) \ArrowLine(50,80)(70,80)
\Photon(70,80)(105,80){1.5}{4} \DashLine(50,80)(50,40){3}
\ArrowLine(70,80)(70,40) \DashLine(70,40)(105,40){3}
\ArrowLine(70,40)(50,40) \ArrowLine(50,40)(15,40)
\Vertex(50,80){1} \Vertex(70,80){1} \Vertex(50,40){1}
\Vertex(70,40){1} \Text(10,85)[]{$b$} \Text(10,35)[]{$\bar b$}
\Text(115,85)[]{\small $W^-$} \Text(115,35)[]{\small $H^+$}
\Text(60,88)[]{$b$} \Text(78,60)[]{$t$} \Text(60,32)[]{$b$}
\Text(43,60)[]{\small $2$} \Text(60,15)[]{$(s)$}
\end{picture}
\hspace{1.0cm}
\begin{picture}(120,120)(0,0)
\ArrowLine(15,80)(50,80) \ArrowLine(50,80)(70,80)
\Photon(70,40)(105,80){-1.5}{8} \DashLine(50,80)(50,40){3}
\ArrowLine(70,80)(70,40) \DashLine(70,80)(105,40){3}
\ArrowLine(70,40)(50,40) \ArrowLine(50,40)(15,40)
\Vertex(50,80){1} \Vertex(70,80){1} \Vertex(50,40){1}
\Vertex(70,40){1} \Text(10,85)[]{$b$} \Text(10,35)[]{$\bar b$}
\Text(115,85)[]{\small $W^-$} \Text(115,35)[]{\small $H^+$}
\Text(60,88)[]{$t$} \Text(77,60)[]{$b$} \Text(60,32)[]{$t$}
\Text(43,60)[]{\small $3$} \Text(60,15)[]{$(t)$}
\end{picture}
\hspace{1.0cm}
\begin{picture}(120,120)(0,0)
\ArrowLine(15,80)(50,80) \DashLine(50,80)(70,80){3}
\Photon(70,80)(105,80){1.5}{4} \ArrowLine(50,80)(50,40)
\DashLine(70,80)(70,40){3} \DashLine(70,40)(105,40){3}
\DashLine(70,40)(50,40){3} \ArrowLine(50,40)(15,40)
\Vertex(50,80){1} \Vertex(70,80){1} \Vertex(50,40){1}
\Vertex(70,40){1} \Text(10,85)[]{$b$} \Text(10,35)[]{$\bar b$}
\Text(115,85)[]{\small $W^-$} \Text(115,35)[]{\small $H^+$}
\Text(60,88)[]{$\tilde b$} \Text(78,60)[]{$\tilde t$}
\Text(60,32)[]{$\tilde b$} \Text(42,60)[]{$\tilde \chi^0$}
\Text(60,15)[]{$(u)$}
\end{picture}

\begin{picture}(120,120)(0,0)
\ArrowLine(15,80)(50,80) \DashLine(50,80)(70,80){3}
\Photon(70,40)(105,80){-1.5}{8} \ArrowLine(50,80)(50,40)
\DashLine(70,80)(70,40){3} \DashLine(70,80)(105,40){3}
\DashLine(70,40)(50,40){3} \ArrowLine(50,40)(15,40)
\Vertex(50,80){1} \Vertex(70,80){1} \Vertex(50,40){1}
\Vertex(70,40){1} \Text(10,85)[]{$b$} \Text(10,35)[]{$\bar b$}
\Text(115,85)[]{\small $W^-$} \Text(115,35)[]{\small $H^+$}
\Text(60,88)[]{$\tilde t$} \Text(78,60)[]{$\tilde b$}
\Text(60,32)[]{$\tilde t$} \Text(42,60)[]{$\tilde \chi^+$}
\Text(60,15)[]{$(v)$}
\end{picture}
\hspace{1.0cm}
\begin{picture}(120,120)(0,0)
\ArrowLine(60,80)(40,60) \ArrowLine(40,60)(60,40)
\DashLine(60,80)(80,60){3} \DashLine(80,60)(60,40){3}
\ArrowLine(20,80)(60,80) \ArrowLine(60,40)(20,40)
\Photon(80,60)(100,80){1.5}{6} \DashLine(40,60)(100,40){3}
\Vertex(60,80){1} \Vertex(40,60){1} \Vertex(80,60){1}
\Vertex(60,40){1} \Text(15,85)[]{$b$} \Text(15,35)[]{$\bar b$}
\Text(115,85)[]{\small $W^-$} \Text(115,35)[]{\small $H^+$}
\Text(38,70)[]{$t$} \Text(38,50)[]{$b$} \Text(75,75)[]{\small $3$}
\Text(72,40)[]{\small $2$} \Text(60,15)[]{$(w)$}
\end{picture}
\hspace{1.0cm}
\begin{picture}(120,120)(0,0)
\DashLine(60,80)(40,60){3} \DashLine(40,60)(60,40){3}
\Line(60,80)(80,60) \Line(80,60)(60,40) \ArrowLine(20,80)(60,80)
\ArrowLine(60,40)(20,40) \Photon(80,60)(100,80){1.5}{6}
\DashLine(40,60)(100,40){3} \Vertex(60,80){1} \Vertex(40,60){1}
\Vertex(80,60){1} \Vertex(60,40){1} \Text(15,85)[]{$b$}
\Text(15,35)[]{$\bar b$} \Text(115,85)[]{\small $W^-$}
\Text(115,35)[]{\small $H^+$} \Text(38,70)[]{$\tilde t$}
\Text(38,50)[]{$\tilde b$} \Text(80,78)[]{\small $\tilde\chi^+$}
\Text(72,40)[]{\small $\tilde \chi^0$} \Text(60,15)[]{$(x)$}
\end{picture}

\begin{figure}[ht]
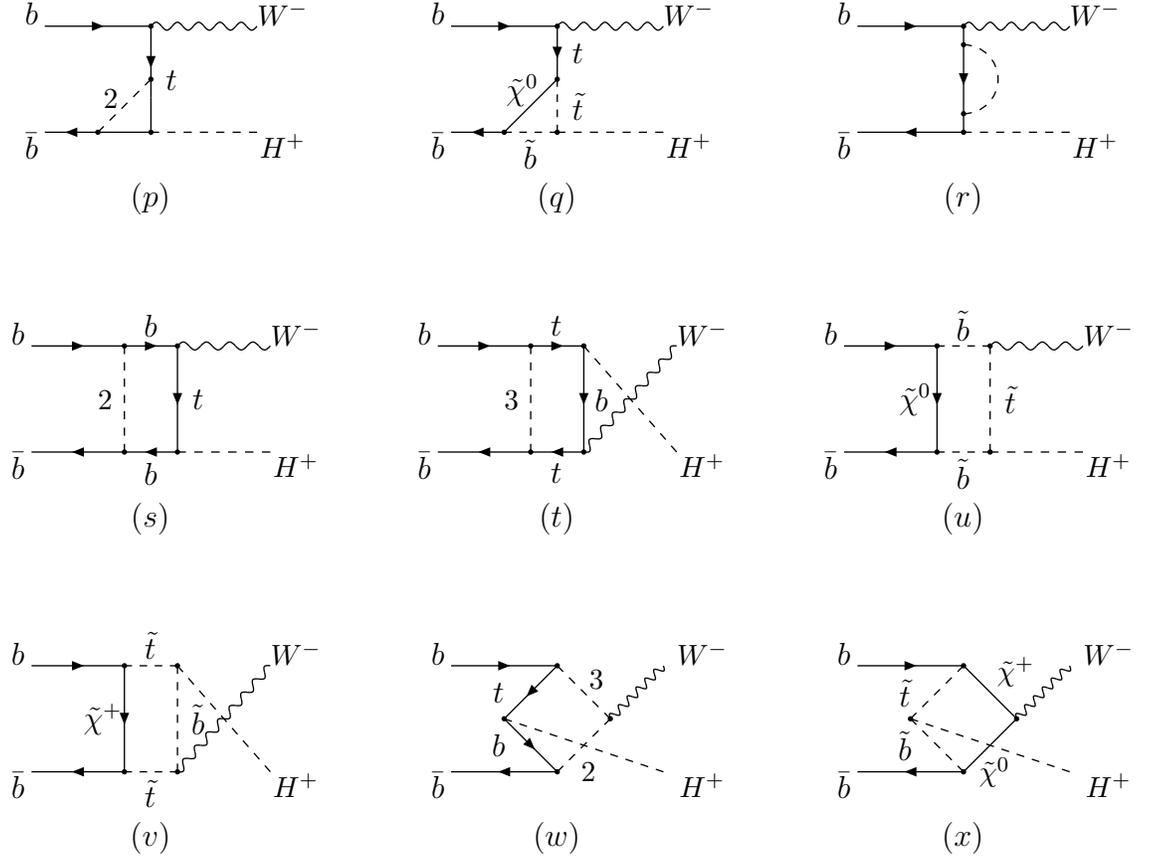
 \caption[]
{\small Feynman diagrams contributing to supersymmetric
electroweak corrections to $b\bar b\rightarrow W^-H^+$: $(a)$ and
$(b)$ are tree level diagrams; $(c)-(x)$ are one-loop corrections.
The dashed line $1$ represents $H,h,A$; the dashed line $2$
represents $H,h,A,G^0$; the dashed line $3$ represents $H^+,G^+$.
For diagram $(r)$, the dashed line in the loop represents
$H,h,A,G^0,H^+,G^+,\tilde t,\tilde b$.}
\end{figure}

\begin{picture}(120,120)(0,0)
\Line(40,60)(80,60) \ArrowLine(20,60)(40,60)
\ArrowLine(80,60)(100,60) \DashCArc(60,60)(20,0,180){3}
\Vertex(40,60){1} \Vertex(80,60){1} \Text(20,50)[]{$t(b)$}
\Text(100,50)[]{$t(b)$} \Text(60,15)[]{$(a)$}
\end{picture}
\hspace{1.0cm}
\begin{picture}(120,120)(0,0)
\Photon(20,60)(40,60){1.5}{3} \Photon(80,60)(100,60){1.5}{3}
\ArrowArc(60,60)(20,0,180) \ArrowArc(60,60)(20,180,360)
\Vertex(40,60){1} \Vertex(80,60){1} \Text(20,50)[]{\small $W^-$}
\Text(105,50)[]{\small $W^-$} \Text(60,88)[]{$t$}
\Text(60,32)[]{$b$} \Text(60,15)[]{$(b)$}
\end{picture}
\hspace{1.0cm}
\begin{picture}(120,120)(0,0)
\Photon(20,60)(40,60){1.5}{3} \Photon(80,60)(100,60){1.5}{3}
\ArrowArc(60,60)(20,0,180) \ArrowArc(60,60)(20,180,360)
\Vertex(40,60){1} \Vertex(80,60){1} \Text(20,50)[]{\small $Z^0$}
\Text(105,50)[]{\small $Z^0$} \Text(60,88)[]{$t(b)$}
\Text(60,32)[]{$t(b)$} \Text(60,15)[]{$(c)$}
\end{picture}

\begin{picture}(120,120)(0,0)
\DashLine(20,60)(40,60){3} \DashLine(80,60)(100,60){3}
\ArrowArc(60,60)(20,0,180) \ArrowArc(60,60)(20,180,360)
\Vertex(40,60){1} \Vertex(80,60){1} \Text(20,50)[]{\small $H_i$}
\Text(105,50)[]{\small $H_i$} \Text(60,88)[]{$t(b)$}
\Text(60,32)[]{$t(b)$} \Text(60,15)[]{$(d)$}
\end{picture}
\hspace{1.0cm}
\begin{picture}(120,120)(0,0)
\DashLine(20,60)(40,60){3} \DashLine(80,60)(100,60){3}
\DashCArc(60,60)(20,0,180){3} \DashCArc(60,60)(20,180,360){3}
\Vertex(40,60){1} \Vertex(80,60){1} \Text(20,50)[]{\small $H_i$}
\Text(105,50)[]{\small $H_i$} \Text(60,88)[]{\small $\tilde
t(\tilde b)$} \Text(60,32)[]{\small $\tilde t(\tilde b)$}
\Text(60,15)[]{$(e)$}
\end{picture}
\hspace{1.0cm}
\begin{picture}(120,120)(0,0)
\DashLine(20,55)(100,55){3} \DashCArc(60,70)(15,0,360){3}
\Vertex(60,55){1} \Text(20,45)[]{\small $H_i$}
\Text(105,45)[]{\small $H_i$} \Text(60,93)[]{\small $\tilde
t(\tilde b)$} \Text(60,15)[]{$(f)$}
\end{picture}

\begin{picture}(120,120)(0,0)
\DashLine(20,60)(40,60){3} \DashLine(80,60)(100,60){3}
\ArrowArc(60,60)(20,0,180) \ArrowArc(60,60)(20,180,360)
\Vertex(40,60){1} \Vertex(80,60){1} \Text(20,50)[]{\small $H^+$}
\Text(105,50)[]{\small $H^+$} \Text(60,88)[]{$b$}
\Text(60,32)[]{$t$} \Text(60,15)[]{$(g)$}
\end{picture}
\hspace{1.0cm}
\begin{picture}(120,120)(0,0)
\DashLine(20,60)(40,60){3} \DashLine(80,60)(100,60){3}
\DashCArc(60,60)(20,0,180){3} \DashCArc(60,60)(20,180,360){3}
\Vertex(40,60){1} \Vertex(80,60){1} \Text(20,50)[]{\small $H^+$}
\Text(105,50)[]{\small $H^+$} \Text(60,88)[]{\small $\tilde b$}
\Text(60,32)[]{\small $\tilde t$} \Text(60,15)[]{$(h)$}
\end{picture}
\hspace{1.0cm}
\begin{picture}(120,120)(0,0)
\DashLine(20,60)(40,60){3} \DashLine(80,60)(100,60){3}
\ArrowArc(60,60)(20,0,180) \ArrowArc(60,60)(20,180,360)
\Vertex(40,60){1} \Vertex(80,60){1} \Text(20,50)[]{\small $H$}
\Text(105,50)[]{$h$} \Text(60,88)[]{$t(b)$} \Text(60,32)[]{$t(b)$}
\Text(60,15)[]{$(i)$}
\end{picture}

\begin{picture}(120,120)(0,0)
\DashLine(20,60)(40,60){3} \DashLine(80,60)(100,60){3}
\DashCArc(60,60)(20,0,180){3} \DashCArc(60,60)(20,180,360){3}
\Vertex(40,60){1} \Vertex(80,60){1} \Text(20,50)[]{\small $H$}
\Text(105,50)[]{$h$} \Text(60,88)[]{\small $\tilde t(\tilde b)$}
\Text(60,32)[]{\small $\tilde t(\tilde b)$} \Text(60,15)[]{$(j)$}
\end{picture}
\hspace{1.0cm}
\begin{picture}(120,120)(0,0)
\DashLine(20,55)(100,55){3} \DashCArc(60,70)(15,0,360){3}
\Vertex(60,55){1} \Text(20,45)[]{\small $H$} \Text(105,45)[]{$h$}
\Text(60,93)[]{\small $\tilde t(\tilde b)$} \Text(60,15)[]{$(k)$}
\end{picture}
\hspace{1.0cm}
\begin{picture}(120,120)(0,0)
\Photon(20,60)(40,60){1.5}{3} \DashLine(80,60)(100,60){3}
\ArrowArc(60,60)(20,0,180) \ArrowArc(60,60)(20,180,360)
\Vertex(40,60){1} \Vertex(80,60){1} \Text(20,50)[]{\small $Z$}
\Text(105,50)[]{\small $A$} \Text(60,88)[]{$t(b)$}
\Text(60,32)[]{$t(b)$} \Text(60,15)[]{$(l)$}
\end{picture}

\begin{picture}(120,120)(0,0)
\Photon(20,60)(40,60){1.5}{3} \DashLine(80,60)(100,60){3}
\DashCArc(60,60)(20,0,180){3} \DashCArc(60,60)(20,180,360){3}
\Vertex(40,60){1} \Vertex(80,60){1} \Text(20,50)[]{\small $Z$}
\Text(105,50)[]{\small $A$} \Text(60,88)[]{\small $\tilde t(\tilde
b)$} \Text(60,32)[]{\small $\tilde t(\tilde b)$}
\Text(60,15)[]{$(m)$}
\end{picture}
\hspace{1.0cm}
\begin{picture}(120,120)(0,0)
\DashLine(20,60)(40,60){3} \Photon(80,60)(100,60){1.5}{3}
\ArrowArc(60,60)(20,0,180) \ArrowArc(60,60)(20,180,360)
\Vertex(40,60){1} \Vertex(80,60){1} \Text(20,50)[]{\small $H^+$}
\Text(105,50)[]{\small $W^+$} \Text(60,88)[]{$b$}
\Text(60,32)[]{$t$} \Text(60,15)[]{$(n)$}
\end{picture}
\hspace{1.0cm}
\begin{picture}(120,120)(0,0)
\DashLine(20,60)(40,60){3} \Photon(80,60)(100,60){1.5}{3}
\DashCArc(60,60)(20,0,180){3} \DashCArc(60,60)(20,180,360){3}
\Vertex(40,60){1} \Vertex(80,60){1} \Text(20,50)[]{\small $H^+$}
\Text(105,50)[]{\small $W^+$} \Text(60,88)[]{\small $\tilde b$}
\Text(60,32)[]{\small $\tilde t$} \Text(60,15)[]{$(o)$}
\end{picture}
\begin{figure}[ht] \caption[]
{\small Feynman diagrams contributing to renormalization
constants: The dashed line represents $H,h,A,G^0,H^+,G^+,\tilde
t,\tilde b$ for diagram (a), and $H_i$ in diagrams $(d)-(f)$
represents $H,h,A$.}
\end{figure}

\begin{figure}
\centerline{\psfig{file=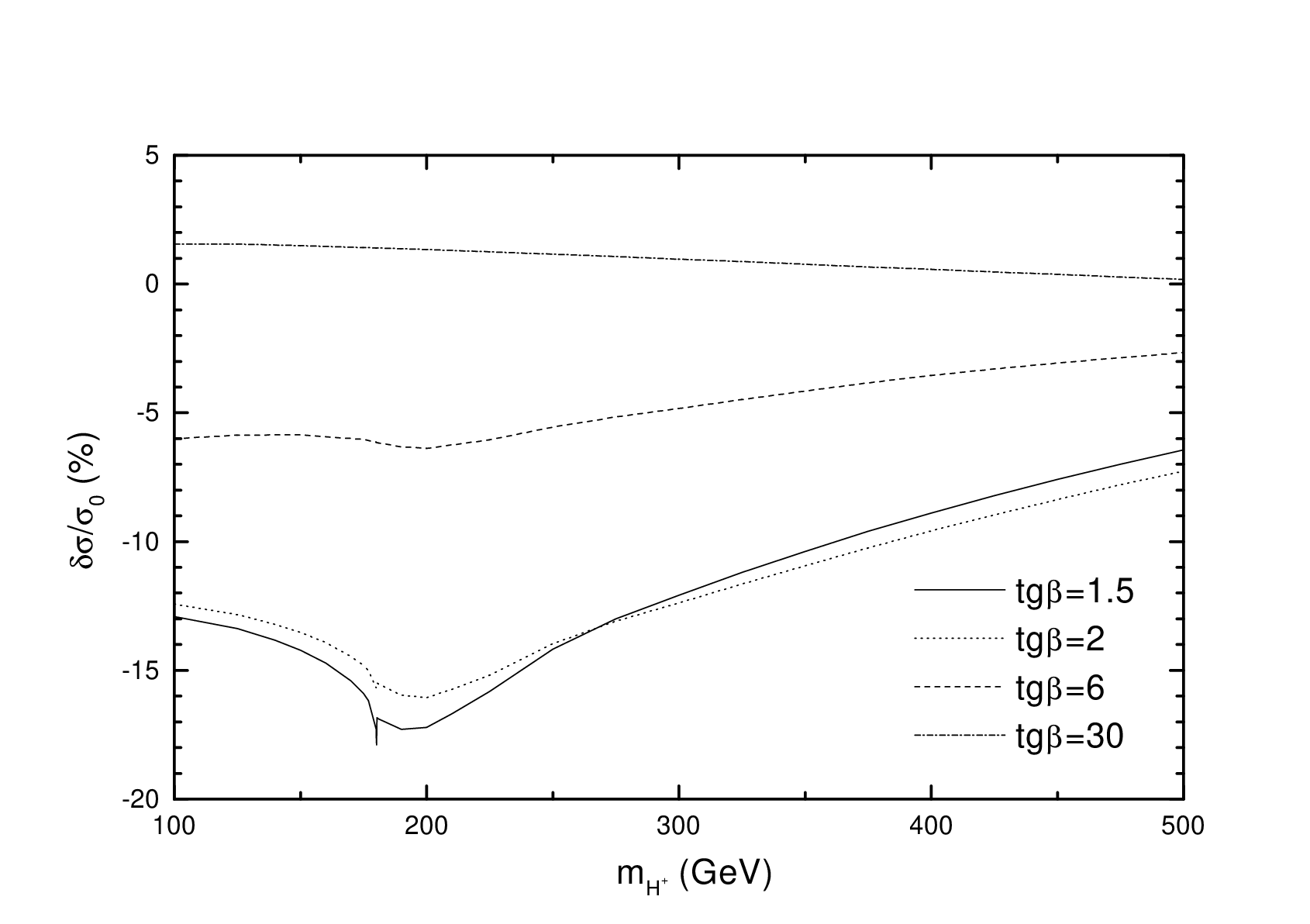, width=250pt}} \caption[]
{\small The Yukawa corrections versus $m_{H^+}$ for
$\tan\beta=1.5$, $2$, $6$ and $30$, respectively.}
\end{figure}

\begin{figure}
\centerline{\psfig{file=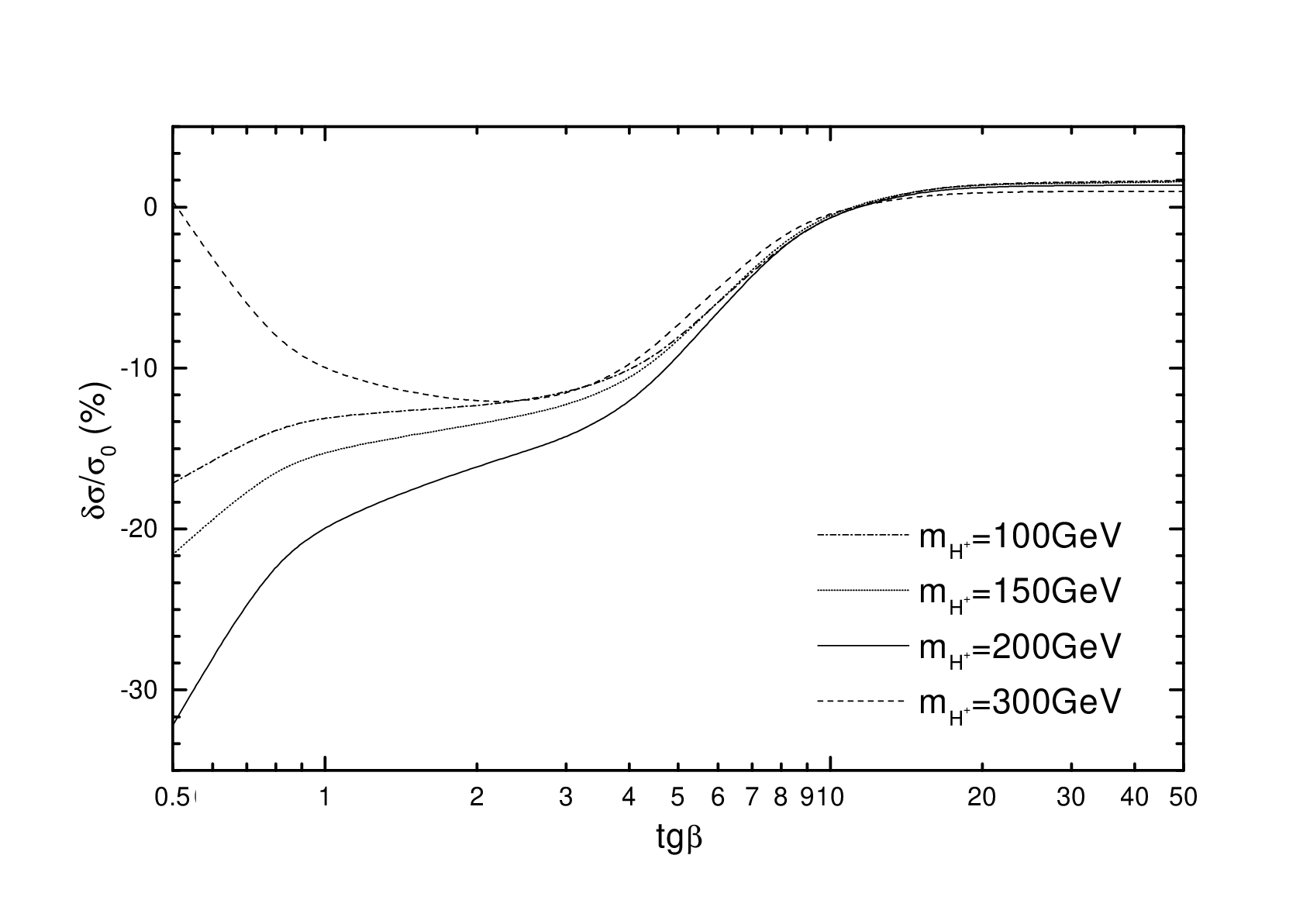, width=250pt}} \caption[]
{\small The Yukawa corrections versus $\tan\beta$ for
$m_{H^+}=100$, $150$, $200$ and $300$GeV, respectively.}
\end{figure}

\begin{figure}
\centerline{\psfig{file=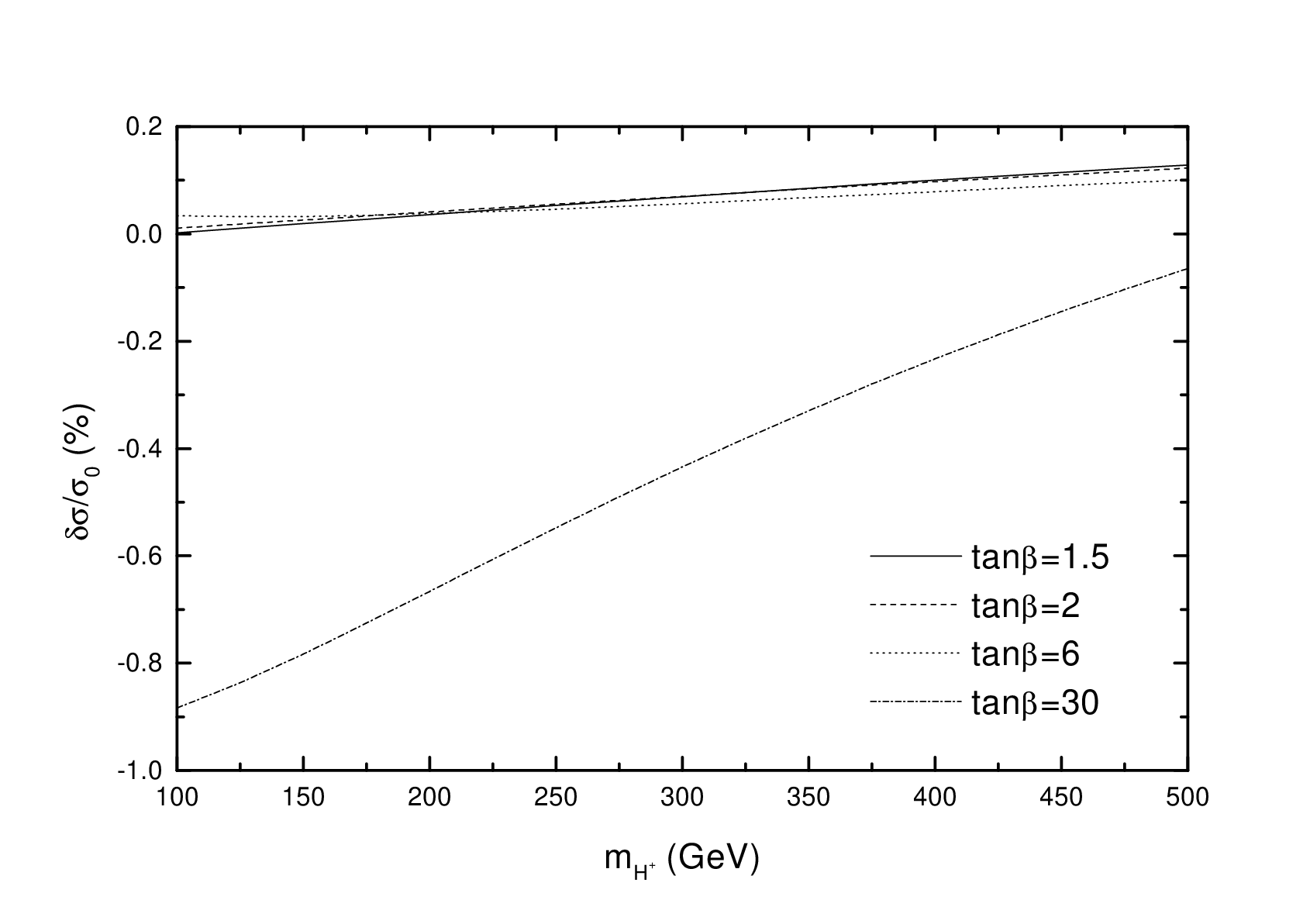, width=250pt}} \caption[]
{\small The genuine SUSY EW corrections versus $m_{H^+}$ for
$\tan\beta=1.5$, $2$, $6$ and $30$, respectively, assuming
$M_2=300$GeV, $\mu=-100$GeV, $A_t=A_b=200$GeV and $M_{\tilde
Q}=M_{\tilde U}=M_{\tilde D}=500$GeV.}
\end{figure}

\begin{figure}
\centerline{\psfig{file=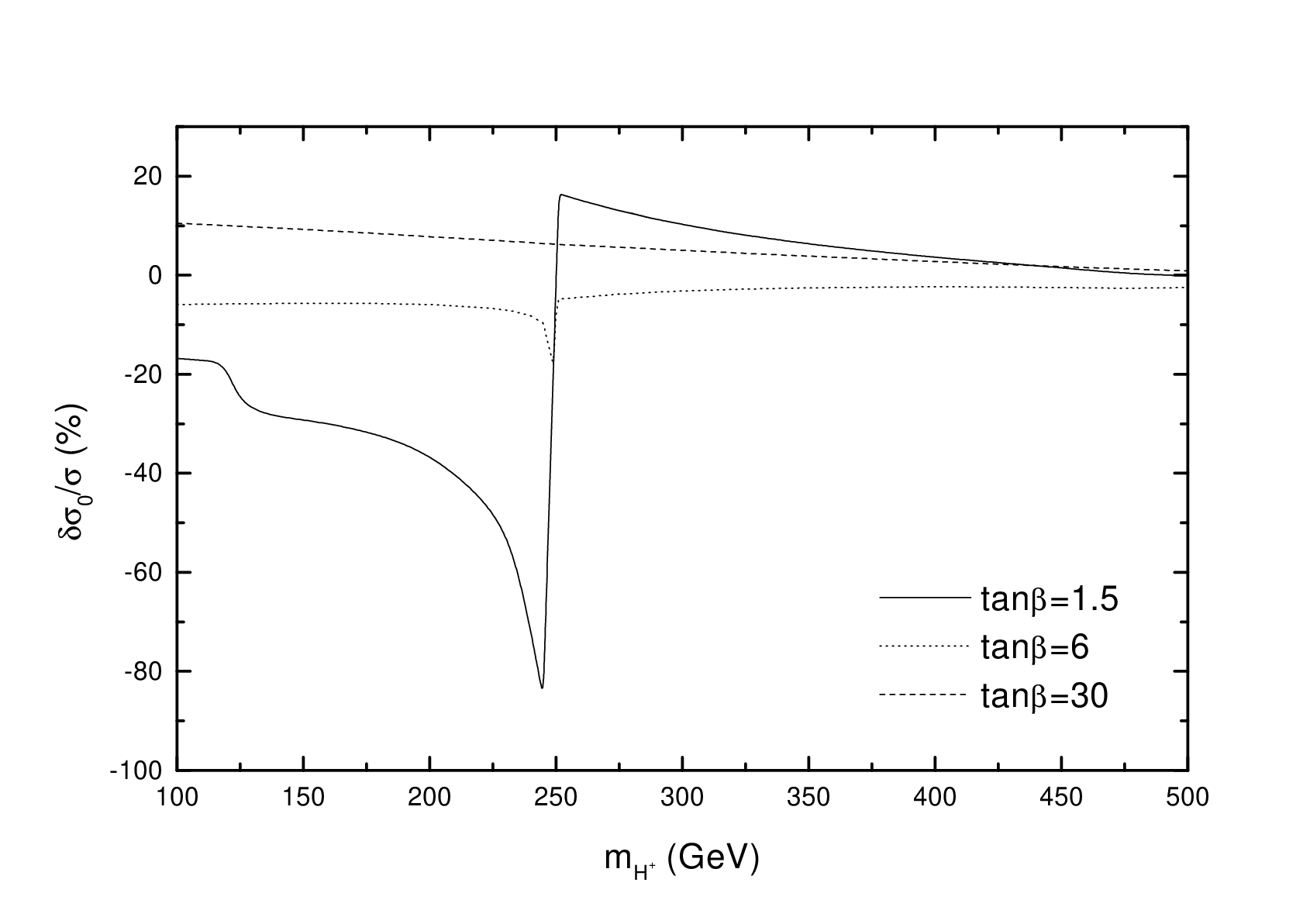, width=250pt}} \caption[]
{\small The genuine SUSY EW corrections versus $m_{H^+}$ for
$\tan\beta=1.5$, $6$ and $30$, respectively, assuming
$M_2=200$GeV, $\mu=100$GeV, $A_t=A_b=1$TeV, $M_{\tilde
Q}=M_{\tilde U}$, $m_{\tilde t_1}=100$GeV and $m_{\tilde
b_1}=150$GeV.}
\end{figure}

\begin{figure}
\centerline{\psfig{file=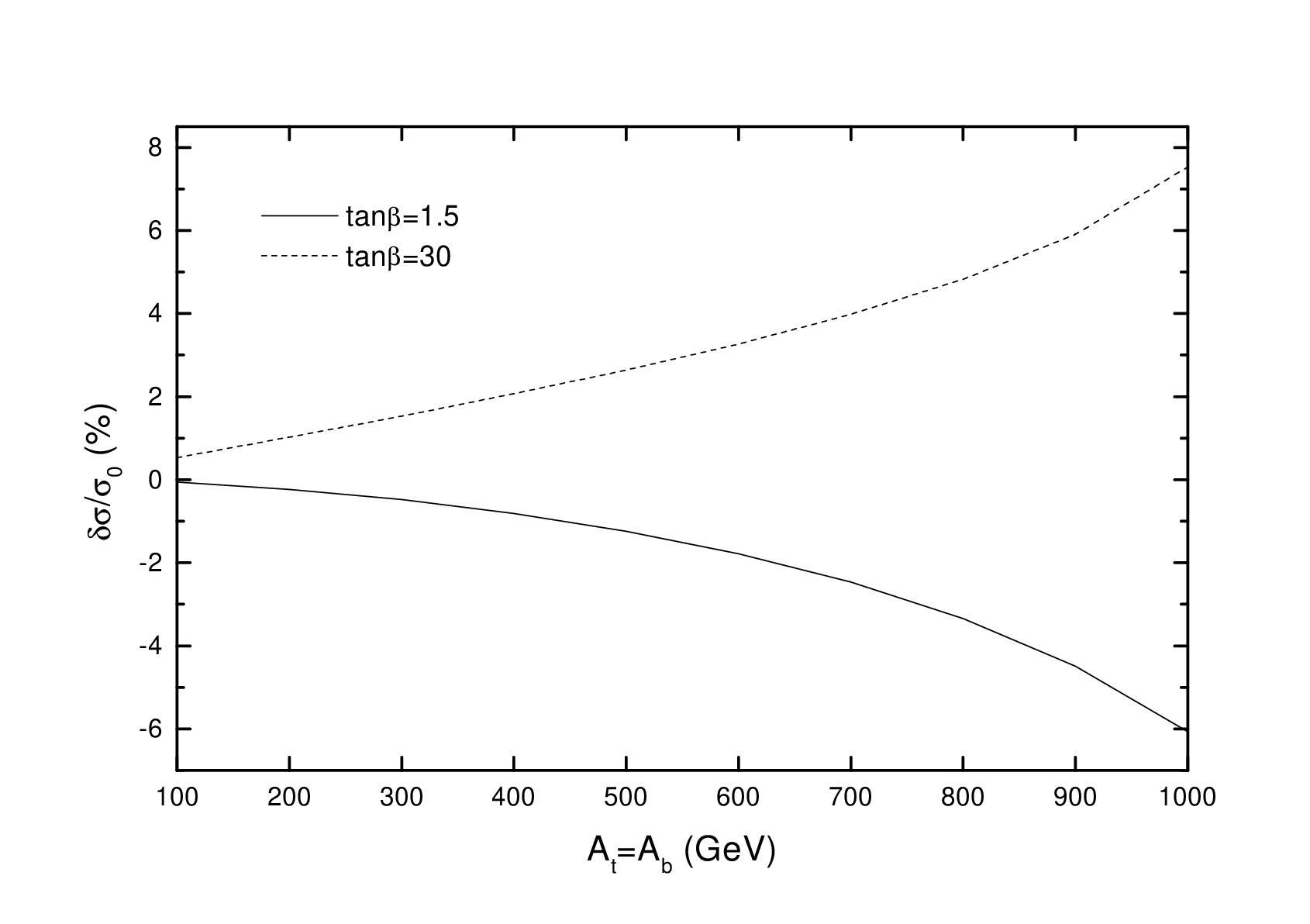, width=250pt}} \caption[]
{\small The genuine SUSY EW corrections versus $A_t=A_b$ for
$\tan\beta=1.5$ and $30$, respectively, assuming $m_{H^+}=200$GeV,
$M_2=300$GeV, $\mu=100$GeV, and $M_{\tilde Q}=M_{\tilde
U}=M_{\tilde D}=400$GeV.}
\end{figure}

\begin{figure}
\centerline{\psfig{file=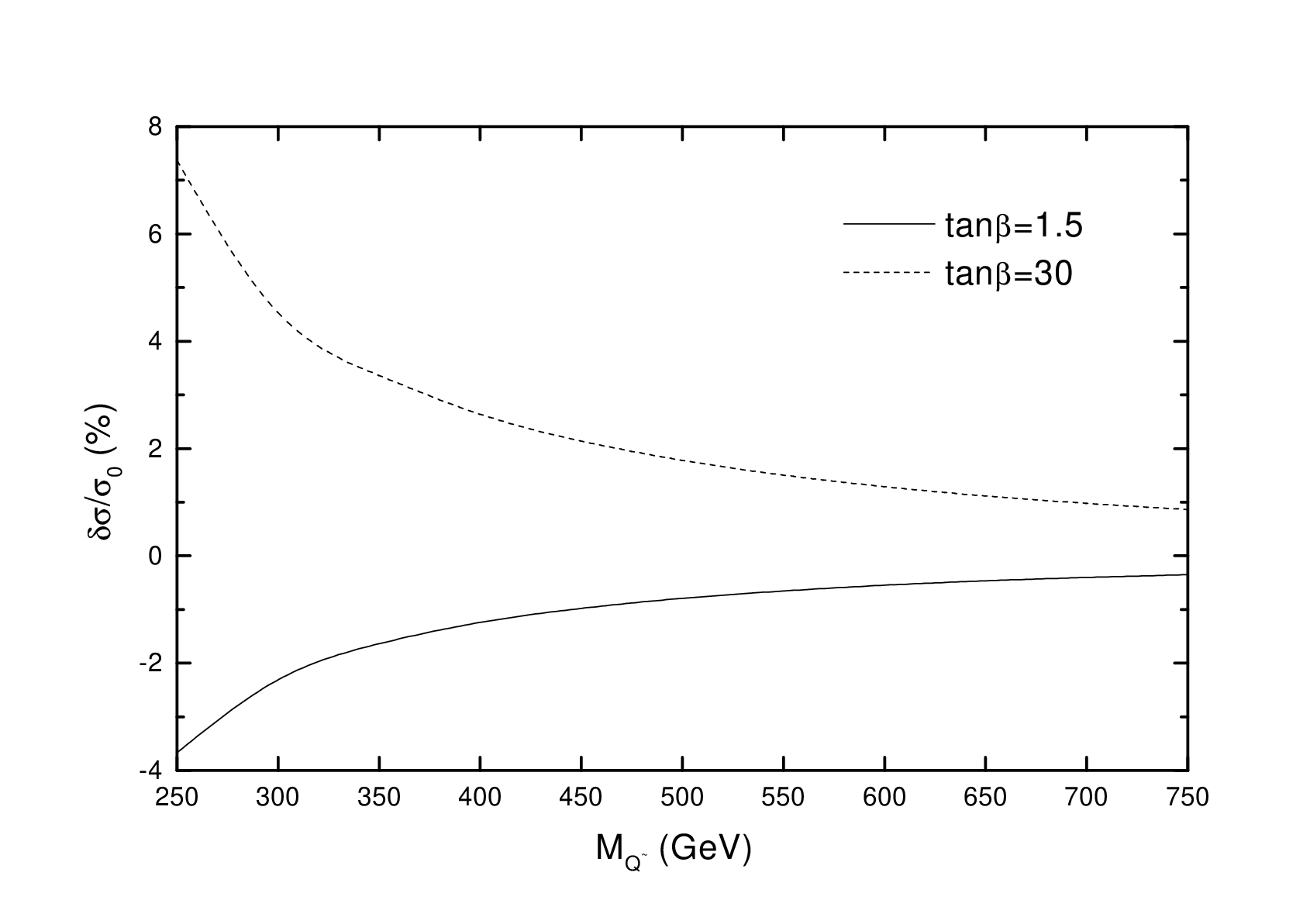, width=250pt}} \caption[]
{\small The genuine SUSY EW corrections versus $M_{\tilde
Q}=M_{\tilde U}=M_{\tilde D}$ for $\tan\beta=1.5$ and $30$,
respectively, assuming $m_{H^+}=200$GeV, $M_2=300$GeV,
$\mu=100$GeV, and $A_t=A_b=500$GeV.}
\end{figure}

\begin{figure}
\centerline{\psfig{file=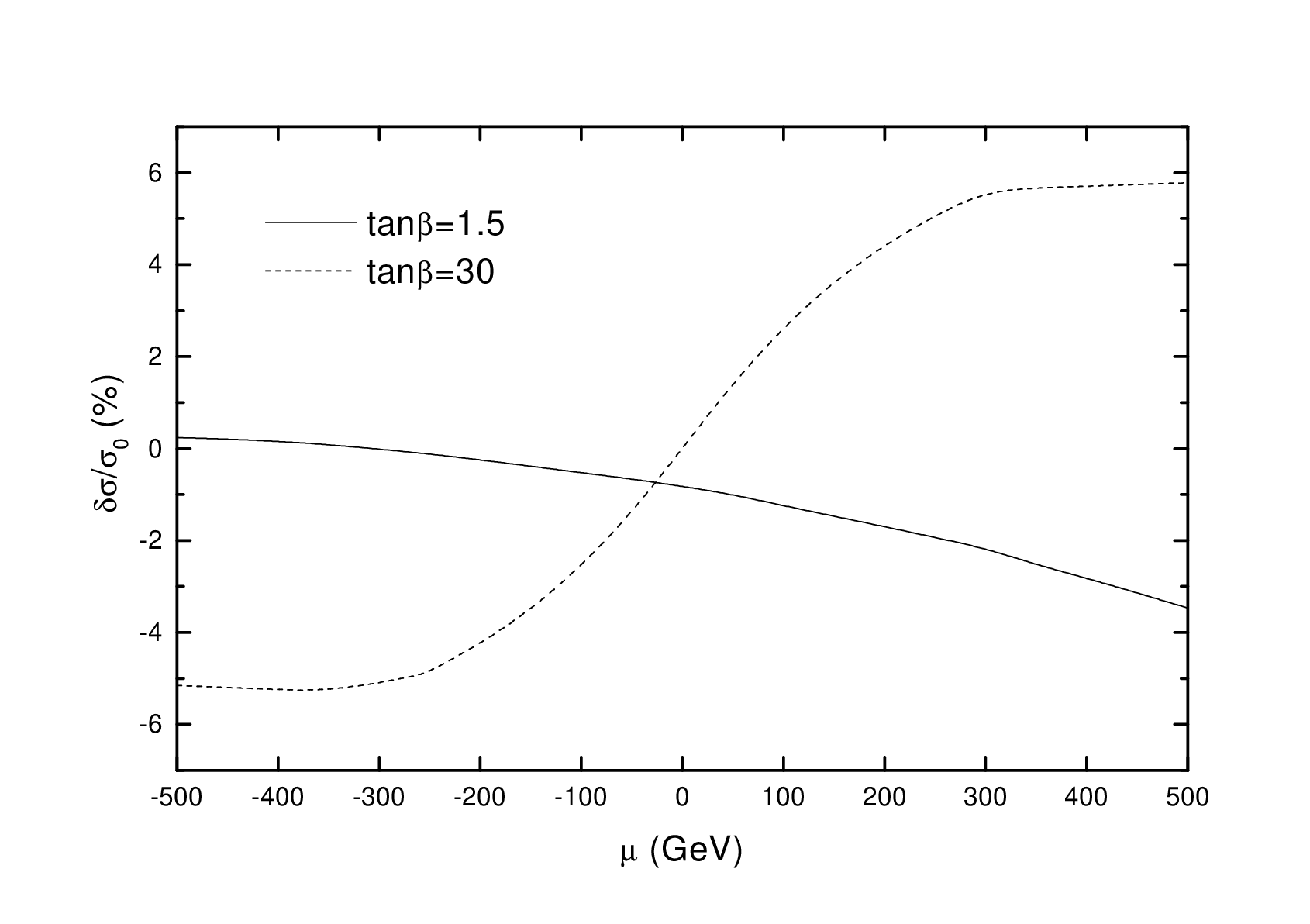, width=250pt}} \caption[]
{\small The genuine SUSY EW corrections versus $\mu$ for
$\tan\beta=1.5$ and $30$, respectively, assuming $m_{H^+}=200$GeV,
$M_2=300$GeV, $A_t=A_b=500$GeV and $M_{\tilde Q}=M_{\tilde
U}=M_{\tilde D}=400$GeV.}
\end{figure}

\setcounter{figure}{0} \setcounter{table}{0}
\setcounter{equation}{0}

\newpage
\part{FCNC Physics}

\section{Preface}

In this part, I will present some works on flavor changing neutral
current (FCNC) processes. As already emphasized, the FCNC
processes are forbidden at tree level in the SM, so it acts as the
ground to test the quantum structure of the SM; at the same time,
it is also the ideal place in searching new physics beyond the SM.

In the first section of this part, $t\bar c$ associated production
in the SM at linear colliders is presented, which corrects some
mistakes in literature; the second section is the $t\bar c $
production in supersymmetrical models, the FCNC process is
mediated by gluino; the third section is $b\bar s$ production in
the SM at linear colliders; the fourth section is about inclusive
process of $B$ meson decay in a CP-softly broken two-Higgs-doublet
model; the last section is given to $B$ exclusive decay to $K$ and
lepton pair in supersymmetrical models.

\newpage
\section{ Top-Charm Associated Production at High Energy
$e^{+}e^{-}$ Colliders in Standard Model}

\begin{footnotesize}
\begin{center}\begin{minipage}{5in}
\begin{center} ABSTRACT\end{center}
The flavor changing neutral current tcV(V=$\gamma$,Z)
couplings in the production vertex
for the process $e^+e^-\rightarrow t\bar c\mbox{ or }\bar t c$ in the standard model are investigated.
The precise calculations keeping all quark masses non-zero are carried out.
áThe total production cross section is found to be  $1.84 \times 10^{-9}$ fb
at $\sqrt s$=200 Gev and  $0.572 \times 10^{-9}$ fb at $\sqrt s$=500 Gev respectively. The result
is  much smaller than that given in
ref.~\cite{clwy} by a factor of $10^{-5}$.
\end{minipage}\end{center}
\end{footnotesize}

Top quark physics has been extensively investigated~\cite{rev}. The advantage of
examining top quark physics than other quark physics is that one can directly determine
the properties of top quark itself and does not need to worry about non-perturbative QCD
effects which are difficult to attack because there exist no top-flavored hadron states
at all. The properties of top quark could reveal information on flavor
physics, electroweak symmetry breaking as well new physics beyond the standard
model(SM).

One of important fields in top physics is to study flavor changing
neutral current (FCNC) coupings. There are no flavor changing
neutral currents at tree-level in the SM. FCNC appear
at loop-levels and consequently offer a good place to test quantum effects of
the fundamental quantum field theory on which SM based. Furthermore, they are
very small at one loop-level due to the unitary of Cabbibo-Kobayashi-Maskawa
 (CKM) matrix. In models beyond SM new particles beyond the particles in SM
may appear in the loop and have significant contributions to flavor
changing transitions. Therefore, FCNC interactions give an ideal place to
search for new physics. Any positive observation of FCNC couplings
deviated from that in SM would unambiguously signal the presence of new
physics. Searching for FCNC is clearly one of important goals of high energy
colliders, in particular, $e^+e^-$ colliders~\cite{pro}.

The flavor changing transitions involving external up-type quarks which are
due to FCNC couplings are much more suppressed than those involving
external down-type quarks in SM. The effects for external up-type quarks
are derived by virtual exchanges of down-type quarks in a loop for which
GIM mechanism~\cite{gim} is much more effective because the mass splittings
between down-type quarks are much less than those between up-type quarks.
Therefore, the tc transition which is studied in the latter
opens a good window to search for new physics.

The FCNC vertices tcV(V=$\gamma$, Z) can be probed either in rare decays
of t quark or via top-charm associated production.
A lot of works have been done in the former case~\cite{rd}. And a number of
papers on the latter case have also appeared~\cite{hh,clwy,many}. In this letter we shall
investigate the latter case in the process \begin{eqnarray}
e^+e^-\rightarrow t\bar{c}\mbox{ or } \bar{t}c.
\end{eqnarray}
Comparing t quark rare decays where the momentum transfer $q^2$ is
limited, i. e., it should be less or equal to mass square of t quark
$m_t^2$, the production process (1) allows the large (time-like) momentum
transfer, which is actually determined by the energies available at
$e^+e^-$ colliders. The reaction (1) has some advantages because of the
ability to probe higher dimension operators at large momenta and striking
kinematic signatures which are straightforward to detect in the clean
environment of $e^+e^-$ collisions. In particular, in some extensions of
SM which induce FCNC there are large underlying mass scales and large
momentum transfer so that these models are more naturally probed via
t$\bar{c}$ associated production than t quark rare decays.

The production cross sections of the process (1) in SM have been calculated in
refs.~\cite{clwy,many}. In the early references~\cite{many} a top quark mass
$m_t\le m_Z$ is assumed and the on-shell Z boson dominance is adopted. The
reference~\cite{clwy} considered a large top quark mass and abandoned the
on-shell Z boson dominance. However, the "self energy" diagrams  have been
omitted in ref.~\cite{clwy}. This is  not legal because the one-loop
contribution for FC transitions is of the leading term of the FC transitions
and must be finite, i.e., although there are some divergences for some
diagrams they should cancel each other in the sum of contributions of all
diagrams. Furthermore, the order of values of cross sections  given in
ref.~\cite{clwy} is not correct.

The order of values of cross sections for the process (1) in SM can easily
be estimated. The differential cross section can be written as
\begin{eqnarray}
\frac{d\sigma}{dcos\theta} ={\frac{N_c}{32\pi s}} (1-\frac{m^2_t}{s})
 {\frac{1}{4}} {\sum_{spins}|M|^2}
\end{eqnarray}
Where $N_c$ is the color factor, $\theta$ is the the angle between
incoming electron $e^{-}$ and outgoing top quark $t$ and M is the amplitude of
the process. In eq.(2) the charm quark mass in kenetic factors has been
omitted. Due to the GIM mechanism, one has
\begin{eqnarray}
\sum_{spins}|M|^2 & = & e^8 |\sum_{j=d,s,b} V_{jt}^{\star}V_{jc} f(x_j, y_j)|^2
 \nonumber \\ & = &  e^8 |V_{tb}^{\star} V_{cb} \frac{m_b^2-m_s^2}{m_w^2}
\frac {\partial f}{\partial x_j}|_{x_j,y_j=0} + ...|^2,
\end{eqnarray}
where $x_j=m_j^2/m_w^2, y_j=m_j^2/s$, and "..." denote the less important
terms for $\sqrt s\ge 200$ Gev. Assuming $\frac {\partial
f}{\partial x_j}|_{x_j,y_j=0} $ = O(1), one obtains from eqs. (2),(3)\\
$$ \sigma \sim 10^{-8}-10^{-9} fb$$
at $\sqrt s$ = 200 Gev. However, the results given in ref.~\cite{clwy} are\\
$$ \sigma = 0.71\times 10^{-2} fb$$
for $m_t$=165 Gev and
$$ \sigma = 4.1\times 10^{-4} fb$$
for $m_t$=190 Gev, which are much larger than the above estimation by a factor
of $10^{5}$. In order to test SM and search for new physics from observations
of some process one needs to know what are the precise results for the relevant
observables of the process in SM. Therefore, it is necessary to calculate
precisely the cross sections in the SM. In this letter we calculate the differential and
total cross sections  of the process (1) in SM.

In SM for the process (1) there are three kinds of Feynman diagram at one
loop, "self enengy" (actually it is a FC transition, not a usual self energy
diagram), triangle and box diagram, which are shown in Fig.1. We carry out
calculations in the Feynman-t'Hooft gauge. The contributions of the neutral
Higgs H and Goldstone bosons $G^{0,\pm}$ which couple to electrons are
neglected since they are proportional to the electron mass and we have put the
mass of electron to zero.
\par
We do the reduction using FeynCalc ~\cite{3} and keep all masses non-zero except
for the mass of electron. To control
the ultraviolet divergence, the dimensional regularization is used. As a
consistent check, we found that all divergences are canceled in the sum.  The
calculations are carried out in the frame of the centre of mass system (CMS)
and Mandelstam variables have been employed: \begin{eqnarray}
s=(p_1 +p_2)^2 =(k_1 +k_2)^2 \hspace{7mm} t=(p_1 -k_1)^2 \hspace{7mm}
u=(p_1 -k_2)^2,
\end{eqnarray}
 where
$p_1,p_2$ are the
momenta of electron and positron respectively, and $k_1,k_2$ are the momenta of
top quark $t$ and anti-charm quark $\bar{c}$ respectively.
\par
The amplitude of process $e^{+}e^{-} \rightarrow t\bar{c}$ can be
expressed as
{\small \begin{eqnarray} M &=&  \sum_{j=d,s,b} 16\pi^2 \alpha^2
V_{cj}^{\star}V_{tj} [ g_1  \bar{u_t}  \gamma^{\mu}  P_L v_c
\bar{v_e}  \gamma_{\mu}  P_R  u_e  + g_2 \bar{u_t}  \gamma^{\mu}
P_L  v_c  \bar{v_e}  \gamma_{\mu}  P_L  u_e  +  g_3 \bar{u_t}  P_L
v_c  \bar{v_e}  \not\!{k_1}  P_R  u_e  +\nonumber\\ && g_4
\bar{u_t}  P_L  v_c  \bar{v_e}  \not\!{k_1}  P_L  u_e  +  g_5
\bar{u_t} \not\!{p_1}  P_L  v_c \bar{v_e}  \not\!{k_1}  P_L  u_e +
  g_6 \bar{v_e} \gamma^{\mu}  P_L  u_e  \bar{u_t}  \gamma_{\mu}  \not\!{p_1}  P_L  v_c  + \nonumber\\
&& g_7 \bar{u_t} \gamma^{\mu}  P_R  v_c  \bar{v_e}  \gamma_{\mu} P_R  u_e  +
g_8 \bar{u_t}  \gamma^{\mu}  P_R  v_c  \bar{v_e}  \gamma_{\mu}  P_L  u_e +  g_9 \bar{u_t}  P_R  v_c  \bar{v_e}  \not\!{k_1}  P_R  u_e  +\nonumber\\
&& g_{10} \bar{u_t}P_R  v_c  \bar{v_e}  \not\!{k_1}  P_L  u_e  +  g_{11} \bar{v_e}  \gamma^{\mu}  P_L  u_e  \bar{u_t}  \gamma_{\mu} \not\!{p_1}  P_R  v_c ]
\end{eqnarray}
} where $\alpha$ is fine structure constant, $V_{ij}$ is CKM
matrix element, $P_L$ is defined as $(1-\gamma^5)/2$, and $P_R$ is
defined as $(1+\gamma^5)/2$. The exact expressions of the
coefficients $g_j(j=1,2,...11)$ are too long to be given. Instead,
in order to show the essential points, we give them in the limit
of $m_i$/m (i=d,s,c, m=$m_w, m_t, s)$ approach to zero.  In the
limit $g_j(j=7,8,9,10,11)$ is zero, and the others are given as
follows.
{\small \begin{eqnarray} g_1  &=&  a_3m_j^2 - 2a_4m_j^2s_w^4 +
6a_4C_2^cm_j^2m_t^2s_w^2 + 6m_w^2(2C_{11}^dm_t^2 + 2C_{22}^ds +
2C_{12}^d(m_t^2 + s) )(a_3 + 2a_4c_w^2s_w^2)  +\nonumber\\ &&
12m_w^2C_2^d(a_3(m_t^2 + s) + a_4( 2sc_w^2s_w^2 + m_t^2c_w^2s_w^2
- m_t^2s_w^4)) - B_0^b(m_j^4 - m_j^2m_t^2 + m_j^2m_w^2 +
\nonumber\\ && 2m_t^2m_w^2 - 2m_w^4)(a_1 + 2a_2s_w^2(3 - 4s_w^2))
+ 12C_{00}^d(a_3(m_j^2 + 6m_w^2) + a_4s_w^2(c_w^2m_j^2 +
12c_w^2m_w^2 - \nonumber\\ && m_j^2s_w^2)) +
6C_1^dm_w^2(2a_3(m_t^2 + s) + 2a_4s_w^2(c_w^2m_t^2 + 2c_w^2s -
m_t^2s_w^2)) - 6C_0^dm_w^2(2a_3(m_j^2  - s) -\nonumber\\ &&
2a_4s_w^2(2c_w^2m_t^2 + 2c_w^2s + 2m_j^2s_w^2 - m_t^2s_w^2
-3m_t^2c_w^2 )) + 2C_0^cm_j^2(a_3(m_j^2 + 2m_w^2 - m_t^2) +
\nonumber\\ && a_4s_w^2(3m_j^2 - 2m_j^2s_w^2 - 4m_w^2s_w^2 +
2m_t^2s_w^2)) - 2(2C_{00}^c + C_{11}^cm_t^2 + C_{22}^cs + C_2^cs +
\nonumber\\ && C_{12}^c(m_t^2 + s))(a_3(m_j^2 + 2m_w^2) +
2a_4s_w^2(3m_w^2 - m_j^2s_w^2 - 2m_w^2s_w^2)) + \nonumber\\ &&
B_0^a( a_1(m_j^2 - m_w^2)(m_j^2 + 2m_w^2) -2a_1m_t^2m_j^2 +
2a_2s_w^2(3m_j^4 - 6m_j^2m_t^2 + 3m_j^2m_w^2 - 6m_w^4 -\nonumber\\
&&  4m_j^4s_w^2 - 4m_j^2m_w^2s_w^2 + 8m_w^4s_w^2 +
8m_t^2m_j^2s_w^2 )) - 2C_1^c(a_3( 2m_w^2s +m_t^2m_j^2) -
\nonumber\\ && a_4s_w^2(3m_j^2m_t^2  + 2m_t^2m_j^2s_w^2 - 6m_w^2s
+ 4m_w^2ss_w^2))
\end{eqnarray}
\begin{eqnarray}
g_2  &=&   a_3m_j^2 + a_4m_j^2(1 - 2s_w^2)(s_w^2 - 3C_2^cm_t^2) + a_5(8D_{00}^e +u(D_1^e +D_2^e +2D_3^e +2D_{12}^e +4D_{13}^e +\nonumber\\
&& 2D_{23}^e +2D_{33}^e) + (2m_t^2D_3^e +2sD_{13}^e +2D_{33}^em_t^2) ) + 6m_w^2(2C_{11}^dm_t^2 + 2C_{22}^ds +\nonumber\\
&& 2C_{12}^d(m_t^2 + s))(a_3 - a_4c_w^2(1 - 2s_w^2))  + 6m_w^2C_2^d(2a_3(m_t^2 +s) - a_4(1-2s_w^2)(m_t^2c_w^2 -m_t^2s_w^2 +\nonumber\\
&& 2sc_w^2)) - B_0^b(m_j^4 - m_j^2m_t^2 + m_j^2m_w^2 + 2m_t^2m_w^2 - 2m_w^4)(a_1 - a_2(3 - 4s_w^2)(1 - 2s_w^2)) + \nonumber\\
&& 6C_{00}^d(2a_3(m_j^2 + 6m_w^2) - a_4(1 - 2s_w^2)(c_w^2m_j^2 + 12c_w^2m_w^2 - m_j^2s_w^2)) + 6C_1^dm_w^2(2a_3(m_t^2 + s) - \nonumber\\
&& a_4(1 - 2s_w^2)(c_w^2m_t^2 + 2c_w^2s - m_t^2s_w^2)) - 6C_0^dm_w^2(2a_3(m_j^2  - s) + a_4(1 - 2s_w^2)(- c_w^2m_t^2 + 2c_w^2s + \nonumber\\
&& 2m_j^2s_w^2 - m_t^2s_w^2)) + C_0^cm_j^2(2a_3(m_j^2 + 2m_w^2 - m_t^2) - a_4(1 - 2s_w^2)(3m_j^2 - 2m_j^2s_w^2 - 4m_w^2s_w^2 +\nonumber\\
&& 2m_t^2s_w^2)) - 2(2C_{00}^c + C_{11}^cm_t^2 + C_{22}^cs + C_2^cs + C_{12}^c(m_t^2 + s))(a_3(m_j^2 + 2m_w^2) - \nonumber\\
&& a_4(1 - 2s_w^2)(3m_w^2 - m_j^2s_w^2 - 2m_w^2s_w^2)) + B_0^a(a_1(m_j^2 - m_w^2)(m_j^2 + 2m_w^2) - 2a_1m_t^2m_j^2 - \nonumber\\
&& a_2(1 - 2s_w^2)(3m_j^4 - 6m_j^2m_t^2 + 3m_j^2m_w^2 - 6m_w^4 - 4m_j^4s_w^2 - 4m_j^2m_w^2s_w^2 + 8m_w^4s_w^2 + 8m_t^2m_j^2s_w^2)) - \nonumber\\
&& C_1^c(2a_3(m_t^2m_j^2 + 2sm_w^2) + a_4(1 - 2s_w^2)(3m_j^2m_t^2  - 6sm_w^2  + 4sm_w^2s_w^2 + 2m_t^2m_j^2s_w^2))
\end{eqnarray}
\begin{eqnarray}
g_3  &=&  12a_4s_w^2m_t(2C_2^dm_w^2 -C_2^cm_j^2) + 4m_tC_0^cm_j^2(a_3 - 2a_4s_w^4) + 24m_tm_w^2(2C_1^d + C_0^d)(a_3 + 2a_4c_w^2s_w^2)  +\nonumber\\
&& 8C_1^cm_t(a_3m_j^2 - 2a_4s_w^4m_j^2) + 12m_t(C_{11}^d +C_{12}^d)(a_3(m_j^2 + 2m_w^2) + a_4s_w^2(c_w^2m_j^2 + 4c_w^2m_w^2 - m_j^2s_w^2)) +\nonumber\\
&& 4m_t(C_{11}^c + C_{12}^c)(a_3(m_j^2 + 2m_w^2) + 2a_4s_w^2(3m_w^2 - m_j^2s_w^2 - 2m_w^2s_w^2))
\end{eqnarray}
\begin{eqnarray}
g_4  &=&  - 2a_5m_t(2D_{23}^e + D_2^e + 2D_{33}^e + 2D_3^e) + 6a_4m_t(C_2^cm_j^2 -2C_2^dm_w^2)(1 - 2s_w^2) +  4C_0^cm_tm_j^2(a_3 + \nonumber\\
&& a_4s_w^2(1 - 2s_w^2)) + 24m_tm_w^2(2C_1^d + C_0^d )(a_3 - a_4c_w^2(1 - 2s_w^2))  +  8C_1^cm_t(a_3m_j^2 + \nonumber\\
&& a_4s_w^2m_j^2(1 - 2s_w^2)) + 6(C_{11}^d + C_{12}^d)m_t(2a_3(m_j^2 + 2m_w^2) - a_4(1 - 2s_w^2)(c_w^2m_j^2 + 4c_w^2m_w^2 - \nonumber\\
&& m_j^2s_w^2)) + 4m_t(C_{11}^c +C_{12}^c)(a_3(m_j^2 + 2m_w^2) - a_4(1 - 2s_w^2)(3m_w^2 - m_j^2s_w^2 - 2m_w^2s_w^2))    \\
g_5  &=&  - 4a_5( D_{12}^e + D_{13}^e )  \\
g_6  &=& a_5m_t( 2D_{12}^e + 2D_{13}^e  + 2D_{23}^e + D_2^e +2D_{33}^e + 2D_3^e)
\end{eqnarray}
} with $m_j^2=m_b^2$(since $m_s,m_d$ have been omitted in the
above expressions of g's),\\ where $a_i (i=1,2,...,5)$ are defined
by
\begin{eqnarray}
a_1=\frac{1}{96s\pi^2s_w^2m_t^2m_w^2},\hspace{3mm} a_2=\frac{1}{768\pi^2c_w^2s_w^4m_t^2m_w^2 (m_z^2-im_z\Gamma{z}-s)},
 \hspace{3mm} a_3=\frac{1}{192s\pi^2s_w^2 m_w^2}  \nonumber\\
 a_4=\frac{1}{384\pi^2c_w^2s_w^4m_w^2 (m_z^2-im_z\Gamma{z}-s)},\hspace{3mm} a_5=\frac{1}{32\pi^2s_w^4} \nonumber
\end{eqnarray}
with $ c_w= cos\theta_w$ and $s_w= sin\theta_w$.
In the presentation of $g_j$ above, we have used the definition of scalar
integrals $Bs$, $Cs$,and $Ds$\cite{3}, and these
functions, $Bs$, $Cs$,and $Ds$, with superscripts a,b,...,e have the arguments
\\
\begin{math}
(0,m_j^2,m_w^2),\hspace{3mm} (m_t^2,m_j^2,m_w^2),\hspace{3mm}
(m_t^2,0,s,m_j^2,m_w^2,m_j^2), \hspace{3mm} (m_t^2,0,s,m_w^2,m_j^2,m_w^2)\\
\hspace{3cm} (0,s,m_t^2,u,0,0,0,m_w^2,m_w^2,m_j^2)\\
\end{math}
respectively. Here $m_j$ denotes the mass of down-type quark b.

In the numerical calculations the following values of the
parameters have been used ~\cite{4}:
\begin{eqnarray}
m_e=0,\hspace{4mm}m_c=1.4Gev,\hspace{4mm}m_t=175Gev, \hspace{4mm}m_d=0.005Gev,\hspace{4mm}m_s=0.17Gev,\hspace{4mm}\nonumber\\
m_b=4.4Gev,\hspace{4mm}m_w=80.41Gev,\hspace{4mm}m_z=91.187Gev,\hspace{4mm} \Gamma_z=2.5Gev,\hspace{4mm} \alpha=\frac{1}{128} \nonumber
\end{eqnarray}
\par
In order to keep the unitary condition of CKM matrix exactly, we employ the
standard parametrization and take the values~\cite{4,ckm}
\begin{eqnarray}
s_{12}=0.220,\hspace{7mm} s_{23}=0.039,\hspace{7mm}
s_{13}=0.0031,\hspace{7mm} \delta_{13}=70^{\circ} . \nonumber
\end{eqnarray}

Numerical results are shown in Figs. 2, 3.
In Fig.2, we show the total cross section $\sigma_{tot}$ of the process
$e^{+}e^{-} \rightarrow t\bar{c}$ as a function of the centre of mass energy
$\sqrt{s}$.  One can see from the figure that the
total cross section is the order of $10^{-10} \sim 10^{-9}$ fb, as expected, and decreases when center-of-mass energy
increases and is large enough ($\ge 250$ Gev ).
We fixed the centre of mass energy $\sqrt{s}$ at $200 Gev$. Differential cross
section of the process at the energy as a function of $\cos{\theta}$
 is shown in Fig.3.
\par
To summarize, we have calculated the production cross sections of the process
$e^{+}e^{-} \rightarrow t\bar{c}$ in SM.  We found that the total cross
section is $1.84\times 10^{-9} fb$ at $\sqrt s$ = 200 Gev and $0.572\times 10^{-9}$ fb at
$\sqrt s$ = 500 Gev. It is too small to be of experimental relevance.
Therefore, this is a remarkable situation that allows for a precise test of the
SM and, in particular, of the GIM mechanism in SM. Even a small number of
$t\bar c$ events, detected at LEP II or a NLC running with a yearly integrated
luminosity of ${\cal L}\ge 10^2 [fb]^{-1}$, will unambiguously indicate new
FCNC dynamics beyond SM.

\begin{figure}
\epsfxsize=6cm
\centerline{\epsffile{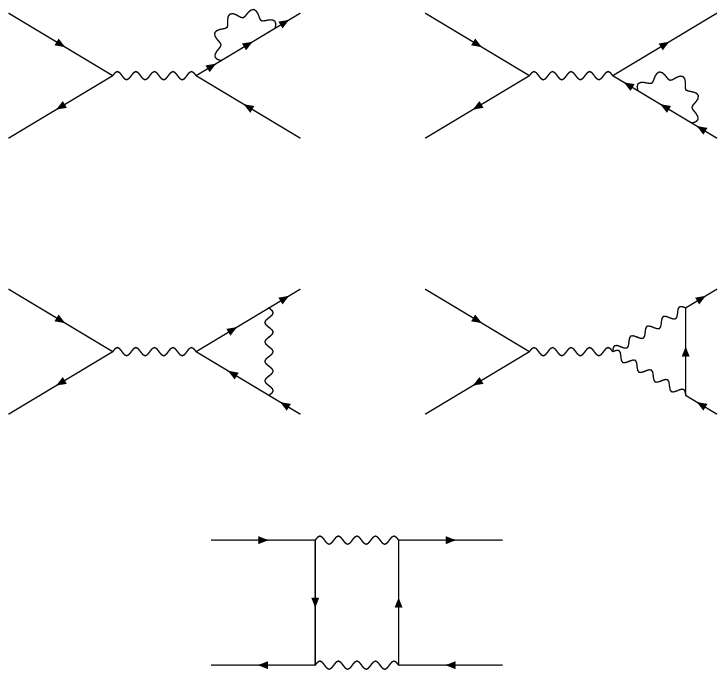}}
\caption[]{Feynman diagrams of
prosess $e^+e^- \rightarrow t \bar c$}
\end{figure}

\begin{figure}
\epsfxsize=6cm
\centerline{\epsffile{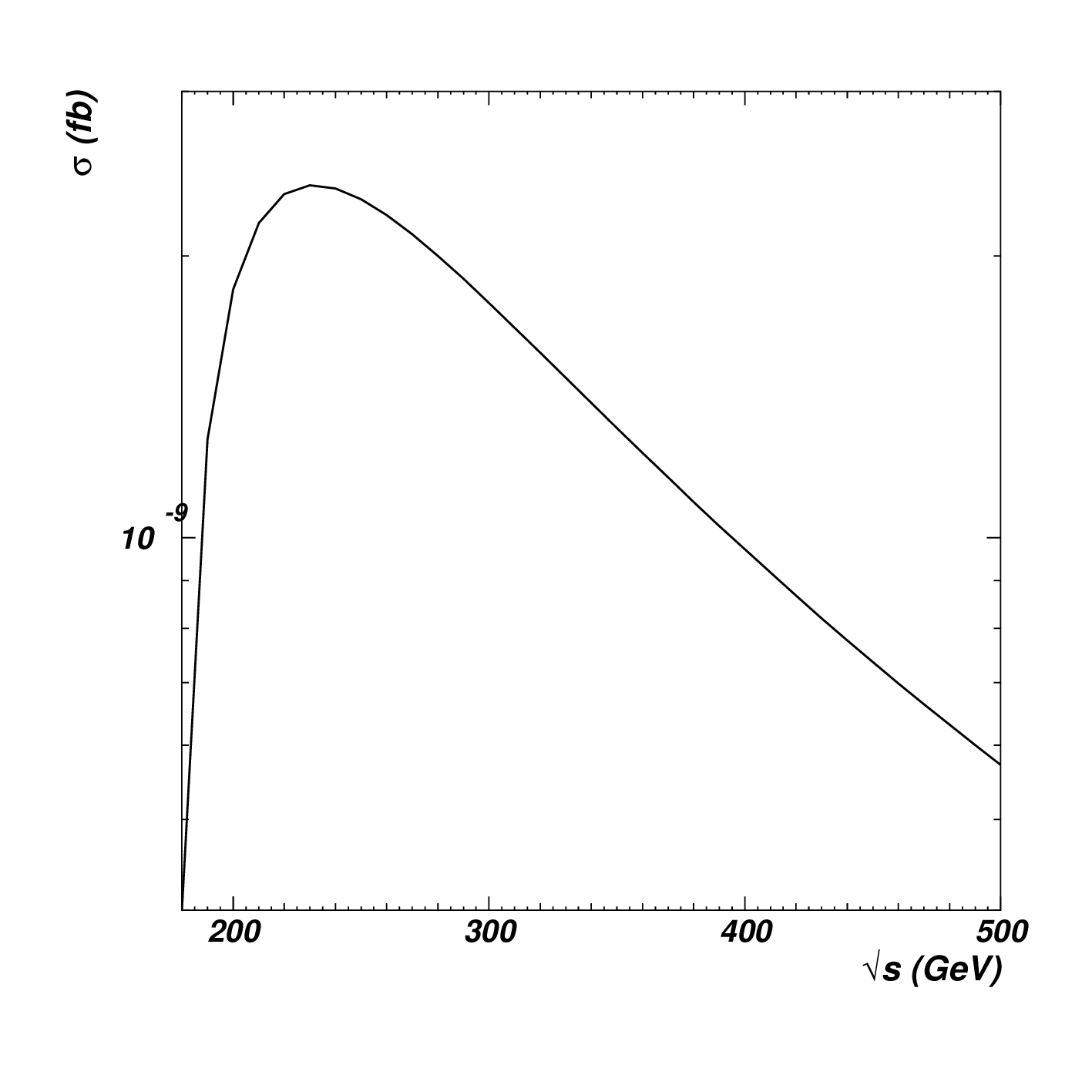}}
\caption[]{ Cross section of the
process $e^+e^- \rightarrow t \bar c$ as a function of
$\sqrt{s}$.}
\end{figure}

\begin{figure}
\epsfxsize=6cm
\centerline{\epsffile{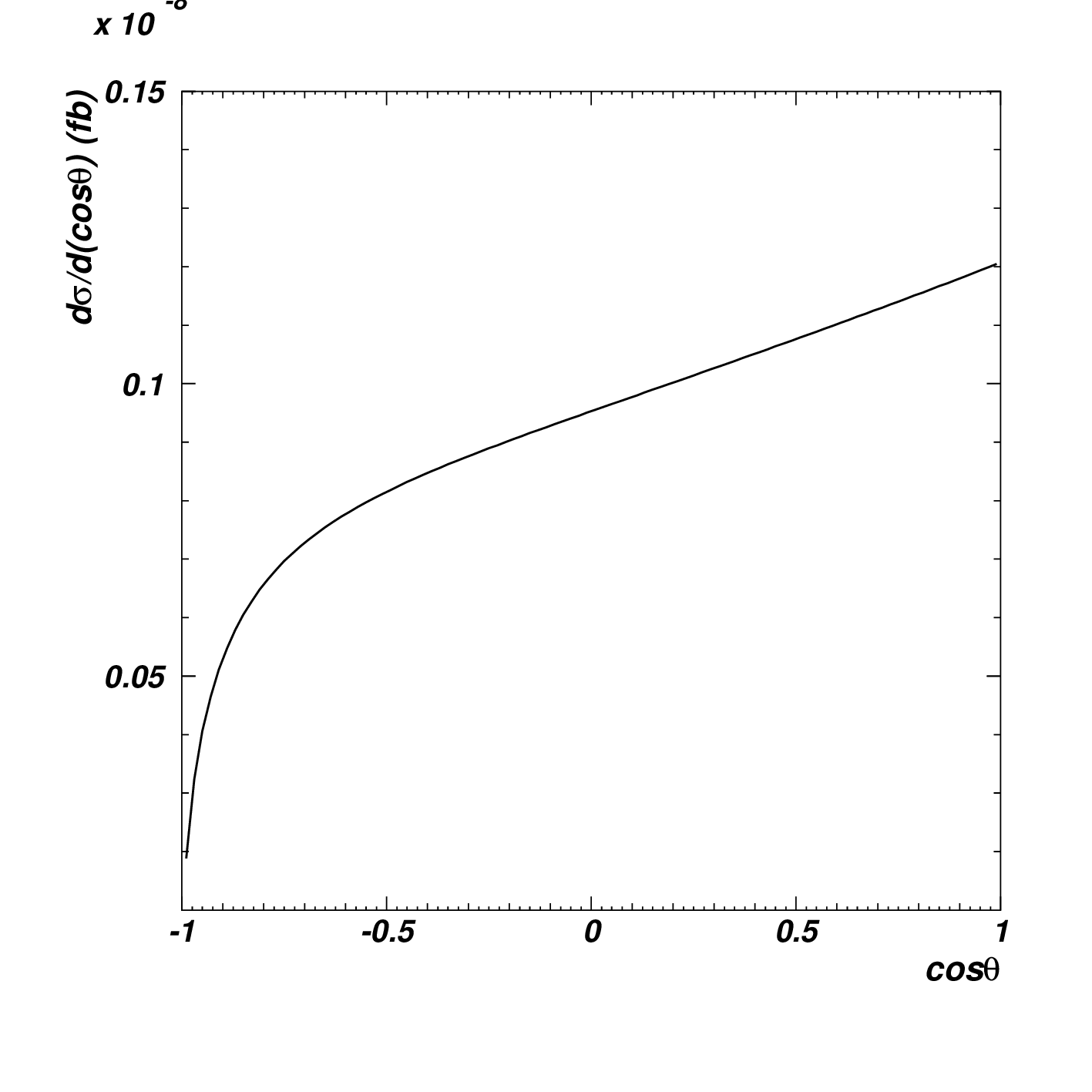}}
\caption[]{  Differential cross section of the
process $e^+e^- \rightarrow t \bar c$, where
$\sqrt{s}=200$ GeV.
}
\end{figure}


\setcounter{figure}{0} \setcounter{table}{0}
\setcounter{equation}{0}

\newpage

\section{SUSY-QCD Effect on Top-Charm Associated Production
at Linear Collider     }

\begin{footnotesize}
\begin{center}\begin{minipage}{5in}
\begin{center} ABSTRACT\end{center}
We evaluate the contribution of SUSY-QCD to top-charm associated
production at next generation linear colliders. Our results show
that the production cross section of
the process $e^+e^-\rightarrow t\bar c\mbox{ or }\bar t c$ could be as
large as
$0.1$ fb, which is larger than the
prediction of the SM by a factor of $10^8$.
\end{minipage}\end{center}
\end{footnotesize}


One of the most important physics in top quark sector is to
probe anomalous flavor changing
neutral current (FCNC) couplings.
In the Standard Model (SM), FCNC couplings are forbidden at
the tree level and much suppressed in loops by the GIM
mechanism. Any signals on FCNC couplings in
the
processes of
top
quark decay and productions or indirectly in loops will
indicate the existence of
new physics beyond the SM. Recently in the framework of
effective lagrangian, Han and Hewett\cite{han}
have examined carefully the possibility of exploring
the FCNC couplings $tcZ/ tc \gamma$ in the production vertex
for the reaction $e^+ e^- \rightarrow t {\bar c} + {\bar t}
c$ and concluded that at higher energy colliders with $0.5
-1$ TeV center-of-mass energy, the resulting sensitivity to
FCNC couplings will be
better than the present constraints \cite{zhang}. In this
paper, in the minimal supersymmetric standard model (MSSM)
we study the
process $e^+ e^- \rightarrow t {\bar c}
+ {\bar t}
c$
and perform an detail calculation of the contribution from
the FCNC
couplings in the vertex
of gluino-squark-quark
to the production cross section. We will point out that
at higher energy $e^+ e^-$ colliders
the cross
section could be as large as 0.1 fb
which is at least eight order of magnitude larger than the
prediction of the SM $\sim 10^{-10} - 10^{-9}$
fb \cite{sm}.

The MSSM is arguably
the most promising candidate for physics beyond the SM.
Beside many attractive features of supersymmetry in
understanding the mass hierarchy, gauge coupling unification,
the weak scale SUSY models in generally lead to a rich flavor
physics. In fact, SUSY models often have arbitrary flavor
mixings and mass parameters in the squark and slepton sectors
and these mass matrices after diagonalization induce FCNC
couplings at tree level in the vertex of
gluino-squark-quark ${\it etc}$.
Phenomenologically one would have to assume certain
symmetries or dynamical
mechanisms to prevent large FCNC among the first and second
generations. On the other hand the
flavor structure, especially among the second and third
generations in the SUSY sector motivates us to seek for
new physics and any experimental observation on the FCNC
processes beyond the SM would undoubtedly shed light on our
understanding for flavor physics. In this paper we take
model of Ref. \cite{Ellis,Duncan2} where the FCNC couplings
relevant to our calculation is given by:
\begin{eqnarray}
{\cal L_{FC}}=-\sqrt{2}g_sT^aK\overline{\tilde g}P_Lq\tilde{q}_L + h.c.
\end{eqnarray}
In (1), K is the supersymmetric version of the
Kobayashi--Maskawa
matrix, which is explicitly expressed as:
\begin{eqnarray}
K_{ij}=\left(\matrix{1&\varepsilon&\varepsilon^2\cr
 -\varepsilon&1&\varepsilon\cr-\varepsilon^2&-\varepsilon&1\cr}\right)
\end{eqnarray}
where $\epsilon$ parameterizes the strength of flavor mixing
and
is shown to be as large as $1/2$ without
contradicting with the low energy experimental
data \cite{Duncan2}.

In Fig.(1) we give the Feynman diagrams for the
process $e^+(p_1) e^-(p_2)
\rightarrow t(k_1) \bar{c}(k_2)$.
 In
calculations,
we have neglected the scalar u-quark contribution since it is
highly suppressed
by $K_{12}K_{13}$; and we use
the dimensional regularization to control the ultraviolet
divergence.
We have checked that all divergences cancel out in the final
result with the
summing up of all of the diagrams.
The calculations are carried out in the frame of the center of mass system (CMS)
and Mandelstam variables have been employed:
\begin{eqnarray}
s=(p_1 +p_2)^2 =(k_1 +k_2)^2 \hspace{7mm} t=(p_1 -k_1)^2 \hspace{7mm}
u=(p_1 -k_2)^2.
\end{eqnarray}

After a straightforward calculations, one obtains for the
amplitudes
\begin{eqnarray}
M&=&{e \over S}\bar{v}(p_1)\gamma_{\mu} u(p_2) \bar{u}(k_1)
V^\mu (tc\gamma) v(k_2) \nonumber \\
&+& {g \over 2 \cos \theta_W (S-M_Z^2)}
\bar{v}(p_1)\gamma_{\mu} (g_V^e-g_A^e \gamma_5) u(p_2) \bar{u}(k_1)
V^\mu (tcZ) v(k_2)
\end{eqnarray}
where,
$g_V^e=1/2-2 \sin^2 \theta_W$, $g_A^e=1/2$, and
$V^\mu (tc\gamma)$ and $V^\mu (tcZ)$ are the on-shell quarks effective vertices
given by \footnote{For simplicity, we only give the results in the
limit of $m_c =0$. However in our numerical calculations, we use the
full formulas. }
\begin{eqnarray}
V^\mu(tc\gamma;Z)&=&
f_1^{\gamma;Z} \gamma_\mu P_R+
f_2^{\gamma;Z} \gamma_\mu P_L+
f_3^{\gamma;Z} k_{1\mu} P_R +
f_4^{\gamma;Z} k_{1\mu} P_L
\nonumber \\
&& +
f_5^{\gamma;Z} k_{2\mu} P_R +
f_6^{\gamma;Z} k_{2\mu} P_L.
\end{eqnarray}
The form factors, $f_i^{\gamma;Z}$ are
\begin{eqnarray}
f_1^\gamma &=&
\sum_{\tilde{q}=
\tilde{c}, \tilde{t}}
{(\pm 1)\epsilon e g_s^2 \cos (\theta_{\tilde{q}}) \sin(\theta_{\tilde{q}})
m_{\tilde{g}}
\over
12 m_t \pi^2 } [
B_0(0, m_{\tilde{g}}^2, m_{\tilde{q}_2}^2)
- B_0(m_t^2, m_{\tilde{g}}^2, m_{\tilde{q}_2}^2)]
+ R.R.
\nonumber \\
f_2^\gamma &=&
\sum_{\tilde{q}=
\tilde{c}, \tilde{t}}
{(\pm 1)\epsilon e g_s^2 \sin^2 (\theta_{\tilde{q}}) \over
24 m_t^2 \pi^2 } [
 (m_{\tilde{g}}^2-m_{\tilde{q}_2}^2) B_0(0, m_{\tilde{g}}^2, m_{\tilde{q}_2}^2)
- (m_{\tilde{g}}^2-m_{\tilde{q}_2}^2+ m_t^2)
B_0(m_t^2, m_{\tilde{g}}^2, m_{\tilde{q}_2}^2)
\nonumber \\
&&+
4 m_t^2 C_{00} ]
+ R.R
\nonumber \\
f_3^\gamma &=&
\sum_{\tilde{q}=
\tilde{c}, \tilde{t}}
{(\mp 1)\epsilon e g_s^2 \sin(\theta_{\tilde{q}})
\cos (\theta_{\tilde{q}})
m_{\tilde{g}}
 \over
12  \pi^2 } [ C_0+2 C_1] + R.R.
\nonumber \\
f_4^\gamma &=&
\sum_{\tilde{q}=
\tilde{c}, \tilde{t}}
{(\pm 1)\epsilon e g_s^2 \sin(\theta_{\tilde{q}})
\cos (\theta_{\tilde{q}})
m_{\tilde{g}}
 \over
12  \pi^2 } [ C_0+2 C_2] + R.R.
\nonumber \\
f_5^\gamma &=&
\sum_{\tilde{q}=
\tilde{c}, \tilde{t}}
{(\pm 1)\epsilon e g_s^2 \sin^2(\theta_{\tilde{q}})
m_{t}
 \over
12  \pi^2 } [ C_0+2 C_{11}] + R.R.
\nonumber \\
f_6^\gamma &=&
\sum_{\tilde{q}=
\tilde{c}, \tilde{t}}
{(\mp 1)\epsilon e g_s^2 \sin^2(\theta_{\tilde{q}})
m_{t}
 \over
12  \pi^2 } [ C_0+2 C_{12}] + R.R.
\end{eqnarray}
\begin{eqnarray}
f_1^Z &=&
\sum_{\tilde{q}=
\tilde{c}, \tilde{t}}
{(\mp 1)\epsilon g g_s^2 \sin^2(\theta_w)
\cos (\theta_{\tilde{q}}) \sin(\theta_{\tilde{q}}) \over
12 m_t \cos (\theta_w) \pi^2 } [
B_0(0, m_{\tilde{g}}^2, m_{\tilde{q}_2}^2)
- B_0(m_t^2, m_{\tilde{g}}^2, m_{\tilde{q}_2}^2)]
+ R.R.
\nonumber \\
f_2^Z &=&
\sum_{\tilde{q}=
\tilde{c}, \tilde{t}}
{(\mp 1)\epsilon g g_s^2 \sin^2 (\theta_{\tilde{q}}) \over
96 m_t^2 \cos (\theta_w) \pi^2 } \{
 (-3+4 \sin^2(\theta_w)) [(m_{\tilde{g}}^2-m_{\tilde{q}_2}^2) B_0(0, m_{\tilde{g}}^2, m_{\tilde{q}_2}^2) \nonumber\\
&& - (m_{\tilde{g}}^2-m_{\tilde{q}_2}^2+ m_t^2)
B_0(m_t^2, m_{\tilde{g}}^2, m_{\tilde{q}_2}^2)]
+ 4 m_t^2 (-3 \sin^2 (\theta_{\tilde{q}})+4 \sin^2(\theta_w)) C_{00} \nonumber \\
&& -12 m_t^2 \cos^2 (\theta_{\tilde{q}})
\hat{C}_{00} \}
+ R.R
\nonumber \\
f_3^Z &=&
\sum_{\tilde{q}=
\tilde{c}, \tilde{t}}
{(\pm 1)\epsilon g g_s^2 \sin(\theta_{\tilde{q}})
\cos (\theta_{\tilde{q}})
m_{\tilde{g}}
 \over
48 \cos (\theta_w) \pi^2 } [
(4 \sin^2(\theta_w)-3 \sin^2 (\theta_{\tilde{q}})) ( C_0+2 C_1) \nonumber \\
&& + 3 \sin^2 (\theta_{\tilde{q}})
( \hat{C}_0+2 \hat{C}_1) ]  + R.R.
\nonumber \\
f_4^Z &=&
\sum_{\tilde{q}=
\tilde{c}, \tilde{t}}
{(\mp 1)\epsilon g g_s^2 \sin(\theta_{\tilde{q}})
\cos (\theta_{\tilde{q}})
m_{\tilde{g}}
 \over
48  \cos (\theta_w) \pi^2 } [
(4 \sin^2(\theta_w)-3 \sin^2 (\theta_{\tilde{q}})) ( C_0+2 C_2) \nonumber \\
&& + 3 \sin^2 (\theta_{\tilde{q}})
( \hat{C}_0+2 \hat{C}_2) ]  + R.R.
\nonumber \\
f_5^Z &=&
\sum_{\tilde{q}=
\tilde{c}, \tilde{t}}
{(\mp 1)\epsilon g g_s^2 \sin^2(\theta_{\tilde{q}})
m_{t}
 \over
48  \cos (\theta_w) \pi^2 } [
(4 \sin^2(\theta_w)-3 \sin^2 (\theta_{\tilde{q}})) ( C_0+2 C_{11}) \nonumber \\
&& - 3 \cos^2 (\theta_{\tilde{q}})
( \hat{C}_0+2 \hat{C}_{11}) ]  + R.R.
\nonumber \\
f_6^Z &=&
\sum_{\tilde{q}=
\tilde{c}, \tilde{t}}
{(\pm 1)\epsilon g g_s^2 \sin^2(\theta_{\tilde{q}})
m_{t}
 \over
48 \cos (\theta_w) \pi^2 } [
(4 \sin^2(\theta_w)-3 \sin^2 (\theta_{\tilde{q}})) ( C_0+2 C_{12}) \nonumber \\
&& - 3 \cos^2 (\theta_{\tilde{q}})
( \hat{C}_0+2 \hat{C}_{12}) ]  + R.R.
\end{eqnarray}
where $R.R.$ represents the replacement of
$\theta_{\tilde{q}} \rightarrow \pi/2+ \theta_{\tilde{q}}$
and $m_{\tilde{q}_1} \leftrightarrow m_{\tilde{q}_2}$.
The variables
of three point
functions $C_{i}$, $C_{ij}$ \cite{denner} and
$\hat{C}_{i}$, $\hat{C}_{ij}$
are
$(m_t^2, S, 0, m_{\tilde{g}}^2,m_{\tilde{q}_2}^2, m_{\tilde{q}_2}^2)$
 and $(m_t^2, S, 0, m_{\tilde{g}}^2,m_{\tilde{q}_2}^2, m_{\tilde{q}_1}^2)$,
 respectively.


 In the MSSM the mass eigenstates of the
squarks $\tilde{q}_1$ and $\tilde{q}_2$ are related to the weak
eigenstates $\tilde{q}_L$ and
$\tilde{q}_R$
by \cite{MSSM}
\begin{eqnarray}
\left(\begin{array}{c}
\tilde{q}_1 \\ \tilde{q}_2\end{array}\right)=
R^{\tilde{q}}\left(\begin{array}{c}
\tilde{q}_L \\ \tilde{q}_R\end{array}\right)\ \ \ \ \mbox{with}\ \ \ \
R^{\tilde{q}}=\left(\begin{array}{cc}
                   \cos\theta_{\tilde{q}} & \sin\theta_{\tilde{q}}\\
                   -\sin\theta_{\tilde{q}} & \cos\theta_{\tilde{q}}
                   \end{array}
                   \right).
\label{eqeq1}
\end{eqnarray}
For the squarks, the mixing angle $\theta_{\tilde{q}}$ and
the masses $m_{\tilde{q}_{1,2}}$ can be calculated by
diagonalizing the following mass
matrices
\begin{eqnarray}
M^2_{\tilde{q}}=\left(\begin{array}{cc}
          M_{LL}^2 & m_q M_{LR}\\
           m_q M_{RL} & M_{RR}^2
           \end{array} \right), \nonumber \\
M_{LL}^2=m_{\tilde{Q}}^2+m_q^2+m_{z}^2\cos 2\beta (I_q^{3L}-e_q\sin^2\theta_w),
\nonumber \\
M_{RR}^2= m_{\tilde{U},\tilde{D}}^2 +m_q^2+m_{z}^2\cos 2\beta e_q\sin^2\theta_w,
 \nonumber \\
M_{LR}= M_{RL}=\left\{ \begin{array}{ll}
                A_t-\mu \cot \beta & (\tilde{q}= \tilde{t})\\
                A_b-\mu \tan \beta & (\tilde{q}= \tilde{b}),
                \end{array}
                \right.
\label{eqeq2}
\end{eqnarray}
where $ m_{\tilde{Q}}^2$, $ m_{\tilde{U},\tilde{D}}^2$ are
 soft SUSY breaking mass terms of the left- and right-handed
 squark, respectively; $\mu$ is the
coefficient of the $H_1H_2$
term in the superpotential;
 $A_t$ and $A_b$ are the
coefficient of the dimension-three tri-linear soft SUSY-breaking
terms; $I_q^{3L}, e_q$ are the weak isospin and electric charge of
the squark $\tilde{q}$. From Eqs. \ref{eqeq1} and \ref{eqeq2}, we
have
\begin{eqnarray}
m^2_{\tilde{t}_{1,2}}&=&{1\over 2}\left[ M^2_{LL}+
M^2_{RR}\mp \sqrt{
(M^2_{LL}-M^2_{RR})^2+4 m^2_t M^2_{LR}}\right] \nonumber \\
\tan\theta_{\tilde{t}}&=&{m^2_{\tilde{t}_1}-M^2_{LL} \over m_t M_{LR}}.
\label{eq3}
\end{eqnarray}

 Now we present the numerical results.
For the SM parameters, we take
\begin{eqnarray}
m_Z=91.187 GeV, ~~~~m_W=80.33 GeV&&, ~~~~ m_t=176.0 GeV, ~~~~
m_c=1.4 GeV
\nonumber \\
\alpha=1/128 &&, ~~~~ \alpha_S=0.118
\end{eqnarray}
For the MSSM parameters, we choose $\mu=-100 GeV$ and
$\epsilon^2 =
1/4$.
To simplify the calculation we have taken
that  $m_{\tilde{U}} = m_{\tilde
{D}}=m_{\tilde{Q}}=A_t=m_S$ (global SUSY).
In Figs. 2-5, we show the cross sections of the process
$e^+e^- \rightarrow t \bar c$ as functions of $m_S$,
$m_{\tilde{g}}$,
$\sqrt{s}$ and $\tan\beta$. One can see that
the production cross section increases as squarks  and
gluino masses decrease, and it could reach $0.1 $ fb for
favorable parameters. This is an enhancement by a factor of
$10^8$
relative to the SM prediction.
 Such enhancement could be easily understood as following:
\begin{eqnarray}
\frac{\sigma_{SUSY}}{\sigma_{SM}}
\sim \left( \frac{\alpha_s \Delta m_{\tilde{q}}^2}{\alpha m_b^2} \right)^2,
\end{eqnarray}
where $\Delta m_{\tilde{q}}^2$ represents the possible mass square difference
among squarks. If $\Delta m_{\tilde{q}}^2$ varies from $100^2 -200^2 (GeV)^2$,
$\frac{\sigma_{SUSY}}{\sigma_{SM}}= 10^7 \sim 10^8$. At the same
time, this kind of enhancement could also be observed
in FCNC decay process of top quark
\cite{decay}.
Due to the
rather clean experimental environment and well-constrained
kinematics, the signal of ${\bar t} c$ or $t {\bar c}$
would be spectacular \cite{han}. We expect the
SUSY-QCD effects studied in this paper be observed at higher
energy $e^+ e^-$ colliders.


\begin{figure}
\epsfxsize=7 cm
\centerline{\epsffile{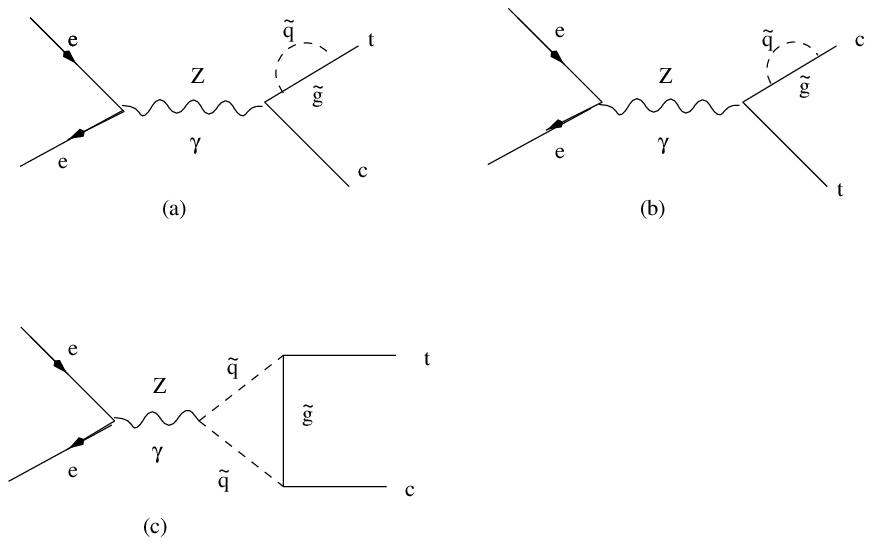}}
\caption[]{
The Feynmann diagrams for the process $e^+ e^- \rightarrow t \bar{c}$.
}
\end{figure}

\begin{figure} 
\epsfxsize=7 cm 
\centerline{\epsffile{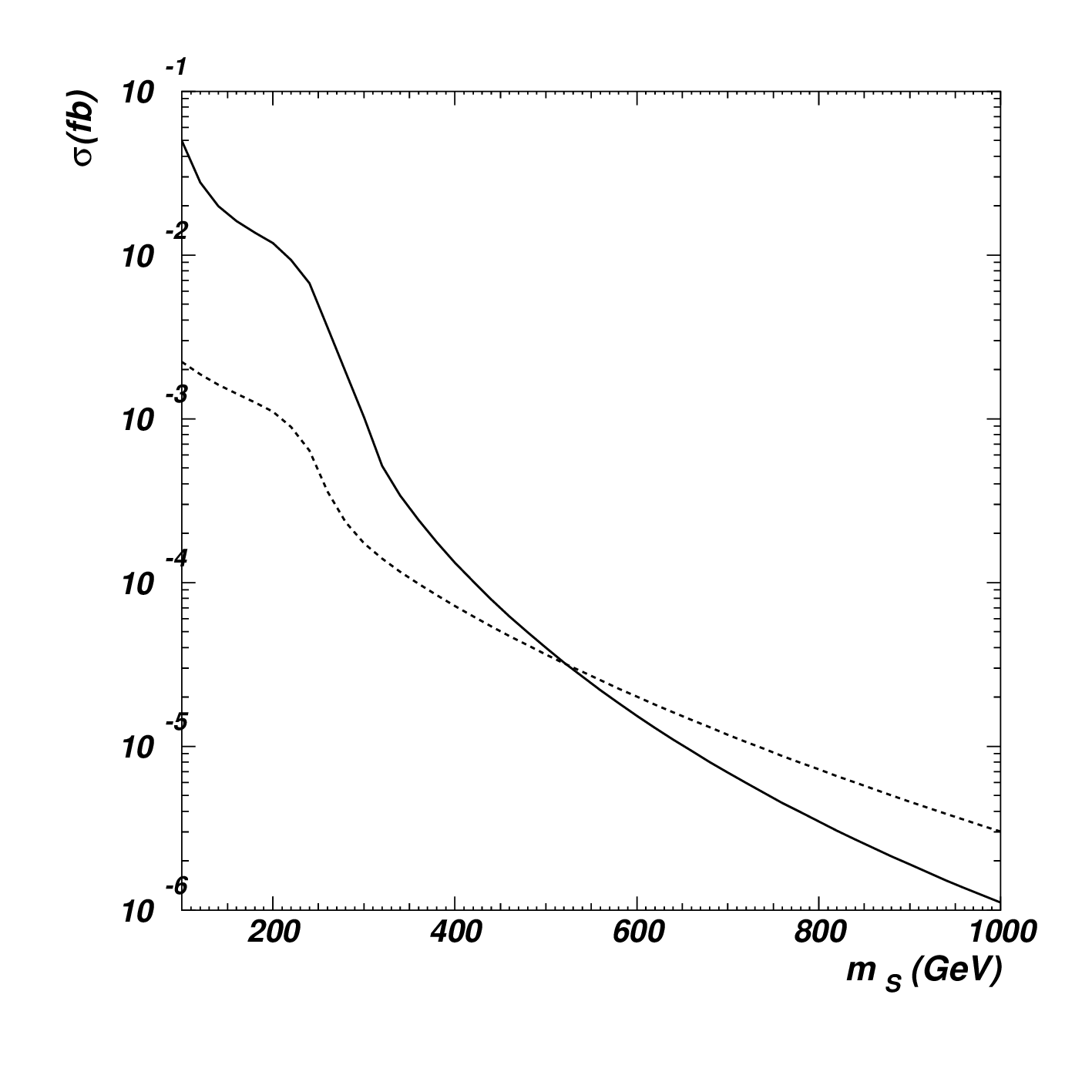}} 
\caption[]{
The cross section for the process $e^+ e^- \rightarrow t \bar{c}$
as 
a function  of  $m_S$,
where $\sqrt{S}= 500 GeV$,  $\tan\beta =2 $,
$\epsilon^2=1/4$ and $\mu =-100 GeV$. The solid  and dashed lines
represent $m_{\tilde{g}}= 100 GeV$ and  $500 GeV$, respectively.
} 
\end{figure} 

\begin{figure} 
\epsfxsize=7 cm 
\centerline{\epsffile{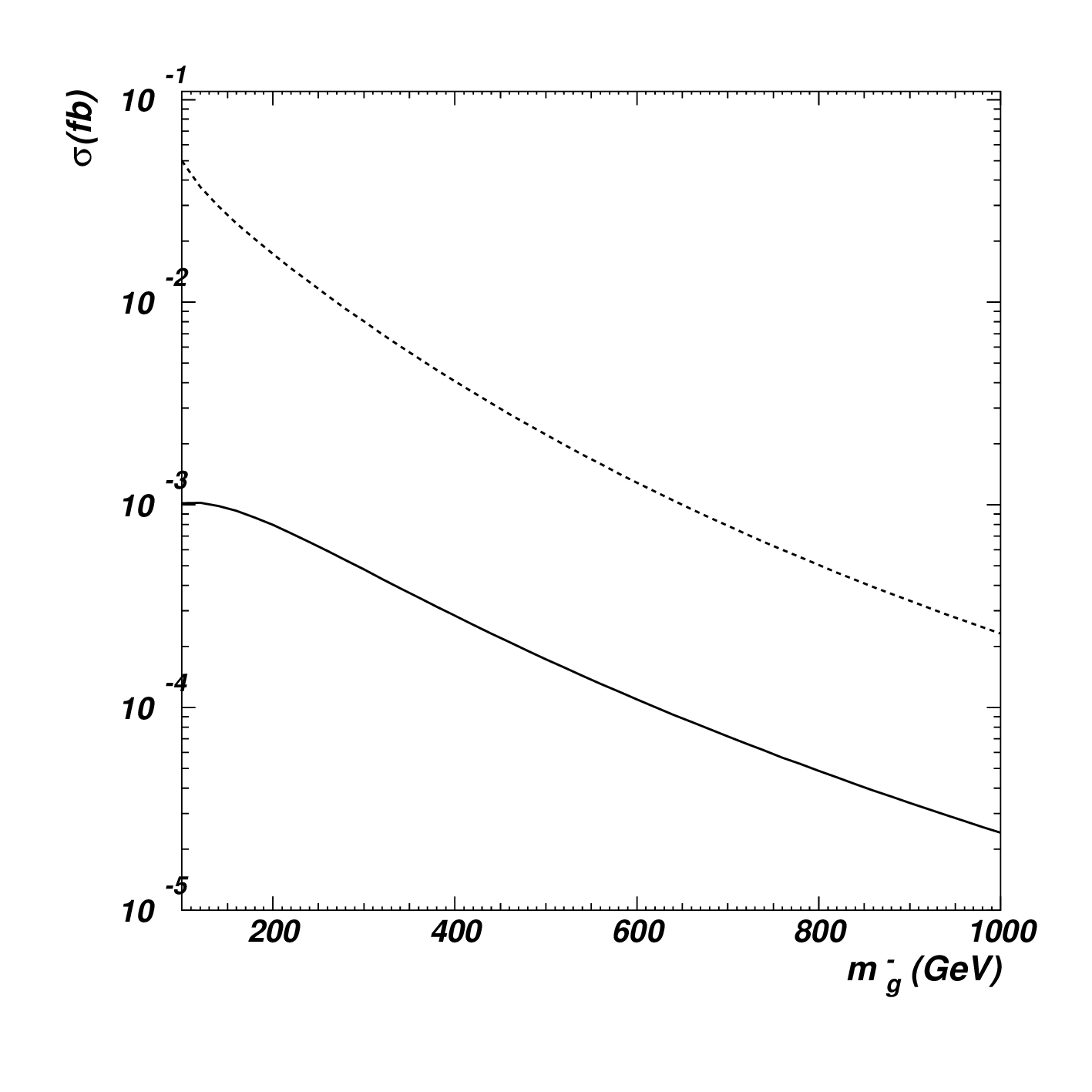}} 
\caption[]{
The cross section for the process $e^+ e^- \rightarrow t \bar{c}$
as 
a function  of  $m_{\tilde{g}}$,
where $\sqrt{S}= 500 GeV$,  $\tan\beta =2 $,
$\epsilon^2=1/4$ and $\mu =-100 GeV$.
The solid  and dashed lines
represent $m_S= 300 GeV$ and  $100 GeV$, respectively.
}
\end{figure} 
 
\begin{figure} 
\epsfxsize=7 cm 
\centerline{\epsffile{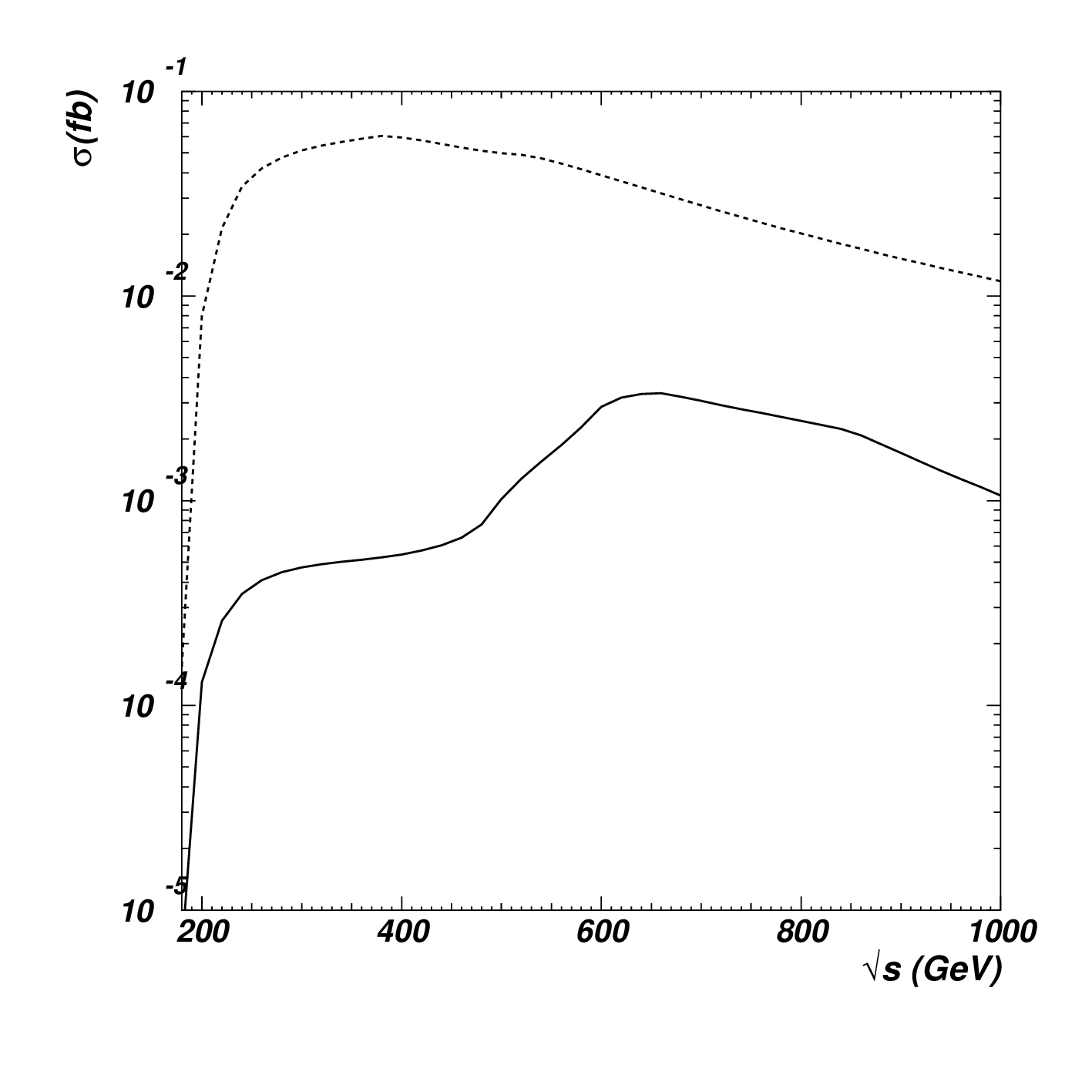}} 
\caption[]{
The cross section for the process $e^+ e^- \rightarrow t \bar{c}$
as 
a function  of  $\sqrt{S}$,
where $m_{\tilde{g}}= 100 GeV$,  $\tan\beta =2 $,
$\epsilon^2=1/4$ and $\mu =-100 GeV$. The solid and dashed lines represent
$m_S= 300 GeV$ and $100 GeV$, respectively.
} 
\end{figure} 

\begin{figure} 
\epsfxsize=7 cm 
\centerline{\epsffile{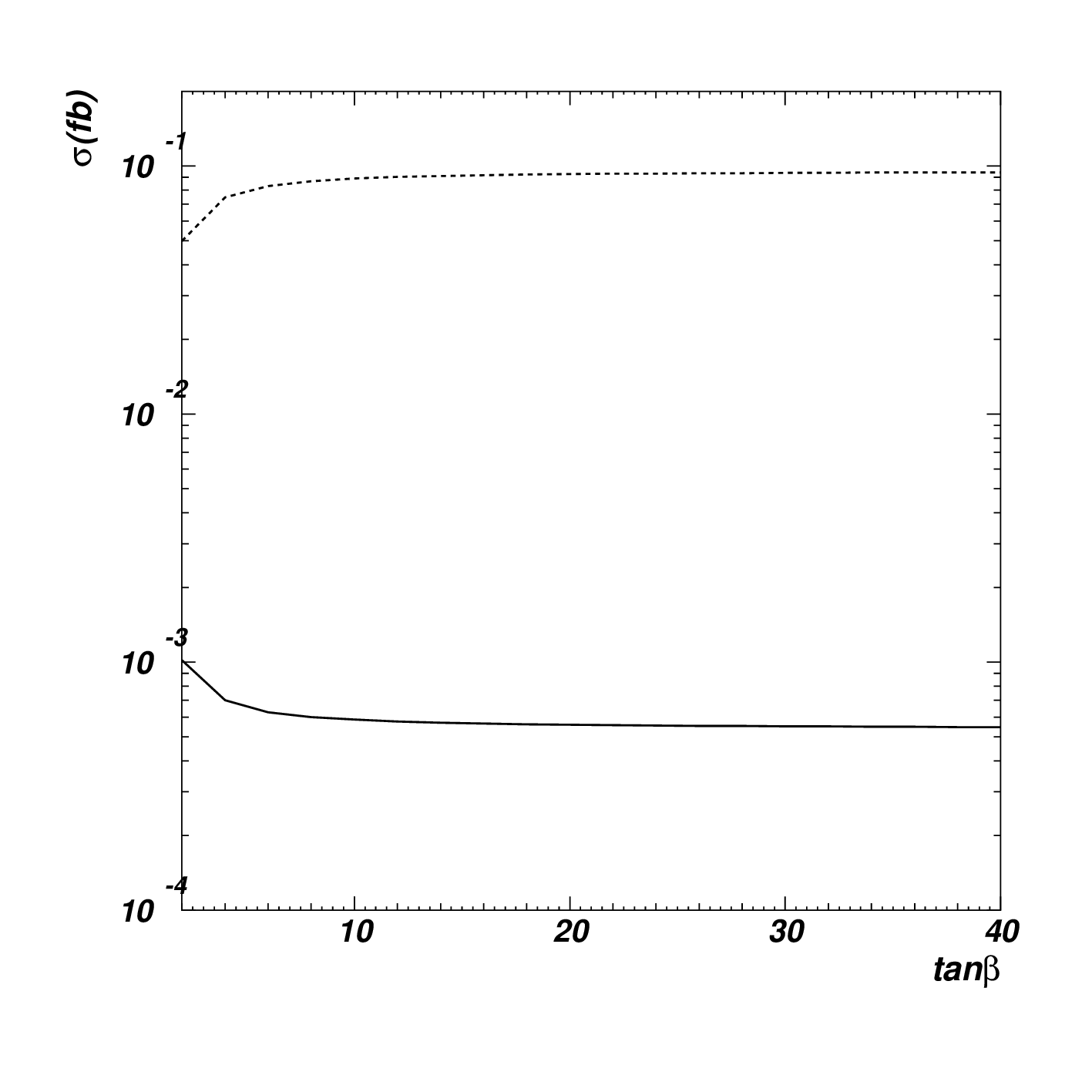}} 
\caption[]{
The cross section for the process $e^+ e^- \rightarrow t \bar{c}$
as 
a function  of  $\tan\beta$,
where $\sqrt{S}= 500 GeV$, $m_{\tilde{g}}= 100 GeV$, 
$\epsilon^2=1/4$ and $\mu =-100 GeV$.
The solid and dashed lines represent
$m_S= 300 GeV$ and $100 GeV$, respectively.
} 
\end{figure} 


\setcounter{figure}{0} \setcounter{table}{0}
\setcounter{equation}{0}

\newpage
\section{Bottom-Strange Associated Production at High Energy
$e^{+}e^{-} $ Colliders in Standard Model}

\begin{minipage}{15cm}
\begin{center} Abstract \end{center}
We investigate the flavor changing neutral current bsV(V=$\gamma$,Z) couplings
in the production vertex  for the process $e^+e^-\rightarrow b\bar s
\mbox{ or } \bar b s$
in the standard model. The precise calculations keeping all
quark masses non-zero are carried out.
Production cross sections are found to be the order of $10^{-3}$ fb at LEP II
and the order of $10^{-1}$ fb when center-of-mass energy is near the mass of
neutral gauge boson Z. \end{minipage}

\subsection{Introduction}

\hspace{4mm}There are no flavor changing neutral currents (FCNC) at
tree-level in the standard model (SM). FCNC appear at loop-levels and
consequently offer a good place to test quantum effects of the
fundamental quantum field theory on which SM based. Furthermore, they are
very small at one loop-level due to the unitarity of Cabbibo-Kobayashi-Maskawa
 (CKM) matrix. In models beyond SM new particles beyond the particles in SM
may appear in the loop and have significant contributions to
flavor changing transitions. Therefore, FCNC interactions give an
ideal place to search for new physics. Any positive observation of
FCNC couplings deviated from that in SM would unambiguously signal
the presence of new physics. Searching for FCNC is clearly one of
important goals of the next generation of high energy
colliders~\cite{pro11}.

The flavor changing transitions involving external up-type quarks
which are due to FCNC couplings are much more suppressed than
those involving external down-type quarks in SM. The effects for
external up-type quarks are derived by virtual exchanges of
down-type quarks in a loop for which GIM mechanism ~\cite{gim1} is
much more effective because the mass splitting between down-type
quarks are much less than those between up-type quarks. Therefore,
for example, the bs transition which is studied in the paper has
larger probability to be observed than that for the tc transition.

The b-hadron system promises to give a fertile ground to test the
SM and probe new physics. The FCNC vertices bsV(V=$\gamma$, Z)
have been extensively examined in rare decays of b-hadron
system~\cite{my,neu,rev1}. The observation of FCNC processes in
both the exclusive $B\rightarrow K^{\star}\gamma$ and inclusive
$B\rightarrow X_s\gamma$ channels has placed the rare B decays on
a new footing and has put a stringent constraint on classes of
models~\cite{hew}. Analyses of the inclusive decay $B \rightarrow
X_s l^+l^-$ show that in the minimal supergravity model(SUGRA)
there are regions in the parameter space where the branching ratio
of $b\rightarrow s l^{+}l^{-}(l=e, {\mu})$ is enhanced by about
50\% compared to the SM~\cite{gost} and the first distinct signals
of SUSY could come from the observation of $B \rightarrow
X_s{\mu}^{+}{\mu}^{-}$ if tan$\beta$ is large ( $\geq 30$ ) and
the mass of the lightest neutral Higgs boson m$_h$ is not too
large (say, less than 150 Gev)~\cite{my}. The B factories
presently under construction will collect some $10^7$---$10^8$ B
mesons per year which can be used to obtain good precision on low
branching fraction modes.

The FCNC vertices bsV(V=$\gamma$, Z) can also be investigated via bottom-strange associated production. In the paper we shall investigate the process
\begin{eqnarray}
e^+e^-\rightarrow b\bar{s} \mbox{ or } \bar{b}s.
\end{eqnarray}
Comparing b quark rare decays where the momentum transfer $q^2$ is
limited, i. e., it should be less or equal to mass square of b
quark $m_b^2$, the production process (1) allows the large
(time-like) momentum transfer, which is actually determined by the
energies available at $e^+e^-$ colliders. The reaction (1) has
some advantages because of the ability to probe higher dimension
operators at large momenta and striking kinematic signatures which
are straightforward to detect in the clean environment of $e^+e^-$
collisions. In particular, in some extensions of SM which induce
FCNC there are large underlying mass scales and large momentum
transfer so that these models are more naturally probed via
b$\bar{s}$ associated production than b quark rare decays.

It has been shown that the cross sections of $e^+e^-\rightarrow
t\bar c$ in SM are too small to be observed at LEP or
NLC~\cite{tc}. As pointed above, in SM the cross sections of
$e^+e^-\rightarrow b \bar s$ should be much larger than those of
tc final states. Are they large enough to be seen at LEP or NLC?
In the paper we would like to address the problem by calculating
cross sections and backward-forward asymmetry of the process (1)
in SM.

\subsection{Analytic calculations} In SM for the process
(1) there are three kinds of Feynman diagram at one loop, self
energy-type, triangle and box diagram, which are shown in Fig.1.
We carry out calculations in the Feynman-t'Hooft gauge. The
contributions of the neutral Higgs H and Goldstone bosons
$G^{0,\pm}$ which couple to electrons are neglected since they are
proportional to the electron mass and we have put the mass of
electron to zero.
\par
We do the reduction using FeynCalc \cite{333} and keep all masses
non-zero except for the mass of electron. To control the
ultraviolet divergence, the dimensional regularization is used. As
a consistent check, we found that all divergences are canceled in
the sum of contributions of all Feynman diagrams. The calculations
are carried out in the frame of the center of mass system (CMS)
and  Mandelstam variables have been employed: \begin{eqnarray}
s=(p_1 +p_2)^2 =(k_1 +k_2)^2 \hspace{7mm} t=(p_1 -k_1)^2
\hspace{7mm} u=(p_1 -k_2)^2,
\end{eqnarray}
 where
$p_1,p_2$ are the
momentum of electron and positron respectively, and $k_1,k_2$ are the momentum of
bottom quark $b$, and anti-strange quark $\bar{s}$ respectively.
\par
The amplitude of process $e^{+}e^{-} \rightarrow \bar b s$ can be
expressed as {\small
\begin{eqnarray}
M &=& \sum_{j=u,c,t} 16\pi^2 \alpha^2 V_{jb}^{\star}V_{js}[g_1 \bar{u_b} \gamma^{\mu}  P_R  v_s  \bar{v_e}  \gamma_{\mu} P_R  u_e +
 g_2 \bar{u_b}  \gamma^{\mu}  P_L v_s  \bar{v_e}  \gamma_{\mu}  P_R  u_e +
g_3 \bar{u_b}  \gamma^{\mu}  P_R  v_s  \bar{v_e}  \gamma_{\mu}  P_L  u_e +\nonumber\\
&& g_4 \bar{u_b}  \gamma^{\mu}  P_L  v_s  \bar{v_e}  \gamma_{\mu}  P_L  u_e +
 g_5 \bar{u_b}  P_R  v_s  \bar{v_e}  \not\!{k_1}  P_R  u_e +
 g_6 \bar{u_b}  P_L  v_s  \bar{v_e}  \not\!{k_1}  P_R  u_e +
g_7 \bar{u_b}  P_R  v_s  \bar{v_e}  \not\!{k_1}  P_L  u_e +\nonumber\\
&&g_8 \bar{u_b}  P_L  v_s  \bar{v_e}  \not\!{k_1}  P_L  u_e +
 g_9 \bar{u_b} \not\!{p_1}  P_L  v_s  \bar{v_e}  \not\!{k_1}  P_L  u_e +
 g_{10} \bar{v_e}  \gamma^{\mu}  P_L  u_e  \bar{u_b}  \gamma_{\mu}  \not\!{p_1}  P_R  v_s + \nonumber\\
&& g_{11} \bar{v_e} \gamma^{\mu}  P_L  u_e  \bar{u_b}  \gamma_{\mu}  \not\!{p_1}  P_L  v_s]
\end{eqnarray}
}
where $\alpha$ is fine structure constant, $V_{ij}$ is CKM matrix element, $P_L$ is
defined as $(1-\gamma^5)/2$, and $P_R$ is defined as $(1+\gamma^5)/2$.
The expressions of the coefficients $g_j(j=1,2,...11)$ can be
found in Appendix.
\par
Having the amplitude M, it is straightforward to obtain the differential
cross section by
\begin{eqnarray}
\frac{d\sigma}{dcos\theta} ={\frac{N_c}{16\pi}}
{\frac{|\vec{k_1}|}{s^{3\over{2}}}} {\frac{1}{4}} {\sum_{spins}|M|^2}
\end{eqnarray}
Where $N_c$ is the color factor and $\theta$ is the angle between
incoming electron $e^{-}$ and outgoing
bottom quark $b$.

\subsection{Numerical results} \hspace{4mm} In the numerical
calculations the following values of the parameters have been used
~\cite{pdg}:
\begin{eqnarray}
m_e=0,\hspace{4mm}m_u=0.005Gev,\hspace{4mm}m_c=1.4Gev,\hspace{4mm}m_t=175Gev,\hspace{4mm}m_s=0.17Gev,\hspace{4mm}\nonumber\\
m_b=4.4Gev,\hspace{4mm}m_w=80.41Gev,\hspace{4mm}m_z=91.187Gev,\hspace{4mm}
\Gamma_z=2.5Gev,\hspace{4mm} \alpha=\frac{1}{128} \nonumber
\end{eqnarray}
\par
In order to keep the unitary condition of CKM matrix exactly, we
employ the standard parameterization and took the values
~\cite{pdg,ckm1}
\begin{eqnarray}
s_{12}=0.220,\hspace{7mm} s_{23}=0.039,\hspace{7mm} s_{13}=0.0031,\hspace{7mm} \delta_{13}= 70^{\circ} \nonumber
\end{eqnarray}
Numerical results are shown in Figs. 2, 3, 4.
\par
In Fig.2, we show the total cross section $\sigma_{tot}$ of the process $e^{+}e^{-} \rightarrow b\bar{s}$ as a function of
the center-of-mass energy $\sqrt{s}$.
There are three peaks, corresponding to the pole of neutral gauge boson
$Z^0$, a pair of charged gauge boson $W$ threshold, and a pair of top quark
$t\bar{t}$ threshold respectively.
In  most of high energy region, total cross section is the order of
$10^{-3}$ fb, which is too small to be seen at LEP II or planning NLC colliders. Therefore, even a small number of bs events, detected at LEP II or NLC, will unambiguously indicate  new FCNC couplings beyond SM. Smallness of the total cross section can easily be understood. One has
\begin{eqnarray}
\sum_{spins}|M|^2 & = & e^8 |\sum_{j=u,c,t} V_{jt}^{\star}V_{jc} f(x_j, y_j)|^2
\nonumber \\ & = &  e^8 |V_{tb}^{\star} V_{ts} \frac{m_t^2-m_c^2}{m_w^2}
\frac {\partial f}{\partial x_j}|_{x_j,y_j=0} + ...|^2,
\end{eqnarray}
due to GIM mechanism, where $x_j=m_j^2/m_w^2, y_j=m_j^2/s$, and "..." denote the less important terms for $\sqrt s\ge 200$ Gev. Assuming $\frac {\partial
f}{\partial x_j}|_{x_j,y_j=0} $ = O(1), one obtains from eqs. (4), (5)\\
$$ \sigma \sim 10^{-3} fb$$
at $\sqrt s$ = 200 Gev.
\par
We fixed the center-of-mass energy $\sqrt{s}$ at $200$ Gev. Differential
cross  section of the process at the energy as a function of $\cos{\theta}$
 is shown in Fig.3.
\par
The Fig.4 is devoted to the backward-forward asymmetry
\begin{equation}
A_{FB}=\frac{\int_{0}^{\pi/2}{d\sigma \over d\theta}d\theta -
\int_{\pi/2}^{\pi}{d\sigma \over d\theta}d\theta }
{\int_{0}^{\pi/2}{d\sigma \over d\theta}d\theta +
\int_{\pi/2}^{\pi}{d\sigma \over d\theta}d\theta }
\end{equation}
\par
as a function of $\sqrt{s}$.
\par
To summarize, we have calculated the process $e^{+}e^{-} \rightarrow b\bar{s}$ in SM.  We found that the total cross
section is of the order of $10^{-3} fb$ in the
 high energy region which is still too small to be seen at LEP II or planning NLC. However, it is worth to note that
 the total cross section at Z resonance may reach as large as $10^{-1}$ fb. Therefore, it is possible to see the process if a luminosity reaches 100-1000 $fb^{-1}$. In addition to that, the process is
 of  a good place to search for new physics.

\subsection{Appendix}

{\small \begin{eqnarray} g_{1} &=&  m_s(B_0^a (m_b^2 - m_s^2)
(m_j^2 - m_w^2) (m_j^2 + 2 m_w^2) + B_0^b m_s^2 (m_b^4 - 2 m_b^2
m_j^2 + m_j^4 + m_b^2 m_w^2 + m_j^2 m_w^2 -\nonumber\\ && 2 m_w^4)
- B_0^c m_b^2 (m_s^4 - 2 m_s^2 m_j^2 + m_j^4 + m_s^2 m_w^2 + m_j^2
m_w^2 - 2 m_w^4)) (a_1 - 4 a_2 s_w^4) + \nonumber\\ && 2 m_b m_s
(2 C_{00}^e + C_{11}^e m_b^2 + C_0^e m_j^2 + C_1^e (m_b^2 + m_j^2
- 2 m_w^2) + C_{22}^e s + C_2^e s +\nonumber\\ && C_{12}^e (m_b^2
- m_s^2 + s)) (a_3 + 6 a_4 s_w^2 - 8 a_4 s_w^4) - 6 C_{00}^d m_b
m_s (a_3 + 4 a_4 s_w^2 (c_w^2 - s_w^2)) +\nonumber\\ && 12 m_b m_s
m_w^2 (C_0^d + C_1^d) (a_3 + 6 a_4 c_w^2 s_w^2 - 2 a_4 s_w^4)
\end{eqnarray}
\begin{eqnarray}
g_{2} &=&  -a_3 m_j^2  + 8 a_4 m_j^2 s_w^4  - 6 m_w^2(C_{11}^d m_b^2 + C_{22}^d s + C_{12}^d (m_b^2 - m_s^2 + s) ) (a_3 + 8 a_4 c_w^2 s_w^2)  + \nonumber\\
&& m_b m_s^2 ( B_0^a (m_b^2 - m_s^2) (m_j^2 - m_w^2) + (m_b^2 m_s^2 - m_b^2 m_j^2 - m_s^2 m_j^2 + m_j^4  + m_j^2 m_w^2 -\nonumber\\
&& 2 m_w^4) (B_0^b -B_0^c) + m_w^2 (B_0^b (2 m_b^2 -m_s^2) -B_0^c (2 m_s^2 - m_b^2) )) (a_1 + 6 a_2 s_w^2 - 4 a_2 s_w^4) - \nonumber\\
&& 6 C_2^d m_w^2 (a_3 (m_b^2 - m_s^2 + s) + 4 a_4 s_w^2 ((m_b^2 - m_s^2) (c_w^2 - s_w^2) + 2 s c_w^2) ) - 6 C_{00}^d (a_3 (m_j^2 + 6 m_w^2) + \nonumber\\
&& 4 a_4 s_w^2 (c_w^2 m_j^2 + 12 c_w^2 m_w^2 - m_j^2 s_w^2)) + 2 (2 C_{00}^e + C_{11}^e m_b^2 + C_{22}^e s + C_{12}^e (m_b^2 - m_s^2 + s) )\nonumber\\
&& (a_3 (m_j^2 + 2 m_w^2) + 4 a_4 s_w^2 (3 m_w^2 - 2 m_j^2 s_w^2 - 4 m_w^2 s_w^2)) + 2 C_0^e m_j^2 (a_3 (m_b^2 + m_s^2 - m_j^2 - 2 m_w^2) +\nonumber\\
&&  2 a_4 s_w^2 (3 m_s^2 - 3 m_j^2 - 4 m_b^2 s_w^2 - 4 m_s^2 s_w^2 + 4 m_j^2 s_w^2 + 8 m_w^2 s_w^2)) + 2 C_2^e (a_3 s (m_j^2 + 2 m_w^2) -\nonumber\\
&& 2 a_4 s_w^2 (3 m_b^2 m_j^2 - 3 m_s^2 m_j^2 - 6 m_w^2 s + 4 m_j^2 s s_w^2 + 8 m_w^2 s s_w^2) )  - 6 C_1^d m_w^2 (a_3 (m_b^2 - m_s^2 + s) +\nonumber\\
&& 4 a_4 s_w^2 (m_b^2 c_w^2 - m_b^2 s_w^2  - 2 c_w^2 m_s^2 + 2 c_w^2 s)) - 6 C_0^d m_w^2 (a_3 (s - m_s^2 - m_j^2) +\nonumber\\
&& 4 a_4 s_w^2 (2 s c_w^2 - m_b^2 + 2 m_j^2 s_w^2  - 2 c_w^2 m_s^2)) + 2 C_1^e (a_3 (2 s m_w^2 + m_b^2 m_s^2 + m_b^2 m_j^2 - 2 m_w^2 m_s^2) + \nonumber\\
&& 2 a_4 s_w^2 (3 m_b^2 m_s^2 - 3 m_b^2 m_j^2 + 6 s m_w^2 - 4 m_b^2 m_s^2 s_w^2 - 4 m_b^2 m_j^2 s_w^2 - 8 s m_w^2 s_w^2 -\nonumber\\
&& 6 m_w^2 m_s^2 + 8 m_w^2 m_s^2 s_w^2))
\end{eqnarray}
\begin{eqnarray}
g_{3} &=&  - a_5 m_b m_s (2 D_{23}^f + D_3^f) + m_s (B_0^a (m_b^2 - m_s^2) (m_j^2 - m_w^2) (m_j^2 + 2 m_w^2) + B_0^b m_s^2 (m_b^4 - 2 m_b^2 m_j^2 +\nonumber\\
&& m_j^4 + m_b^2 m_w^2 + m_j^2 m_w^2 - 2 m_w^4) - B_0^c m_b^2 (m_s^4 - 2 m_s^2 m_j^2 + m_j^4 + m_s^2 m_w^2 + m_j^2 m_w^2 - 2 m_w^4) ) (a_1 +\nonumber\\
&& 2 a_2 s_w^2 - 4 a_2 s_w^4) + 2 m_b m_s (2 C_{00}^e + C_{11}^e m_b^2 + C_0^e m_j^2 + C_1^e (m_b^2 + m_j^2 - 2 m_w^2) + C_{22}^e s + C_2^e s +\nonumber\\
&& C_{12}^e (m_b^2 - m_s^2 + s) ) (a_3 - 3 a_4 + 10 a_4 s_w^2 - 8 a_4 s_w^4) - 6 C_{00}^d m_b m_s (a_3 - 2 a_4 (1 - 2 s_w^2)^2) +\nonumber\\
&& 12 m_b m_s m_w^2 (C_0^d + C_1^d) (a_3 - a_4 (3 - 4 s_w^2) (1 - 2 s_w^2))
\end{eqnarray}
\begin{eqnarray}
g_{4} &=&  - a_5 (2 D_{00}^f - (2 D_{13}^f + D_1^f) (m_b^2 - t) + 2 D_{23}^f t + D_3^f t) - a_3 m_j^2  - 4 a_4 m_j^2 s_w^2 (1 - 2 s_w^2)  -\nonumber\\
&& 6 m_w^2 (C_{11}^d m_b^2 + C_{22}^d s + C_{12}^d (m_b^2 - m_s^2 + s) ) (a_3 - 4 a_4 c_w^2 (1 - 2 s_w^2) ) + \nonumber\\
&& m_b m_s^2 (B_0^a (m_b^2 - m_s^2) (m_j^2 - m_w^2) + B_0^b (m_b^2 m_s^2 - m_b^2 m_j^2 - m_s^2 m_j^2 + m_j^4 + 2 m_b^2 m_w^2 - m_s^2 m_w^2 +\nonumber\\
&& m_j^2 m_w^2 - 2 m_w^4) - B_0^c (m_b^2 m_s^2 - m_b^2 m_j^2 - m_s^2 m_j^2 + m_j^4 - m_b^2 m_w^2 + 2 m_s^2 m_w^2 + m_j^2 m_w^2 - 2 m_w^4))\times \nonumber\\
&& (a_1 - 3 a_2 + 8 a_2 s_w^2 - 4 a_2 s_w^4) - 6 C_2^d m_w^2 (a_3 (m_b^2 - m_s^2 + s) + 2 a_4 (c_w^2 m_s^2 - c_w^2 m_b^2 - 2 c_w^2 s +\nonumber\\
&& m_b^2 s_w^2 + 2 c_w^2 m_b^2 s_w^2 - m_s^2 s_w^2 - 2 c_w^2 m_s^2 s_w^2 + 4 c_w^2 s s_w^2 - 2 m_b^2 s_w^4 + 2 m_s^2 s_w^4) ) - 6 C_{00}^d (a_3 (m_j^2 + 6 m_w^2) +\nonumber\\
&& 2 a_4 (m_j^2 s_w^2 - c_w^2 m_j^2 - 12 c_w^2 m_w^2 + 2 c_w^2 m_j^2 s_w^2 + 24 c_w^2 m_w^2 s_w^2 - 2 m_j^2 s_w^4) ) +\nonumber\\
&& 2 (2 C_{00}^e + C_{11}^e m_b^2 + C_{22}^e s)(a_3 (m_j^2 + 2 m_w^2) + 2 a_4 (2 m_j^2 s_w^2 - 3 m_w^2 + 10 m_w^2 s_w^2 - 4 m_j^2 s_w^4 - 8 m_w^2 s_w^4) ) +\nonumber\\
&& 2 C_0^e m_j^2 (a_3 (m_b^2 + m_s^2 - m_j^2 - 2 m_w^2) + a_4 (3 m_j^2 - 3 m_s^2 + 4 m_b^2 s_w^2 + 10 m_s^2 s_w^2 - 10 m_j^2 s_w^2 - 8 m_w^2 s_w^2 -\nonumber\\
&& 8 m_b^2 s_w^4 - 8 m_s^2 s_w^4 + 8 m_j^2 s_w^4 + 16 m_w^2 s_w^4) ) + 2 C_2^e (a_3 s (m_j^2 + 2 m_w^2) + a_4 (3 m_b^2 m_j^2 - 3 m_s^2 m_j^2 -\nonumber\\
&& 6 m_w^2 s - 6 m_b^2 m_j^2 s_w^2 + 6 m_s^2 m_j^2 s_w^2 + 4 m_j^2 s s_w^2 + 20 m_w^2 s s_w^2 - 8 m_j^2 s s_w^4 - 16 m_w^2 s s_w^4) ) + \nonumber\\
&& 2 C_{12}^e (m_b^2 - m_s^2 + s) (a_3 (m_j^2 + 2 m_w^2) - 2 a_4 (3 m_w^2 - 2 m_j^2 s_w^2 - 10 m_w^2 s_w^2 + 4 m_j^2 s_w^4 + 8 m_w^2 s_w^4) ) -\nonumber\\
&& 6 C_1^d m_w^2 (a_3 (m_b^2 -m_s^2 + s) + 2 a_4 (m_b^2 s_w^2 - 3 c_w^2 m_b^2 + 6 c_w^2 m_b^2 s_w^2 - 2 m_b^2 s_w^4 + 2 c_w^2 t - 4 c_w^2 s_w^2 t +\nonumber\\
&& 2 c_w^2 u - 4 c_w^2 s_w^2 u) ) - 6 C_0^d m_w^2 (a_3 (s - m_s^2 - m_j^2) + 2 a_4 (m_b^2 s_w^2 + c_w^2 m_b^2 - 2 c_w^2 m_b^2 s_w^2 - 2 m_j^2 s_w^2 -\nonumber\\
&& 2 m_b^2 s_w^4 + 4 m_j^2 s_w^4 + 2 m_s^2 c_w^2 - 4 m_s^2 c_w^2 s_w^2 - 2 s c_w^2 + 4 s c_w^2 s_w^2) ) + 2 C_1^e (a_3 (m_b^2 m_s^2 + m_b^2 m_j^2 +\nonumber\\
&& 2 m_b^2 m_w^2 - 2 m_w^2 t - 2 m_w^2 u) + a_4 (3 m_b^2 m_j^2 - 3 m_b^2 m_s^2 + 6 m_s^2 m_w^2 + 10 m_b^2 m_s^2 s_w^2 - 2 m_b^2 m_j^2 s_w^2 -\nonumber\\
&& 20 m_s^2 m_w^2 s_w^2 - 8 m_b^2 m_s^2 s_w^4 - 8 m_b^2 m_j^2 s_w^4 + 16 m_s^2 m_w^2 s_w^4 - 6 s m_w^2 + 20 s m_w^2 s_w^2 - 16 s m_w^2 s_w^4) )
\end{eqnarray}
\begin{eqnarray}
g_{5} &=&  -24 a_4 m_s s_w^2 (C_2^e m_j^2 - 2 C_2^d m_w^2) - 4 m_s (C_{11}^e m_b^2 + C_0^e m_j^2 + C_1^e (m_b^2 + m_j^2 - 2 m_w^2) ) (a_3 + 6 a_4 s_w^2 -\nonumber\\
&& 8 a_4 s_w^4) - 6 m_s (C_{11}^d m_b^2 + 2 C_0^d m_w^2 + C_1^d (m_b^2 - m_j^2 + 2 m_w^2) ) (a_3 + 4 a_4 c_w^2 s_w^2 - 4 a_4 s_w^4) - \nonumber\\
&& 6 C_{12}^d m_s (a_3 (m_b^2 - m_j^2 - 2 m_w^2) + 4 a_4 s_w^2 (m_b^2 c_w^2 - c_w^2 m_j^2 - 4 c_w^2 m_w^2 - m_b^2 s_w^2 + m_j^2 s_w^2)) -\nonumber\\
&& 4 C_{12}^e m_s (a_3 (m_b^2 - m_j^2 - 2 m_w^2) + 2 a_4 s_w^2 (3 m_b^2 - 6 m_w^2 - 4 m_b^2 s_w^2 + 4 m_j^2 s_w^2 + 8 m_w^2 s_w^2))
\end{eqnarray}
\begin{eqnarray}
g_{6} &=&  24 a_4 m_b s_w^2 (C_2^e m_j^2 - 2 C_2^d m_w^2) - 12 m_b m_w^2 (C_0^d + 2 C_1^d) (a_3 + 8 a_4 c_w^2 s_w^2) -\nonumber\\
&& 4 m_b m_j^2 (C_0^e + 2 C_1^e) (a_3 - 8 a_4 s_w^4) - 6 C_{11}^d m_b (a_3 (m_j^2 + 2 m_w^2) + 4 a_4 s_w^2 (c_w^2 m_j^2 + 4 c_w^2 m_w^2 - m_j^2 s_w^2) ) +\nonumber\\
&& 6 C_{12}^d m_b (a_3 (m_s^2 - m_j^2 - 2 m_w^2) + 4 a_4 s_w^2 (c_w^2 m_s^2 - c_w^2 m_j^2 - 4 c_w^2 m_w^2 - m_s^2 s_w^2 + m_j^2 s_w^2) ) - \nonumber\\
&& 4 C_{11}^e m_b (a_3 (m_j^2 + 2 m_w^2) + 4 a_4 s_w^2 (3 m_w^2 - 2 m_j^2 s_w^2 - 4 m_w^2 s_w^2) ) + 4 C_{12}^e m_b (a_3 (m_s^2 - m_j^2 - 2 m_w^2) +\nonumber\\
&& 2 a_4 s_w^2 (3 m_s^2 - 6 m_w^2 - 4 m_s^2 s_w^2 + 4 m_j^2 s_w^2 + 8 m_w^2 s_w^2) )
\end{eqnarray}
\begin{eqnarray}
g_{7} &=&  2 a_5 m_s (D_{23}^f + D_3^f) + 12 a_4 m_s (C_2^e m_j^2 - 2 C_2^d m_w^2) (1 - 2 s_w^2) - 4 m_s (C_{11}^e m_b^2 + C_0^e m_j^2 + C_1^e (m_b^2 +\nonumber\\
&& m_j^2 - 2 m_w^2) ) (a_3 - 3 a_4 + 10 a_4 s_w^2 - 8 a_4 s_w^4) - 6 m_s (C_{11}^d m_b^2 + 2 C_0^d m_w^2 +\nonumber\\
&& C_1^d (m_b^2 - m_j^2 + 2 m_w^2) ) (a_3 - 2 a_4 (s_w^2 - c_w^2)^2 ) - 6 C_{12}^d m_s (a_3 (m_b^2 - m_j^2  - 2 m_w^2) + 2 a_4 (c_w^2 m_j^2 -\nonumber\\
&& c_w^2 m_b^2  + 4 c_w^2 m_w^2 + m_b^2 s_w^2 + 2 c_w^2 m_b^2 s_w^2 - m_j^2 s_w^2 - 2 c_w^2 m_j^2 s_w^2 - 8 c_w^2 m_w^2 s_w^2 - 2 m_b^2 s_w^4 + 2 m_j^2 s_w^4) ) - \nonumber\\
&& 4 C_{12}^e m_s (a_3 (m_b^2 - m_j^2 - 2 m_w^2) + a_4 (6 m_w^2 - 3 m_b^2 + 10 m_b^2 s_w^2 - 4 m_j^2 s_w^2 - 20 m_w^2 s_w^2 - 8 m_b^2 s_w^4 +\nonumber\\
&& 8 m_j^2 s_w^4 + 16 m_w^2 s_w^4) )
\end{eqnarray}
\begin{eqnarray}
g_{8} &=&  -2 a_5 m_b (D_{22}^f - D_{23}^f + D_2^f) - 12 a_4 m_b (C_2^e m_j^2 - 2 C_2^d m_w^2) (1 - 2 s_w^2) - 12 m_b m_w^2 (C_0^d + 2 C_1^d) (a_3 -\nonumber\\
&& 4 a_4 c_w^2 (1 - 2 s_w^2) ) - 4 m_b m_j^2 (C_0^e + 2 C_1^e) (a_3 + 4 a_4 s_w^2 (1 - 2 s_w^2) ) - 6 C_{11}^d m_b (a_3 (m_j^2 + 2 m_w^2) -\nonumber\\
&& 2 a_4 (c_w^2 m_j^2 + 4 c_w^2 m_w^2 - m_j^2 s_w^2 - 2 c_w^2 m_j^2 s_w^2 - 8 c_w^2 m_w^2 s_w^2 + 2 m_j^2 s_w^4) ) + 6 C_{12}^d m_b (a_3 (m_s^2 - m_j^2 -\nonumber\\
&& 2 m_w^2) - 2 a_4 (c_w^2 m_s^2 - c_w^2 m_j^2 - 4 c_w^2 m_w^2 - m_s^2 s_w^2 - 2 c_w^2 m_s^2 s_w^2 + m_j^2 s_w^2 + 2 c_w^2 m_j^2 s_w^2 + 8 c_w^2 m_w^2 s_w^2 +\nonumber\\
&& 2 m_s^2 s_w^4 - 2 m_j^2 s_w^4) ) - 4 C_{11}^e m_b (a_3 (m_j^2 + 2 m_w^2) + 2 a_4 (2 m_j^2 s_w^2 - 3 m_w^2  + 10 m_w^2 s_w^2 - 4 m_j^2 s_w^4 -\nonumber\\
&& 8 m_w^2 s_w^4) ) + 4 C_{12}^e m_b (a_3 (m_s^2 - m_j^2 - 2 m_w^2) + a_4 (6 m_w^2 - 3 m_s^2 + 10 m_s^2 s_w^2 - 4 m_j^2 s_w^2 -\nonumber\\
&& 20 m_w^2 s_w^2 - 8 m_s^2 s_w^4 + 8 m_j^2 s_w^4 + 16 m_w^2 s_w^4) ) \\
g_{9} &=& 2 a_5 (D_{12}^f - 2 D_{13}^f + D_{22}^f - D_{23}^f + D_2^f) \\
g_{10} &=& - a_5 m_s (2 D_{13}^f + 2 D_{23}^f + D_3^f)  \\
g_{11} &=& a_5 m_b (2 D_{13}^f + D_1^f)
\end{eqnarray}
}
where $a_i$ is defined as
\begin{eqnarray}
a_1 =\frac{1}{192\pi^2sm_bm_s^2m_w^2(m_b^2 - m_s^2)s_w^2}, \hspace{3mm}a_2 =\frac{1}{768\pi^2m_bm_s^2m_w^2(m_b^2 - m_s^2)(m_z^2 -im_z\Gamma{z} - s)c_w^2s_w^4}\nonumber\\
a_3 =\frac{1}{96\pi^2sm_w^2s_w^2}, \hspace{3mm}a_4 =\frac{1}{768\pi^2m_w^2(m_z^2-im_z\Gamma{z}-s)c_w^2s_w^4},\hspace{3mm} a_5 =\frac{1}{32\pi^2s_w^4}\nonumber
\end{eqnarray}
where $c_w= cos\theta_w$ and $s_w= sin\theta_w$. In the
presentation of $g_j$ above, we have used the definition of scalar
integrals $Bs$, $Cs$,and $Ds$\cite{333},and these functions, $Bs$,
$Cs$,and $Ds$, with superscripts a,b,...,f  have the arguments \\
\begin{math}
(0,m_j^2,m_w^2),\hspace{3mm} (m_b^2,m_j^2,m_w^2),\hspace{3mm} (m_s^2,m_j^2,m_w^2), \hspace{3mm} (m_b^2,m_s^2,s,m_w^2,m_j^2,m_w^2)\\
 \hspace{3mm} (m_b^2,m_s^2,s,m_j^2,m_w^2,m_j^2), \hspace{3cm} (0,m_b^2,m_s^2,0,t,s,0,m_w^2,m_j^2,m_w^2)\\
\end{math}
respectively. Here $m_j$ denotes the mass of up-type quark  $u,c,t$.

\begin{figure}
\epsfxsize=12cm \centerline{\epsffile{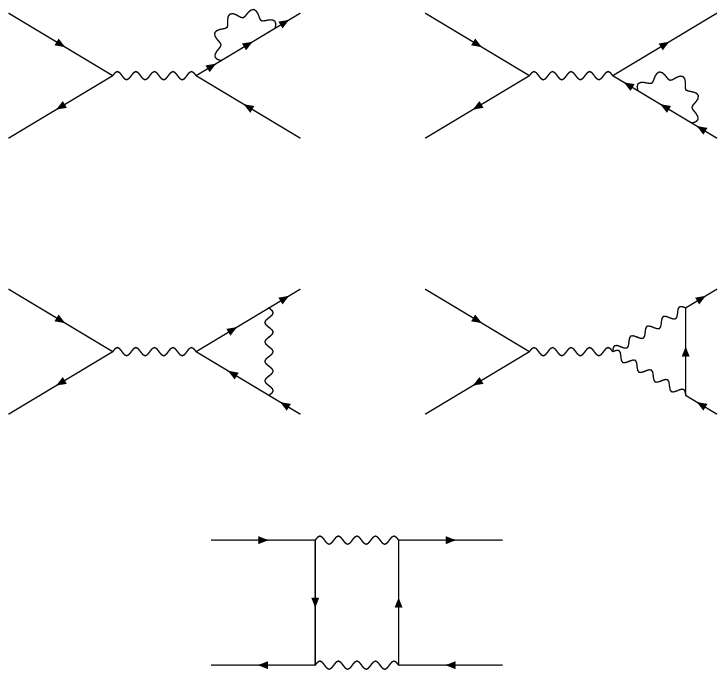}}
\caption[]{Typical Feynman diagram of prosess $e^+e^- \rightarrow
b \bar s$}
\end{figure}

\begin{figure}
\epsfxsize=18cm \centerline{\epsffile{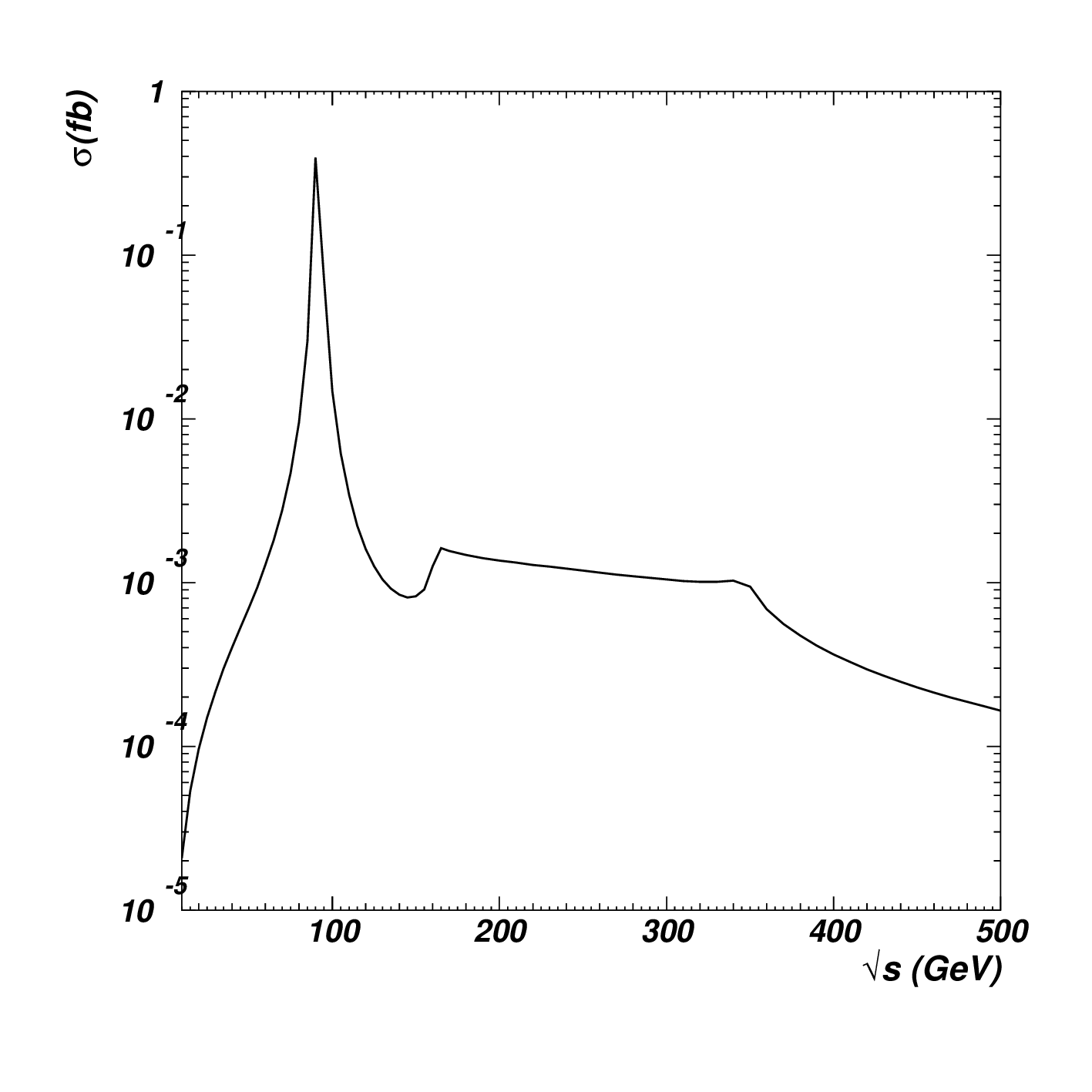}}
\caption[]{Cross section of the process $e^+e^- \rightarrow b \bar
s$ as a function of $\sqrt{s}$.}
\end{figure}

\begin{figure}
\epsfxsize=18cm \centerline{\epsffile{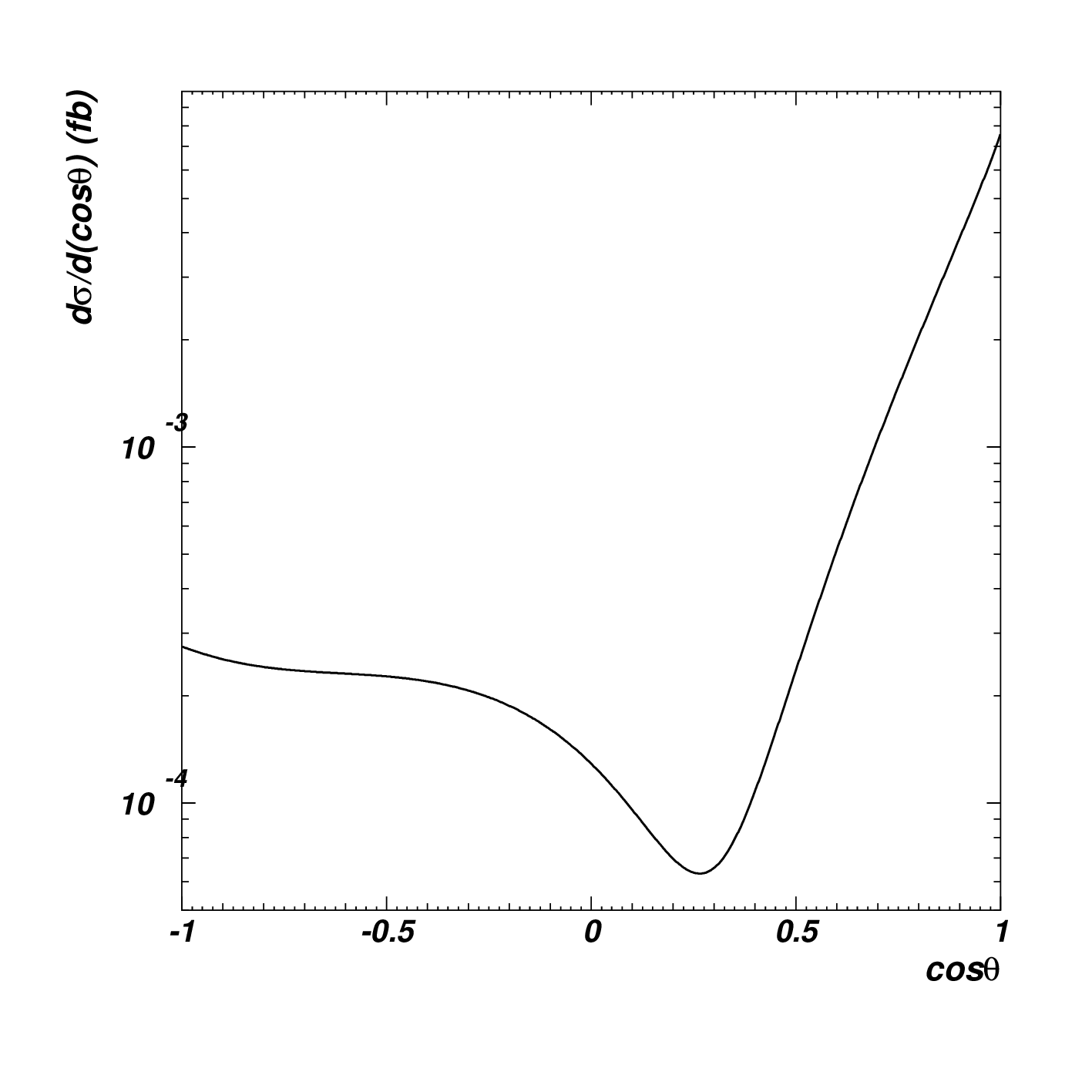}} \caption[]{
Differential cross section of the process $e^+e^- \rightarrow b
\bar s$, where $\sqrt{s}=200$ GeV. }
\end{figure}

\begin{figure}
\epsfxsize=18cm \centerline{\epsffile{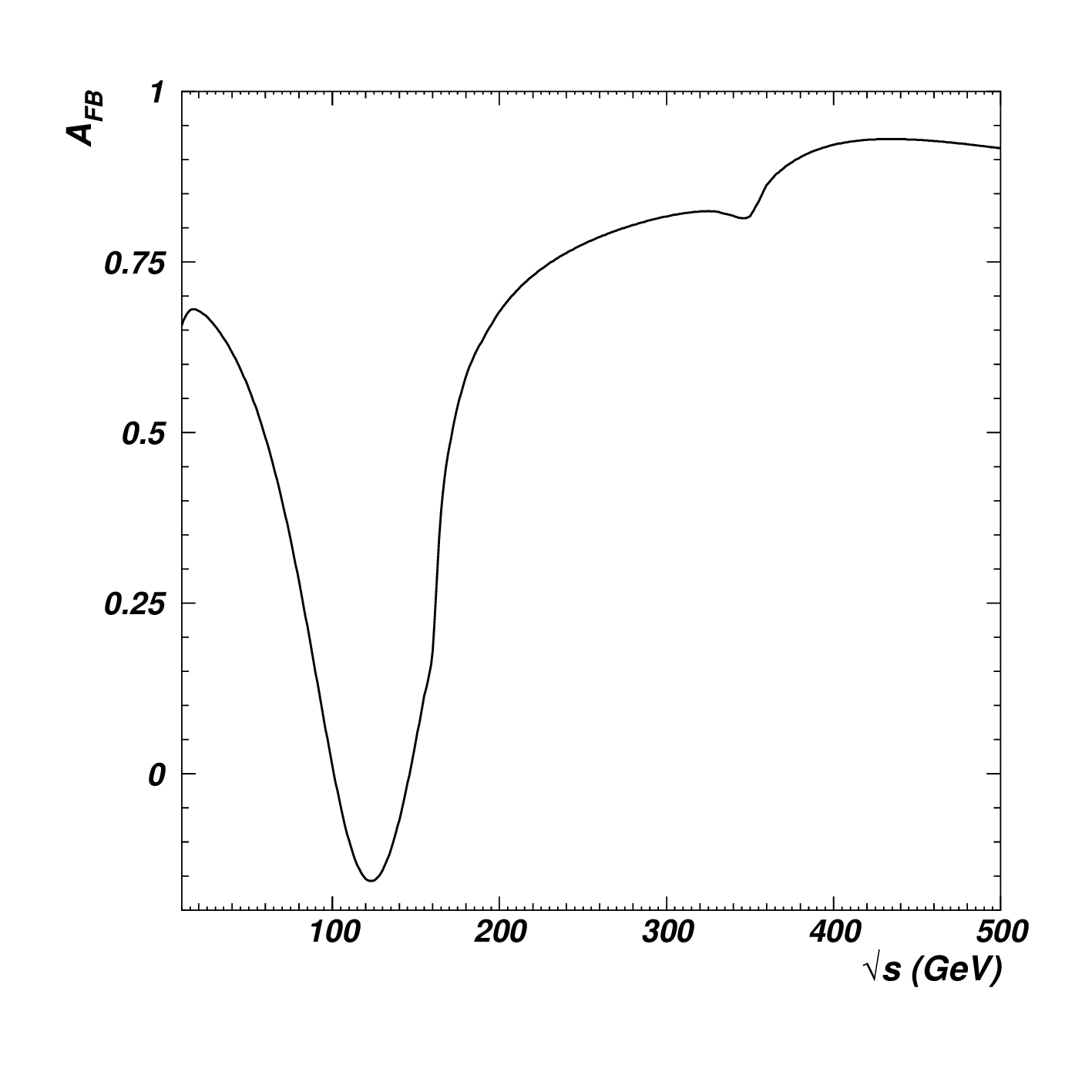}} \caption[]{
$A_{FB}$ of the process $e^+e^- \rightarrow b \bar s$ as a
function of $\sqrt{s}$. }
\end{figure}



\setcounter{figure}{0} \setcounter{table}{0}
\setcounter{equation}{0}

\newpage
\def\beq{\begin{equation}}
\def\eeq{\end{equation}}
\def\bea{\begin{eqnarray}}
\def\eea{\end{eqnarray}}
\def\ve{\vert}
\def\vel{\left|}
\def\ver{\right|}
\def\nnb{\nonumber}
\def\ga{\left(}
\def\dr{\right)}
\def\aga{\left\{}
\def\adr{\right\}}
\def\rar{\rightarrow}
\def\nnb{\nonumber}
\def\la{\langle}
\def\ra{\rangle}
\def\ba{\begin{array}}
\def\ea{\end{array}}
\def\tep{$B \rar K \ell^+ \ell^-$}
\def\tepm{$B \rar K \mu^+ \mu^-$}
\def\tept{$B \rar K \tau^+ \tau^-$}
\def\ds{\displaystyle}

\def\bos{\lower 0.5cm\hbox{{\vrule width 0pt height 1.3cm}}}
\def\aaa{\lower 0.cm\hbox{{\vrule width 0pt height .8cm}}}
\def\dol{\lower 0.6cm\hbox{{\vrule width 0pt height .8cm}}}

\section{$B\rightarrow X_s\tau^+\tau^-$ in a CP softly broken
two Higgs doublet model}

\begin{footnotesize}
\begin{center}\begin{minipage}{5in}
\begin{center} ABSTRACT\end{center}

 The differential branching ratio, forward-backward asymmetry,
CP asymmetry and
lepton polarization
for a B-meson to decay to strange hadronic final states and a
$\tau^+\tau^-$ pair in a CP softly broken
two Higgs doublet model are computed.
It is shown that
contributions of neutral Higgs bosons  to the decay are
quite significant when $\tan\beta$ is large.
And it is proposed to measure the direct
CP asymmetry in back-forward asymmetry.

\end{minipage}\end{center}
\end{footnotesize}


\subsection{Introduction}

The origin of the CP violation has been one of main issues
in high energy physics since the discovery of the CP violation in
the $K_0-\overline{K}_0$ systerm in 1964 \cite{new1}.
The measurements of electric dipole moments of the neutron and
electron and the matter-antimatter asymmetry in the universe
indicate that one needs new sources of CP violation in addition
to the CP violation come from CKM matrix, which has been one
of motivations to search new theoretical models beyond the
standard model (SM).

The minimal extension of the SM is to enlarge the Higgs sectors
of the SM \cite{new2}. It has been shown that if one adheres to the
natural flavor conservation (NFC) in the Higgs sector, then
a minimum of three Higgs doublets are necessary in order to have
spontaneous CP violations \cite{new3}. However, the constraint
can be evaded if one allows the real and image parts of $\phi_1^+
\phi_2$ have different self-couplings and adds a linear term of Re($\phi^+_1\phi_2$)
in the Higgs potential (see below Eq. (\ref{eq2}) with $m_4^2=0$).
Then, one can construct a CP spontaneously broken two Higgs
doublet (2HDM), which is the minimal and the most "economical" one \footnote{
Comparing the Model III  2HDM \cite{new4}, in which  CP is explicitly
violated, the CP spontaneously broken 2HDM has only two new
parameters besides the masses of the Higgs bosons in the
large $\tan\beta$ limit (see below). In this sense it is the most
"economical".
}
among the extensions of the SM that provide  new source of CP violation. Furthermore,
in addition to the above terms, if one adds a linear term of Im($\phi^+_1\phi_2$), then
one has a CP softly broken 2HDM~\cite{Georgi}.

Flavor changing neutral current (FCNC) transitions $B\rightarrow X_s\gamma$ and
$B\rightarrow X_sl^+l^-$ provide testing grounds for the SM at
the loop level and sensitivity to new physics.
Rare decays $B\rightarrow X_sl^+l^-(l=e,\mu)$ have been
extensively investigated in both SM and the beyond
 \cite{gswg,ex11}.
In these processes contributions from exchanging neutral
Higgs bosons (NHB)
can be safely neglected because of smallness of $\frac{m_l}{m_W}
(l=e,\mu)$.
The inclusive decay $B\rightarrow X_s\tau^+
\tau^-$ has also been investigated in the SM, the model II 2HDM and
SUSY models  with and without
including the contributions of NHB \cite{new7,new8,Dai,add2}.
In this note we investigate the inclusive decay $B\rightarrow X_s\tau^+\tau^-$
 with emphasis on CP violation effect in a CP softly broken
2HDM, which we shall call Model IV hereafter for the sake of
simplisity. We consider the Model IV in which the up-type quarks
get masses from Yukawa couplings to the one Higgs doublet $H_2$
and down-type quarks and leptons get masses from Yukawa couplings
to the another Higgs doublet $H_1$. The Higgs boson couplings to
down-type quarks and leptons depend on only the CP violated phase
$\xi$ which comes from the expectation value of Higgs and the
ratio $tg\beta =\frac{v_2}{v_1}$ in the large $tg\beta$ limit (see
next subsection), which are the free parameters in the model.
Because the couplings of the charged Higgs to fermions in Model IV
are the same as those in the model II, the constraints on
$\tan\beta$ due to effects arising from the charged Higgs are the
same as those in the model II. Constraints on $tg\beta$ from
$K-\bar{K}$ and $B-\bar{B}$ mixing, $\Gamma(b\rightarrow s\gamma)
$,$\Gamma(b\rightarrow c\tau\bar{\nu}_{\tau})$ and $R_b$ have been
given \cite{15}
\begin{equation}
0.7\le tg\beta\le 0.52(\frac{m_{H^{\pm}}}{1Gev})
\end{equation}
(and the lower limit $m_{H^{\pm}}\ge 200Gev$ has also been given in the ref.
\cite{15}). It is obvious that the contributions from exchanging neutral Higgs
bosons now is enhanced roughly by a factor of $tg^2\beta$ and can compete with
those from exchanging $\gamma,~Z$ when $tg\beta$ is large enough.
 Because the CP violation effects in $B\rightarrow X_s \tau^+\tau^-$ come
 from the couplings of NHB to leptons and quarks,
 we shall be interested in the large $\tan\beta$ limit in this note.
 The constraints on $\xi$ can be obtained from the electric dipole moments
 (EDM) of the neutron and electron, which will be analysed in the next
 subsection.

\subsection{Model description}
Consider two complex $y=1$, $SU(2)_w$ doublet scalar fields, $\phi_1$ and
$\phi_2$. The Higgs potential which spontaneously breaks $SU(2)\times U(1)$
down to $U(1)_{EM}$ can be written in the following form \cite{Georgi}:
\begin{eqnarray}
V(\phi_1,\phi_2) &=&
\sum_{i=1,2} [m_i^2 \phi_i^+ \phi_i +\lambda_i  (\phi_i^+ \phi_i)^2] \nonumber\\
&&+ m_3^2 Re(\phi_1^+\phi_2) + m_4^2 Im(\phi_1^+ \phi_2) \nonumber\\
&&+ \lambda_3 [ (\phi_1^+ \phi_1)(\phi_2^+ \phi_2) ]
+ \lambda_4 [ (\phi_1^+ \phi_2) (\phi_2^+ \phi_1) ]
\nonumber \\
&&
+ \lambda_5
 [ \mbox{Re}(\phi_1^+ \phi_2)]^2
+ \lambda_6
 [ \mbox{Im}(\phi_1^+ \phi_2) ]^2
 \label{eq2}
\end{eqnarray}
Hermiticity requires that all parameters are real.
The potential is CP softly broken due to the presence of the term $m_4^2 Im(\phi_1^+ \phi_2)$.
It is easy to see that the minimum of the potential is at
\begin{eqnarray}
<\phi_1>=\left( \begin{array}{c}
0 \\
v_1
\end{array}
\right), \ \ \ \
<\phi_2>=\left( \begin{array}{c}
0 \\
v_2 e^{i\xi}
\end{array}
\right),
\end{eqnarray}
thus breaking $SU(2)\times U(1)$ down to $U(1)_{EM}$ and
simutaneously breaking CP, as desired. It should be noticed that
only for  $\lambda_5 \neq \lambda_6$, the phase $\xi$ can't
rotated away as usual, which breaks the CP-conservation. If
$m_4^2$=0 in (2) then the potential is CP invarint. It has been
shown that the CP spontaneously breaking happens at
(3)~\cite{leb}. We limit ourself to the case of $m_4^2 \neq 0$ in
the paper and shall investigate the $m_4^2$=0 case in a separate
paper ~\cite{hz1}.

In the following we will work out the mass spectrum of the Higgs
boson. For charged components, the mass-squared matrix for negtive states is
\begin{eqnarray}
\lambda_4 \left(
\begin{array}{cc}
v_1^2 & -v_1 v_2 e^{i\xi} \\
-v_1 v_2 e^{-i\xi} & v_2^2
\end{array}
\right),
\end{eqnarray}
Diagonalizing the mass-squared matrix results in one
zero-mass Goldstone state:
\begin{eqnarray}
G^-=e^{i\xi} \sin\beta \phi_2^- +\cos\beta \phi_1^-,
\end{eqnarray}
and one massive charged Higgs boson state:
\begin{eqnarray}
H^-=e^{i\xi} \cos\beta \phi_2^- - \sin \beta \phi_1^-,
\\
m_{H^-}=|\lambda_4| (v_1^2+v_2^2),
\end{eqnarray}
where $\tan\beta = v_2/v_1$.
Correspondingly we could also get the positive states $G^+$ and $H^+$ with the
same masses zero and $|\lambda_4| (v_1^2+v_2^2)$, respectively.

For neutral Higgs components, because CP-conservation is breaking, the
mass-squared matrix is $4\times 4$, which could not be simply
separated into
two $2\times 2$ matrices as usual.
However, in
the case of large $\tan\beta$ which is we intrested in,
the neutral parts can be
written as separately two $2\times 2$ matrices and one of them is
\begin{eqnarray}
v_2^2 \left(
\begin{array}{cc}
\frac{\lambda_5+ \lambda_6+(\lambda_6-\lambda_5) \cos(2\xi)}{2} &
- \frac{(\lambda_6-\lambda_5) \sin(2\xi)}{2} \\
- \frac{(\lambda_6-\lambda_5) \sin(2\xi)}{2} &
\frac{\lambda_5+ \lambda_6+(\lambda_5-\lambda_6) \cos(2\xi)}{2}
\end{array}
\right).
\end{eqnarray}
Diagonalizing the Higgs boson mass-squared matrix results in two
eigenstates:
\begin{eqnarray}
\left(
\begin{array}{c}
H^0_1\\
H^0_2
\end{array}
\right) =
\sqrt{2} \left(
\begin{array}{cc}
c_\xi & -s_\xi\\
s_\xi & c_\xi
\end{array}
\right)
\left(
\begin{array}{c}
{\rm Im} \phi_1^0 \\
{\rm Re} \phi_1^0
\end{array}
\right)
\label{eq9}
\end{eqnarray}
with masses
\begin{eqnarray}
m_{H^0_1}^2=\lambda_5 v_2^2 \nonumber \\
m_{H^0_2}^2=\lambda_6 v_2^2,
\label{eq10}
\end{eqnarray}
where $c_\xi=\cos\xi$ and $s_\xi=\sin\xi$. The diagonalizing of the
$4\times 4$ neutral Higgs mass-squared matrix has been analytically
carried out under some assumptions in Ref. \cite{Vend} and the results
reduce to Eq. (\ref{eq9}) and (\ref{eq10}) in the case of large
$\tan\beta$.

The another $2\times 2$ matrix can be similarly deal with.
Because the couplings of the third physical neutral Higgs boson
and neutral Goldstone to down-type quarks and leptons
are not enhanced for large $\tan\beta$ case in which we are interested,
we do not show the explicit results.

Now, we turn to the discussion of the Higgs-fermion-fermion couplings.
After completing the transformation from the weak states to the mass
states, the couplings of neutral Higgs to fermions which are
relevant to our analysis are
\begin{eqnarray}
H^0_1 \bar f f:\ \ \ \ &&  \frac{i g m_f}{2 m_w \cos\beta}
(s_\xi- i c_\xi \gamma_5)
\nonumber\\
H^0_2 \bar f f: \ \ \ \
 &&  -\frac{i g m_f}{ 2 m_w \cos\beta} (c_\xi+ i s_\xi \gamma_5)
 \label{eq11}
\end{eqnarray}
where $f$ represents down-type quarks and leptons. And the
couplings of the charged Higgs bosons to fermions are the same as
those in the
CP-conservative 2HDM (model II, for examples see Ref. \cite{19}).
This is in contrary with the model III in which the couplings
of the charged Higgs to fermions are quite different from model II.
It is easy to see from Eq. (\ref{eq11}) that the contributions
come from exchanging NHB is proportional to
$\sqrt{2} G_F s_\xi c_\xi  m_f^2/\cos^2\beta$, so that the
constaints due to EDM translate into the constraints
on $\sin 2\xi \tan^2\beta$ ($1/\cos\beta \sim \tan\beta$ in the large
$\tan\beta$ limit). According to the analysis in Ref. \cite{new6},
we have the constraint
\begin{eqnarray}
\sqrt{|\sin 2\xi|}\tan\beta < 50
\label{eqa}
\end{eqnarray}
from the neutron EDM. And the constraint from the electron EDM
is not stronger than Eq. (\ref{eqa}). It is obvious from Eq. (\ref{eqa}) that there
is a constraint on $\xi$ only if $\tan \beta >  50$ and the stringent constraint on $\tan \beta$
comes out and is $\tan \beta < 50$ when $\xi = \pi $/4.

\subsection{Formula for $B \rightarrow X_{s} \tau^{+} \tau^{-}$ }
Inclusive decay rates of heavy hadrons can be calculated in heavy quark
effective theory (HQET) \cite{17} and it has been shown that the
leading terms in $1/m_Q$
expansion turn out to be the decay of a free (heavy) quark and corrections stem
from the order $1/m_Q^2$ \cite{18}. In what follows we shall
calculate the leading term.
The transition rate for
$b\rightarrow s\tau^+\tau^-$ can be computed in the framework of the QCD
corrected effective weak hamiltonian, obtained by integrating out the top quark,
Higgs bosons and $W^{\pm},Z$ bosons
\begin{equation}\label{ham}
H_{eff}=\frac{4G_F}{\sqrt{2}}V_{tb}V^*_{ts}(\sum_{i=1}^{10}C_i(\mu)O_i(\mu)
+\sum_{i=1}^{10}C_{Q_i}(\mu)Q_i(\mu))
\end{equation}
where $O_i(i=1,\cdots ,10)$ is the same as that given in the
ref.\cite{gswg}, $Q_i$'s come from exchanging the neutral Higgs
bosons and are defined in Ref. \cite{Dai}. The explicit
expressions of the operators governing $B\rightarrow
X_s\tau^+\tau^-$ are given as follows:
\begin{eqnarray}
O_7 &=& (e/16\pi^2) m_b (\bar{s}_{L\alpha} \sigma^{\mu\nu}
b_{R\alpha}) F_{\mu\nu}, \nonumber \\
O_8 &=& (e/16\pi^2) (\bar{s}_{L\alpha} \gamma^{\mu}
b_{L\alpha}) \bar\tau \gamma_{\mu} \tau, \nonumber \\
O_9 &=& (e/16\pi^2) (\bar{s}_{L\alpha} \gamma^{\mu}
b_{L\alpha}) \bar\tau \gamma_{\mu} \gamma_5 \tau, \nonumber \\
Q_1 &=& (e^2/16\pi^2) (\bar{s}_{L\alpha} b_{R\alpha})
 (\bar\tau \tau), \nonumber \\
Q_2 &=& (e^2/16\pi^2) (\bar{s}_{L\alpha} b_{R\alpha})
 (\bar\tau \gamma_5 \tau).
\end{eqnarray}

At the renormalization point $\mu=m_W$ the coefficients $C_i$'s in
the effective hamiltonian have been given in the ref.\cite{gswg}
and $C_{Q_i}$'s are (neglecting the $O(tg\beta)$ term)
\begin{eqnarray}
C_{Q_1}(m_W)&=&\frac{m_bm_{\tau}tg^2\beta x_t}{2 sin^2\theta_W}
\{
\sum_{i=H_1,H_2} \frac{ A_{i}}{m_{i}^2} (f_1 B_{i}+f_2 E_i) \},
\nonumber \\
C_{Q_2}(m_W)&=&\frac{m_bm_{\tau}tg^2\beta x_t}{2 sin^2\theta_W}
\{
\sum_{i=H_1,H_2} \frac{ D_{i}}{m_{i}^2} (f_1 B_{i}+f_2 E_i) \},
\nonumber \\
C_{Q_3}(m_W)&=&\frac{m_be^2}{m_{\tau}g_s^2}(C_{Q_1}(m_W)+C_{Q_2}(m_W)),
\nonumber
\\
C_{Q_4}(m_W)&=&\frac{m_be^2}{m_{\tau}g_s^2}(C_{Q_1}(m_W)-C_{Q_2}(m_W)), \nonumber
\\
C_{Q_i}(m_W)&=&0, ~~~~i=5,\cdots, 10
\label{eq1}
\end{eqnarray}
where
\begin{eqnarray}
A_{H_1}&=-s_\xi,  \  D_{H_1}&=i c_\xi,
\nonumber \\
A_{H_2}&= c_\xi, \  D_{H_2}&=i s_\xi,
\nonumber \\
B_{H_1}&=\frac{ i c_\xi-s_\xi}{2}, \
B_{H_2}&=\frac{c_\xi+i s_\xi}{2},
\nonumber
\end{eqnarray}
\begin{eqnarray}
f_1&=& \frac{x_t ln x_t}{x_t-1}-
\frac{x_{H^\pm} ln x_{H^\pm}-x_t ln x_t }{x_{H^\pm}-x_t},
\nonumber \\
f_2&=& \frac{x_t ln x_t}{(x_t-1)(x_{H^\pm}-1)}-
\frac{x_{H^\pm} ln x_{H^\pm} }{(x_{H^\pm}-x_t)(x_{H^\pm}-1)}
\end{eqnarray}
with $ x_i=m_i^2/m_w^2$.
In Eq. (\ref{eq1}), $E_i$ are given by
\begin{eqnarray}
E_{H_1}&=& \frac{1}{2} (-s_\xi c_1+ c_\xi c_2),
\nonumber \\
E_{H_2}&=& \frac{1}{2} (c_\xi c_1+ s_\xi c_2),
\nonumber \\
c_1&=& -x_{H^\pm}+c_\xi x_{H_1} (c_\xi+i s_\xi)
+s_\xi x_{H_2} (s_\xi-i c_\xi),
\nonumber \\
c_2&=& i\left(- x_{H^\pm}+s_\xi x_{H_1} (s_\xi-i c_\xi)
+c_\xi x_{H_2} (c_\xi+i s_\xi) \right).
\end{eqnarray}

Neglecting the strange quark mass, the effective hamiltonian (\ref{ham}) leads
to the following matrix element for $b\rightarrow s\tau^+\tau^-$
\begin{eqnarray}\label{matrix}\nonumber
M&=&\frac{G_F\alpha}{\sqrt{2}\pi}V_{tb}V^*_{ts}[C^{eff}_8\bar{s}_L\gamma_{\mu}
b_L\bar{\tau}\gamma^{\mu}\tau+C_9\bar{s}_L\gamma_{\mu}b_L\bar{\tau}\gamma^{\mu}
\gamma^5\tau\\
&+&2C_7m_b\bar{s}_Li\sigma^{\mu\nu}\frac{q^{\nu}}{q^2}b_R\bar{\tau}\gamma^{\mu}
\tau+C_{Q_1}\bar{s}_Lb_R\bar{\tau}\tau+C_{Q_2}\bar{s}_Lb_R\bar{\tau}\gamma^5
\tau],
\end{eqnarray}
where \cite{gswg,new7,10}
\begin{eqnarray}\label{coeff}\nonumber
C^{eff}_8&=&C_8+\{g(\frac{m_c}{m_b},\hat{s})\\
&+&\frac{3}{\alpha^2}k\sum_{V_i=
\psi^{\prime}, \psi^{\prime\prime}...}
\frac{\pi M_{V_i}\Gamma(V_i\rightarrow\tau^+\tau^-)}{M^2_{V_i}-q^2
-iM_{V_i}\Gamma_{V_i}}\}(3C_1+C_2),
\end{eqnarray}
with $\hat{s}=q^2/m_b^2,~~q=(p_{\tau^+}+p_{\tau^-})^2$. In
(\ref{coeff}) $g(\frac{m_c}{m_b},\hat{s})$ arises from the
one-loop matrix element of the four-quark operators and can be
found in Refs. \cite{gswg,dba}.
 The second term
in braces in (\ref{coeff}) estimates the long-distance
contribution from the intermediate, $\psi^{\prime}$,
$\psi^{\prime\prime}$ ... \cite{gswg,10}. In our numerical
calculations, we choose $ k (3C_1+C_2)=-0.875$ \cite{PDG}.

The QCD corrections to coefficients $C_i$ and $C_{Q_i}$ can be incooperated
in the standard way by using the renormalization group equations.
Although the $C_i$ at the scale $\mu=O(m_b)$
have been given in the next-to-leading order
approximation (NLO) and without including mixing with $Q_i$, we
use the values of $C_i$ only in the leading order approximation (LO)
since no $C_{Q_i}$ have been calculated in NLO.
The $C_i$ and $C_{Q_i}$ with LO QCD corrections have been
given in Ref. \cite{Dai}.
\begin{eqnarray}\label{c7}
C_7(m_b)&=&\eta^{-16/23}\left[ C_7(m_W)-[\frac{58}{135}(\eta^{10/23}-1)
+\frac{29}{189}(\eta^{28/23}-1)]C_2(m_W) \right.
\nonumber \\
&&\left. -0.012C_{Q_3}(m_W)\right],
\label{eq18}
\end{eqnarray}
\begin{eqnarray}
C_8(m_b)&=&C_8(m_W)+\frac{4\pi}{\alpha_s(m_W)}[-\frac{4}{33}(1-\eta^{-11/23})
+\frac{8}{87}(1-\eta^{-29/23})]C_2(m_W),\\
C_9(m_b)&=&C_9(m_W),\\
C_{Q_i}(m_b)&=&\eta^{-\gamma_Q/\beta_0}C_{Q_i}(m_W),~~i=1,2,
\end{eqnarray}
where $\gamma_Q=-4$ \cite{21} is the anomalous dimension of $\bar{s}_Lb_R$,
$\beta_0=11-2 n_f/3$, and $\eta=\alpha_s(m_b)/\alpha_s(m_W)$.

After a straightforward calculation,  we obtain the
invariant dilepton mass distribution \cite{Dai}
\begin{eqnarray}
\frac{{\rm d}\Gamma(B\rightarrow X_s\tau^{+}\tau^{-})}{{\rm d}s}
 &=& B(B\rightarrow X_c l {\bar \nu}) \frac{{\alpha}^2}
 {4 \pi^2 f(m_c/m_b)} (1-s)^2(1-\frac{4t^2}{s})^{1/2}
 \frac{|V_{tb}V_{ts}^{*}|^2}{|V_{cb}|^2} D(s) \nonumber \\
 D(s) &=& |C_8^{eff}|^2(1+\frac{2t^2}{s})(1+2s)
      + 4|C_7|^2(1+ \frac{2t^2}{s})(1+\frac{2}{s}) \nonumber \\
    & &  + |C_9|^2 [ ( 1 + 2s) + \frac{2t^2}{s}(1-4s)]
      +12 {\rm Re}(C_7 C_{8}^{eff*})(1+\frac{2t^2}{s}) \nonumber \\
  & & + \frac{3}{2}|C_{Q_1}|^2 (s-4t^2) + \frac{3}{2}|C_{Q_2}|^2s
      + 6{\rm Re}(C_9 C_{Q_2}^{*}) t
\label{eq22}
\end{eqnarray}
where s=$q^2/m_b^2$, t=$m_{\tau}/m_{b}$,
$B(B\rightarrow X_c l {\bar \nu})$ is the branching ratio,
$f$ is the phase-space factor and f(x)=$1-8 x^2+8 x^6
-x^8-24 x^4 \ln ~ x$.

 The CP asymmetry for the $B \rightarrow X_s l^+ l^-$ and
$\overline{ B} \rightarrow \overline{ X}_s l^+ l^-$ is defined as
\begin{eqnarray}
A_{CP}^1(s)=\frac{{\rm d}\Gamma/{\rm d}s -  {\rm d}\overline{\Gamma}/{\rm d}s}{
{\rm d}\Gamma/{\rm d}s +  {\rm d}\overline{\Gamma}/{\rm d}s}.
\end{eqnarray}
We also give the forward-backward asymmetry
\begin{eqnarray}
A(s)=\frac{\int^{1}_{0}dz \frac{d^2\Gamma}{ds dz} -
\int^{0}_{-1}dz \frac{d^2\Gamma}{ds dz}}{
\int^{1}_{0}dz \frac{d^2\Gamma}{ds dz} +
\int^{0}_{-1}dz \frac{d^2\Gamma}{ds dz}}
=\frac{E(s)}{D(s)}
\end{eqnarray}
where $z=\cos\theta$ and $\theta$ is the angle between the momentum
of the B-meson and that of $l^+$ in the center of mass frame of the
dileptons $\tau^+\tau^-$. Here,
\begin{eqnarray}
E(s)={\rm Re} (C_8^{eff} C_9^* s+2 C_7 C_9^*+
C_8^{eff} C_{Q1}^* t+ 2 C_7 C_{Q2}^* t).
\label{eq26}
\end{eqnarray}
The CP asymmetry in the forward-backward asymmetry
for $B \rightarrow X_s \tau^+ \tau^-$ and
$\overline{ B} \rightarrow \overline{X}_s \tau^+ \tau^-$
is defined as
\begin{eqnarray}
A_{CP}^2(s)=
\frac{ A(s)- \overline{A} (s)}{ A(s)+ \overline{A}(s)}.
\end{eqnarray}
It is easy to see from Eq. (\ref{eq22})  that  the CP asymmetry
$A_{CP}^1$ is very small because the weak phase difference in
$C_7 C_8^{eff}$ arises from the small mixing of $O_7$ with
$Q_3$ (see Eq. (\ref{eq18})). In contrast with it, $A_{CP}^2$
can reach a large value when $\tan\beta$ is large, as can be seen
from Eq. (\ref{eq26}) and (\ref{eq1}).
Therefore, we propose to measure $A_{CP}^2$ in order to search for
new CP violation sources.

Let us now discuss the lepton polarization effects. We define three
orthogonal unit vectors:
\bea
\vec{e}_L &=& \frac{\vec{p}_1}{\vel \vec{p}_1 \ver}~, \nnb \\
\vec{e}_N&=& \frac{\vec{p}_{s} \times \vec{p}_1}
{\vel \vec{p}_{s} \times \vec{p}_1 \ver}~, \nnb \\
\vec{e}_T &=& \vec{e}_N \times \vec{e}_L~, \nnb
\eea
where $\vec{p}_1$ and $\vec{p}_{s}$ are the three momenta of the
$\ell^-$ lepton
and the $s$ quark, respectively, in the center of mass of the
$\ell^+~\ell^-$ system. The differential decay rate for any given spin
direction $\vec{n}$ of the $\ell^-$ lepton, where $\vec{n}$ is a unit vector
in the $\ell^-$ lepton rest frame, can be written as
\bea
\frac{d \Gamma \ga \vec{n} \dr}{{\rm d} s} =
\frac{1}{2} \ga \frac{d \Gamma}{{\rm d} s} \dr_{\!\!\! 0}
\Big[ 1 + \ga P_L\, \vec{e}_L + P_N\, \vec{e}_N + P_T\, \vec{e}_T \dr \cdot
\vec{n} \Big]~,
\label{eq27}
\eea
where the subscript "0" corresponds to the unpolarized case, and $P_L,~P_T$,
and $P_N$, which correspond to the longitudinal, transverse and normal
projections of the lepton spin, respectively, are functions of $s$.
From Eq. (\ref{eq27}), one has
\bea
P_i (s) = \frac{ {\displaystyle{\frac{d \Gamma}{d s}
\ga \vec{n}=\vec{e}_i \dr -
\frac{d \Gamma}{ds}\ga \vec{n}=-\vec{e}_i \dr}} }
{ {\displaystyle{\frac{d \Gamma}{ds}\ga \vec{n}=\vec{e}_i \dr +
\frac{d \Gamma}{ds}\ga \vec{n}=-\vec{e}_i \dr}} } ~.
\eea

The calculations for the $P_i$'s (i = $L,~T,~N$) lead to the following
results:
\bea
P_L &=&  (1-\frac{4 t^2}{s})^{1/2} \frac{D_L(s)}{D(s)},
\nonumber \\
P_N&=& \frac{3 \pi}{4 s^{1/2}} (1-\frac{4 t^2}{s})^{1/2}
\frac{D_N(s)}{D(s)},
\nonumber \\
P_T&=& -\frac{3 \pi t}{2 s^{1/2}}
\frac{D_T(s)}{D(s)},
\eea
where
\bea
D_L(s) &=& {\rm Re}\left(
 2 (1+2 s) C_8^{eff} C_9^*+12 C_7 C_9^*- 6 t C_{Q_1} C_9^*-
3 s C_{Q_1} C_{Q_2}^* \right), \nonumber \\
D_N(s) &=&  {\rm Im} \left(
2 s  C_{Q_1} C_7^*+s C_{Q_1} C_8^{eff *}+s C_{Q_2} C_9^*+
4 t C_9 C_7^*+ 2 t s C_8^{eff\ *} C_9 \right),
 \nonumber \\
D_T(s) &=&   {\rm Re}\left(
-2 C_7 C_9^*+ 4 C_8^{eff} C_7^* +\frac{4}{s} |C_7|^2-
C_8^{eff} C_9^* \right. \nonumber \\
&& \left. +s |C_8^{eff}|^2 -\frac{s-4 t^2}{2 t} C_{Q_1} C_9^*
-\frac{s}{t} C_{Q_2} C_7^*-\frac{s}{2 t} C_8^{eff} C_{Q_2}^* \right).
\label{eq30}
\eea
$P_i$ (i=L, T, N) have been given
in the ref. [9], where there are some errors in $P_T$ and they gave only
two terms in $D_N$, the numerator of $P_N$.
We remind that $P_N$ is the CP-violating projection of the lepton spin
onto the normal of the decay plane. Because $P_N$ in $B \rightarrow
X_s l^+ l^-$ comes from both the quark and lepton sectors,
purely hadronic and leptonic CP-violating observables,
such as $d_n$ or $d_e$, do not necessarily strongly constrain $P_N$
\cite{add1}. So it is advantageous to use $P_N$ to investigate
CP violation effects in some extensions of SM \cite{add3}.
In the model IV 2HDM, as pointed out above, $d_n$ and $d_e$ constrain
$\sqrt{|\sin 2\xi|}\tan\beta$ and consequently $P_N$ through
$C_{Q_i}$ ($i=1,2$) (see Eq. (\ref{eq30})).

\subsection{Numerical results}
The following parameters have been used in the numerical calculations:
$$
m_t=175Gev,~m_b=5.0Gev,~m_c=1.6Gev,~m_{\tau}=1.77Gev,~\eta=1.724,
$$
$$
m_{H_1}=100 Gev, ~m_{H_2}=m_{H^\pm}=200 Gev.
$$

Numerical results are shown in Figs. 1-9. From Figs. 1 and 2, we can see
that the contributions of NHB to the differential branching ratio
$d\Gamma/ds$ are significant when $\tan\beta$ is not smaller than 30 and the
masses of NHB are in the reasonable region, and the forward-backward asymmetry
$A(s)$ is more sensitive to $\tan\beta$
than $d\Gamma/ds$, which is similar to the case of the normal
2HDM without CP violation \cite{Dai}.

The direct CP violation $A_{CP}^i$ ($i=1,2$) and CP-violating polarization
$P_N$ of $B \rightarrow X_s \tau^+\tau^-$ are presented in Figs. 3-7,
respectively. As expected, $A_{CP}^1$ is about $0.1$\% and hard to be
measured. However, $A_{CP}^2$ can reach about $10$\%.
$A_{CP}^2$ is strongly dependent of the CP violation
phase $\xi$ and comes mainly from exchanging NHBs as expected.
From Figs. 6 and 7, one can see that $P_N$ is also strongly dependent of
the CP violation phase $\xi$ and can be as large as 5\% for some
values of $\xi$, which should be within the luminosity reach of
coming B factories, and comes mainly from NHB contributions in the most of range
of $\xi$.

Figs. 8 and 9 show the longitudinal and transverse polarizations
respectively. It is obviously that the contributions of NHB can change
the polarization greatly, especially when $\tan\beta$ is large,
 and the dependence of $P_L$ on CP violation phase $\xi$ is not significant
in the most of range of $\xi$.
The longitudinal  polarization of $B \rightarrow X_s \tau^+\tau^-$
has been calculated in SM and several new physics scenarios
\cite{new7}. Switching off the NHB contributions, our results
are in agreement with those in Ref. \cite{new7}.

In summary, we have calculated the differential braching ratio,
back-forward asymmetry, lepton polarizations and some CP violated
observables for $B\rightarrow X_s \tau^+\tau^-$ in the model IV 2HDM.
As the main features of the model, NHB play an important role
in inducing CP violations,
in particular, for large $\tan\beta$.
We propose to measure $A_{CP}^2$, the direct CP asymmetry in back-forward
asymmetry, in stead of $A_{CP}^1$, the usual direct CP violation in
branching ratio, because the former could be observed if $\tan\beta$
is large enough (say, $\geq 30$) and the latter is too small to be
observed. It is possible to discriminate the
model IV from the other 2HDMs by measuring the CP-violated observables
such as $A_{CP}^2$, $P_N$ if the nature chooses large $\tan\beta$.




\begin{figure}
\epsfxsize=18 cm \centerline{\epsffile{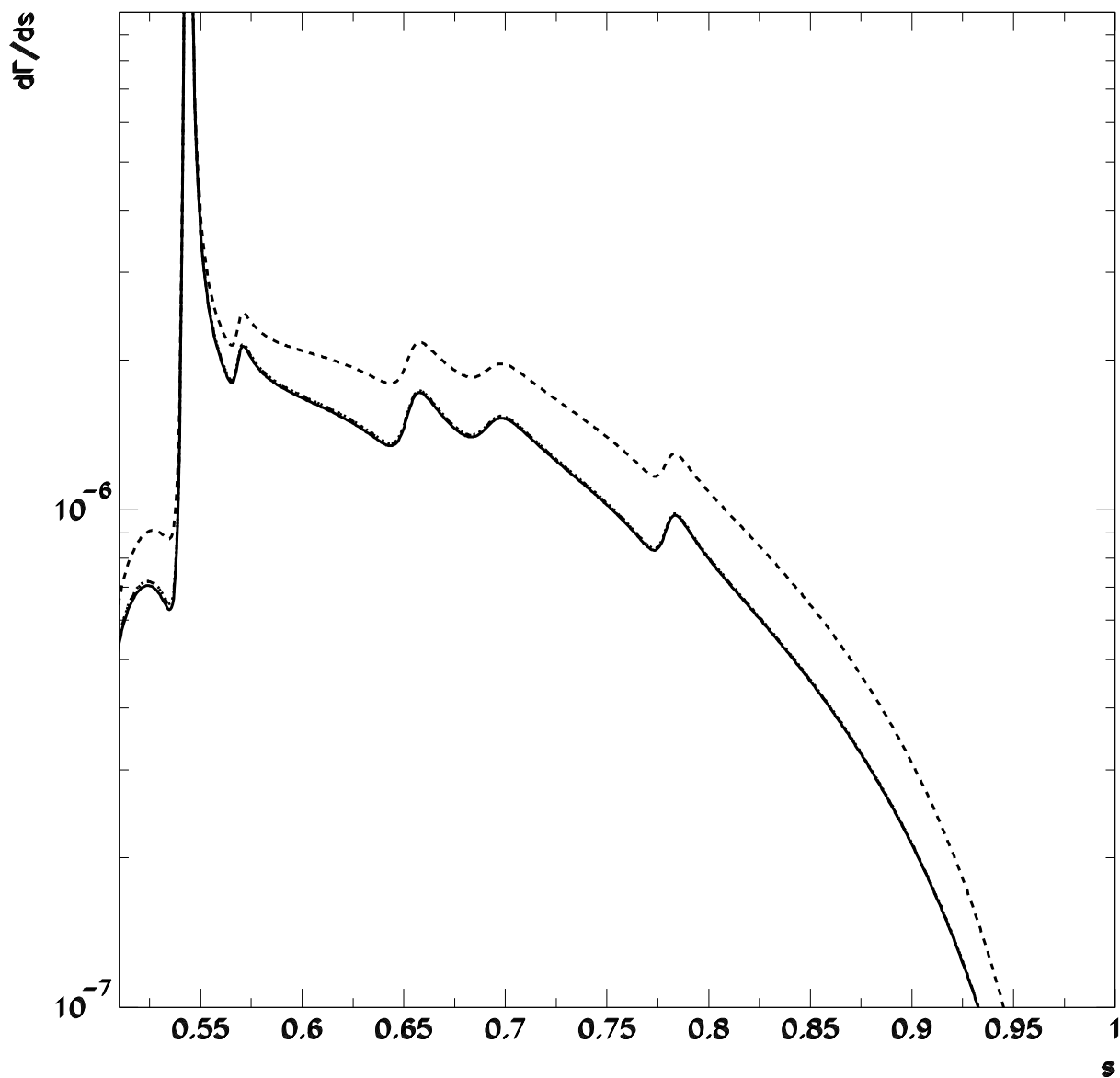}}
\caption[]{ Differential branching ratio as function of $s$, where
$\xi=\pi/4$,  solid and dashed lines represent $\tan\beta=10$ and
$30$, dot-dashed line represents the case of switching off
$C_{Q_i}$ contributions.}
\end{figure}

\begin{figure}
\epsfxsize=18 cm \centerline{\epsffile{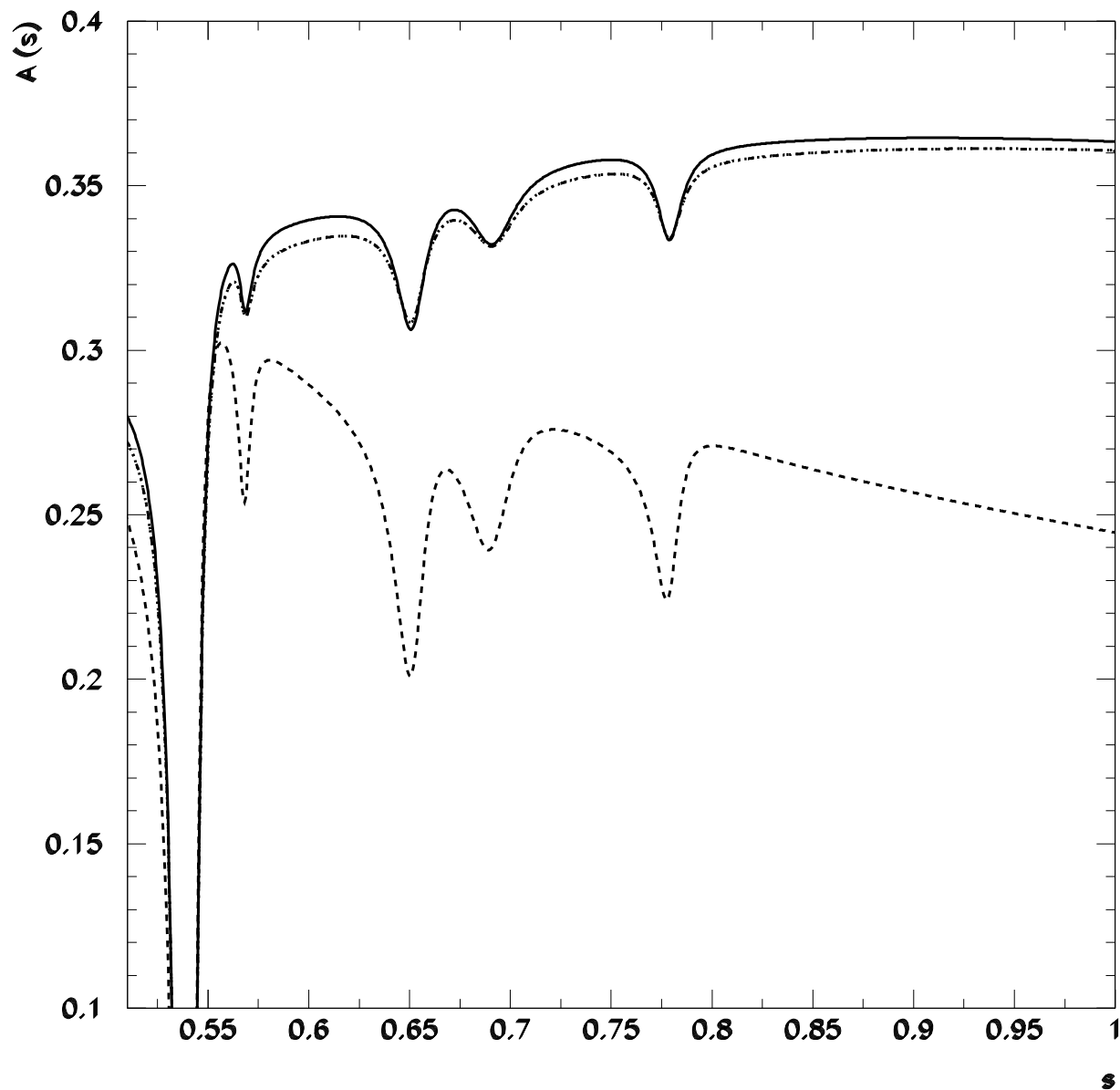}} \caption[]{
Backward-forward asymmetry as function of $s$, where $\xi=\pi/4$,
solid and dashed lines represent $\tan\beta=10$ and $30$,
dot-dashed line represents the case of switching off $C_{Q_i}$
contributions.}
\end{figure}

\begin{figure}
\epsfxsize=18 cm \centerline{\epsffile{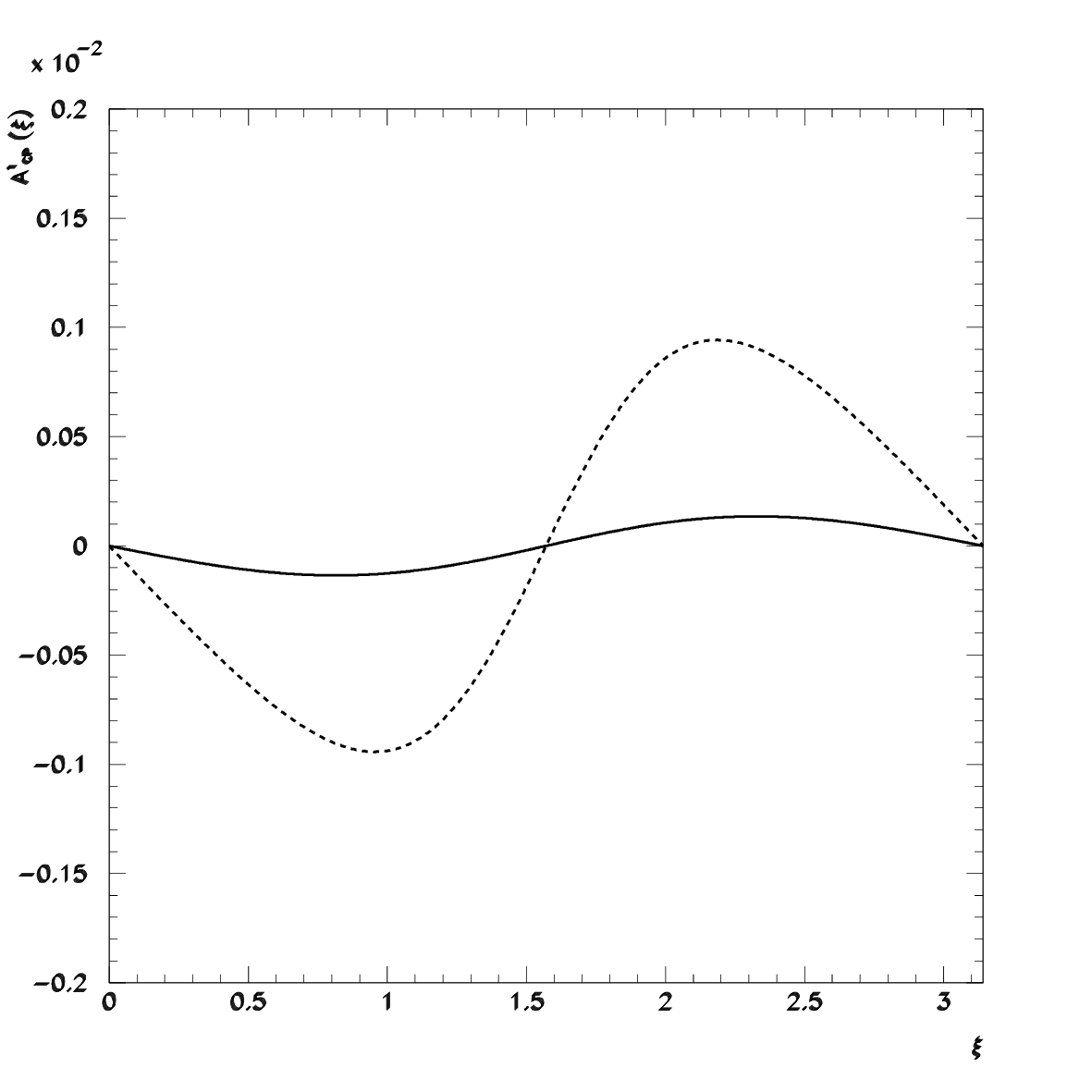}} \caption[]{
$A^1_{CP}$ as function of $\xi$, where $s=0.8$,  solid and dashed
lines represent $\tan\beta=10$ and $30$.}
\end{figure}


\begin{figure}
\epsfxsize=18 cm \centerline{\epsffile{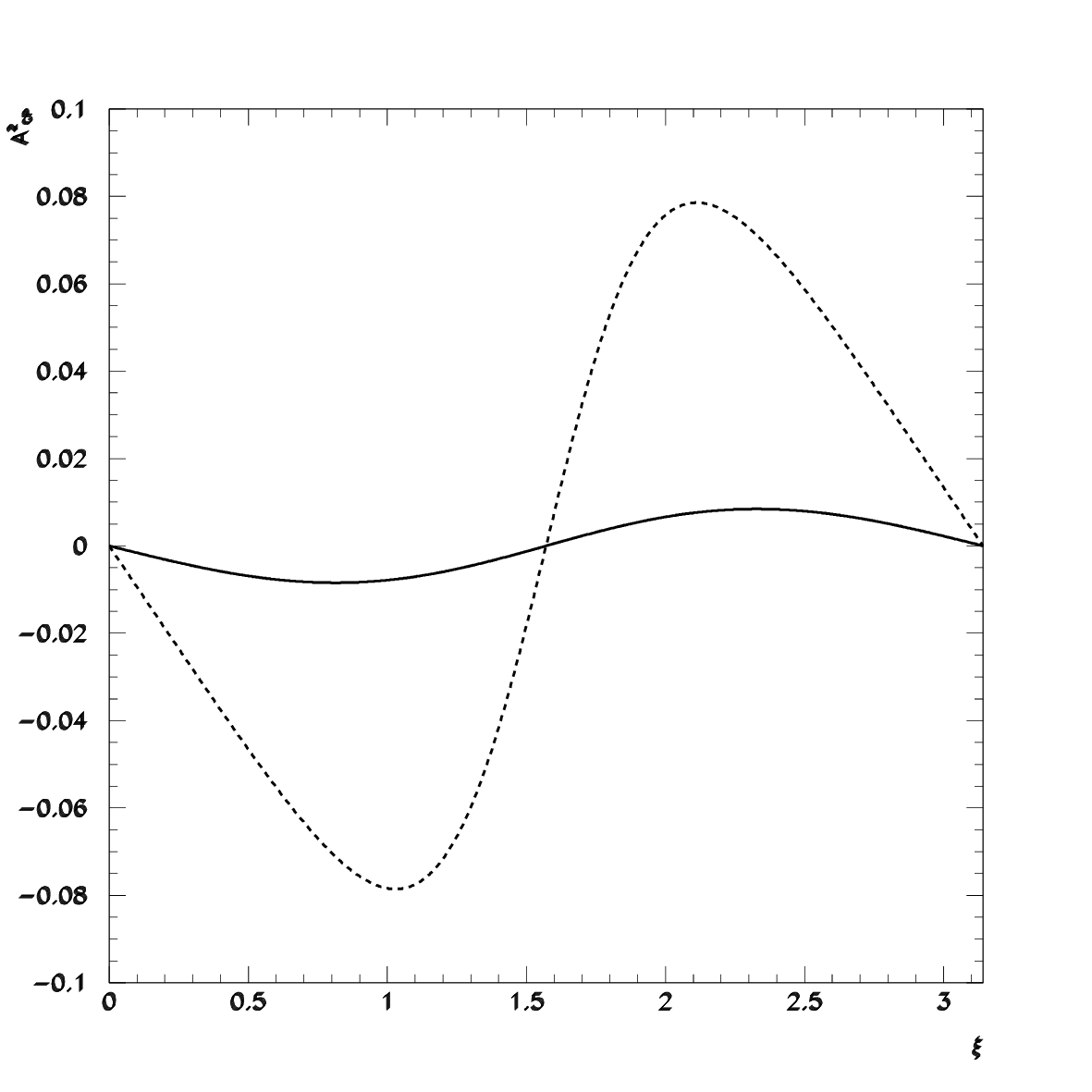}} \caption[]{
$A^2_{CP}$ as function of $\xi$, where $s=0.8$,  solid and dashed
lines represent $\tan\beta=10$ and $30$.}
\end{figure}

\begin{figure}
\epsfxsize=18 cm \centerline{\epsffile{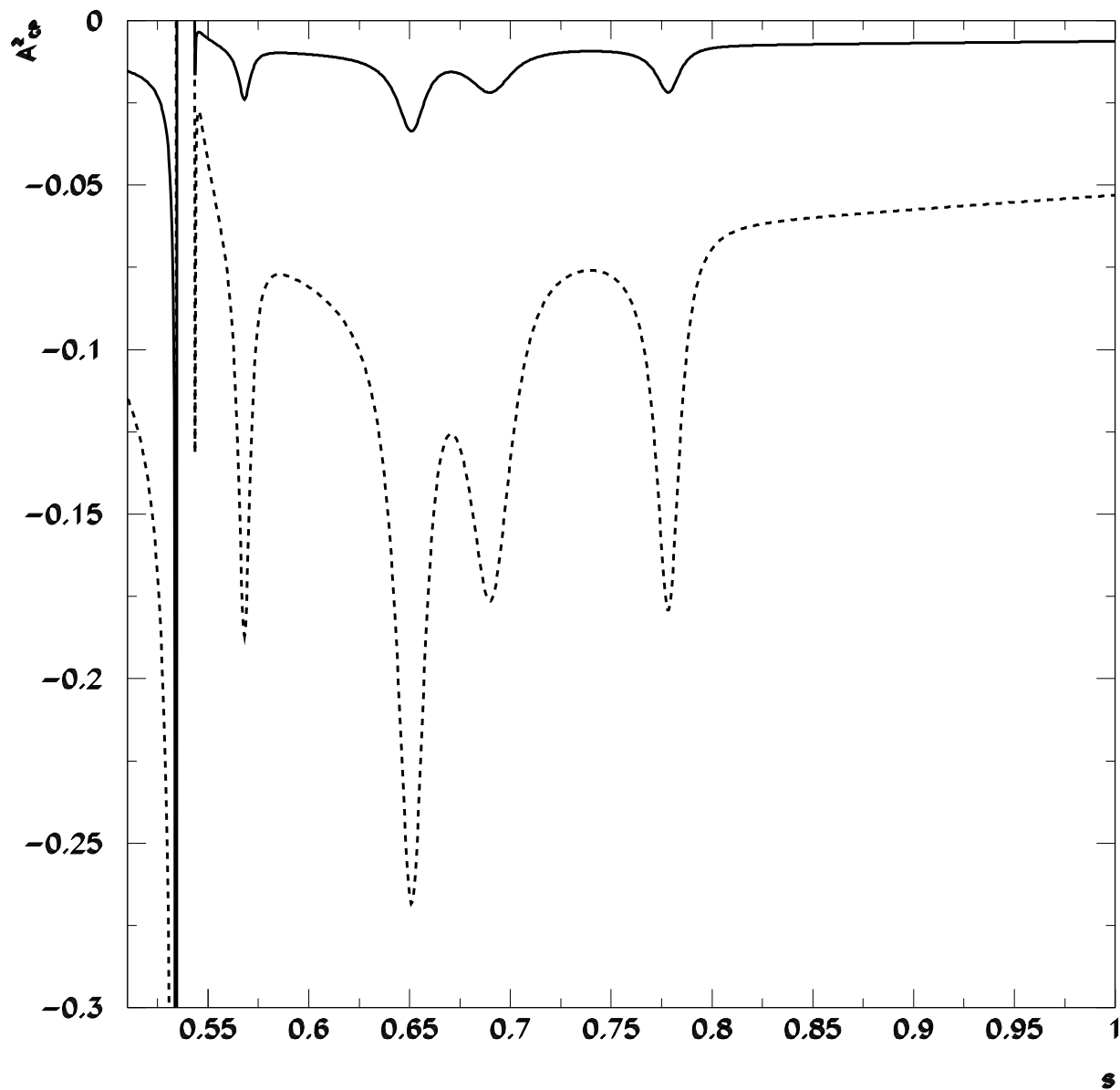}} \caption[]{
$A^2_{CP}$ as function of $s$, where $\xi=\pi/4$,  solid and
dashed lines represent $\tan\beta=10$ and $30$.}
\end{figure}


\begin{figure}
\epsfxsize=18 cm \centerline{\epsffile{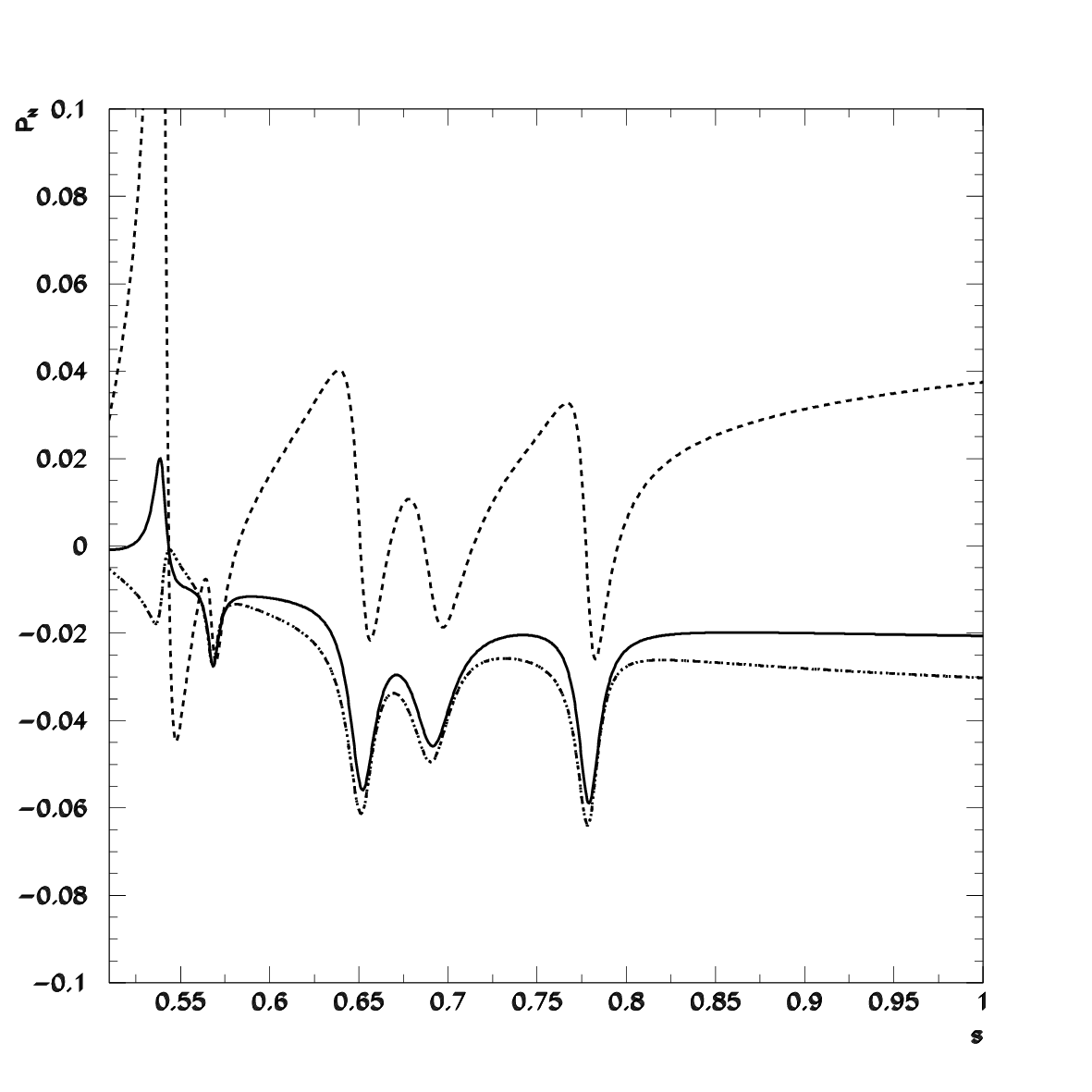}} \caption[]{
$P_N$ as function of $s$, where $\xi=\pi/4$,  solid and dashed
lines represent $\tan\beta=10$ and $30$, dot-dashed line
represents the case of switching off $C_{Q_i}$ contributions.}
\end{figure}

\begin{figure}
\epsfxsize=18 cm \centerline{\epsffile{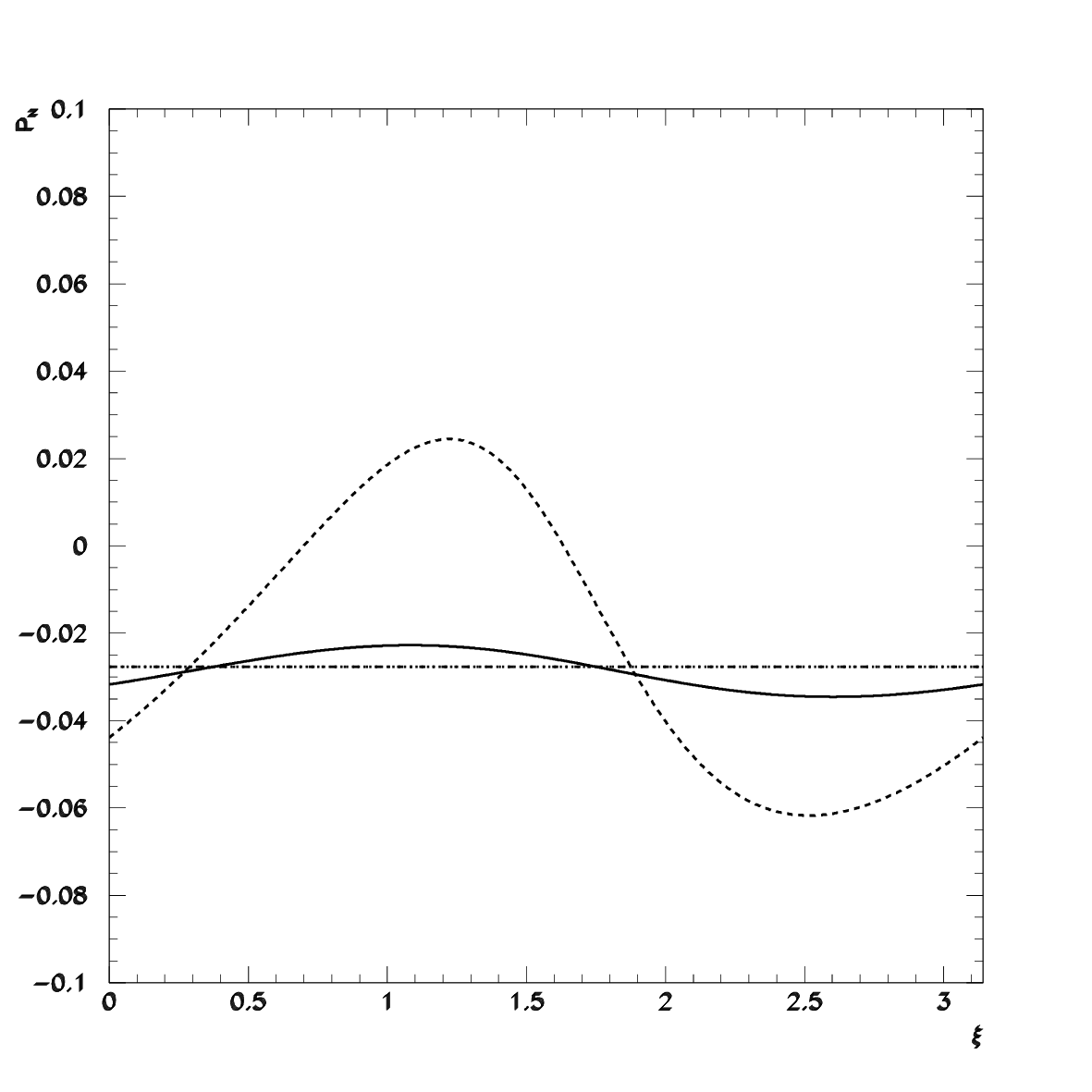}} \caption[]{
$P_N$ as function of $\xi$, where $s=0.8$,  solid and dashed lines
represent $\tan\beta=10$ and $30$, dot-dashed line represents the
case of switching off $C_{Q_i}$ contributions.}
\end{figure}


\begin{figure}
\epsfxsize=18 cm \centerline{\epsffile{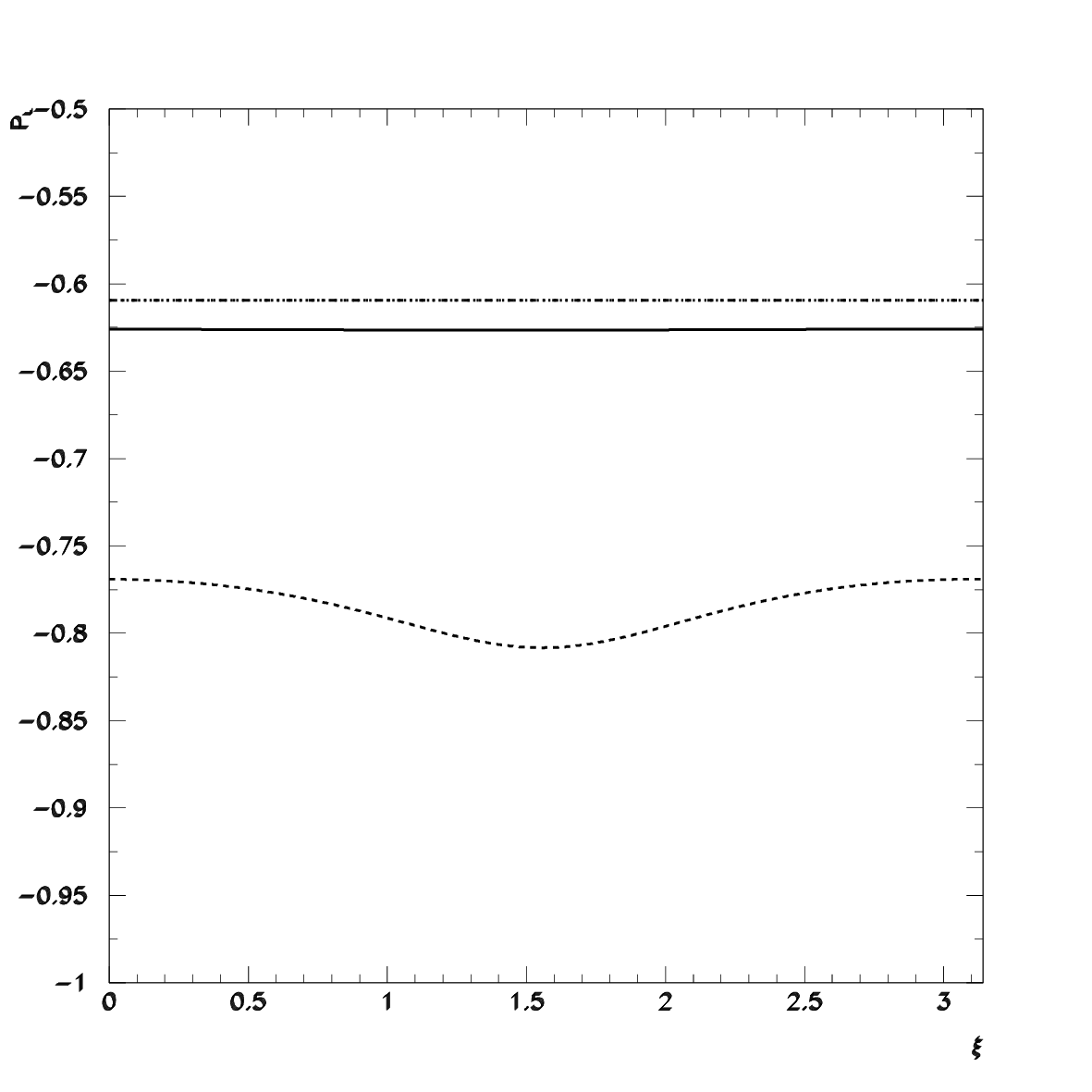}} \caption[]{
$P_L$ as function of $\xi$, where $s=0.8$,  solid and dashed lines
represent $\tan\beta=10$ and $30$, dot-dashed line represents the
case of switching off $C_{Q_i}$ contributions.}
\end{figure}

\begin{figure}
\epsfxsize=18 cm \centerline{\epsffile{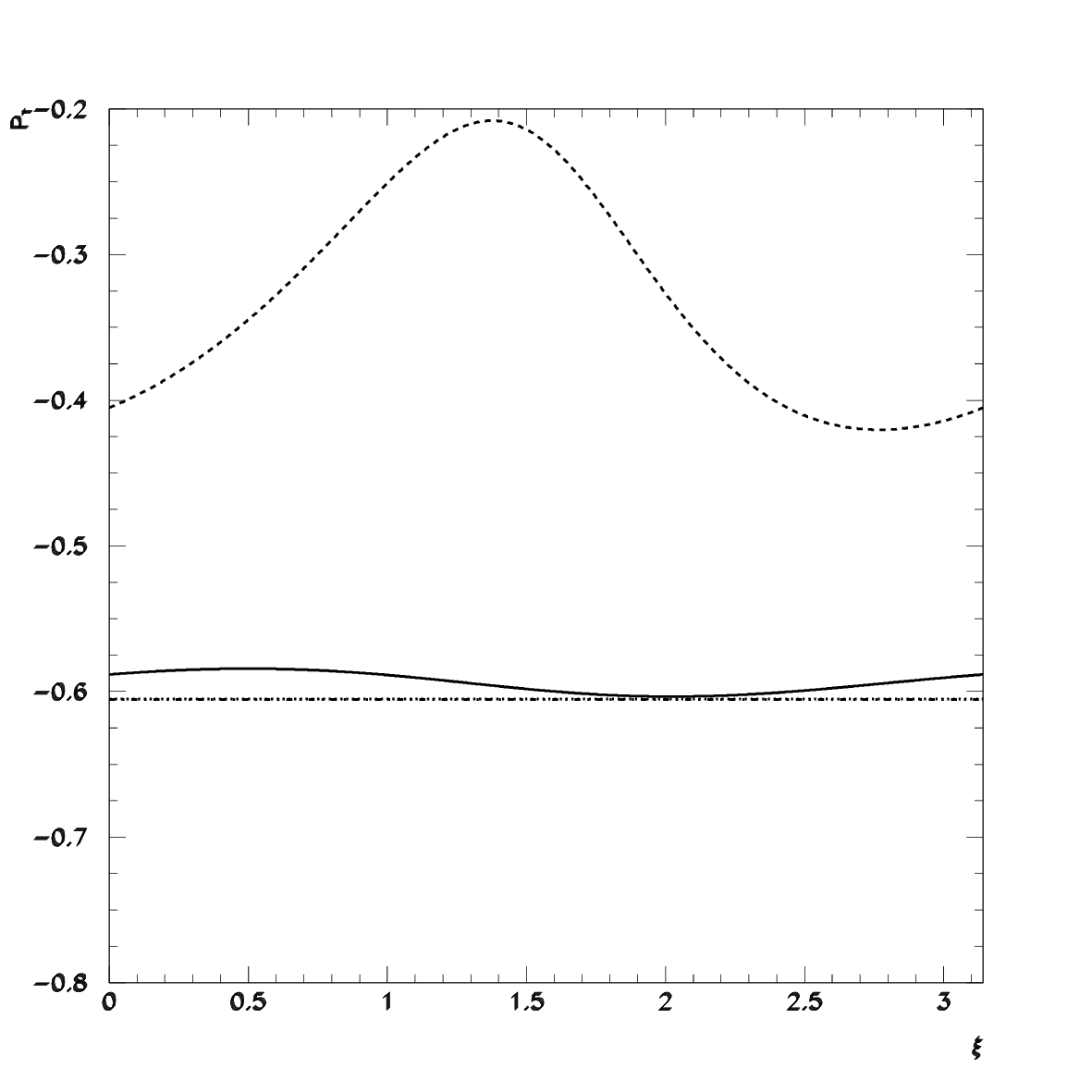}} \caption[]{
$P_T$ as function of $\xi$, where $s=0.8$,  solid and dashed lines
represent $\tan\beta=10$ and $30$, dot-dashed line represents the
case of switching off $C_{Q_i}$ contributions.}
\end{figure}



\setcounter{figure}{0} \setcounter{table}{0}
\setcounter{equation}{0}

\newpage



\renewcommand{\textfraction}{0}
\renewcommand{\theequation}{\arabic{subsection}.\arabic{equation}}

\def\journal#1#2#3#4{{\it #1} {\bf #2} (#3) #4}
\def\epj{Euro. Phys. Jour.}
\def\prl{Phys. Rev. Lett.}
\def\pl{Phys. Lett.}
\def\np{Nucl. Phys.}
\def\ptp{Prog. Theor. Phys.}
\def\mpl{Mod. Phys. Lett.}
\def\zp{Z. Phys.}
\def\pr{Phys. Rev.}
\def\prp{Phys. Rep.}
\def\nc{Nuovo Cim.}
\def\jhep{JHEP}
\def\yf{Yad. Fiz.}
\def\tmf{Teo. Mat. Fiz.}
\def\jetp{JETP Lett.}
\def\ijmp{Int. Jour. Mod. Phys.}
\def\ppnp{Prog. Part. Nucl. Phys.}
\def\ctp{Commun. Theor. Phys.}
\def\rmp{Rev. Mod. Phys.}
%
\def\ml{{\hat{m_{\ell}}}}
\def\ddp{{D^\prime}}
\def\mc{{\hat{m_c}}}

\def\o{{\cal O}}
\def\c{C}
\def\cs{{\c_7}}
\def\cn{{\c_9}}
\def\ct{{\c_{10}}}
\def\cne{\cn^{\rm eff}}
\def\cse{\cs^{\rm eff}}
\def\m{{\cal M}}
\def\gl{\Gamma}
\def\g{\gamma}
\def\l{\ell}
\def\ks{{K^\ast}}
\def\lb{\bar{\l}}
\def\cb{\bar{c}}
\def\ub{\bar{u}}
\def\bxll{B \rightarrow X \, \l^+ \, \l^-}
\def\bkll{B \rightarrow K \, \l^+ \, \l^-}
\def\bksll{B \rightarrow \ks \, \l^+ \, \l^-}
\def\brholl{B \rightarrow \rho \, \l^+ \, \l^-}
\def\bpill{B \rightarrow \pi \, \l^+ \, \l^-}
\def\bpll{B \rightarrow P \, \l^+ \, \l^-}
\def\bvll{B \rightarrow V \, \l^+ \, \l^-}
\def\bsg{B \rightarrow X_s \, \g}
\def\bxg{B \rightarrow X \, \g}
\def\bxlv{B \rightarrow X \, \l \, \nu}
\def\bkm{B \rightarrow K \mu^+ \mu^-}
\def\bkt{B \rightarrow K \tau^+ \tau^-}
\def\bksm{B \rightarrow K^* \mu^+ \mu^-}
\def\bkst{B \rightarrow K^* \tau^+ \tau^-}
\def\b{{\cal B}}
\def\afb{{{\cal A}_{\rm FB}}}
\def\alp{{{\cal A}_{\rm LP}}}
\def\acp{{{\cal A}_{\rm CP}}}
\def\mqx{{m_{q_X}}}
\def\ql{{q_l}}
\def\qx{{q_X}}
\def\qxb{{\bar{q}_X}}
\def\qv{{q_V}}
\def\qp{{q_P}}
\def\qvb{{\bar{q}_V}}
\def\qpb{{\bar{q}_P}}
\def\bb{\bar{b}}
\def\qlb{{\bar{q}_l}}
\def\pb{{p_B}}
\def\px{{p_X}}
\def\pp{{p_K}}
\def\pv{{p_{K^*}}}
\def\fx{{f_X}}
\def\fp{{f_P}}
\def\fv{{f_V}}
\def\ph{{p_h}}
\def\kl{{k_\l}}
\def\klp{{k_{\l^+}}}
\def\klm{{k_{\l^-}}}
\def\he{{\cal H}_{\rm eff}}
\def\qs{q^2}
\def\qq{q^4}
\def\mb{{m_B}}
\def\mmp{{m_P}}
\def\mv{{m_V}}
\def\mx{{m_X}}
\def\d{{\rm d}}
\def\D{{\cal D}}
\def\e{{\rm e}}
\def\t{{\rm T}}
\def\bm{{M^2}}
\def\xp{{X^\prime}}
\def\qu{{q_u}}
\def\qub{{\bar{q}_u}}
\def\oc{\tilde{\o}}
\def\mh{\hat{m}}
\def\mbh{\mh_b}
\def\msh{\mh_s}
\def\mph{\mh_K}
\def\mvh{\mh_{K^*}}
\def\mxh{\mh_X}
\def\mlh{\mh_\l}
\def\qh{\hat{q}}
\def\pvh{\hat{p}_{K^*}}
\def\pph{\hat{p}_K}
\def\pbh{\hat{p}_B}
\def\ph{\hat{p}}
\def\sh{\hat{s}}
\def\h{{\cal H}}
\def\t{{\cal T}}
\def\a{{\cal A}}
\def\s{{\cal S}}
\def\cqb{C_{Q1}}
\def\cqc{C_{Q2}}
\def\ep{{\epsilon^\ast}}
\def\ap{{A^\prime}}
\def\bp{{B^\prime}}
\def\cp{{C^\prime}}
\def\rp{{D^\prime}}
\def\uh{{\hat{u}}}
\def\la{{\lambda}}
\def\lal{{\lambda_\l}}
\def\as{{\alpha_s}}
\def\be{\begin{equation}}
\def\ee{\end{equation}}
\def\ba{\begin{eqnarray}}
\def\ea{\end{eqnarray}}
\def\nnb{\nonumber}
\renewcommand{\thefootnote}{\fnsymbol{footnote}}
\section{ Exclusive Semileptonic Rare Decays $B \to (K,K^*)
\ell^+ \ell^-$ in Supersymmetric Theories}
\begin{footnotesize}
\begin{center}\begin{minipage}{5in}
\begin{center} ABSTRACT\end{center}

The invariant mass spectrum, forward-backward asymmetry, and
lepton  polarizations of the exclusive processes $B\rightarrow
K(K^*)\ell^+ \ell^-, \ell=\mu,~\tau$ are analyzed under
supersymmetric context. Special attention is paid to the effects
of neutral Higgs bosons (NHBs). Our analysis shows that the
branching ratio of the process $\bkm$ can be quite largely
modified by the effects of neutral Higgs bosons and the
forward-backward asymmetry would not vanish. For the process
$\bksm$,  the lepton transverse polarization is quite sensitive to
the effects of NHBs, while the invariant mass spectrum, forward-
backward asymmetry, and lepton longitudinal polarization are not.
For both $\bkt$ and $\bkst$, the effects of NHBs are quite
significant. The partial decay widths of these processes are also
analyzed, and our analysis manifest that even taking into account
the theoretical uncertainties in calculating weak form factors,
the effects of NHBs could make SUSY shown up.

\end{minipage}\end{center}
\end{footnotesize}

\renewcommand{\thefootnote}{\arabic{footnote}}
\setcounter{footnote}{0}

\subsection{Introduction}

The inclusive rare processes $b\rightarrow X_s \ell^+ \ell^-, \ell=e,~\mu,~\tau$
have been intensively studied in
literatures\cite{gsw,ex1,dhh,hly,fky,fkmy,kkl,goto96,lmss99,ht96,gln,ks}.
As one of flavor changing neutral current processes, it
is sensitive to fine structure of the standard model and to the possible
new physics as well, and is expected to shed light on the existence of new
physics before the possible new particles are produced at colliders.

It is well known that invariant mass spectrum, forward-backward asymmetries, and lepton
polarizations are important observables to probe new physics, while the first two
observables are mostly analyzed. About lepton polarizations, it is known that due to the smallness
of the mass of it, therefore electron polarizations are very difficult to be measured
experimentally. So only the lepton polarizations of muon and tau are
considered in literatures~\cite{ht96,ks,ch,hz}. The longitudinal polarization of tau in $B \rightarrow X_s \tau^+
 \tau^-$ has been calculated in standard model (SM) and several new physics scenarios~\cite{ht96}.  For
$B\rightarrow X_s l^+ l^-$ ($l =\mu, \tau$), the polarizations of lepton in SM are
analyzed in \cite{ks} and it is pointed out that for the $\mu$ channel, the only
significant component is $P_L$, while all three components are sizable in the
$\tau$ channel.The analysis has been extended to supersymmetric models (SUSY) and a CP softly broken two Higgs doublet
model in refs. ~\cite{ch} and \cite{hz} respectively. The reference \cite{fky} also gives a general model-independent
analysis of the lepton polarization asymmetries in the process $B \rightarrow
X_s \tau^+ \tau^-$ and it is found that the contribution from
$C_{LRLR}+C_{LRRL}$ is much larger than other scalar-type interactions.

Compared with the inclusive processes $b\rightarrow X_s \ell^+ \ell^-,
\ell=e,~\mu,~\tau$, the theoretical study of the exclusive processes
$B\rightarrow K(K^*) \ell^+ \ell^-$ is relatively hard. For inclusive semileptonic decays of B, the decay
rates can be calculated in heavy quark effective theory (HQET) \cite{hqet}. However,
for exclusive semileptonic decays of B, to make theoretical predictions, additional
knowledge of decay form factors is needed, which is related with the
calculation of hadronic transition matrix elements. Hadronic transition matrix
elements depend on the non-perturbative properties of QCD, and can only
be reliably calculated by using a nonperturbative method. The form factors for B decay into $K^{(*)}$ have been computed with
different methods such as quark models~\cite{jw}, SVZ  QCD sum rules~\cite{cfss}, light cone sum rules
(LCSRs)~ \cite{bbk,chzh,brau,kr,ball}. Compared to the lattice approach which
mainly deal with the form factors at small recoil, the QCD sum rules on the
light-cone can complementarily provide information of the form factors
at smaller values of $\sh$. And they are consistent with perturbative QCD and
the heavy quark limit. In this work, we will use the weak decay form factors
calculated by using the technique of the light cone QCD sum rules and given in
\cite{abhh}.

A upper limit on the branching ratio of $B^0\rightarrow K^{0*} \mu^+\mu^-$  has been recently given by CLEO \cite{cl}:
\begin{equation}
BR ( B^o \rightarrow K^{0*} \mu^+\mu^-) < 4.0 \times 10^{-6} ,
\end{equation}
and  they will be
precisely measured at B factories, these exclusive processes are quite worthy
of intensive study and have attracted many attentions \cite{abhh,dt,ex,dl,liu,akk,ju,mns,giw,as,asok,gk}.
In reference \cite{abhh}, by
using improved theoretical calculations of the decay dorm factors in the light
cone QCD sum rule approach, dilepton invariant mass spectra and the
forward-backward asymmetry of these exclusive decays are analyzed in the
standard model and a number of popular variants of the supersymmetric models.
However, as the author claimed, the effects of neutral Higgs exchanges are
neglected. For exclusive processes, as pointed out in
\cite{gk}, the polarization asymmetries of $\mu$ and $\tau$ for $B \rightarrow
K^(*) \mu^+ \mu^-$ and $ B \rightarrow K^(*) \tau^+ \tau^-$ are also accessible
at the B-factories under construction.  In reference \cite{as}, the lepton polarizations and CP violating effects in
 $B \rightarrow K^* \tau^+ \tau^-$ are analyzed in SM and two Higgs doublet models.

As pointed in refs.~\cite{dhh,hly}, in two-Higgs-doublet models and SUSY models, neutral Higgs boson could contribute
largely to the inclusive processes $b\rightarrow X_s \ell^+ \ell^-,
\ell=~\mu,~\tau$ and greatly modify the branching ratio and forward-backward asymmetry in the large tan$\beta$
case. The effects of
neutral Higgs in the 2HDM to polarizations of $\tau$ in $B\rightarrow K \tau^+
\tau^-$ are analyzed in \cite{asok}, and it was found that polarizations of the charged
final lepton are very sensitive to the tan$\beta$.

In this paper, we will investigate the exclusive decay $B\rightarrow K(K^*) \ell^+ \ell^-, \ell=\mu,~\tau$
in SUSY models. We shall evaluate branching ratios and forward-backward asymmetries (FBA) with emphasis on the
effects of neutral Higgs  and analyze lepton polarizations in MSSM.
According to the analysis of \cite{liu}, different sources of the vector current
could manifest themselves in different regions of phase space, for the very
low $\sh$ the photonic penguin dominates, while the Z penguin and W box
becomes important towards high $\sh$. In order to search the regions of $\sh$ where
neutral Higgs bosons could contribute large, we analyze the partial
decay widths of these two processes. Beside that they
are accessible to B factories, our motivation also bases on the fact that to
the inclusive processes $B\rightarrow X_s \ell^+ \ell^-, \ell=\mu, ~\tau$,
neutral Higgs could make quite a large contributions at certain large tan$\beta$ regions
of parameter space in SUSY models, since part of supersymmetric contributions
is proportional to tan$^3\beta$~ \cite{hly}. Such regions considerably exist in SUGRA
and M-theory inspired models~\cite{hllyz}.
We also analyze the effects of neutral
Higgs to the position of the zero value of the forward-backward asymmetry.
Our results show that the branching ratio of the process $\bkm$
can be quite largely modified by the effects of neutral Higgs bosons and the forward-backward asymmetry
would not vanish. Because the FBA for $\bkll (\ell=\mu,~\tau)$ vanishes
if there are no the contributions of NHBs and the contributionsof NHBs can
be large enough to be observed only in SUSY and/or 2HDM with large tan$\beta$, a non-zero FBA for $\bkll$
would signal the existance of new physics.
For the process $\bksm$,  the lepton transverse polarization is quite sensitive
to the effects of NHBs, while the invariant mass spectrum, forward-
backward asymmetry, and lepton longitudinal polarization are not. For both $\bkt$ and $\bkst$, the effects of NHBs are
quite significant. Our analysis manifest that even taking into account the theoretical uncertainties in calculating
weak form factors, the effects of NHBs could show SUSY up. In a word, our analysis manifest
that effects of NHBs is quite remarkable in some regions of parameter space
of SUSY, even for the process $\bkm$.

The paper is organized as follows. In the subsection 2, the
effective Hamiltonian is presented and the form factors given by
using light cone  sum rule method are briefly discussed. Basic
formula of observables are introduced in subsection 3. Section 4
is devoted to the numerical analysis. In subsection 5 we make
discussions and conclusions.

\subsection{Effective Hamiltonian and Form Factors}
\setcounter{equation}{0}

By integrating out the degrees of heavy freedom from the
full theory, MSSM, at electroweak(EW) scale, we can get the effective Hamiltonian
describing the rare semileptonic decay  $b \to s  \ell^+ \ell^-$:
\begin{equation}
\he  =  -\frac{4G_{F}}{\sqrt{2}} V_{tb} V_{ts}^{*} (\sum_{i=1}^{10}
    C_{i}(\mu) O_{i}(\mu) + \sum_{i=1}^{10} C_{Q_i}(\mu) Q_i(\mu))\; ,
        \label{eq:he}
\end{equation}

where the first ten operators and Wilson coefficients (WC) at EW scale can be found in
\cite{goto96,bbmr}\footnote{In our previous papers, e.g., \cite{dhh,hly}, we follow the convention of
ref. \cite{gsw} for the indices of operators as well as Wilson coefificients. In this paper.
we use more popular conventions (see, e.g., \cite{efh}). That is, $O_8\rightarrow O_9$
and $O_9\rightarrow O_{10}$.}, and last ten operators and WC which represent the contributions of
neutral Higgs can be found in \cite{hly}.

With the renormalization group equations to resum the QCD corrections, WCs at
energy scale $\mu=m_b$ are evaluated. Theoretical uncertainties
related to renormalization-scale can be substantially reduced when the
next-leading-logarithm corrections are included \cite{efh}.

The above Hamiltonian leads to the following free quark decay amplitude:
\begin{eqnarray}         \m(b\to s\ell^+\ell^-) & = & \frac{G_F
\alpha}{\sqrt{2}  \pi} \,                  V_{t s}^\ast V_{tb} \, \left\{
             \cne  \left[ \bar{s}  \g_\mu  L  b \right] \,
       \left[ \lb  \g^\mu  \l \right]                 + \ct  \left[ \bar{s}
\g_\mu  L  b \right] \,                           \left[ \lb  \g^\mu  \g_5  \l
\right]                 \right. \nonumber \\         & & \left.
                - 2 \mbh  \cse  \left[ \bar{s}  i  \sigma_{\mu \nu}
                        \frac{\qh^{\nu}}{\sh}  R  b \right]
                        \left[ \lb  \g^\mu  \l \right]
                +\cqb \left[ \bar{s}  R  b \right] \,
                         \left[ \lb \l \right]
                +\cqc \left[ \bar{s}  R  b \right] \,
                         \left[ \lb \g_5 \l \right]\right\} \; .
        \label{eq:m}
\end{eqnarray}

where $C^{eff}_9$ is defined as \cite{agm,bm}
\begin{equation}
\cne (\mu, \hat{s}) = C_9(\mu) + Y(\mu, \sh) + \frac{3 \pi}{\alpha^2} C (\mu)
         \sum_{V_i = \psi(1s),..., \psi(6s)} \kappa_i
      \frac{\Gamma(V_i \rightarrow \ell^+ \ell^-)\, m_{V_i}}{
      {m_{V_i}}^2 - \sh \, {m_B}^2 - i m_{V_i} \Gamma_{V_i}}
\label{eqn:cni}
\end{equation}
where $\hat{s}=s/m_b^2$,s=$q^2$, $C(\mu)=\left(3 \, C_1 + C_2 + 3 \, C_3
                + C_4 + 3 \, C_5 + C_6 \right)$, and

\ba
        {Y}(\mu, \sh) & = & g(\mc,\sh) C(\mu) \nonumber \\
         &&  -\frac{1}{2} g(1,\sh)
                \left( 4 \, C_3 + 4 \, C_4 + 3 \,
                C_5 + C_6 \right)
         - \frac{1}{2} g(0,\sh) \left( C_3 +
                3 \, C_4 \right) \nonumber \\
        &&-     \frac{2}{9} \left( 3 \, C_3 + C_4 +
                3 \, C_5 + C_6 \right) \; .
\label{eq:yert}
\ea
where the function $g(\mc, \sh)$ comes from one loop contributions of four-quark operators and is defined by
\begin{equation}
g(z,\hat{s})=-\frac{4}{9} lnz^2+ \frac{8}{27}+ \frac{16}{9}
\frac{z^2}{\hat{s}}-\left\{ \begin{array}{ll}
 \frac{2}{9} \sqrt{1-\frac{4 z^2}{\hat{s}}} (2+\frac{4 z^2}{\hat{s}})
 \bigg[ ln ( \frac{1+\sqrt{1-4 z^2/ \hat{s}}} {1-\sqrt{1-4 z^2/ \hat{s}}})
 +i \pi \bigg], & \textrm{$4 z^2 < \hat{s}$} \\
 \frac{4}{9} \sqrt{\frac{4 z^2}{\hat{s}}-1} (2+\frac{4 z^2}{\hat{s}})
arctan \bigg( \frac{1}{\sqrt{4 z^2/ \hat{s} -1}} \bigg),
 & \textrm{$4 z^2 > \hat{s}$}
\end{array} \right.
\end{equation}
The last terms in
(\ref{eqn:cni}) are nonperturbative effects from $(\bar{c} c)$ resonance contributions,
while the phenomenological factors $\kappa_i$ can be fixed from the processes \cite{abhh}
$B\to K^{(\ast)} V_i \to K^{(\ast)} \ell^+ \ell^-$ and as given in the Table. \ref{tab:kappa}.
\begin{table}[b]
        \begin{center}
        \begin{tabular}{|c|c|c|}
 \hline
    \multicolumn{1}{|c|}{$\kappa$}
      & \multicolumn{1}{|c|}{$J/\Psi$}
      & \multicolumn{1}{|c|}{$\Psi^\prime$} \\
        \hline
  $K$           &2.70                   & 3.51 \\
  $K^\ast$      &1.65                   & 2.36 \\
        \hline
        \end{tabular}
        \end{center}
\caption{\it Fudge factors in $B\to K^{(\ast)} J/\Psi, \Psi^\prime \to
K^{(\ast)} \ell^+ \ell^- $ decays calculated using the LCSR form factors.}
\label{tab:kappa}
\end{table}

Exclusive decays $B\to (K,K^*) \ell^+ \ell^-$ are described in terms
of matrix elements of the quark operators in Eq. (\ref{eq:m}) over meson
states, which can be parametrized in terms of form factors.

For the process $B\to K \ell^+ \ell^-$, the non-vanishing matrix
elements are ($q=p_B-p$)
\begin{equation}\label{eq:ff1}
\langle K(p) | \bar s \gamma_\mu b | B(p_B)\rangle  =  f_+(s) \left\{
(p_B+p)_\mu - \frac{m_B^2-m_K^2}{s} \, q_\mu \right\} +
\frac{m_B^2-m_K^2}{s} \, f_0(s)\, q_\mu,
\end{equation}
and
\begin{eqnarray}
\langle K(p) | \bar s \sigma_{\mu\nu} q^\nu (1+\gamma_5) b | B(p_B)\rangle
& = &  \langle K(p) | \bar s \sigma_{\mu\nu} q^\nu b |
B(p_B)\rangle\nonumber\\
& = & i\left\{ (p_B+p)_\mu s - q_\mu (m_B^2-m_K^2)\right\} \,
  \frac{f_T(s)}{m_B+m_K}.\label{eq:ff2}
\end{eqnarray}
While for $B\to K^* \ell^+ \ell^-$, related transition matrix elements are
\begin{eqnarray}
\langle K^*(p) | (V-A)_\mu | B(p_B)\rangle & = & -i \epsilon^*_\mu
(m_B+m_{K^*}) A_1(s) + i (p_B+p)_\mu (\epsilon^* p_B)\,
\frac{A_2(s)}{m_B+m_{K^*}}\nonumber\\
\lefteqn{+ i q_\mu (\epsilon^* p_B) \,\frac{2m_{K^*}}{s}\,
\left(A_3(s)-A_0(s)\right) +
\epsilon_{\mu\nu\rho\sigma}\epsilon^{*\nu} p_B^\rho p^\sigma\,
\frac{2V(s)}{m_B+m_{K^*}}\,.}\hspace*{2cm}\label{eq:ff3}
\end{eqnarray}
and
\begin{eqnarray}
\langle {K^*} | \bar s \sigma_{\mu\nu} q^\nu (1+\gamma_5) b |
B(p_B)\rangle & = & i\epsilon_{\mu\nu\rho\sigma} \epsilon^{*\nu}
p_B^\rho p^\sigma \, 2 T_1(s)\nonumber\\
& & {} + T_2(s) \left\{ \epsilon^*_\mu
  (m_B^2-m_{{K^*}}^2) - (\epsilon^* p_B) \,(p_B+p)_\mu \right\}\nonumber\\
& & {} + T_3(s)
(\epsilon^* p_B) \left\{ q_\mu - \frac{s}{m_B^2-m_{{K^*}}^2}\, (p_B+p)_\mu
\right\}\label{eq:T}
\end{eqnarray}

Where $\epsilon_\mu$ is polarization vector of the vector meson $K^*$. By means of the equation of motion,
one obtains  several relations between form factors
\begin{eqnarray}  A_3(s) & = &
\frac{m_B+m_{K^*}}{2m_{K^*}}\, A_1(s) - \frac{m_B-m_{K^*}}{2m_{K^*}}\,
A_2(s),\nonumber\\ A_0(0) & = & A_3(0), \nonumber\\
\langle K^* |\partial_\mu A^\mu | B\rangle & = & 2 m_{K^*}
(\epsilon^* p_B) A_0(s),\nonumber \\
T_1(0) &= & T_2(0).
\label{eq:exq}
\end{eqnarray}
All signs are defined in such a way as to render the form factors real and positive. The physical
range in $\sh$ extends from $\sh_{\rm min} = 4 \mlh^2$ to $\sh_{\rm max} = (1-{\hat m}_{K,K^*})^2$.

The calculation of the form factors given above is a real task, and one has to
rely on certain approximate methods. We use the results calculated by using technique
of LCSRs and given in \cite{abhh}. And form factors can be parametrized as
\begin{equation}\label{eq:para}
F(\hat{s}) = F(0) \exp ( c_1 \hat{s} + c_2 \hat{s}^2 + c_3 \hat{s}^3).
\end{equation}
The parameterization formula works within 1\% accuracy for $s<15\,$GeV$^2$ and
can avoid the spurious singularities at $s=m^2_{B}$. Related parameters is
given in the Table. 4 of \cite{abhh}

\subsection{Formula of Observables}
\setcounter{equation}{0}

In this subsection we provide formula of experimental observables,
which include dilepton invariant mass spectrum, forward-backward
asymmetry, and  lepton polarizations.

>From eqs. (2.2 - 2.8), it is straightforward to obtain the matrix element of $B\rightarrow K (K^{*}) l^+l^-$ as follows.
\begin{equation}
        \m  =  \frac{G_F  \alpha}{2 \sqrt{2} \pi} \,
                V_{ts}^\ast  V_{tb}  m_B \, \left[
                  \t_\mu^1 \, \left( \lb \, \g^\mu \, \l \right)
                + \t_\mu^2 \,
                  \left( \lb \, \g^\mu \, \g_5 \, \l \right)
                +\s  \left( \lb \l \right ) \right] \; ,
        \label{eq:med}
\end{equation}
where for $B\to K\ell^+\ell^-$,
\begin{eqnarray}
  \t_\mu^1 & = & \ap(\sh) \, \ph_\mu,
   \label{eq:t1bpll}\\
  \t_\mu^2 & = & \cp(\sh) \, \ph_\mu + \rp(\sh) \, \qh_\mu \; ,
     \label{eq:t2bpll}\\
  \s & = & \s_1(\sh)
     \label{eq:sbpll}\,
\end{eqnarray}
and for $B\to K^*\ell^+\ell^-$,
\begin{eqnarray}
  \t_\mu^1 & = &
    A(\sh) \, \epsilon_{\mu\rho\alpha\beta} \ep^\rho \pbh^\alpha
    \pvh^\beta
    - i B(\sh) \, \ep_\mu
    + i C(\sh) \, (\ep \cdot \hat{p}_B) \ph_\mu  \; ,
    \label{eq:t1bvll}\\
  \t_\mu^2 & = &
  E(\sh) \, \epsilon_{\mu\rho\alpha\beta} \ep^\rho \pbh^\alpha \pvh^\beta
  - i F(\sh) \, \ep_\mu
    + i G(\sh) \, (\ep \cdot \hat{p}_B) \ph_\mu
    + i H(\sh) \, (\ep \cdot \hat{p}_B) \qh_\mu \; ,
    \label{eq:t2bvll}\\
  \s & = &i 2 \mvh (\ep \cdot \hat{p}_B) \s_2(\sh)
\end{eqnarray}
with $p \equiv p_B + p_{K,K^*}$. Note that, using the equation of
motion for lepton fields, the terms in $\hat{q}_\mu$
in ${\cal T}^1_\mu$ vanish.

The auxiliary functions above are
defined as
\begin{eqnarray}
  \ap(\sh) & = & \cne(\sh) \, f_+(\sh)
         + \frac{2 \mbh}{1 + \mph} \cse f_T(\sh) \;, \label{eq:axb} \\
  \cp(\sh) & = & \ct \, f_+(\sh) \; , \\
  \rp(\sh) & = & \ct \, f_-(\sh) -\frac{1-\mph^2}{2 \mlh (\mbh-\msh)} \cqc
f_0(\sh) \; , \\
  \s_1(\sh)  &=& \frac{1-\mph^2}{(\mbh-\msh)} \cqb f_0(\sh)
\;,\\   A(\sh) & = & \frac{2}{1 + \mvh} \cne(\sh) V(\sh)
         + \frac{4 \mbh}{\sh} \cse T_1(\sh) \; , \\
  B(\sh) & = & (1 + \mvh) \left[ \cne(\sh) A_1(\sh)
         + \frac{2 \mbh}{\sh} (1 - \mvh) \cse T_2(\sh) \right] \; , \\
  C(\sh) & = & \frac{1}{1 - \mvh^2} \left[
         (1 - \mvh) \cne(\sh) A_2(\sh)
         + 2 \mbh \cse \left(
           T_3(\sh) + \frac{1 - \mvh^2}{\sh} T_2(\sh) \right) \right] \; , \\
  E(\sh) & = & \frac{2}{1 + \mvh} \ct V(\sh) \; , \label{eq:dt}\\
  F(\sh) & = & (1 + \mvh) \ct A_1(\sh) \; , \\
  G(\sh) & = & \frac{1}{1 + \mvh} \ct A_2(\sh) \; , \\
  H(\sh) & = & \frac{\ct}{\sh} \left[
       (1 + \mvh) A_1(\sh) - (1 - \mvh) A_2(\sh) - 2 \mvh A_0(\sh) \right]
\nonumber \\
&&+\frac{\mvh}{\mlh (\mbh+\msh)} A_0(\sh) \cqc \; , \\
\s_2(\sh)&=&-\frac{1}{(\mbh+\msh)} A_0(\sh) \cqb \;. \label{eq:axd}
\end{eqnarray}
where
\begin{eqnarray}
f_0(\sh)&=&\frac{1}{1-\mph^2} [\sh f_-(\sh)+(1-\mph^2) f_+(\sh)]
\end{eqnarray}
and to get the auxiliary functions given above, we have used equations of motion
\begin{eqnarray}
q^{\mu} (\bar \psi_1 \g_\mu \psi_2)&=&(m_1-m_2) \bar \psi_1 \psi_2, \\
q^{\mu} (\bar \psi_1 \g_\mu \g_5 \psi_2)&=&-(m_1+m_2) \bar \psi_1 \g_5 \psi_2.
\end{eqnarray}
The contributions of NHBs have been incorporated in the terms of $\s_1(\sh)$, $D^{\prime}(\sh)$,
$H(\sh)$ and $\s_2(\sh)$. It is remarkable that the contributions of NHBs in
$D^{\prime}(\sh)$ and $H(\sh)$ are proportional to the inverse mass of the lepton, and
for the case $l=\mu$, the effects of NHBs can be manifested through these terms.

A phenomenological effective Hamiltonian is recently given in \cite{akk}.
If ignoring tensor type interactions in the phenomenological Hamiltonian (it is shown that physical observables
are not sensitive to the presence of tensor type interactions~\cite{fkmy}), it is easy to verify that  the matrix
element of $B\rightarrow K^{(*)}\ell^+\ell^-$ can always be
expressed as the form of the equation (\ref{eq:med}) with the auxiliary functions  defined as
\ba
\ap(\sh) & = & w_{c_1} \, f_+(\sh)
         - \frac{w_{c_9}+w_{c_{10}}}{1+\mph} f_T(\sh) \; , \label{eq:auxb} \\
  \cp(\sh) & = & w_{c_2} \, f_+(\sh)\; , \\
  \rp(\sh) & = & w_{c_2} \, f_-(\sh) -\frac{1-\mph^2}{2 \mlh (\mbh-\msh)} w_{c_6}
f_0(\sh) \; , \\
  \s_1(\sh)  &=& \frac{1-\mph^2}{(\mbh-\msh)} w_{c_5} f_0(\sh)
\;,\\   A(\sh) & = & \frac{2}{1 + \mvh} w_{c_1} V(\sh)
         - \frac{2)}{\sh} (w_{c_9}+w_{c_{10}}) T_1(\sh) \; , \\
  B(\sh) & = & -(1 + \mvh) \left[ w_{c_3} A_1(\sh)
         + \frac{1}{\sh} (1 - \mvh) (w_{c_9}+w_{c_{10}})  T_2(\sh) \right] \; , \\
  C(\sh) & = & -\frac{1}{1 - \mvh^2} \left[
         (1 - \mvh) w_{c_3} (\sh) A_2(\sh) \right .\nnb \\
&&\left .+ (w_{c_9}-w_{c_{10}}) \left(
           (1+\mvh) T_3(\sh) + \frac{1 - \mvh^2}{\sh} T_2(\sh) \right) \right] \; , \\
  E(\sh) & = & \frac{2}{1 + \mvh} w_{c_2} V(\sh) \; , \\
  F(\sh) & = & -(1 + \mvh) w_{c_4} A_1(\sh) \; , \\
  G(\sh) & = & -\frac{1}{1 + \mvh} w_{c_4} A_2(\sh) \; , \\
  H(\sh) & = & -\frac{2 \mvh}{\sh}\,w_{c_4}\, \left( A_3(\sh)-A_0(\sh) \right)+\frac{\mvh}{\mlh (\mbh+\msh)} w_{c_8} A_0(\sh)\; , \\
 \s_2(\sh)&=&-\frac{1}{(\mbh+\msh)} w_{c_7} A_0(\sh)  \;,
\label{eq:auxd}
\ea
where
\ba
w_{c_1} &=& \frac{1}{4} (C_{LL}+C_{LR}+C_{RL}+C_{RR}) \;, \\
w_{c_2} &=& \frac{1}{4} (-C_{LL}+C_{LR}-C_{RL}+C_{RR})\;, \\
w_{c_3} &=& \frac{1}{4} (-C_{LL}-C_{LR}+C_{RL}+C_{RR})\;, \\
w_{c_4} &=& \frac{1}{4} (C_{LL}-C_{LR}-C_{RL}+C_{RR}) \;, \\
w_{c_5} &=& \frac{1}{4} (C_{LRLR}+C_{RLLR}+C_{LRRL}+C_{RLRL})  \;, \\
w_{c_6} &=& \frac{1}{4} (C_{LRLR}+C_{RLLR}-C_{LRRL}-C_{RLRL})\;, \\
w_{c_7} &=& \frac{1}{4} (C_{LRLR}-C_{RLLR}+C_{LRRL}-C_{RLRL}) \;, \\
w_{c_8} &=& \frac{1}{4} (C_{LRLR}-C_{RLLR}-C_{LRRL}+C_{RLRL}) \;, \\
w_{c_9} &=& m_b C_{BR}    \;, \\
w_{c_{10}} &=& m_s C_{SL}   \;, \\.
\ea
In the above equations $C_{LL}, C_{LR}$ etc. are defined in ref. ~\cite{fkmy}.
Therefore our formula given below can also be used to make  model independent phenomenological analysis,
if using Eqs. ((\ref{eq:auxb})-(\ref{eq:auxd})) in stead of Eqs. ((\ref{eq:axb})-(\ref{eq:axd})).

Keeping the lepton mass, we find the double differential
decay widths $\gl^K$ and $\gl^{K^*}$ for the decays $B\to
K\ell^+\ell^-$ and $B\to K^*\ell^+\ell^-$, respectively, as
\begin{eqnarray}
  \frac{\d^2 \gl^K}{\d\sh \d\uh} & = &
  \frac{G_F^2  \alpha^2  m_B^5}{2^{11}  \pi^5}
      \left| V_{ts}^\ast  V_{tb} \right|^2  \nonumber \\
& & \times\left\{
(|\ap|^2 +|\cp|^2)
(\la -\uh^2 ) +|\s_1|^2 (\sh-4 \mlh^2)+Re(\s_1 \ap^{*}) 4 \mlh  \uh
\right. \nonumber \\
&& + \left. |\cp|^2 4 \ml^2 (2+2 \mph^2-\sh)
+Re( \cp \ddp^{*}) 8 \ml^2 (1-\mph^2)
+|\ddp|^2 4 \ml^2 \sh \right\} \; ,    \label{eq:ddwbpll} \\
  \frac{\d^2 \gl^{K^*}}{\d\sh \d\uh} & = &
  \frac{G_F^2 \, \alpha^2 \, m_B^5}{2^{11} \pi^5}
      \left| V_{ts}^\ast \, V_{tb} \right|^2
        \nonumber \\
  & &  \times\Bigg\{
  \frac{|A|^2}{4}
   \left(\sh (\la + \uh^2) + 4 \mlh^2 \la  \right)
  + \frac{|E|^2}{4} \left(\sh (\la + \uh^2) - 4 \mlh^2 \la  \right)
  +|\s_2|^2 (\sh-4\mlh^2) \la
        \Bigg.
        \nonumber \\
  & & \Bigg.
  + \frac{1}{4 \mvh^2} \left[
  |B|^2 \left( \la - \uh^2 + 8 \mvh^2 (\sh +2 \mlh^2 ) \right)
  + |F|^2 \left( \la - \uh^2 + 8 \mvh^2 (\sh -4 \mlh^2) \right) \right]
        \Bigg.
        \nonumber \\
  & & \Bigg.
  - 2 \sh \uh \left[ {\rm Re}(BE^\ast) + {\rm Re}(AF^\ast)
\right]+\frac{2 \mlh \uh}{\mvh}[{\rm Re}   (\s_2 B^*) (\sh+\mvh^2-1) + {\rm
Re}(\s_2 C^*) \la ]        \Bigg.
        \nonumber \\
  & & \Bigg.
  + \frac{\la}{4 \mvh^2} \left[ |C|^2 (\la - \uh^2)
    + |G|^2 \left( \la - \uh^2 + 4 \mlh^2 (2 + 2 \mvh^2 - \sh) \right) \right]
        \Bigg.
        \nonumber \\
  & & \Bigg.
  - \frac{1}{2 \mvh^2} \left [
  {\rm Re}(BC^\ast) (1 - \mvh^2 - \sh)(\la - \uh^2)
  \right.
  \nonumber \\
  & & \left. \; \; \; \; \; \; \; \; \; \; \; \;
  + {\rm Re}(FG^\ast)
\left( (1 - \mvh^2 - \sh)(\la - \uh^2) + 4 \mlh^2 \la \right) \right]
        \Bigg.
        \nonumber \\
  & & \Bigg.
  - 2 \frac{\mlh^2}{\mvh^2} \la \left[
    {\rm Re}(FH^\ast)
    - {\rm Re}(GH^\ast) (1 - \mvh^2) \right]
 + |H|^2 \frac{\mlh^2}{\mvh^2} \sh \la
  \Bigg\} \; .
   \label{eq:ddwbvll}
\end{eqnarray}
Here the kinematic variables $(\sh,\uh)$ are defined as
\begin{eqnarray}
  \sh & = & \qh^2 = (\ph_+ + \ph_-)^2 \; , \\
  \uh & = & (\pbh - \ph_-)^2 - (\pbh - \ph_+)^2  \;
 \end{eqnarray}
which are bounded as
\begin{eqnarray}
  (2 \mlh)^2 \leq & \sh & \leq (1 - \hat{m}_{K,K^*})^2  \; ,
  \label{eq:sbound}\\
  -\uh(\sh) \leq & \uh & \leq \uh(\sh) \; ,
  \label{eq:ubound}
\end{eqnarray}
with $\ml=m_{\ell}/m_B$ and
\begin{eqnarray}
  \uh(\sh) & =& \sqrt{\la (1-4 \frac{\mlh^2}{\sh})}  \; , \\
 \la & =& 1+\hat{m}_{K,K*}^4+\sh^2-2 \sh-2 \hat{m}_{K,K*}^2(1+\sh) \;, \\
\D&=&\sqrt{1-\frac{4 \mlh^2}{s}}\;.
\end{eqnarray}
Note that the variable $\uh$ corresponds to $\theta$, the angle
between the momentum of the $B$-meson and the positively charged lepton
$\ell^+$  in the dilepton CMS frame, through the relation $\uh = -\uh(\sh) \cos \theta$
\cite{amm91}.

Integrating over $\uh$ in the kinematic region
given in Eq. (\ref{eq:ubound}) we get the formula of dilepton invariant mass
spectra (IMS)
\begin{eqnarray}
  \frac{\d \gl^K}{\d\sh} & = &
  \frac{G_F^2  \alpha^2  m_B^5}{2^{10} \pi^5}
      \left| V_{ts}^\ast  V_{tb} \right|^2  \uh(\sh)  D^{K}\\
D^{K}&=& (|\ap|^2 +|\cp|^2)
( \la- \frac{\uh(\sh)^2}{3} )+ |\s_1|^2 (\sh-4 \mlh^2) \nonumber \\
& & +  |\cp|^2 4 \ml^2 (2+2 \mph^2-\sh)
+ Re( \cp \ddp^{*}) 8 \ml^2 (1-\mph^2)
+|\ddp|^2 4 \ml^2 \sh \; ,\label{eq:dwbpll}\\
  \frac{\d \gl^{K^*}}{\d\sh} & = &
  \frac{G_F^2 \, \alpha^2 \, m_B^5}{2^{10} \pi^5}
      \left| V_{ts}^\ast  V_{tb} \right|^2 \, \uh(\sh) D^{K^*}\\
D^{K^*}  &= & \frac{|A|^2}{3} \sh \la (1+2 \frac{\mlh^2}{\sh})
+|E|^2 \sh \frac{\uh(\sh)^2}{3} + |\s_2|^2 (\sh-4 \mlh^2) \la
        \nonumber \\
  & & + \frac{1}{4 \mvh^2} \left[
|B|^2 (\la-\frac{\uh(\sh)^2}{3} + 8 \mvh^2 (\sh+ 2 \mlh^2) )
          + |F|^2 (\la -\frac{ \uh(\sh)^2}{3} + 8 \mvh^2 (\sh- 4 \mlh^2))
\right]       \nonumber \\
  & & +
   \frac{\la }{4 \mvh^2} \left[ |C|^2 (\la - \frac{\uh(\sh)^2}{3})
 + |G|^2 \left(\la -\frac{\uh(\sh)^2}{3}+4 \mlh^2(2+2 \mvh^2-\sh) \right)
\right]      \nonumber \\
  & & -  \frac{1}{2 \mvh^2}
\left[ {\rm Re}(BC^\ast) (\la -\frac{ \uh(\sh)^2}{3})(1 - \mvh^2 - \sh)
\nonumber  \right. \\
& & +    \left.   {\rm Re}(FG^\ast) ((\la -\frac{ \uh(\sh)^2}{3})(1 -
\mvh^2 - \sh) +  4 \mlh^2 \la) \right]
        \nonumber \\
  & & -  2 \frac{\mlh^2}{\mvh^2} \la  \left[ {\rm Re}(FH^\ast)-
 {\rm Re}(GH^\ast) (1-\mvh^2) \right] +\frac{\mlh^2}{\mvh^2} \sh \la |H|^2
 \; .
   \label{eq:dwbvll}
\end{eqnarray}

Both distributions agree with the ones obtained in \cite{abhh,gk},
if $C_{Q_1,2}$ are set to zero.

The differential forward-backward-asymmetry (FBA) is defined as
$$ A_{FB} (s) = \frac{\displaystyle{- \int_0^{u(\sh)} dz \frac{d \Gamma}{ds du}+
\int_{-u(\sh)}^0du
\frac{d \Gamma}{ds du}}}{\displaystyle{\int_0^{u(\sh)} dz \frac{d \Gamma}{ds
du}+
\int_{-u(\sh)}^0du\frac{d \Gamma}{ds du}}}~.$$
For $B\to K\ell^+\ell^-$ decays it reads as follows
\begin{eqnarray}
  \frac{\d \a_{\rm FB}^{K}}{\d \sh} D^K& =& - 2 \mlh \uh(\sh){\rm Re}(\s_1 \ap^*)
\label{eqn:bkas}
\end{eqnarray}
For $B\to K^*\ell^+\ell^-$ decays it reads as follows
\begin{eqnarray}
  \frac{\d \a_{\rm FB}^{K^*}}{\d \sh} D^{K^*}& =& \uh(\sh)
      \Bigg \{ \sh \left [ {\rm Re}(B E^\ast) + {\rm Re}(AF^\ast)\right ]
\Bigg. \; \nonumber \\
&&\Bigg. +\frac{\mlh}{\mvh} \left [{\rm Re}(\s_2
B^*) (1-\sh-\mvh^2) -{\rm Re}(\s_2 C^*)\la \right ]
\Bigg \}
\label{eqn:bksas}
\end{eqnarray}

We can read from (\ref{eqn:bkas}), the FB asymmetry of the process $B\rightarrow K
\ell^+ \ell^-$ does not vanish when the contributions of NHB are taken into
account. With it, our analysis below also show the contributions of NHBs can even be accessible
in B factories.

The lepton polarization can be defined as follows
\ba
\frac{d \Gamma ( \vec{n} )}{d s} &=&
\frac{1}{2} ( \frac{d \Gamma}{d s} )_0
\Big[ 1 + ( P_L\, \vec{e}_L + P_N\, \vec{e}_N + P_T\, \vec{e}_T ) \cdot
\vec{n} \Big]~,
\ea
where the subscript "0" corresponds to the unpolarized amplitude, and
$P_L$, $P_T$, and $P_N$, correspond to the longitudinal, transverse and
normal components of the polarization vector, respectively.

For the process $B\rightarrow K \ell^- \ell^+$, the $P_L^K$, $P_T^K$, and
$P_N^K$, are derived respectively as
\ba
P_L^K D^K&=&\frac{4}{3} \D \Bigg \{ \la {\rm Re}(\ap \cp^{\ast})-3 \mlh
(1-\mph^2) {\rm Re}(\cp^{\ast} \s_1) -3 \mlh \sh {\rm Re}(\rp^*
\s_1) \Bigg \}, \label{eqn:bklp} \\
P_N^K D^K&=&\frac{\pi \sqrt{\sh} \uh(\sh)}{2} \Bigg \{-{\rm Im}(\ap \s_{1}^*)+ 2 \mlh {\rm Im}(\cp \rp^*)\Bigg \},\\
P_T^K D^K&=&\frac{-\pi \sqrt{\la}}{\sqrt{\sh}} \Bigg \{\mlh \left [(1-\mph^2) {\rm
Re}(\ap \cp^*)\Bigg. \right. \nonumber \\
&&\Bigg.\left. +\sh {\rm Re}(\ap \rp^{\ast})\right ] + \frac{(\sh-4\mlh^2)}{2}{\rm
Re}(\cp \s_{1}^{\ast}) \Bigg \}.      \label{eqn:bktp}\ea

$D^K$ is defined in Eq. (\ref{eq:dwbpll}). For the process $B\rightarrow K^* \ell^- \ell^+$, the $P_L^{K^*}$,
$P_T^{K^*}$, and $P_N^{K^*}$, are derived respectively as
\ba
P_L^{K^*} D^{K^*}&=&\D \Bigg \{\frac{2 \sh \la}{3} {\rm Re}(A E^*) +
\frac{(\la+12 \mvh^2)}{3 \mvh^2}{\rm Re}(B F^*) \Bigg. \nonumber\\
&&\Bigg.-\frac{\la (1-\mvh^2-\sh)}{3
\mvh^2}{\rm Re}(B G^*+C F^*) +\frac{\la^2}{3 \mvh^2} {\rm Re}
(C G^*)\Bigg. \nonumber \\
&&\Bigg. +\frac{2 \mlh \la}{\mvh} \left[{\rm Re}(F \s_2^*) - \sh {\rm Re}(H
\s_2^*)-(1-\mvh^2){\rm Re}(G \s_2^*)\right ] \Bigg \},   \label{eqn:bkslp} \\
P_N^{K^*} D^{K^*}&=&\frac{- \pi \sqrt{\sh} \uh(\sh)}{4 \mph}
\Bigg \{\frac{\mlh}{\mvh}\left [ {\rm Im}(F G^*)(1+3 \mvh^2-s)\right.\Bigg.\nonumber \\
&&\left.\Bigg.+{\rm Im}(F H^*)(1-\mvh^2-s)-{\rm Im}(G H^*)\la\right ]\Bigg.\nonumber \\
&&\Bigg.+2 \mvh \mlh
\left [{\rm Im} (B E^*)+{\rm Im}(A F^*)\right ] \Bigg.\nonumber \\
&&\Bigg. -(1-\mvh^2-\sh) {\rm Im} (B \s_2^*)+\la {\rm Im}(C
\s_2^*)\Bigg \}, \\
P_T^{K^*} D^{K^*}&=&\frac{\pi \sqrt{\la} \mlh}{4 \sqrt{\sh}}\Bigg \{ 4 \sh {\rm Re}(A B^*) \Bigg.\nonumber \\
&&\Bigg.+\frac{(1-\mvh^2 -\sh)}{\mvh^2}\left[-{\rm Re}(B F^*)+(1-\mvh^2){\rm Re}(B G^*)+\sh{\rm
Re}(B H^*)\right ]\Bigg.
\nonumber \\
&&\Bigg.+\frac{\la}{\mvh^2} \left [{\rm Re} (C F^*) - (1-\mvh^2){\rm Re}(C G^*)-\sh
{\rm Re}(C H^*) \right ] \Bigg.\nonumber \\
&&\Bigg.+\frac{(\sh-4 \mlh^2)}{\mvh \mlh} \left[(1-\mvh^2-\sh){\rm Re} (F \s_2^*)
-\la {\rm Re}(G \s_2^*)\right]\Bigg \}.
\label{eqn:bkstp}
\ea
$D^{K^*}$ is defined by Eq.(\ref{eq:dwbvll}).
\subsection{Numerical analysis}
Parameters used in our analysis are list in Table \ref{tab:para}. Considering that the branching ratios of
$\bkll$ and $\bksll$ are not very sensitive to the mass of $m_b$, we neglect the difference between the pole mass
and running mass of b quark.
\begin{table}
        \begin{center}
        \begin{tabular}{|l|l|}
        \hline
        $m_b$                   & $4.8$ GeV \\
        $m_c$                   & $1.4$ GeV \\
        $m_s$                   & $0.2$ GeV   \\
        $m_{mu}$                & $0.11$ GeV \\
        $m_{tau}$               & $1.78$ GeV \\
        $M_B$                   & $5.28$ GeV \\
        $M_K$                   & $0.49$ GeV \\
        $M_{K^*}$                   & $0.89$ GeV \\
        $M_{J/psi}(M_{\psi'})$      & $3.10(3.69)$   GeV\\
        $\Gamma_B$                   & $4.22\times10^{-13}$ GeV \\
        $\Gamma_{J/\psi}(\Gamma_{\psi'})$      & $8.70(27.70)\times 10^{-5}$   GeV\\
        $\Gamma(J/psi\rightarrow\ell^+\ell^-)$      & $5.26\times 10^{-6}$   GeV\\
        $\Gamma(\psi'\rightarrow\ell^+\ell^-)$      & $2.14\times 10^{-6}$   GeV\\
        $G_F$                         & $1.17\times 10^{-5}$ GeV$^{-2}$\\
        $\alpha^{-1}$     & 129           \\
        $|V^\ast_{ts} V_{tb}|$ & 0.0385 \\
        \hline
        \end{tabular}
        \end{center}
\caption{\it Values of the input parameters used in our numerical analysis.}
\label{tab:para}
\end{table}

The Wilson coefficients in the SM used in the numerical analysis is given in the Table \ref{tab:wcsm}.
$C_7^{eff}$ is defined as
\ba
C_7^{eff} &=& C_7 -C_5/3 -C_6\;.
\ea
\begin{table}
        \begin{center}
        \begin{tabular}{|c|c|c|c|c|c|c|c|c|c|}
        \hline
        \multicolumn{1}{|c|}{ $C_1$}       &
        \multicolumn{1}{|c|}{ $C_2$}       &
        \multicolumn{1}{|c|}{ $C_3$}       &
        \multicolumn{1}{|c|}{ $C_4$}       &
        \multicolumn{1}{|c|}{ $C_5$}       &
        \multicolumn{1}{|c|}{ $C_6$}       &
        \multicolumn{1}{|c|}{ $C_7^{\rm eff}$}       &
        \multicolumn{1}{|c|}{ $C_9$}       &
        \multicolumn{1}{|c|}{$C_{10}$} &
        \multicolumn{1}{|c|}{ $C$ }     \\
        \hline
        $-0.248$ & $+1.107$ & $+0.011$ & $-0.026$ & $+0.007$ & $-0.031$ &
   $-0.313$ &   $+4.344$ &    $-4.669$    & $+0.362$     \\
        \hline
        \end{tabular}
        \end{center}
\caption{ \it Wilson coefficients of the SM used in the numerical
          analysis.}
\label{tab:wcsm}
\end{table}

$C_{Q_{1,2}}$ come from  exchanging  NHBs and are proportional to tan$^3\beta$ in some regions of the
parameter space in SUSY models.
According to the analysis in \cite{hly,hllyz}, the necessary conditions for the large contributions of NHBs include: ({\it i})
the ratio of vacuum expectation value, tan$\beta$, should be large, ({\it ii}) the mass values of the lighter
chargino and the lighter stop should not be too large (say less than 120 GeV), ({\it iii})mass splitting of charginos
and stops should be large, which also indicate large mixing between stop sector and chargino sector.
As the conditions are satisfied, the process $B\rightarrow X_s \gamma$ will impose a constraint on $C_7$.
It is well known that this process puts a very stringent constraint on the possible new physics and that
SUSY can contribute destructively when the signature of the Higgs mass term $\mu$ is minus.
There exist considerable regions of SUSY parameter space in which NHBs can largely contribute
to the process $b\rightarrow s \ell^+ \ell^-$ while the constraint of $b\rightarrow s \gamma$ is respected
(i.e., the signature of the Wilson coefficient $C_7$ is changed from positive to negative).
When the masses of SUSY particle are relatively heavy (say, 450 Gev),
there are still significant regions in the parameter space of SUSY models in which NHBs could contribute largely.
However, at these cases $C_7$ does not change its sign, because contributions of charged Higgs and
charginos cancel with each other. We will see it is hopeful to distinguish these two kinds of regions of
SUSY parameter space through observing $B\rightarrow K^{(*)}\ell^+\ell^-$.

As pointed out in \cite{dhh,hly}, the contribution of NHBs is proportional to the lepton mass, therefore for $\ell=e$,
contributions of NHBs can be safely neglected. While for cases $\ell=\mu$ and $\ell=\tau$, the contributions of NHBs
can be considerably large. To investigate the effects of NHBs in SUSY models, we take typical
values of $C_{7,9,10}$ and  $C_{Q_{1,2}}$ as given in Table \ref{tab:susy}. The SUSY model without
considering the effects of NHBs (SUSY I in Table 4) is given as a reference frame so that could
the effects of NHBs be shown in high relief.
\begin{table}
        \begin{center}
        \begin{tabular}{|c|c|c|c|c|c|}
        \hline
        \multicolumn{1}{|c|}{ SUSY models} &
        \multicolumn{1}{c|}{ $R_7$}       &
        \multicolumn{1}{c|}{ $R_9$}       &
        \multicolumn{1}{c|}{ $R_{10}$}       &
        \multicolumn{1}{c|}{ $C_{Q_1}$}       &
        \multicolumn{1}{c|}{ $C_{Q_2}$}      \\
        \hline
        SUSY I  & $-1.2$ & $1.1$ & $0.8$ & $0.0$ & $0.0$ \\
        \hline
        SUSY II  & $-1.2$ & $1.1$ & $0.8$ & $6.5(16.5)$ & $-6.5(-16.5)$ \\
        \hline
        SUSY III  & $1.2$ & $1.1$ & $0.8$ & $1.2(4.5)$ & $-1.2(-4.5)$ \\
        \hline
        \end{tabular}
        \end{center}
\caption{ \it Wilson coefficients of the SUSY used in our numerical
        analysis. $R_i$ means $C_i$/$C_i^{SM}$. SUSY I corresponds to the regions where SUSY can destructively
        contribute and can change the sign of $C_7$, but the contributions of NHBs are neglected.
        SUSY II corresponds to the regions where tan$\beta$ is large and the masses of superpartners
        are relatively small. SUSY III corresponds to the regions where tan$\beta$ is large but the masses of
        superpartners are relatively large. In the last two cases the effects of NHBs are taken into account.
        The contributions of NHBs are settled to be different for both the case $\ell=\mu$ and $\ell=\tau$, since $C_{Q_{1,2}}$
        are proportional to the mass of lepton. The values in bracket are for the case $\ell=\tau$}
\label{tab:susy}
\end{table}
\begin{center}
\begin{figure}
\begin{minipage}[t]{8.2cm}
     \epsfig{file=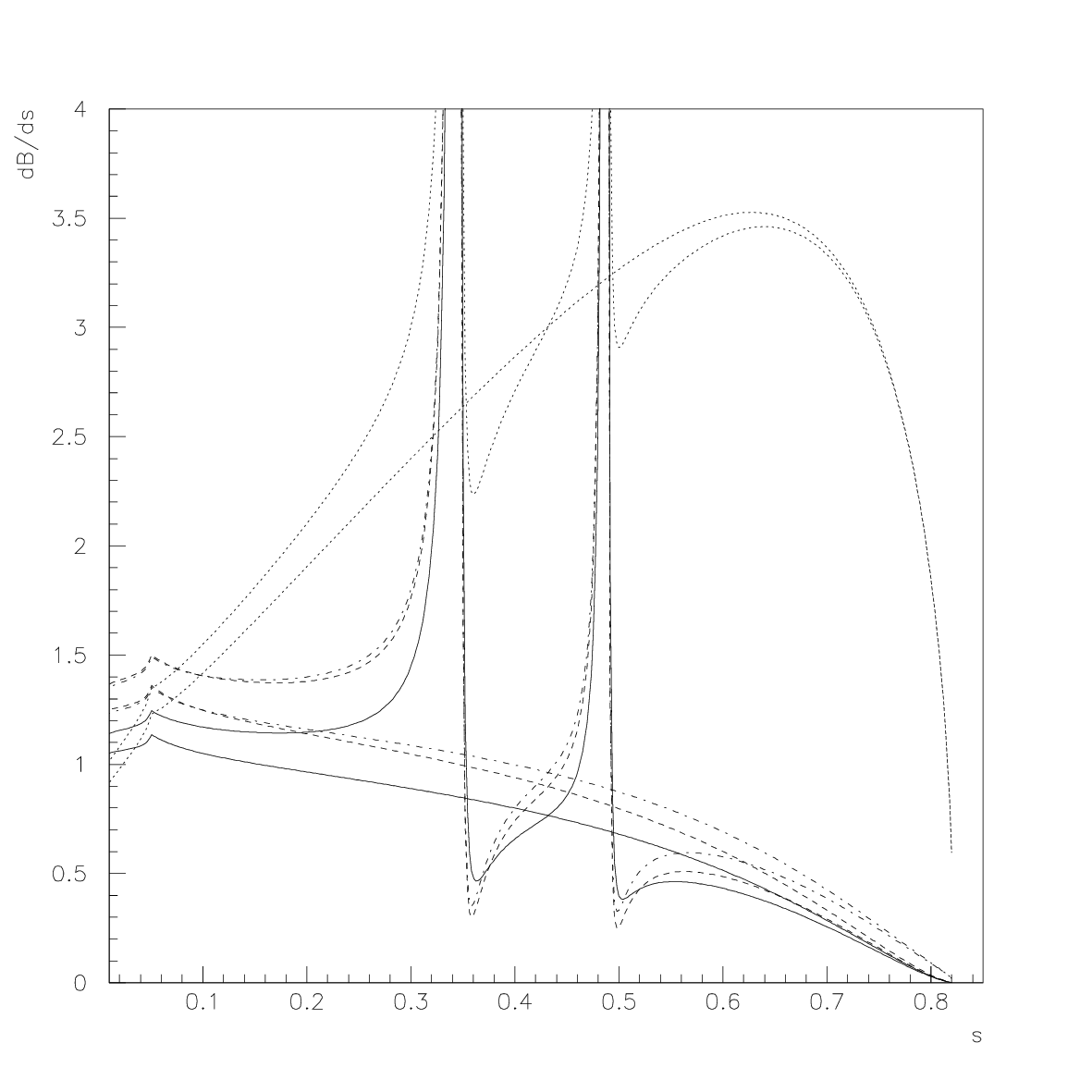,width=8.2cm}
     \mbox{ }\hfill\hspace{1cm}(a)\hfill\mbox{ }
     \end{minipage}
     \hspace{-0.4cm}
     \begin{minipage}[t]{8.2cm}
     \epsfig{file=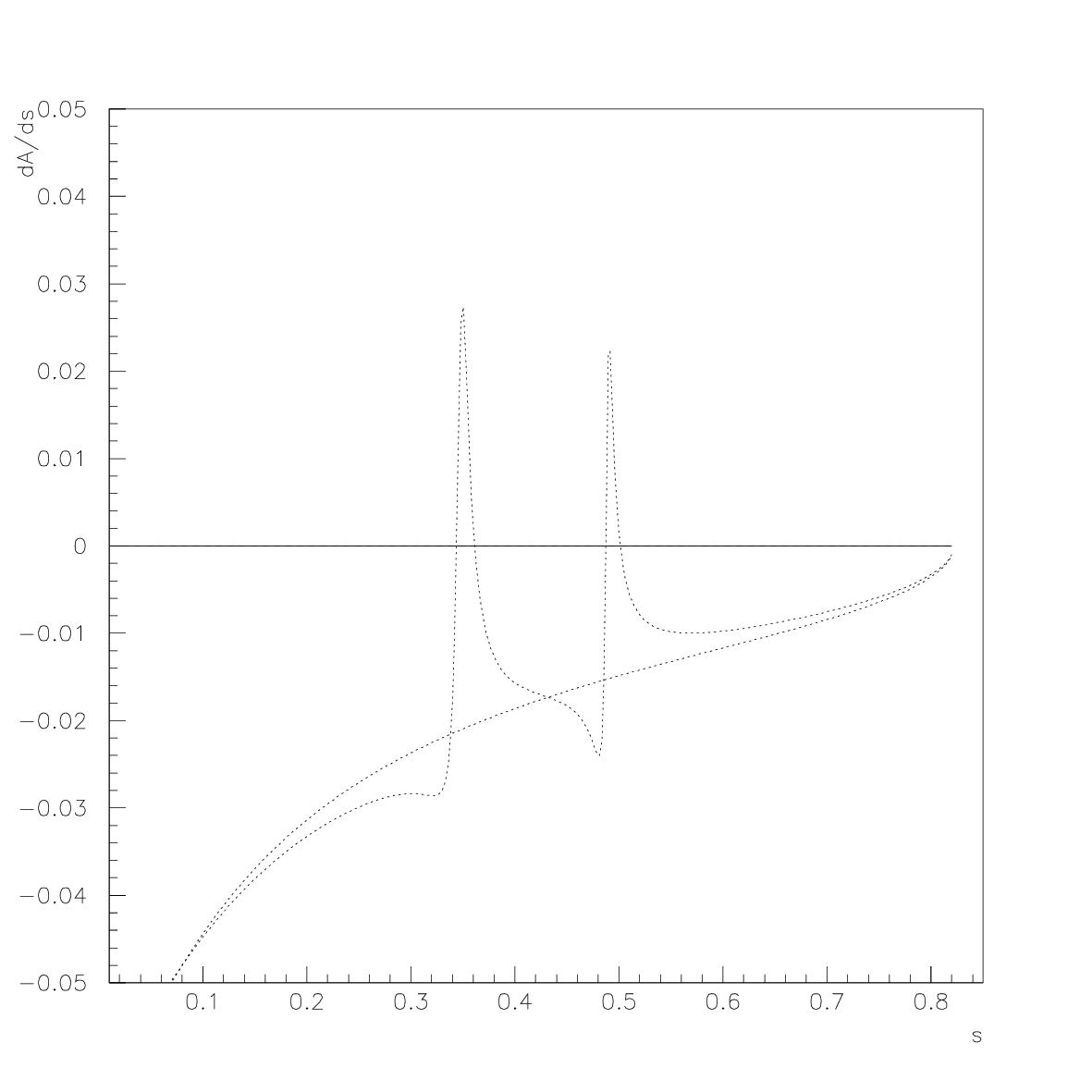,width=8.2cm}
     \mbox{ }\hfill\hspace{1cm}(b)\hfill\mbox{ }
     \end{minipage}
\vskip 0.05truein
     \begin{minipage}[t]{8.2cm}
     \epsfig{file=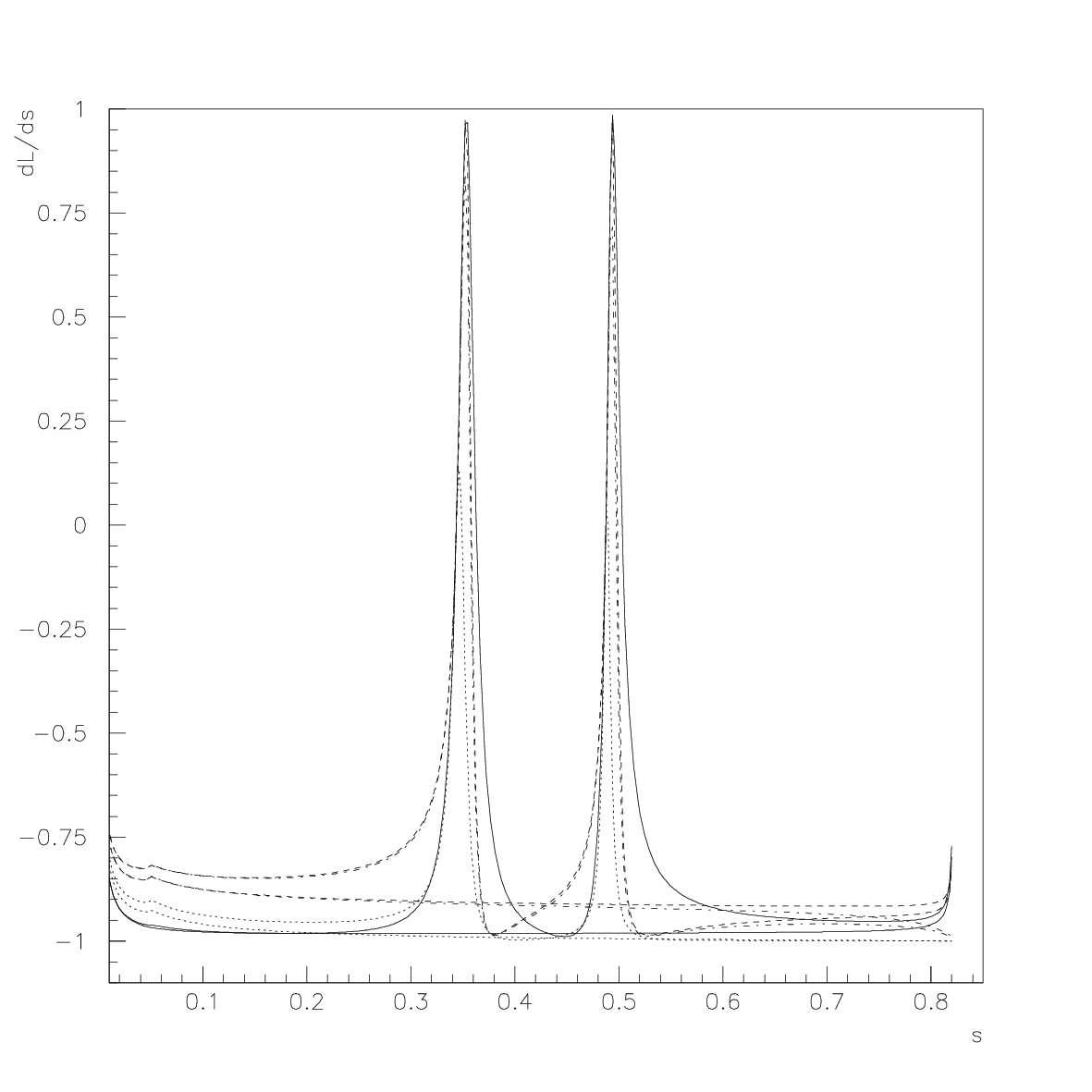,width=8.2cm}
     \mbox{ }\hfill\hspace{1cm}(c)\hfill\mbox{ }
     \end{minipage}
     \hspace{-0.4cm}
     \begin{minipage}[t]{8.2cm}
     \epsfig{file=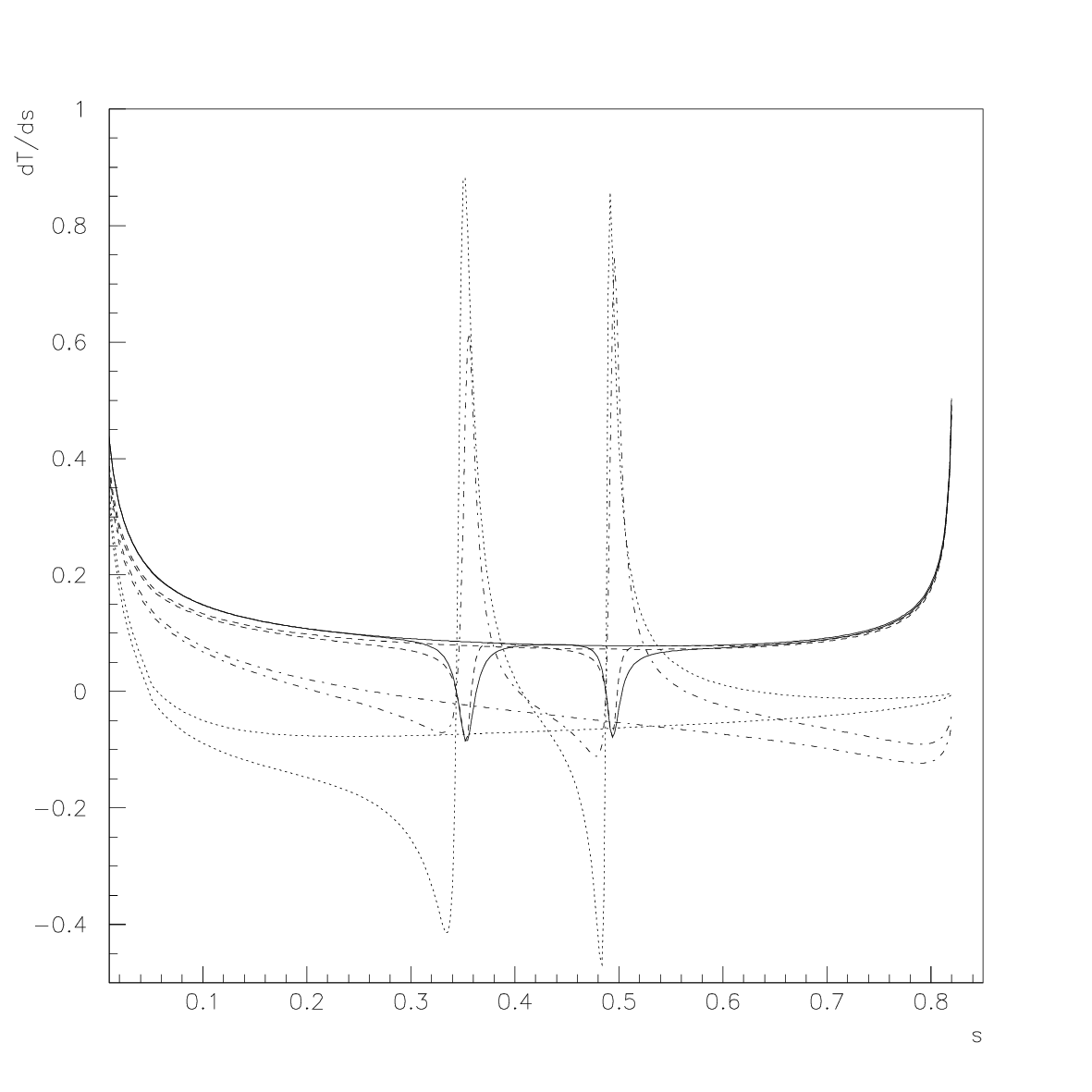,width=8.2cm}
     \mbox{ }\hfill\hspace{1cm}(d)\hfill\mbox{ }
     \end{minipage}
     \caption{\it The IMS(a), FBA(b), LP(c), and TP(d) of the process $\bkm$. The solid line, dashed line, dot line and dashed-dot line represent
   the SM, SUSY I, SUSY II, SUSY III respectively. Both the total (SD+LD) and the pure SD contributions are shown in order to compare.}
\label{fig:kman}
\end{figure}
\end{center}
Numerical results are shown in Figs. 1-4. In Fig.
\ref{fig:kman}(a), the IMS of $B\rightarrow K\mu^+\mu^-$ is
depicted. We see that at the high $\sh$ regions, NHBs greatly
modify the spectrum. While at the low $\sh$ region, the effects of
NHBs become weak. In Fig. \ref{fig:kman}(b), the FB asymmetry of
the $B\rightarrow K \mu^+ \mu^-$ is presented. Fig.
\ref{fig:kman}(b) shows that the average FB asymmetry in
$B\rightarrow K \mu^+ \mu^-$ 0.02. To measure an asymmetry $A$ of
a decay with the branching ratio $Br$ at the $n\sigma$ level, the
required number of events is $N=n^2/(Br A^2)$. For $B \rightarrow
K\mu^+ \mu^-$, the average FB asymmetry is 0.02 or so, the
required number of events is $10^{-12}$ or so. Therefore it is
hard to observe the derivation of FB asymmetry from the SM. In
Fig. \ref{fig:kman}(c) and Fig. \ref{fig:kman}(d), the
longitudinal and transverse polarizations are given. The effect of
NHBs to the longitudinal polarization is weak but the effect to
the transverse is remarkable.
\begin{center}\begin{figure}
\begin{minipage}[t]{8.2cm}
     \epsfig{file=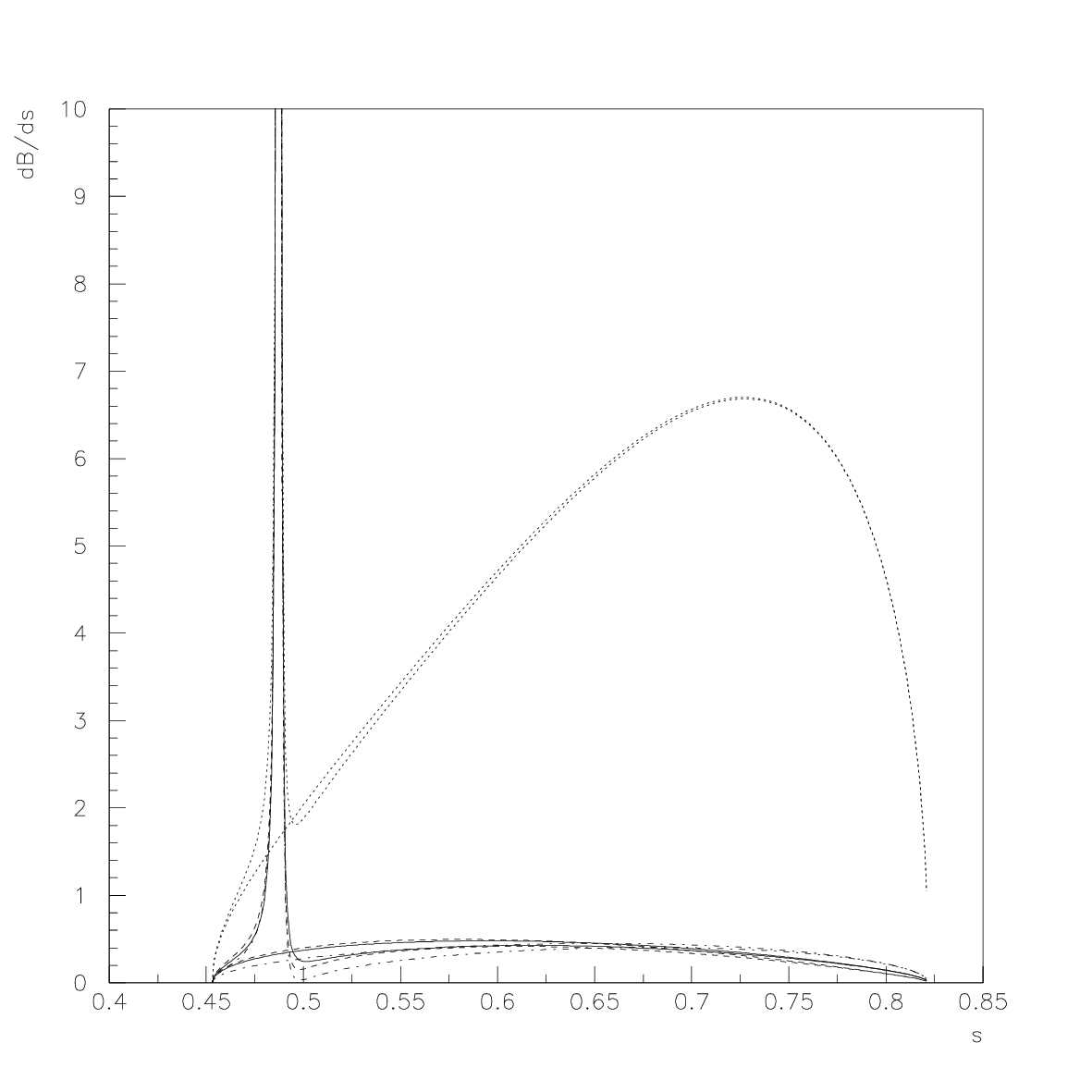,width=8.2cm}
     \mbox{ }\hfill\hspace{1cm}(a)\hfill\mbox{ }
     \end{minipage}
     \hspace{-0.4cm}
     \begin{minipage}[t]{8.2cm}
     \epsfig{file=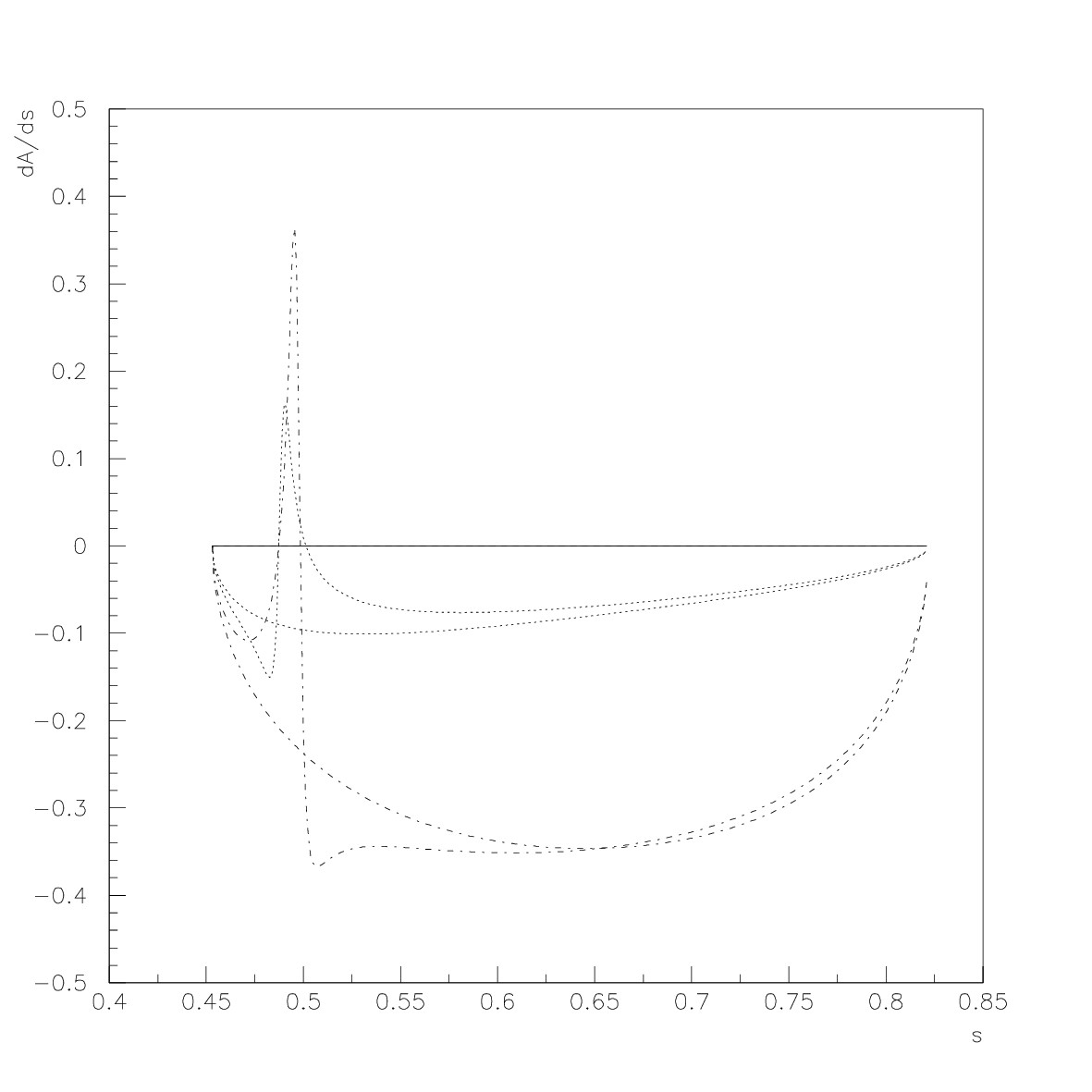,width=8.2cm}
     \mbox{ }\hfill\hspace{1cm}(b)\hfill\mbox{ }
     \end{minipage}
\vskip 0.05truein
     \begin{minipage}[t]{8.2cm}
     \epsfig{file=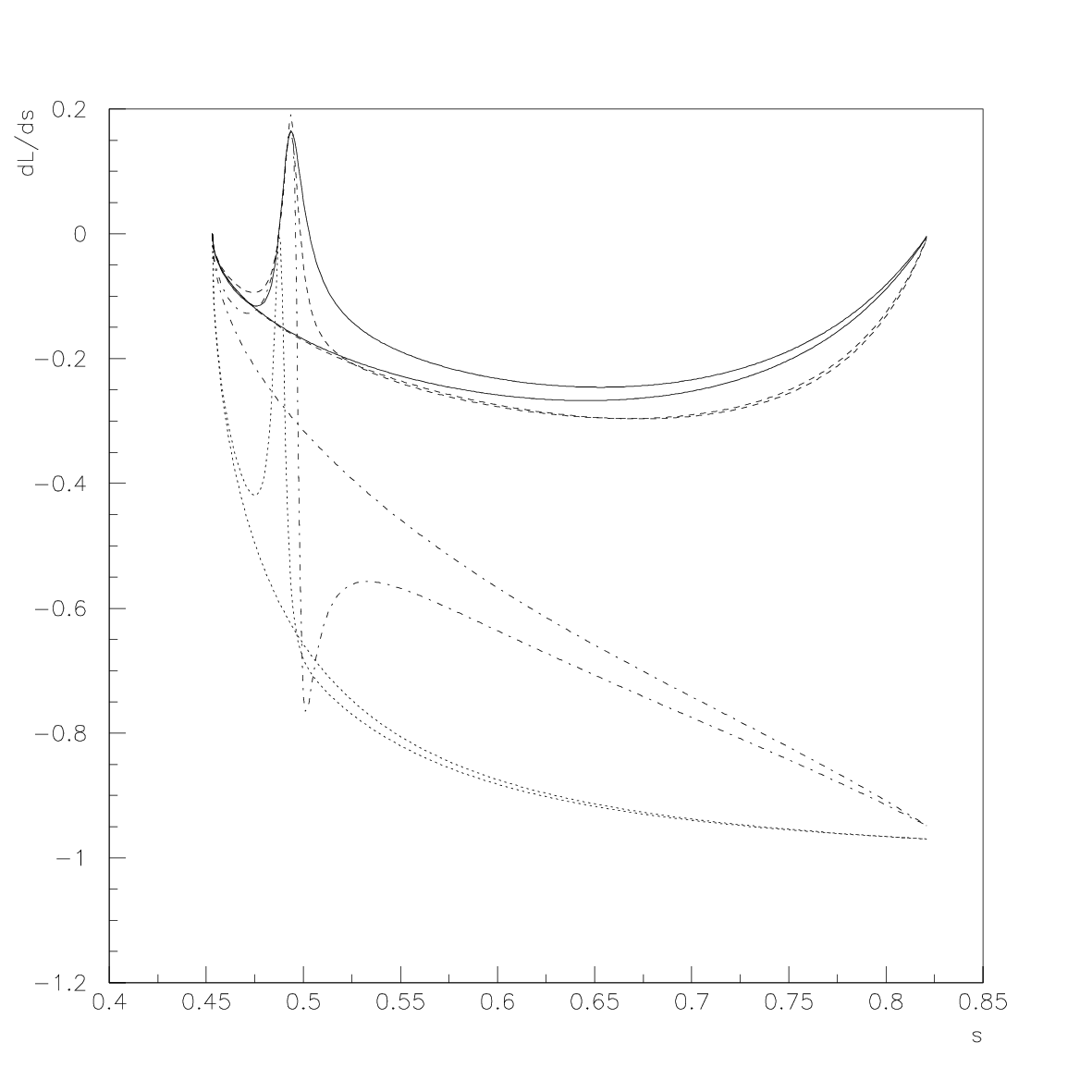,width=8.2cm}
     \mbox{ }\hfill\hspace{1cm}(c)\hfill\mbox{ }
     \end{minipage}
     \hspace{-0.4cm}
     \begin{minipage}[t]{8.2cm}
     \epsfig{file=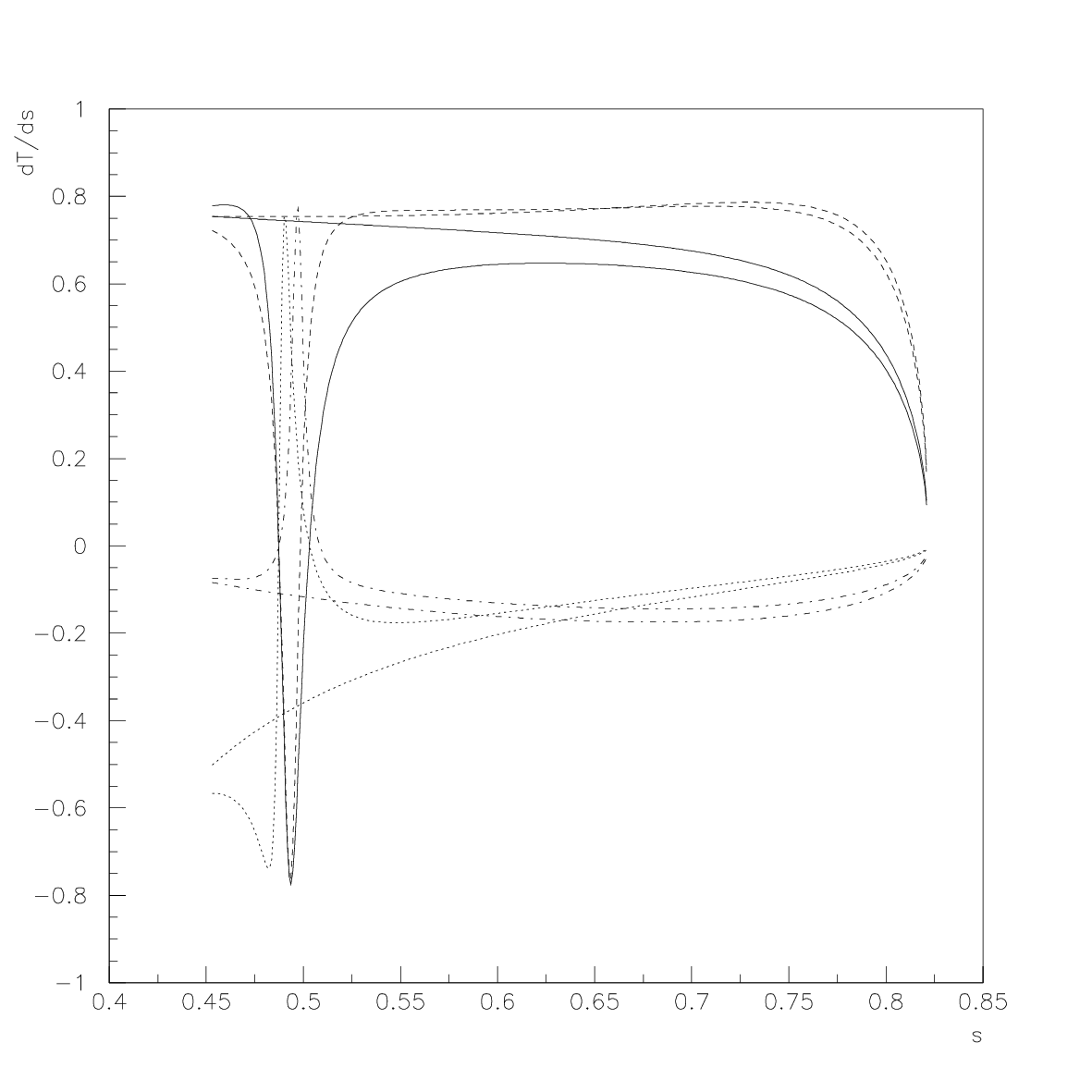,width=8.2cm}
     \mbox{ }\hfill\hspace{1cm}(d)\hfill\mbox{ }
     \end{minipage}
     \caption{\it The IMS(a), FBA(b), LP(c), and TP(d) of the process $\bkt$. The line conventions are the
same as given in the legend of Fig 1.}
\label{fig:ktan}
\end{figure}
\end{center}

In Fig. \ref{fig:ktan}(a) and Fig. \ref{fig:ktan}(b) the IMS and FB asymmetry of $B\rightarrow K\tau^+\tau^-$ are presented respectively.
For SUSY II, the effects of NHBs to IMS is quite manifest, and the average FB asymmetry can reach 0.1.
For SUSY III, the average FB asymmetry can reach $0.3$. Therefore, in order to observe FBA, the required number
of events should be $10^{-9}$ or so and  $10^{-8}$, respectively, so that in B factories, say LHCB, these two
cases are accessible.
In Fig. \ref{fig:ktan}(c) and Fig. \ref{fig:ktan}(d), the longitudinal and transverse polarizations are drawn respectively. The effects of
NHBs are also very obvious.

Figs. 3 and 4 are devoted to the decay $B\rightarrow K^{*}
l^+l^-$. In Fig. \ref{fig:ksman}, the IMS, FB asymmetry, and
polarizations of $B\rightarrow K^* \mu^+ \mu^-$ are given. We see
that this process is not as much as sensitive to the effect of NHB
as $\bkm$. However, the contribution of NHBs will increase the
part with positive FB asymmetry and will be helpful to determine
the zero point of FB asymmetry.  Fig. \ref{fig:ksman}(d) depicts
the transverse polarization of the $\bksm$, and the effect of NHBs
is quite obvious. The zero point of the FB asymmetry can be
slightly modified as shown in Figure \ref{fig:ksman}(b) due to the
contributions of NHBs. \begin{center}\begin{figure}
\begin{minipage}[t]{8.2cm}
     \epsfig{file=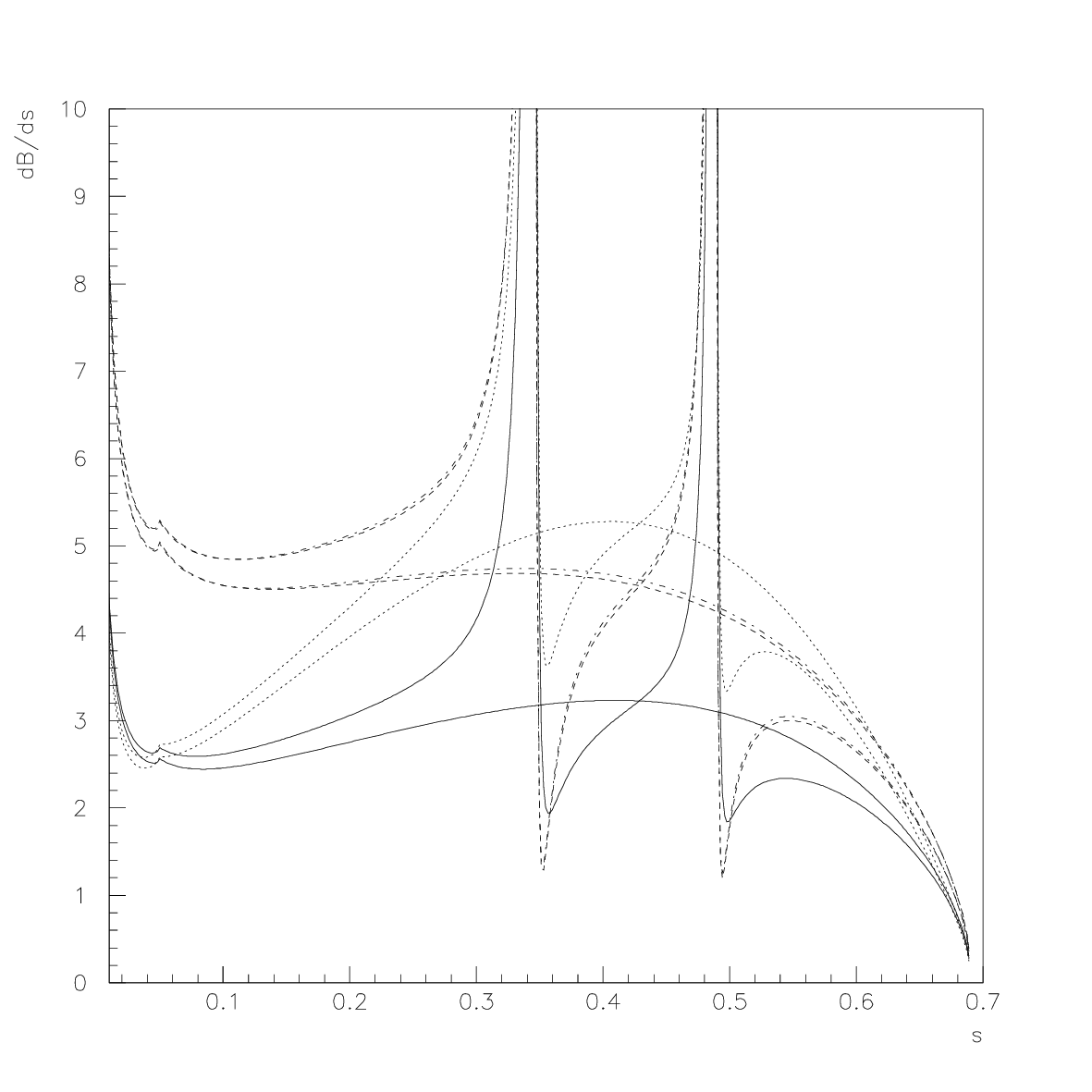,width=8.2cm}
     \mbox{ }\hfill\hspace{1cm}(a)\hfill\mbox{ }
     \end{minipage}
     \hspace{-0.4cm}
     \begin{minipage}[t]{8.2cm}
     \epsfig{file=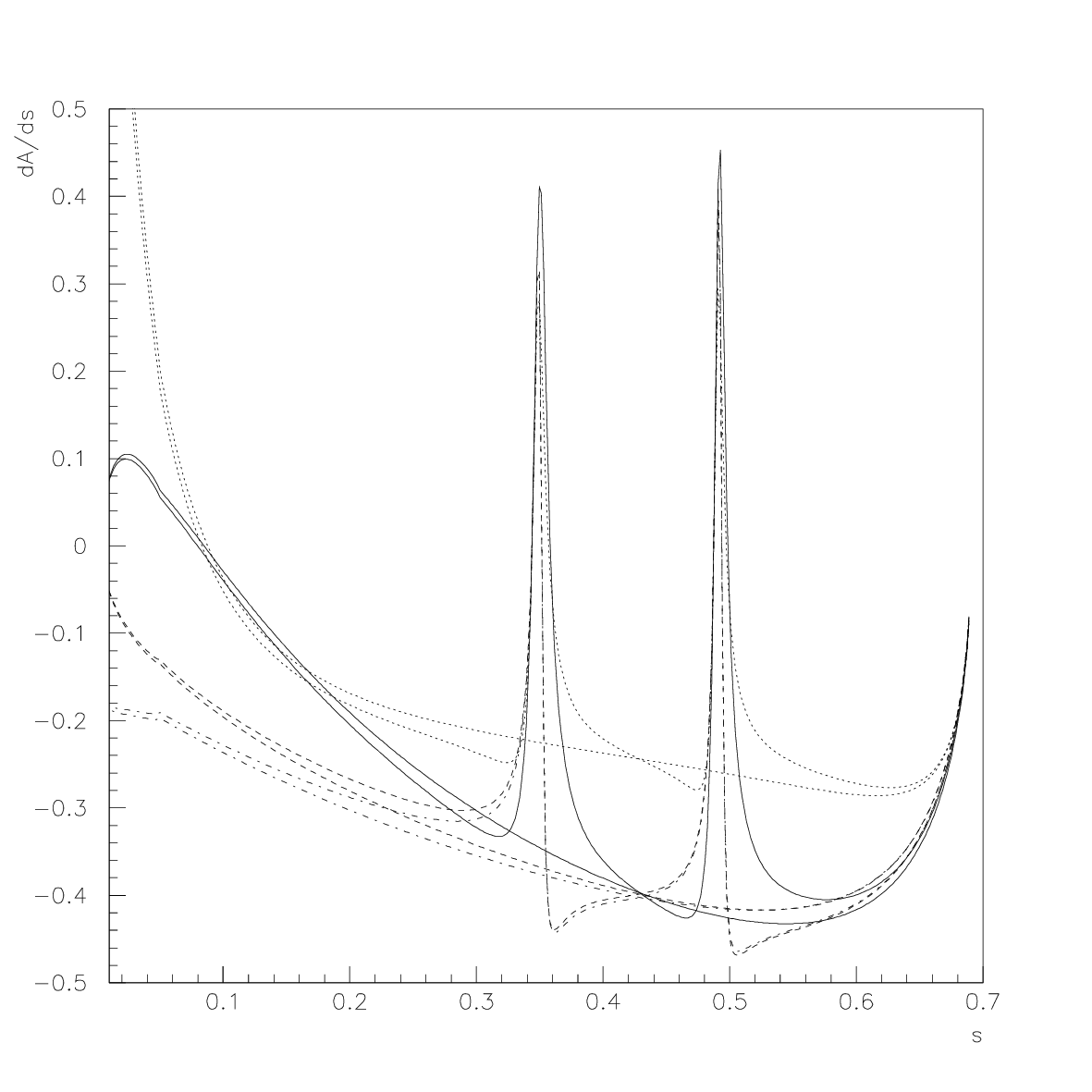,width=8.2cm}
     \mbox{ }\hfill\hspace{1cm}(b)\hfill\mbox{ }
     \end{minipage}
\vskip 0.05truein
     \begin{minipage}[t]{8.2cm}
     \epsfig{file=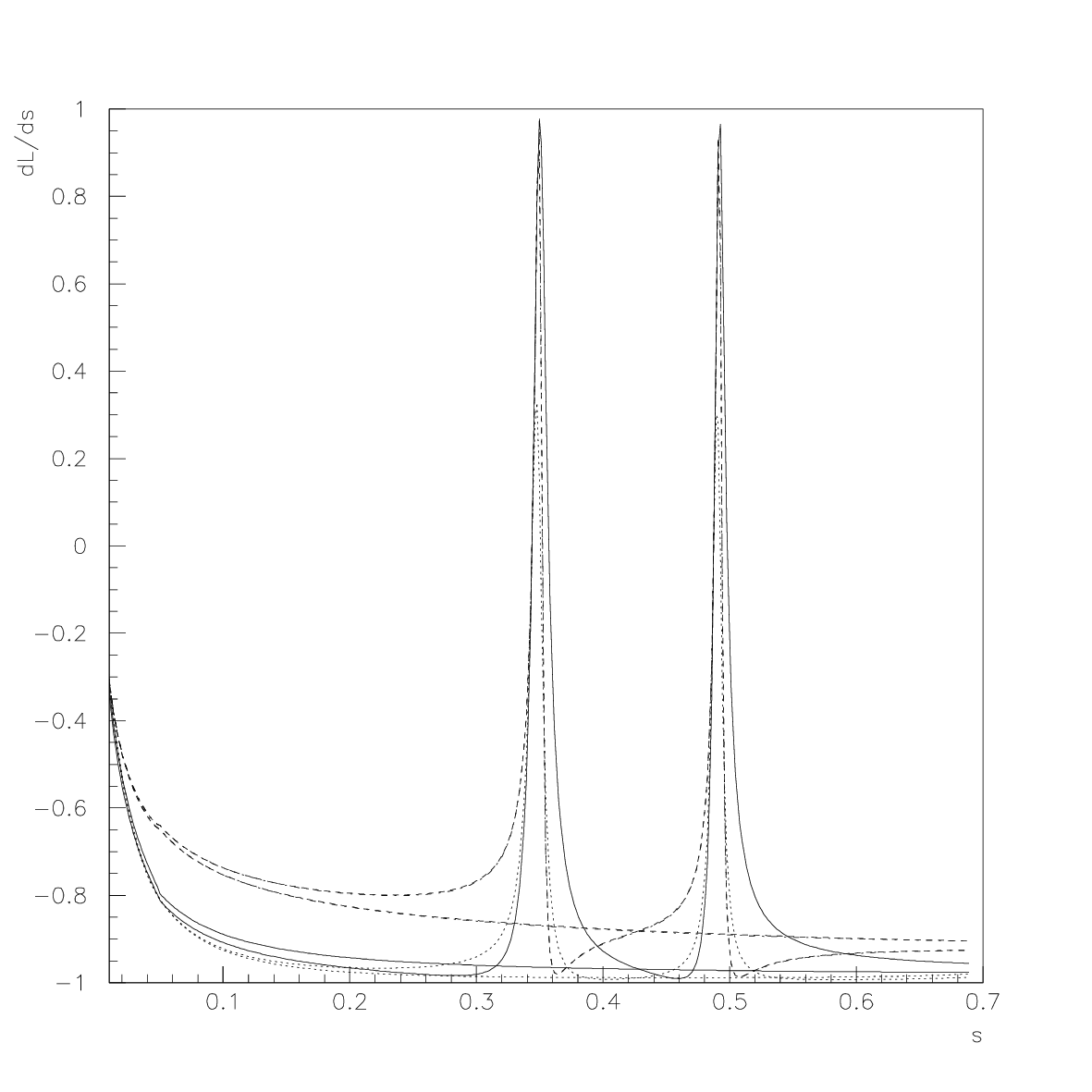,width=8.2cm}
     \mbox{ }\hfill\hspace{1cm}(c)\hfill\mbox{ }
     \end{minipage}
     \hspace{-0.4cm}
     \begin{minipage}[t]{8.2cm}
     \epsfig{file=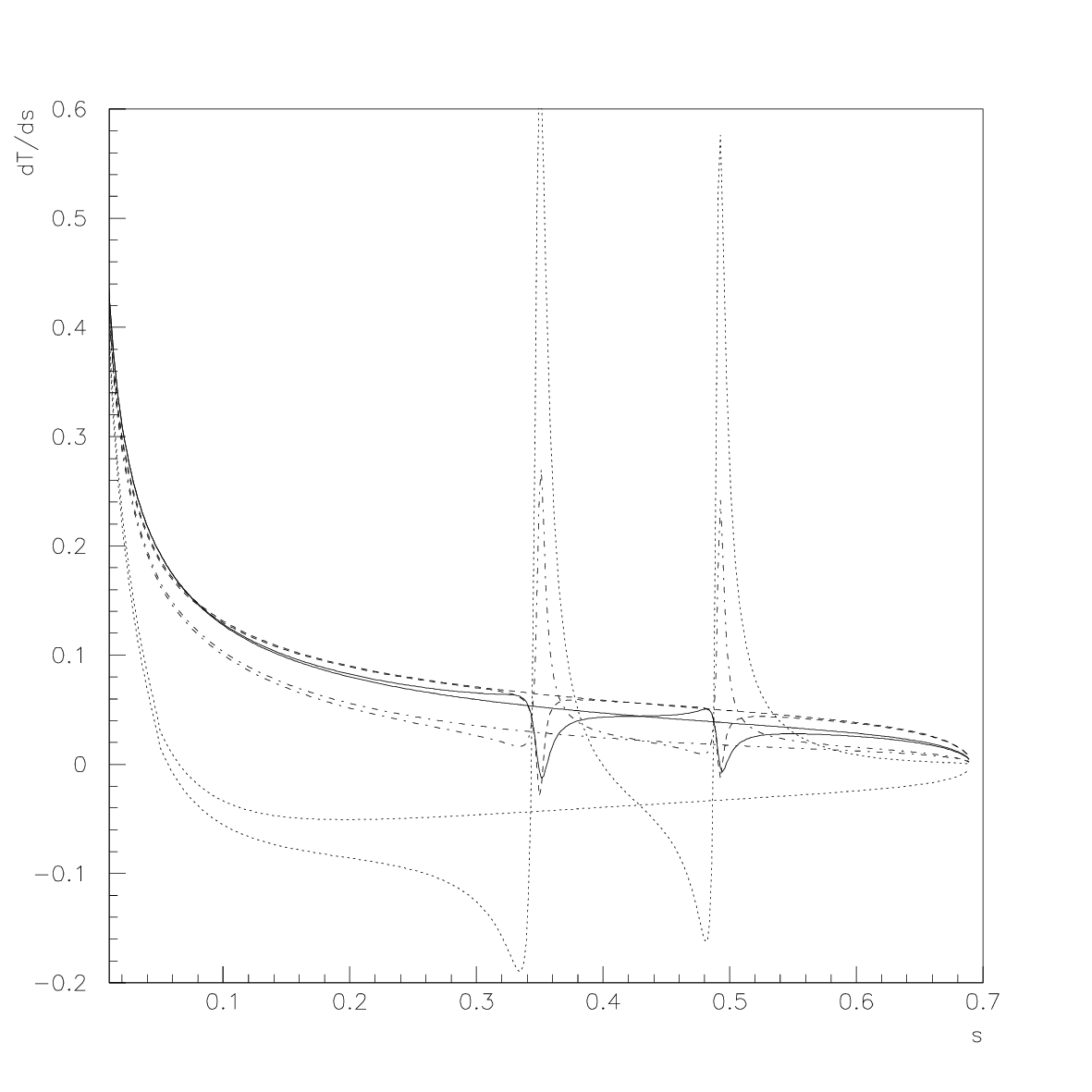,width=8.2cm}
     \mbox{ }\hfill\hspace{1cm}(d)\hfill\mbox{ }
     \end{minipage}
     \caption{\it The IMS(a), FBA(b), LP(c), and TP(d) of the process $\bksm$. The line conventions are the
same as given in the legend of Fig 1.}
\label{fig:ksman}
\end{figure}

\begin{figure}
\begin{minipage}[t]{8.2cm}
     \epsfig{file=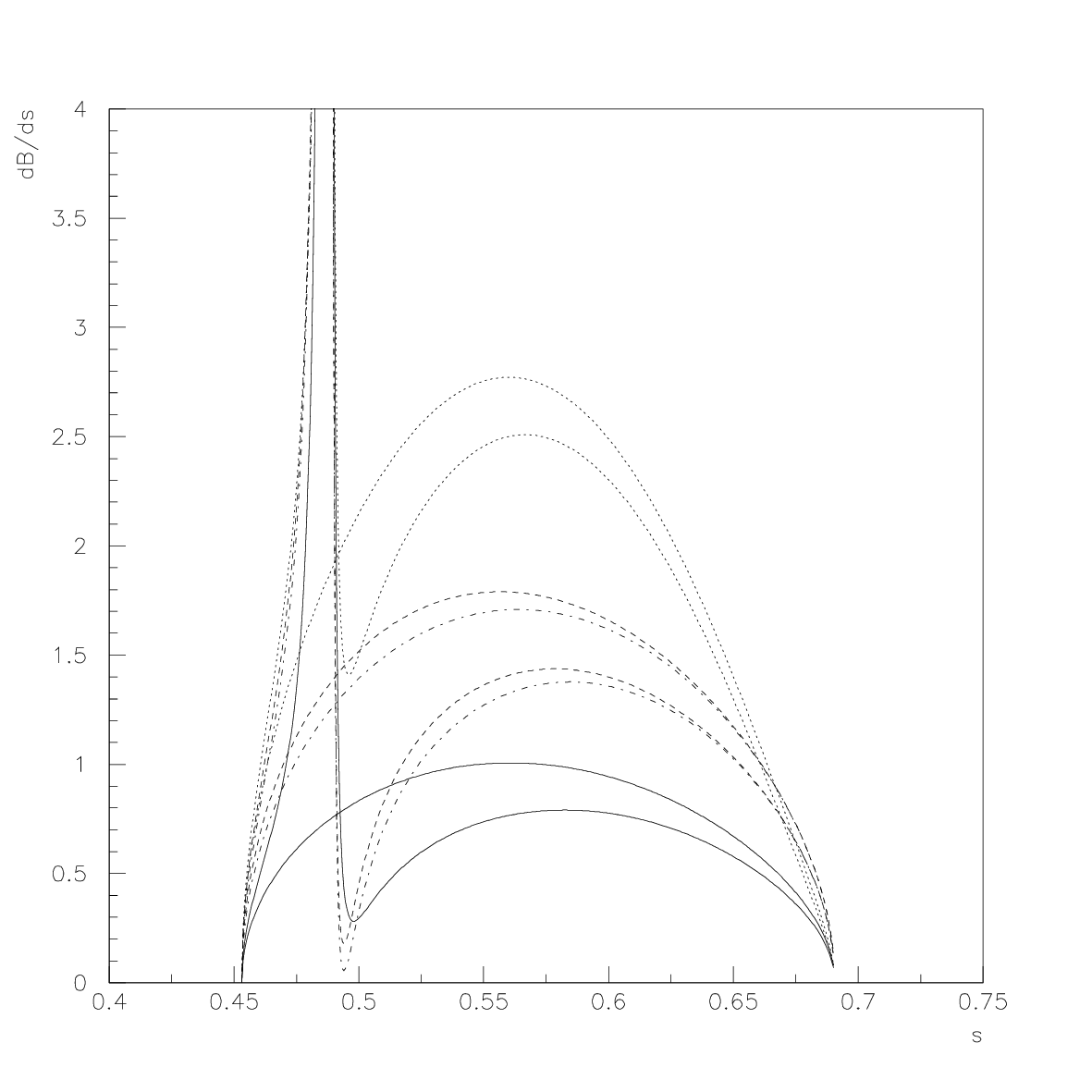,width=8.2cm}
     \mbox{ }\hfill\hspace{1cm}(a)\hfill\mbox{ }
     \end{minipage}
     \hspace{-0.4cm}
     \begin{minipage}[t]{8.2cm}
     \epsfig{file=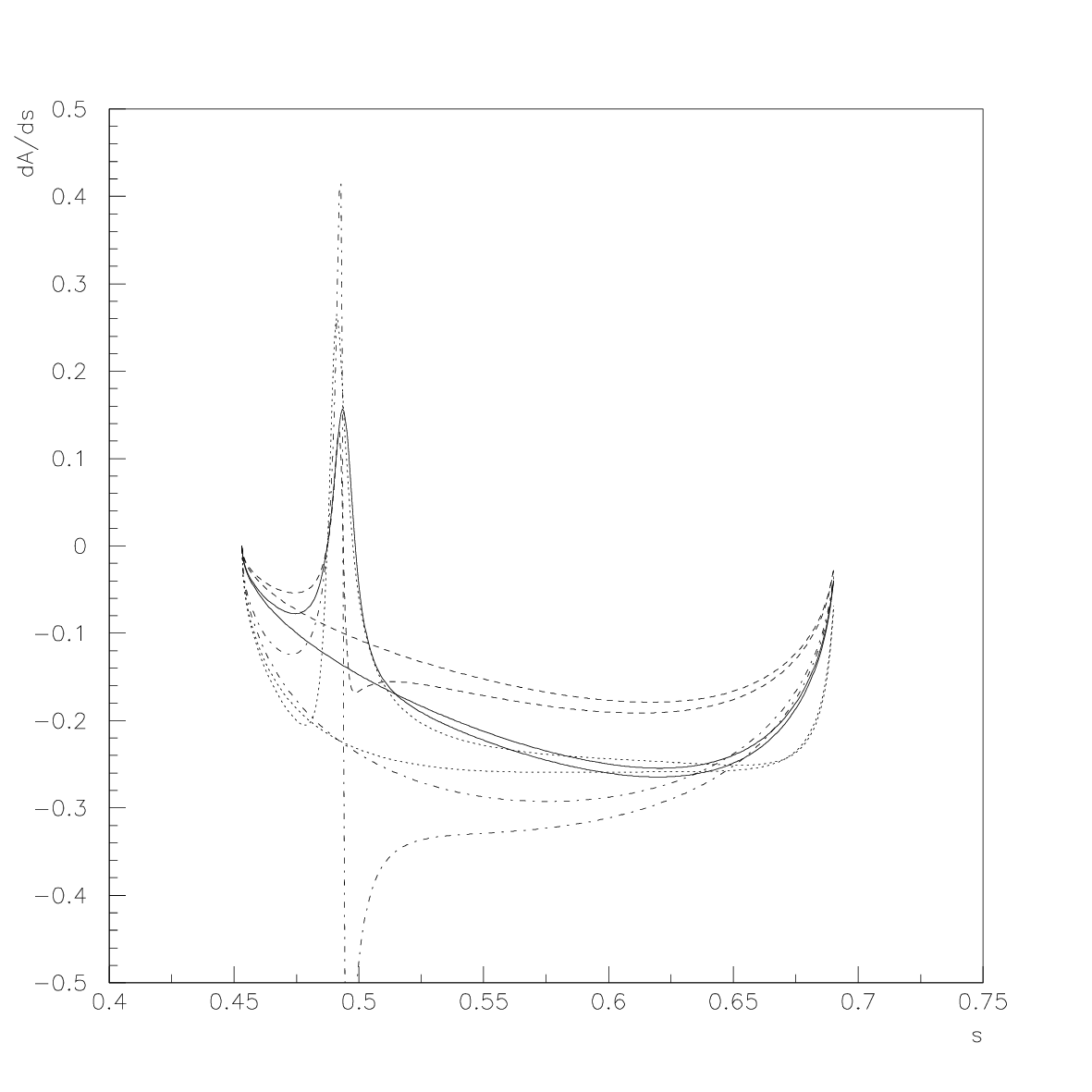,width=8.2cm}
     \mbox{ }\hfill\hspace{1cm}(b)\hfill\mbox{ }
     \end{minipage}
\vskip 0.05truein
     \begin{minipage}[t]{8.2cm}
     \epsfig{file=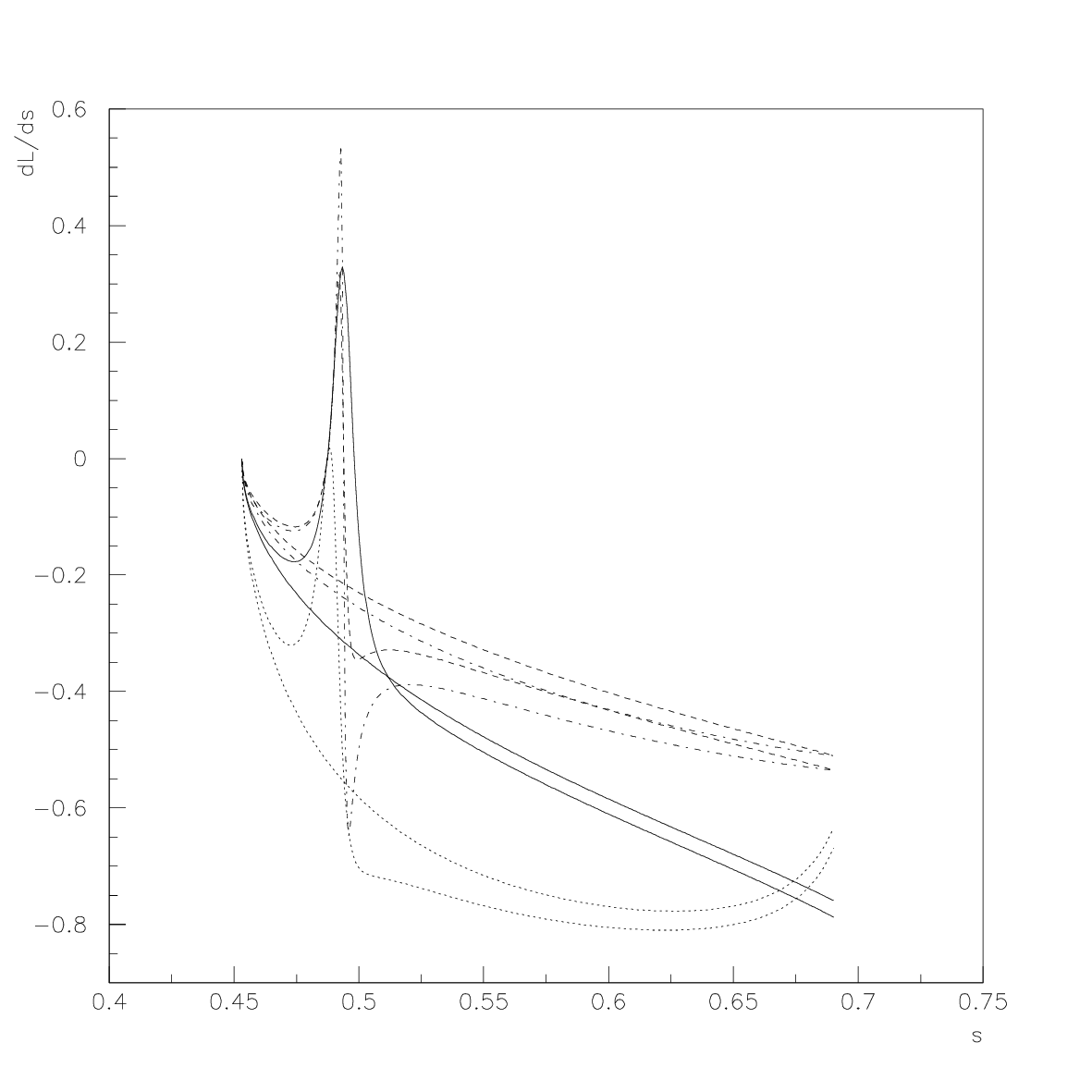,width=8.2cm}
     \mbox{ }\hfill\hspace{1cm}(c)\hfill\mbox{ }
     \end{minipage}
     \hspace{-0.4cm}
     \begin{minipage}[t]{8.2cm}
     \epsfig{file=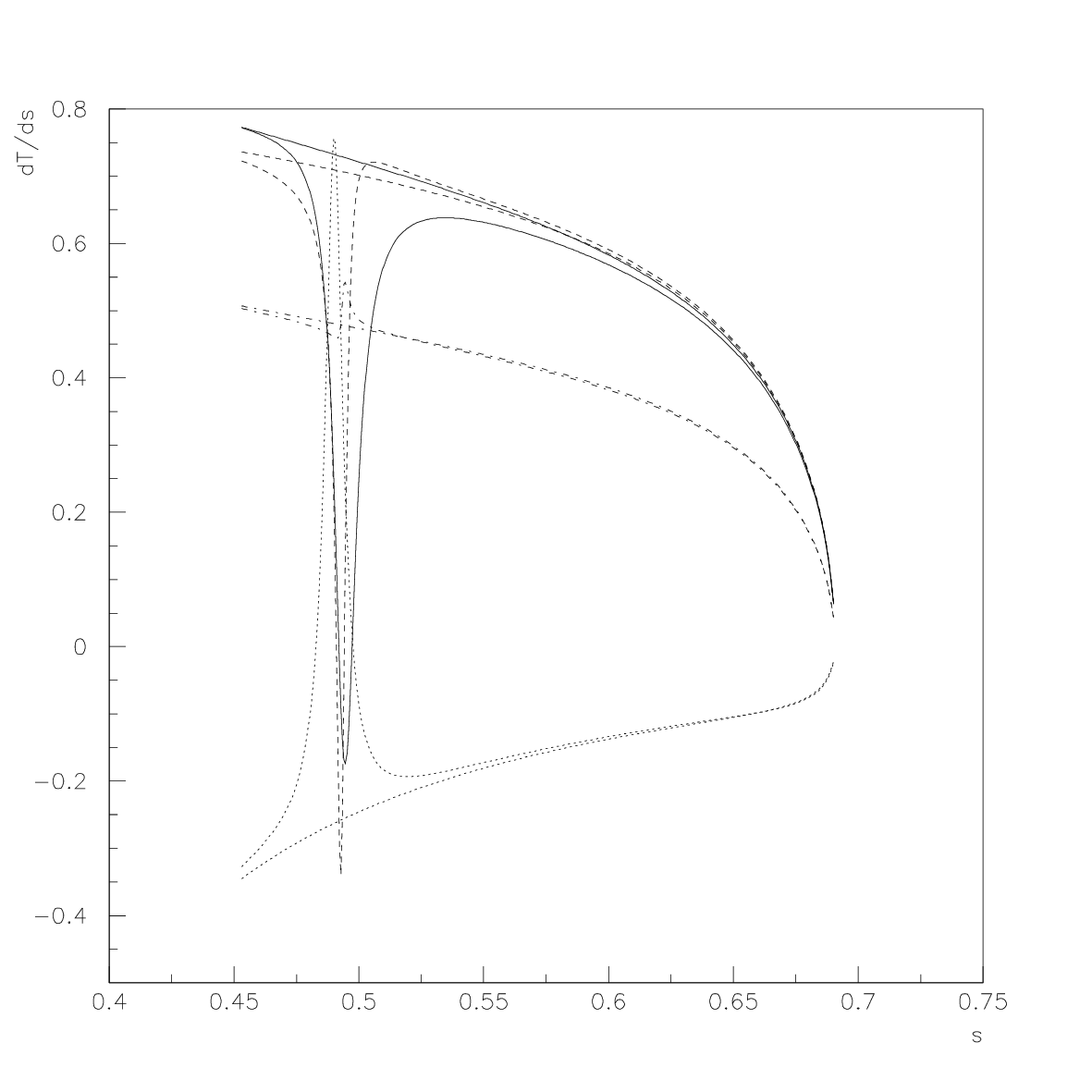,width=8.2cm}
     \mbox{ }\hfill\hspace{1cm}(d)\hfill\mbox{ }
     \end{minipage}
     \caption{\it The IMS(a), FBA(b), LP(c), and TP(d) of the process $\bkst$. The line conventions are the
same as given in the legend of Fig 1.}
\label{fig:kstan}
\end{figure}
\end{center}

In Fig. \ref{fig:kstan}, the IMS, FB asymmetry, longitudinal and transverse polarizations of the
$\bkst$ are depicted. The effect of NHBs does show in great relief.  It is worth to note that
IMS, FBA, and lepton polarizations for $\bksll$ in MSSM without including the contributions of NHBs are also significantly
diffident from those in SM, while for $\bkll$ they have little differences from those in SM. Therefore, compared to the
process $\bkll$, more precise measurements for $\bksll$ are needed  in order to single out the contributions of NHBs.

Normal polarizations for both $\bkll$ and $\bksll$ are small and can be neglected because the imaginary
parts of Wilson coefficients are small in SUSY models without CP violating phases which are implicitly
assumed in the paper.

The behavior of IMS(a), FBA(b), LP(c), and TP(d) shown Figs 1-4 can be understood with the
formula given in the Section 3.
With Eqs. (\ref{eq:dwbpll}), (3.10) and (3.11), we see that the contributions
of NHBs are contained in the terms of $\s_1$ and $D^{\prime}$. At the high $\sh$ regions,
it is these two terms which are important. This explained the behavior of IMS
given in (a) of Fig. 1 and Fig. 2.
The Eq. (\ref{eqn:bkas}) shows that the FBA is proportional to the mass of the lepton.
For the case $\bkm$, due to smallness of the mass $\mu$, the FBA does not vanish but is hard to
be measured. While for the case $\bkt$, the mass $\tau$ is quite large and observing FBA
is relatively easy. For SUSY II, though the numerator of
FBA is comparatively large, the large IMS suppresses the value of FBA;
for SUSY III, the numerator is relatively small, but the FBA do demonstrate the
effects of NHBs more manifestly, as shown in Fig. 2(b) due to smallness of IMS.
The Eqs. (3.63) and (3.64)  show that for the case $\ell=\mu$, the contributions of NHBs
to $P_N, P_T$ are
suppressed by the mass of $\mu$. But for the case $\ell=\tau$, the contributions
of NHBs become quite manifest both for SUSY II and SUSY III.
The term with $\rp$ in Eq. (\ref{eqn:bktp}) will change its sign
when there exists relatively not too small contributions of NHBs, the fact
deduced from Eq. (3.10), that explains why the sign of TP is changed. The difference
between the case SUSY II and SUSY III is small, the reason is just the same as stated in
the analysis of FBA.

Since the terms incorporating the contributions of NHBs is proportional to
$\la$ as shown in Eq. (\ref{eq:dwbvll}), which approaches zero at high $\sh$
regions; while at small $\sh$ regions, the effects
of NHBs are dwarfed by the other contributions. Therefore, only when
 $C_{Q_i}$ are quite large could effects of NHBs be manifest,
as shown in Fig. 3(a) and Fig. 4(a). According to the
Eq. (\ref{eqn:bksas}), at high $\sh$ regions, the effects of NHBs would be
suppressed by $\la$ and $1-\sh-\mvh^2$. The same suppression mechanism exists
for LP. This suppression mechanism explains the fact that the processes $\bksll$
are not sensitive to the effects of NHBs. However, when there exist large contributions of NHBs, the
sign of TP will be changed, as indicated in both Fig. 3(d) and Fig. 4(d).

\begin{table}
\begin{center}
\begin{tabular}{|c|c|c|c|c|c|c|c|c|}
\hline
\multicolumn{2}{|c|}{{\scriptsize model}}&
\multicolumn{1}{|c|}{{\scriptsize A}} &
\multicolumn{1}{c|}{{\scriptsize B}} &
\multicolumn{1}{c|}{{\scriptsize C}} &
\multicolumn{1}{c|}{{\scriptsize D}} &
\multicolumn{1}{c|}{{\scriptsize E}} &
\multicolumn{1}{c|}{{tot(SD)}} &
\multicolumn{1}{c|}{{tot(SD+LD)}} \cr
\hline
{SM} &LCSR&$0.353$ &$54.707$ &$0.032$ &$4.566$ &$0.076$ &$0.573$ &$59.736$ \\
\cline{2-8}\cline{9-9}
&SVZ&$0.215$ &$22.918$ &$0.015$ &$1.593$ &$0.026$ &$0.299$ &$24.767$\\
\hline
{SUSY I} &LCSR&$0.425$ &$54.723$ &$0.037$ &$4.576$ &$0.086$ &$0.675$ &$59.847$ \\
\cline{2-8}\cline{9-9}
&SVZ&$0.179$ &$22.910$ &$0.011$ &$1.586$ &$0.019$ &$0.236$ &$24.704$\\
\hline
{SUSY II} &LCSR&$0.556$ &$54.865$ &$0.131$ &$4.833$ &$0.849$ &$2.067$ &$61.233$ \\
\cline{2-8}\cline{9-9}
&SVZ&$0.348$ &$23.009$ &$0.068$ &$1.726$ &$0.321$ &$1.002$ &$25.473$\\
\hline
{SUSY III} &LCSR&$0.429$ &$54.727$ &$0.040$ &$4.584$ &$0.109$ &$0.717$ &$59.889$ \\
\cline{2-8}\cline{9-9}
&SVZ&$0.181$ &$22.912$ &$0.012$ &$1.590$ &$0.028$ &$0.255$ &$24.723$\\
\hline
\end{tabular}
\caption{Partial decay widths for $\bkm$. LCSR means the approach
light-cone QCD sum rules, SVZ means the SVZ QCD sum rule
\cite{cfss}. Character A means the region ($\sh_0 , {({\hat
m_\psi} -{\hat \delta} )}^2 $), B (${({\hat m_\psi} -{\hat
\delta})}^2 , {({\hat m_\psi} +{\hat \delta} )}^2 $), C (${({\hat
m_\psi} +{\hat \delta})}^2 ,  {({\hat m_{\psi'}} -{\hat \delta}
)}^2 $), D(${({\hat m_{\psi'}} -{\hat \delta})}^2 , {({\hat
m_{\psi'}} +{\hat \delta} )}^2 $) and E (${({\hat m_{\psi'}}
+{\hat \delta})}^2 ,  \sh_{\rm max}^2$). The unit is $\Gamma_B
\times 10^{-6}$, which is $4.22\times 10^{-19}$ GeV. $\delta$ is
selected to be $0.2$ GeV. ${\hat \delta}$ is normalized with
$M_B$}
\end{center}
\label{tab:pdw1}
\end{table}

\begin{table}
\begin{center}
\begin{tabular}{|c|c|c|c|c|c|c|c|c|}
\hline
\multicolumn{2}{|c|}{{\scriptsize model}}&
\multicolumn{1}{|c|}{{\scriptsize A}} &
\multicolumn{1}{c|}{{\scriptsize B}} &
\multicolumn{1}{c|}{{\scriptsize C}} &
\multicolumn{1}{c|}{{\scriptsize D}} &
\multicolumn{1}{c|}{{\scriptsize E}} &
\multicolumn{1}{c|}{{tot(SD)}} &
\multicolumn{1}{c|}{{tot(SD+LD)}} \cr
\hline
{SM} &LCSR&$0.930$ &$83.257$ &$0.141$ &$9.976$ &$0.258$ &$1.882$ &$94.562$ \\
\cline{2-8}\cline{9-9}
&SVZ&$2.943$ &$111.278$ &$0.147$ &$7.504$ &$0.137$ &$3.639$ &$122.008$\\
\hline
{SUSY I} &LCSR&$1.627$ &$83.402$ &$0.198$ &$10.085$ &0.330$$ &$2.915$ &$95.64$ \\
\cline{2-8}\cline{9-9}
&SVZ&$4.517$ &$111.423$ &$0.183$ &$7.552$ &$0.149$ &$5.291$ &$123.825$\\
\hline
{SUSY II} &LCSR&$1.178$ &$83.431$ &$0.234$ &$10.164$ &$0.352$ &$2.677$ &$95.360$ \\
\cline{2-8}\cline{9-9}
&SVZ&$2.801$ &$111.292$ &$0.156$ &$7.525$ &$0.145$ &$3.522$ &$121.918$\\
\hline
{SUSY III} &LCSR&$1.631$ &$83.407$ &$0.201$ &$10.092$ &$0.334$ &$2.938$ &$95.664$ \\
\cline{2-8}\cline{9-9}
&SVZ&$4.518$ &$111.425$ &$0.184$ &$7.553$ &$0.150$ &$5.296$ &$123.830$\\
\hline
\end{tabular}
\caption{Partial decay widths for $\bksm$. Other conventions can be found in Table 5.}
\end{center}
\label{tab:pdw2}
\end{table}

\begin{table}
\begin{center}
\begin{tabular}{|c|c|c|c|c|c|}
\hline
\multicolumn{1}{|c|}{{\scriptsize model}}&
\multicolumn{1}{|c|}{{\scriptsize }}&
\multicolumn{1}{|c|}{{\scriptsize A'}} &
\multicolumn{1}{c|}{{\scriptsize B'}} &
\multicolumn{1}{c|}{{tot(SD)}} &
\multicolumn{1}{c|}{{tot(SD+LD)}} \cr
\hline
{SM} &LCSR&$1.884$ &$0.094$ &$0.132$ &$1.978$ \\
\cline{2-6}
&SVZ&$0.659$ &$0.036$ &$0.054$ &$0.695$ \\
\hline
{SUSY I} &LCSR&$1.884$ &$0.086$ &$0.131$ &$1.970$ \\
\cline{2-6}
&SVZ&$0.655$ &$0.025$ &$0.038$ &$0.680$ \\
\hline
{SUSY II} &LCSR&$2.022$ &$1.496$ &$1.674$ &$3.519$ \\
\cline{2-6}
&SVZ&$0.726$ &$0.552$ &$0.637$ &$1.278$ \\
\hline
{SUSY III} &LCSR&$1.874$ &$0.094$ &$0.129$ &$1.968$ \\
\cline{2-6}
&SVZ&$0.651$ &$0.026$ &$0.035$ &$0.677$ \\
\hline
\end{tabular}
\caption{Partial decay widths of $\bkt$.
Character A' means ($\sh_0, {({\hat m_\psi} -{\hat \delta})}^2$),
B' means (${({\hat m_{\psi'}}+{\hat \delta})}^2, \sh_{\rm max}$).
The unit is $\Gamma_B \times 10^{-6}$, which is $4.22\times 10^{-19}$ GeV.}
\label{tab:pdw3}
\end{center}
\end{table}

\begin{table}
\begin{center}
\begin{tabular}{|c|c|c|c|c|c|}
\hline
\multicolumn{1}{|c|}{{\scriptsize model}}&
\multicolumn{1}{|c|}{{\scriptsize }}&
\multicolumn{1}{|c|}{{\scriptsize A'}} &
\multicolumn{1}{c|}{{\scriptsize B'}} &
\multicolumn{1}{c|}{{tot(SD)}} &
\multicolumn{1}{c|}{{tot(SD+LD)}} \cr
\hline
{SM} &LCSR&$4.045$ &$0.096$ &$0.183$ &$4.141$ \\
\cline{2-6}
&SVZ&$3.029$ &$0.048$ &$0.102$ &$3.076$ \\
\hline
{SUSY I} &LCSR&$4.088$ &$0.173$ &$0.327$ &$4.261$ \\
\cline{2-6}
&SVZ&$3.052$ &$0.072$ &$0.159$ &$3.124$ \\
\hline
{SUSY II} &LCSR&$4.148$ &$0.266$ &$0.460$ &$4.413$ \\
\cline{2-6}
&SVZ&$3.054$ &$0.084$ &$0.167$ &$3.138$ \\
\hline
{SUSY III} &LCSR&$4.078$ &$0.168$ &$0.312$ &$4.246$ \\
\cline{2-6}
&SVZ&$3.050$ &$0.071$ &$0.156$ &$3.121$ \\
\hline
\end{tabular}
\caption{Partial decay widths of $\bkst$. Other conventions can be found at
Table 7.}
\end{center}
\label{tab:pdw4}
\end{table}

The partial decay widths (PDW) are listed in Tables. 5,6,7,and 8. We see that at the high $\sh$ region, for the process
$B\rightarrow K l^+ l^-$, l=$\mu,\tau$, the
contributions of NHBs do show up, as expected. For $B\rightarrow K^{*} l^+l^-$, the effects of NHBs in the high $\hat{s}$
region is signifiacnt when l=$\tau$ while they are small for l=$\mu$.
It can be read out from these four table that
the results are consistent with the Fig. \ref{fig:kman}(a),\ref{fig:ktan}(a),
\ref{fig:ksman}(a), and \ref{fig:kstan}(a).
In order to estimate the theoretical uncertainty brought by the methods
calculating the weak form factors, we use the form factors calculated with LCSR and SVZ QCD sum
rules (SVZ) method \cite{cfss}.
For $\bkll$, PDWs calculated with form factors obtained by SVZ method
is 50\% of those by LCSR approach; while for $\bksll$, PDWs increase 100\% or so.
We see that at low $\sh$ regions the theoretical uncertainty can reach from 100\% to 200\%.
Another point worthy of mention is that the contribution of resonences domainate the integerated decay width, as had been pointed out in \cite{liu}.

\subsection{Conclusion}
We have calculated invariant mass spectrum, back-forward asymmetry, and lepton polarizations for $\bkll$ and $\bksll$
l=$\mu,\tau$ in SUSY theories. In particular, we have analyzed the effects of NHBs to these processes. It is shown that the effects
of the NHBs  to $\bkt$ and $\bkst$ in some regions of parameter space of SUSY models
are considerable and remarkable. The reason lies in the mass of the $\tau$, which can
magnify the effects of NHBs and can be see through from the related formula.
The numerical results imply that there still exist possiblities to observe the effects of NHB in $\bkm$ and $\bksm$ through
IMS, FB asymmetry and lepton polarizations of these processes. In particular, for $B\rightarrow K \mu^+\mu^-$
in the case of SUSY II,   the partial width in the high $\hat{s}$ where short distance physics dominants  can be enhanced by a
factor of 12 compared to SM. Our analysis also show that
the theoretical uncertainties brought in calculating of weak form factors
are quite large. But the effects of NHBs will not be washed out and can stand out in some regions of the parameter space in
MSSM. If only partial widths are
measured, it is difficult to observe the effects of NHBs except for the decay $B\rightarrow K \tau^+\tau^-$. However, the conbined
analysis of IBS, FBA, and lepton polarizations can provide usefull knowledge to look for SUSY.
Finally, we would like to point out that FBA for $B\rightarrow K l^+l^-$ vanishes (or, more precisely, is neglegiblly small)
in SM and it does not vanish in 2HDM and SUSY models with large tan$\beta$ due to the contributions of NHBs. However, only
in SUSY models and for l=$\tau$ it is large enough to be observed in B factories in the near future.


\setcounter{figure}{0} \setcounter{table}{0}
\setcounter{equation}{0}

\newpage
\part{Acknowledgement} It is my pleasure to thank  Prof. C. S.
Huang, Prof. C. S. Li, as well as their respectively group
members. The closely collaboration and the stimulating discussions
with them encourage me to live with High Energy Physics. Their
name are Dr. Q.S. Yan,
 X.H. Wu, W. Liao, Y.S. Yang and L.G. Jin. Prof. R. Oakes and G.
Eilam are also deserved special acknowledgement. The financial
support of China Postdoctoral Foundation and K. C. Wong Education
Foundation, Hong Kong are both gratefully acknowledged. Finally, I
should give my special thanks to Alexander von Humboldt Foundation
and Prof. W. Hollik, their kindly support make me to continue my
research works smoothly.

\newpage
\part{Publication List}

\begin{enumerate}
\item{Higgs Physics}

\begin{enumerate}

\item
Shou Hua Zhu, Chong Sheng
Li and Chong Shou Gao, Lightest neutral Higgs boson pair
production in photon-photon
collisions in the minimal supersymmetric extension of standard model,
Phys. Rev. {\bf D58}, 015006 (1998) (22 pages).

\item
Shou Hua Zhu, Chong Sheng Li and Chong Shou Gao,
Squarks loop corrections to the charged Higgs boson pair production
in photon-photon collisions,
Phys. Rev. {\bf D58}, 055007 (1998) (12 pages).

\item
Shou Hua Zhu,
Pseudoscalar Higgs boson pair production in photon-photon collisions,
Journal of Physics G (nuclear and particle physics) 24, 1703-1721
(1998) (18 pages).

\item
Chong Sheng Li, Shou Hua Zhu and Cong Feng
Qiao,
Radiative Higgs boson decay beyond the standard model,
Phys. Rev. {\bf D57}, 6928-6933 (1998) (6 pages).

\item
Shou Hua Zhu, Chong Sheng Li and Chong Shou Gao,
Single Higgs boson production in gamma-gamma collision in minimal
supersymmetric extension of standard model, Chinese Physics Letters Vol
15, No. 2 (1998) 89 (4 pages).

\item
Chong Sheng Li
and Shou Hua Zhu,
Top quark loop corrections to the neutral Higgs boson production at
the Fermilab Tevatron, Phys. Lett. {\bf B444}, 224(1998) (6 pages).

\item
Qing Hong Cao, Chong Sheng Li and Shou Hua Zhu,
Leading Electroweak Corrections to the Neutral Higgs Boson
Production at the Fermilab Tevatron, Comm. Theor. Phys. 32, 275 (2000),
hep-ph/9810458.

\item
Chao-shang Huang and Shou Hua Zhu,
Supersymmetrical Higgs bosons discovery potential at hadron collider
through bg channel, Phys. Rev. D60, 075012 (1999) (4 pages).

\item
Shou Hua Zhu,
 Charged Higgs associated production with W at linear collider,
 hep-ph/9901221.

\item
Li Gang Jin, Chong Sheng Li, Robert J.
  Oakes, Shou Hua Zhu,
   Yukawa Corrections to Charged Higgs Boson Production in Association with
     a Top Quark at Hadron Colliders,
     Eur. Phys. J. C14, 91-101 (2000), hep-ph/9907482.

\item
Li Gang Jin, Chong Sheng Li, Robert J.
  Oakes, Shou Hua Zhu,
   Supersymmetrical  Electroweak Corrections to Charged Higgs
Boson Production in Association with
     a Top Quark at Hadron Colliders,
     to appear in PRD (2000), hep-ph/0003159.

\item
Ya-sheng Yang, Li Gang Jin, Chong Sheng Li,
Shou Hua Zhu,
 Supersymmetrical  Electroweak Corrections to Charged Higgs
Boson Production in Association with
     W through $b \bar{b}$ channel,
submitted to PRD, hep-ph/0004248.

\end{enumerate}

\item FCNC

\begin{enumerate}
\item
Chao-shang Huang,
Xiao-hong Wu and Shou Hua Zhu,
Top-charm Associated production at high energy $e^+e^- $colliders in
standard model, Phys. Lett. {\bf B452}, 143 (1999) (7 pages).

\item
Chao-shang Huang,
Xiao-hong Wu and Shou Hua Zhu,
Bottom stange associated production at high energy $e^+e^-$ colliders in
standard model, J. Phys. {\bf G 25}, 2215-2223 (1999) (9 pages).

\item
Chong Sheng Li,
Xin-min Zhang, Shou Hua Zhu,
SUSY QCD effect on top charm associated production at linear
colliders, Phys. Rev. D60, 077702 (1999) (4 pages).

\end{enumerate}

\item  B Physics

\begin{enumerate}
\item
Qi-shu Yan, Chao-shang Huang,
Wei Liao and Shou Hua Zhu,
Exclusive semileptonic rare decays $B \rightarrow (K, K^*) l^+ l^-$
in supersymmetric theories,
hep-ph/0004262, submitted to PRD.

\item
Chao-shang Huang,
Shou-Hua Zhu,
 $B \rightarrow X(S) \tau^+ \tau^-$ in a cp spontaneous broken
 two higgs doublet
 model, Phys. Rev. {\bf D61}, 015011 (2000), hep-ph/9905463.

\end{enumerate}

\item  M-theory phenomenology

\begin{enumerate}
\item
Chao-shang Huang, Tian-jun Li,
Wei Liao, Qi-shu Yan and Shou Hua Zhu,
Scales, Couplings Revisited and Low Energy Phenomenology in M-theory
on $S^1/Z(2)$, submitted to EPJC, hep-ph/9810412.

\item
Chao-shang Huang, Tian-jun Li,
Wei Liao, Qi-shu Yan and Shou Hua Zhu,
M-theory Low Energy Phenomenology,
Commun. Theor. Phys., {\bf 32}, 499-506 (1999).

\end{enumerate}

\item  Top physics

\begin{enumerate}
\item
Cong Feng Qiao and Shou Hua Zhu,
Supersymmetric QCD corrections to top quark semi-leptonic decay,
Phys. Lett. {\bf B451}, 93 (1999) (5 pages).

\item
Lian-you Shan and Shou Hua Zhu,
Top decays into light stop and gluino,  to appear in PRD (2000),
hep-ph/9811430.

\end{enumerate}
\end{enumerate}

\end{document}